\documentclass[a4paper,fleqn,usenatbib]{mnras}

\usepackage{newtxtext,newtxmath}

\usepackage[T1]{fontenc}
\usepackage{ae,aecompl}

\usepackage{graphicx}	
\usepackage{amsmath}	
\usepackage{amssymb}	
\usepackage{multicol}        
\usepackage{pdflscape}	

\usepackage{verbatim}

\title[Fundamental disc parameters of 54 SMC Be stars]{The life cycles of Be viscous decretion discs: \\
fundamental disc parameters of 54 SMC Be stars}

\author[L. R. R\'{\i}mulo et al.]{
L. R. R\'{\i}mulo,$^{1}$\thanks{E-mail: lrrimulo@usp.br}
A. C. Carciofi,$^{1}$
R. G. Vieira,$^{1}$
Th. Rivinius,$^{2}$
D. M. Faes,$^{1}$
 \newauthor
A. L. Figueiredo,$^{1}$
J. E. Bjorkman,$^{3}$
C. Georgy,$^{4}$
M. R. Ghoreyshi,$^{1}$
I. Soszy{\'n}ski$^{5}$
\\
$^{1}$Instituto de Astronomia, Geof\'{\i}sica e Ci\^encias Atmosf\'ericas, Universidade de S\~ao Paulo, \\Rua do Mat\~ao 1226, Cidade Universit\'aria, 05508-900 S\~ao Paulo, SP, Brazil\\
$^{2}$ESO, European Organization for Astronomical Research in the Southern Hemisphere, Chile\\
$^{3}$Ritter Observatory, Department of Physics \& Astronomy, Mail Stop 113, University of Toledo, Toledo, OH 43606, US\\
$^{4}$Observatoire de Gen\`eve, Chemin des Maillettes 51, Sauverny, CH-1290 Versoix, Switzerland\\
$^{5}$Warsaw University Observatory, Al. Ujazdowskie 4, 00-478 Warsaw, Poland
}  

\date{Accepted XXX. Received YYY; in original form ZZZ}

\pubyear{2017}

\begin{document}
\label{firstpage}
\pagerange{\pageref{firstpage}--\pageref{lastpage}}
\maketitle

\begin{abstract}
Be stars are main-sequence massive stars with emission features in their spectrum, which originates 
in circumstellar gaseous discs. Even though the viscous decretion disc (VDD) model can satisfactorily explain most observations, two important physical ingredients, namely the magnitude of the viscosity ($\alpha$) and the disk mass injection rate, remain poorly constrained. The light curves of Be stars that undergo events of disc formation and dissipation offer an opportunity to constrain these quantities.
A pipeline was developed to model these events that uses a grid of synthetic light curves, 
computed from coupled hydrodynamic and radiative transfer calculations. 
A sample of 54 Be stars from the OGLE survey of the Small Magellanic Cloud (SMC) was selected for this study.
Because of the way our sample was selected (bright stars with clear disc events), it likely represents the densest discs  in the SMC. Like their siblings in the Galaxy, the mass of the disc in the SMC increases with the stellar mass. 
The typical mass and angular momentum loss rates associated with the disk events are of the order of $\sim$$10^{-10}\, M_\odot\,\mathrm{yr^{-1}}$ and $\sim$$5\times 10^{36}\, \mathrm{g\, cm^{2}\, s^{-2}}$, respectively. 
The values of $\alpha$ found in this work are typically of a few tenths, consistent with recent results in the literature and with the ones found in dwarf novae, but larger than current theory predicts.
Considering the sample as a whole, the viscosity parameter is roughly two times larger at build-up ($\left\langle\alpha_\mathrm{bu}\right\rangle = 0.63$) than at dissipation ($\left\langle\alpha_\mathrm{d}\right\rangle = 0.26$). 
Further work is necessary to verify whether this trend is real or a result of some of the model assumptions. 
\end{abstract}


\section{Introduction}


%
%

In a classical, observational, and quite broad definition, a Be star is a hot, massive star, with a B spectral type (mass ranging roughly from 3 to 17 $M_{\odot}$), non-supergiant, whose spectrum has, or had at some time, one or more Balmer lines in emission \citep{1981BeSN....4....9J,1987pbes.coll....3C}. 
In a more modern and theoretically-oriented definition, a Be star is a very rapidly rotating and non-radially pulsating B star that forms a geometrically thin \emph{viscous decretion disc} (VDD) composed of an outwardly diffusing, viscosity driven gaseous Keplerian disc that is fed by mass ejected from the central star \citep{2013A&ARv..21...69R}, 
and a possibly non-negligible line-driven wind \citep{2016MNRAS.458.2323K}. 
There is no evidence of large scale magnetic fields in Be stars \citep{2012ASPC..464..405W,2016MNRAS.456....2W}. Fast stellar rotation lowers the effective gravity near the stellar equator and a second mechanism, likely to be stellar pulsation \citep{1998A&A...333..125R,2016A&A...588A..56B,2016A&A...593A.106R}, is responsible for pushing this near equatorial matter into orbit in the inner disc.
Once in orbit, a viscous mechanism takes place, diffusing matter and angular momentum outwards, thus making the disc grow.

Be stars are usually quite variable in all observables and in several timescales (days, weeks, months or even years). The variability observed in Be stars indicates that the injection of matter and angular momentum into the disc is frequently quite erratic, with sudden outbursts of mass injection and periods of no or negligible mass injection. 
\citet{2016A&A...593A.106R} propose a terminology, which will be used here, in which a star that possesses a disc is said to be active and, conversely, when there is no detectable disc, the star is inactive. Two additional terms are used to distinguish the phases of active disc formation (outbursting Be star) and dissipation (dissipating Be star).
It has been demonstrated \citep[e.g.,][]{2012ApJ...756..156H} that a Be disc fed roughly at a constant rate, and for a sufficiently long time (a few to several years, depending on the value of the viscosity), reaches a quasi-steady state in which the density is nearly constant in time. If the gas temperature is properly taken into consideration, the radial density profile is typically a complicated function of the distance from the star \citep[e.g.,][]{2008ApJ...684.1374C}. However, a usual approximation is to consider the gas to be isothermal, in which case the density profile assumes a power-law form.
This simple steady-state VDD has been successful in describing the main observed features of individual Be discs \citep{2006ApJ...652.1617C, 2007ApJ...671L..49C, 2008ApJ...687..598J, 2009A&A...504..915C, 2015A&A...584A..85K, klement17} 
and samples of Be stars \citep{2010ApJS..187..228S, 2011ApJ...729...17T, 2017MNRAS.464.3071V}. 
However, despite its great success, there are several open and intriguing theoretical questions about the VDD model. 
Besides the fact that a good description of the mechanism responsible for putting stellar material into orbit is still needed, a complete physical understanding of the forces that drive these discs is also lacking. 
It has been commonly assumed that the forces operating on the discs are the gravity from the central star and the forces that come from the gradient of pressure and from viscosity, the latter being the one capable of producing torque. 
Recently, \citet{2016MNRAS.458.2323K} showed that  radiative line forces may also generate a non-negligible torque, at least for gaseous discs with Solar metallicity. 
{
In this work we will proceed with the assumption that line-driven forces are negligible. We mitigate this potential issue by choosing to study  Be stars in the SMC, whose low metallicity will greatly decrease the strength of the line forces.
}

In the alpha-disc formalism, the kinematic viscosity is scaled with the $\alpha$ parameter, defined such that the $R\phi$ component of the stress tensor is proportional to the gas pressure: $W_{R\phi}=-\alpha P$.
The most reliable and direct way of estimating $\alpha$ is to study the time-dependent disc behaviour, where the diffusive effect of viscosity will have clear observational counterparts. 
Therefore, light curves of long temporal coverage, such as the ones given by microlensing or planetary transit surveys, are excellent instruments to study the dynamical processes in action on the disc, as it builds-up and dissipates.

Dynamical studies of Be star viscous discs are still quite scarce. 
\citet{2008MNRAS.386.1922J}, using a 1-D time-dependent treatment of the alpha-disc and a non-LTE radiative transfer code, studied the temperature and density profiles of a dynamical disc and their respective H$\alpha$ line profiles.
\citet{2012ApJ...756..156H} studied the theoretical photometric effects of time variable mass injection rates on the structure of the disc also using a 1-D time-dependent treatment of the alpha-disc, associated with the Monte Carlo radiative transfer code {\tt HDUST} \citep{2006ApJ...639.1081C,2008ApJ...684.1374C}. \citet{2012ApJ...744L..15C}, by fitting these dynamical models to a dissipating portion in the light curve of the Be star 28 CMa (which passed from an outbursting phase, that lasted from 2001 to 2003, to a dissipating
 phase at the end of 2003), estimated the value of the $\alpha$ parameter for the Be disc of 28 CMa to be $\alpha=1.0\pm 0.2$. Later, however, it was realised that a proper consideration of the previous history of the disc must be taken into consideration even when fitting the dissipating portion of the light curve, and this quite high value has been revisited to be closer to $\alpha=0.2$ \citep{2017ASPC..508..323G}.

Another intriguing issue regarding Be stars is how they acquired such high rotation rates \citep[typically 80\% of break-up, ][]{2013A&ARv..21...69R}.
As the rotating B star evolves, core contraction and internal angular momentum redistribution generally tends to enhance surface angular rotation \citep{2008A&A...478..467E,2013A&A...553A..25G}.
Another scenario \citep[e.g.,][]{1991A&A...241..419P} would involve a past mass-transfer phase in a binary system, during which the primary donates mass and angular momentum to the secondary. The left-over of such a system would be a fast-spinning Be star (the former mass gainer) and a subdwarf O or B star (sdO/sdB, the former mass donor).
Regardless of how they were spun-up, it has been proposed \citep{2011A&A...527A..84K} that the discs of Be stars may provide natural mechanisms for removing large quantities of angular momentum from the fast rotating stars, preventing them to reach the rotation critical limit. 
The evolutionary models of \citet{2013A&A...553A..25G} assumed the appearance of completely formed viscous discs every time their models reached a near-critical rotation. The mass density and the rate of angular momentum loss  of their discs were roughly similar to the ones estimated by \citet{2017MNRAS.464.3071V}, 
who modelled the spectral energy distribution (SED) of 80 Be stars using the VDD model, provided that values of $\alpha$ of at least a few tenths were assumed in both approaches.

The main objective of this paper is to build upon the previous dynamical studies of Be discs and provide, for the first time, a detailed study of the temporal evolution of a large sample of Be stars from the Small Magellanic Cloud (SMC). By studying a large sample of stars, we may begin to answer several open questions related to Be stars and their discs, namely: i) What is the typical value of the viscosity in these discs? ii) Is there any significant evidence for a dependence of $\alpha$ with parameters such as the density of the disc, the spectral type of the star, etc.? iii) What are the typical rates of mass and angular momentum loss in these stars?
To reach these goals we developed a new method for modelling the light curves of Be stars, described in Sects.~\ref{vdd_sect} and \ref{models_bumps}. 
The sample of studied light curves is described in Sect.~\ref{ogle_sect}, and the model results are discussed in Sect.~\ref{sect_results}, followed by the conclusions.

\section{Viscous decretion discs around Be stars}
\label{vdd_sect}

\begin{figure*}
\centering{
\includegraphics[width=1.0\linewidth]{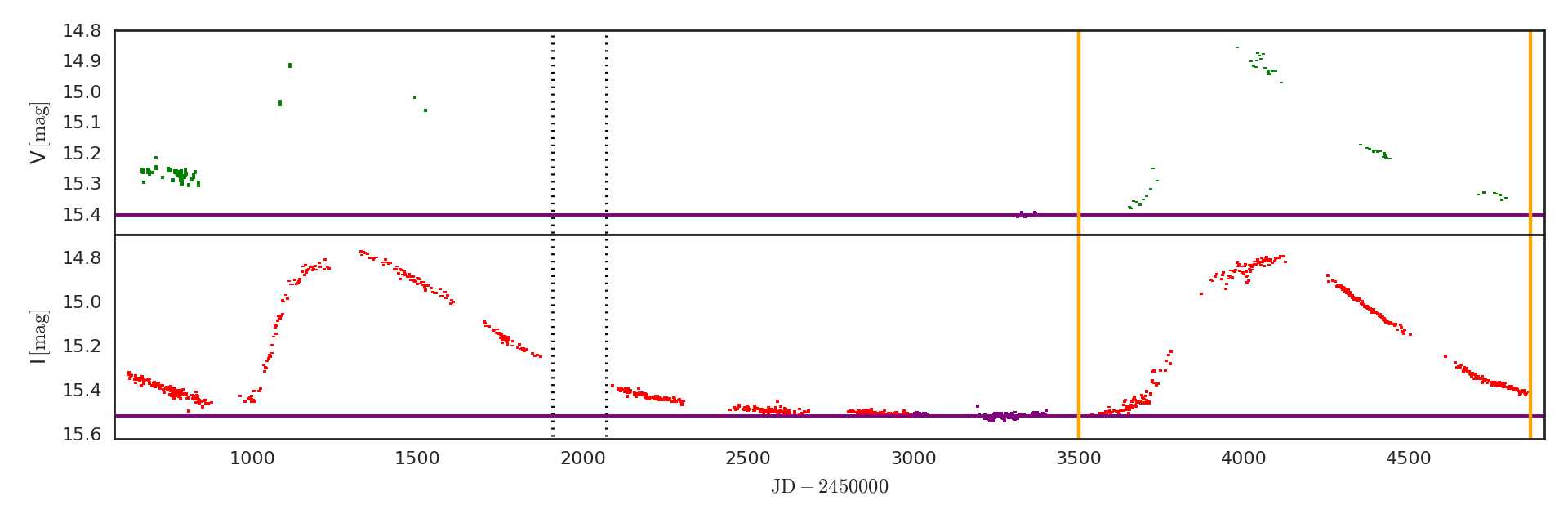}
\includegraphics[width=1.0\linewidth]
{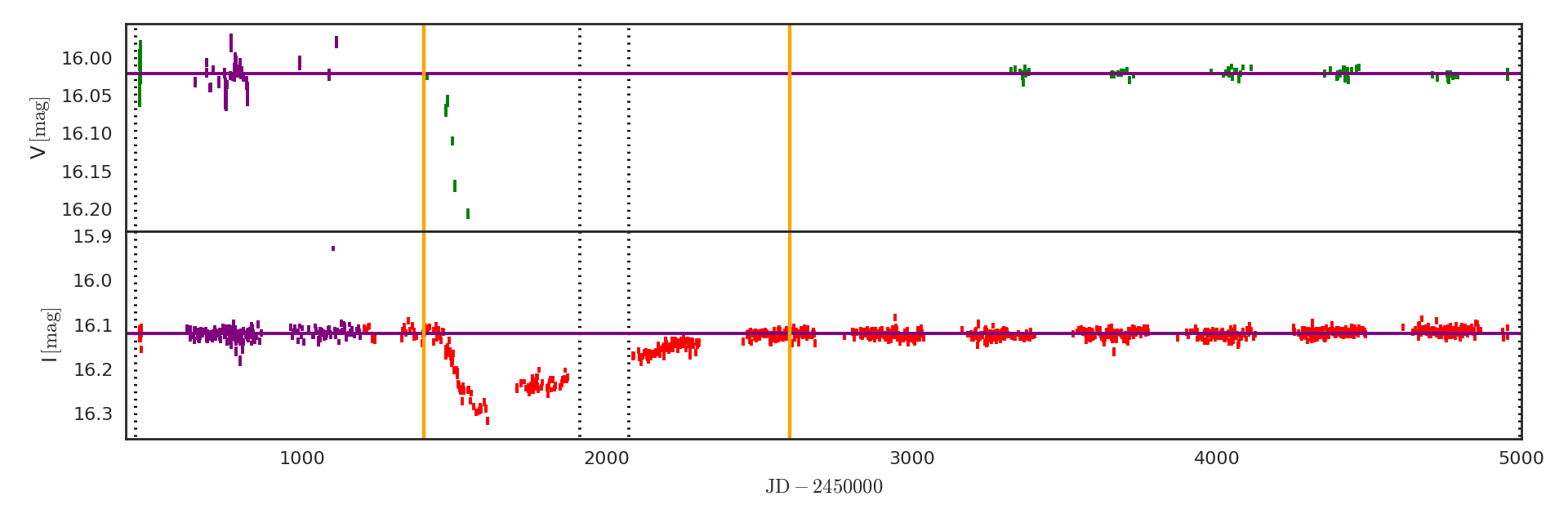}
}
\caption[]
{
Two light curves, in photometric bands $V$ (green), and $I$ (red), selected from the OGLE-II and OGLE-III photometric surveys. {\it Above}: light curve of SMC\_SC1 75701, showing two bumps. {\it Below}: light curve of SMC\_SC6 128831, showing a dip. The pair of vertical dotted straight lines near $\mathrm{JD}-2450000=2000$ separates OGLE-II from OGLE-III data. The measurements shown in purple are assumed to represent the inactive (discless) brightness level of the Be star. Their mean is given by the horizontal purple straight lines. The pairs of vertical orange straight lines bracket our visually selected bumps. These bumps are modelled in Figs.~\ref{example_bb1} and \ref{example_bb2}.
}
\label{example_a}
\end{figure*}
%


The optical light curves of early-type Be stars (with spectral type ranging roughly from B0 to B4) are usually quite variable in  timescales of days to years, with amplitudes of up to tenths of a magnitude \citep{2013A&ARv..21...69R}. The majority of them show very irregular variability. Most present clear single bump-like features, characterised by a fast rise in brightness followed by a slower fading. Frequently, between the brightening and the fading phases some sort of plateau of nearly constant brightness is seen. 
Sometimes these bumps are reversed, so that an initial fast fading is followed by a slow recovering of the stellar brightness.
Two examples of light curves showing these dips are presented in Fig.~\ref{example_a}, taken from OGLE-II \citep{1997AcA....47..319U} and OGLE-III \citep{2008AcA....58...69U} data, for two Be star candidates from the SMC, {based on the selection made by \citet{2002A&A...393..887M}.}
Object {SMC\_SC1 75701} shows two bumps, while {SMC\_SC6 128831} shows a dip.

These bumps and dips resemble the photometric features shown in \citet[][e.g., their Fig. 14]{2012ApJ...756..156H}, where a circumstellar viscous disc builds up as a result of a mass injection into the disc at a constant rate, and then dissipates after the mass injection ceases.
\citet{2012ApJ...756..156H} studied several disc feeding scenarios (constant, cyclic, and outburst) and their photometric counterparts, and demonstrated that the bumps are disc formation/dissipation events of active Be stars seen at near pole-on inclination angles ($i\lesssim 70\deg$), while the dips are associated with near edge-on Be stars (often called shell stars, $i\gtrsim 70\deg$). The inclination angle plays an important role on how the stellar brightness is modified by the presence of disc because, in the second case (edge-on), the disc is seen projected against the stellar disc, thus causing an attenuation of the stellar radiation. This attenuation does not happen for the pole-on case, where the net effect of the disc is to increase the optical brightness as a result of free-bound and free-free radiation from the gas \citep{1974ApJ...191..675G,2015MNRAS.454.2107V}.




In this section, {we describe the basic hydrodynamical concepts of gaseous VDDs, with a focus on how to model the aforementioned events of disc construction and dissipation responsible for the bumps seen in the light curves of active Be stars.}
The evolution of the surface density, 
\begin{equation}
\Sigma(R,t)=\int_{-\infty}^\infty\rho(R,z,t)\mathrm{d}z\,,
\label{sigma_def}
\end{equation}
 of thin circumstellar axisymmetric discs of Be stars 
is described by the following equation \citep{1995ARA&A..33..505P}
\begin{equation}
\frac{\partial\Sigma}{\partial t}
+\frac{1}{R}\frac{\partial}{\partial R}\left(
R\Sigma v_R\right)
=
S_\Sigma
\,,
\label{conservmass1}
\end{equation}
%
where $R$ and $z$ are cylindrical coordinates, $R\Sigma v_R$ is the mass flux crossing radius $R$ per azimuthal angle, $S_\Sigma$ {represents the sources and sinks of mass in the disc per unit of area} (see below), and $\rho$ is the mass density. 
In Be discs with azimuthal symmetry, as assumed in this work, the mass flux at a given radius ($2\pi R\Sigma v_R$) can switch between a positive (decretion) and negative (accretion) value, in response to changes in $S_\Sigma$.
In the circumstellar alpha-disc formulation, the mass flux is given by
\begin{equation}
2\pi R\Sigma 
v_R
=
-4\pi\left(\frac{R}{GM}\right)^\frac{1}{2}
\frac{\partial}{\partial R}\left(R^2\alpha c_{s}^2\Sigma\right)
\,,
\label{fluxofmass}
\end{equation}
where $c_s^2=kT_\mathrm{disc}/\mu m_H$ is the isothermal sound speed and $\alpha$ is the viscosity parameter. 

The orbital velocity of the disc, $v_\phi$, 
is assumed to be Keplerian ($v_\phi = v_K = v_\mathrm{orb}\tilde{R}^{-{1}/{2}}$, with $v_\mathrm{orb}=(GM/R_\mathrm{eq})^{1/2}$ and $\tilde{R}=R/R_\mathrm{eq}$, where $R_\mathrm{eq}$ is the stellar equatorial radius). This assumption holds as long as the gravitational force is much larger than the force due to the pressure gradient, which is generally true for distances from the star of dozens to a few hundreds of stellar radii \citep{2001PASJ...53..119O}, as long as $(c_s/v_K)^2\ll 1$. 


The function $S_\Sigma$ is the rate of mass injected into (or removed from) the disc per unit of area. It represents the variable mass exchange between the star, the disc and the outer medium. We consider that, during outbursts, mass is put into orbit along a very thin ring of radius $R_\mathrm{inj}$ close to the stellar equator. 
Therefore, we assume 
\begin{equation}
S_\Sigma=\dot M_{\rm inj}(t)\frac{\delta(R-R_\mathrm{inj})}{2\pi R}+\mathrm{boundaries}
\,,
\label{sourceterm}
\end{equation}
where $\dot M_{\rm inj}(t)$ is the mass injection rate from the star into the disc at $R_\mathrm{inj}$.

In addition, mass can flow away from the disc through its boundaries. Mass can fall back into the star through the inner boundary at $R_\mathrm{eq}$, or it can leave the system at an outer boundary, $R_\mathrm{out}$. 
The outer boundary can be interpreted as the limiting radius of the disc due to a binary companion \citep[e.g.,][]{2002MNRAS.337..967O} or due to the photoevaporation of the disc \citep[e.g.,][]{2001PASJ...53..119O}.
We consider that all mass that eventually reaches the stellar equator $R_\mathrm{eq}$ is totally absorbed. The same is assumed for the mass that eventually reaches an outer radius $R_\mathrm{out}$
(see below for the definition of $R_\mathrm{out}$).
Therefore, the boundaries consist of $\Sigma(R_\mathrm{eq},t) = \Sigma(R_\mathrm{out},t) = 0$.

For the hydrodynamical simulations, we assume the disc to be isothermal, with $T_\mathrm{disc}=0.6T_\mathrm{eff}$, following \citet{2006ApJ...639.1081C}. We further assume
that the $\alpha$ parameter is constant with $R$, but we allow it to possibly be time-dependent, as there is evidence that it might happen in discs of Be stars \citep{2017ASPC..508..323G}. Consequently, Eqs.~(\ref{conservmass1})--(\ref{sourceterm}) can be scaled in the following way
\begin{equation}
\frac{\partial\Sigma}{\partial t}
=
\frac{1}{\tau}\left\{\frac{2}{\tilde{R}}\frac{\partial}{\partial \tilde{R}}\left[
\tilde{R}^\frac{1}{2}
\frac{\partial}{\partial \tilde{R}}\left(
\tilde{R}^2\Sigma\right) \right]+\tau S_\Sigma\right\}
\,,
\label{singlebe_eq1}
\end{equation}
%
where we introduce the {\emph{timescale parameter}}, $\tau(t)$, given by
\begin{equation}
\tau(t)=\frac{1}{\alpha(t)}\left(\frac{R_\mathrm{eq}^3}{GM}\right)^\frac{1}{2}\frac{v_\mathrm{orb}^2}{c_{s}^2}
\,.
\label{tau}
\end{equation}
%
The timescale parameter is proportional to the viscous timescale at the stellar equator, which is given by $t_\mathrm{vis}=R^2/\nu$, where the viscosity $\nu$ is given by $\nu=(2/3)\alpha c_s^2 R/v_K$. 
Since we are in the thin disc limit ($c_s^2/v_K^2\ll 1$), we see that $\tau$ is much larger than the orbital period at the stellar equator, given by $2\pi(R_\mathrm{eq}^3/GM)^{1/2}$.

The solution of Eq.~\ref{singlebe_eq1} is scaled in time by the timescale parameter, 
which controls how fast matter is distributed throughout the disc and, consequently, its observational counterparts.
It follows that by fitting observed light curves of Be stars, this parameter can be estimated, and once other parameters are known ($R_\mathrm{eq}$, $M$, $T_\mathrm{disc}$), the $\alpha$ value can be determined \citep[e.g.,][]{2012ApJ...744L..15C}.

In order to generate models that do not depend on the time-dependent form of $\alpha(t)$ and also on the parameters $R_\mathrm{eq}$, $M$ and $T_\mathrm{disc}$, which are related to the central star, we define a \emph{time parameter}, $\tilde{\tau}(t)$, such that
%
\begin{equation}
\mathrm{d}\tilde{\tau}=\frac{\mathrm{d}t}{\tau(t)}
\,,
\label{timepar1}
\end{equation}
%
which allows us to solve Eq.~(\ref{singlebe_eq1}) in terms of $\tilde{\tau}$ instead of the physical time $t$. 
The advantage of using this new variable is that it separates the problem of the time dependency of $\alpha(t)$ from the problem of solving Eq.~(\ref{singlebe_eq1}). Consequently, it allows us to create a grid of solutions of that equation that is independent of $\alpha(t)$, $R_\mathrm{eq}$, $M$ and $T_\mathrm{disc}$ (see Sect.~\ref{models_bumps})



In this work, 
Eq.~(\ref{singlebe_eq1}) is solved numerically by the 1-D thin-disc code {\tt SINGLEBE} \citep{2007ASPC..361..230O}.
At selected time parameters $\tilde{\tau}$, 
the solution $\Sigma(\tilde{R},\tilde{\tau})$ enters as  input for the Monte Carlo 3-D radiative transfer code {\tt HDUST} \citep{2004ApJ...604..238C,2006ApJ...639.1081C,2008ApJ...684.1374C} that calculates the emergent spectrum of the  star+disc system. To convert between surface density and mass density (Eq.~\ref{sigma_def}), it is assumed that the disc is vertically sustained by hydrostatic pressure, in which case the vertical density profile is a Gaussian, and 
\begin{equation}
\rho(\tilde{R},z,\tilde{\tau}) = \frac{\Sigma(\tilde{R},\tilde{\tau})}{\sqrt{2\pi}H} e^{-\frac{z^2}{2H^2}}
\,,
\label{rho_sigma1}
\end{equation}
where 
\begin{equation}
\frac{H}{R_\mathrm{eq}}=\left(\frac{c_s}{v_\mathrm{orb}}\right)\tilde{R}^\frac{3}{2}
\label{scaleheight}
\end{equation}
is the scale height. 

Before moving on to the modelling of bump-like events such as the ones of Fig.~\ref{example_a}, 
we introduce in the next section some important parameters of the dynamical discs implied in the equations shown above.

\subsection{Dynamical disc parameters}
\label{more_params}

The radius of mass injection divides the disc in two regions: a narrow region between the inner boundary and the radius of mass injection ($1\leq \tilde{R}\leq \tilde{R}_\mathrm{inj}$), and the much wider region between the radius of mass injection and the outer boundary ($\tilde{R}_\mathrm{inj}\leq \tilde{R}\leq \tilde{R}_\mathrm{out}$). 
The steady-state solution of Eq.~(\ref{singlebe_eq1}) corresponds to the limiting case of a disc that has been fed at a constant rate for an infinitely long time. It is obtained by setting $\partial \Sigma/\partial t = 0$ and assuming that $\alpha$ and $\dot M_\mathrm{inj}$ are time-independent. 
The surface density in the steady-state is given by
\begin{eqnarray}
\Sigma_\mathrm{steady}(\tilde{R})
=
\left\{
\begin{array}{ll}
\frac{\Sigma_0}{\tilde{R}^2} \left(\frac{\tilde{R}^\frac{1}{2}-1}{\tilde{R}_\mathrm{inj}^\frac{1}{2}-1}\right)\Upsilon, & 1 \leq \tilde{R} < \tilde{R}_\mathrm{inj}
\\
\frac{\Sigma_0}{\tilde{R}^2} \left(\frac{\tilde{R}_\mathrm{out}^\frac{1}{2}-\tilde{R}^\frac{1}{2}}{\tilde{R}_\mathrm{out}^\frac{1}{2}-1}\right), & \tilde{R}_\mathrm{inj} \leq \tilde{R} \leq \tilde{R}_\mathrm{out}
\end{array}
\right.
\,,
\label{sigma_stat}
\end{eqnarray}
where $\Upsilon$
$~=~$
$(\tilde{R}_\mathrm{out}^{1/2} -\tilde{R}_\mathrm{inj}^{1/2})/$
$(\tilde{R}_\mathrm{out}^{1/2}-1)$ is a number usually very close to 1 for any Be disc.

The physical quantity $\Sigma_0$ represents the surface density at $R_\mathrm{eq}$, obtained by extrapolating  $\Sigma_\mathrm{steady}$  in the domain $\tilde{R}_\mathrm{inj} \leq \tilde{R} \leq \tilde{R}_\mathrm{out}$ to $R_\mathrm{eq}$. We will refer to this physical quantity as the 
{\emph{asymptotic surface density}}, as it is the asymptotic value reached after an infinitely long disc build-up under a constant $\dot M_\mathrm{inj}$. It is easily shown that $\Sigma_0$ is related to $\dot M_\mathrm{inj}$ by the following equation
%
\begin{equation}
2\pi R_\mathrm{eq}\Sigma_0
\left(\frac{R_\mathrm{eq}}{\tau}\right)
=
\dot M_\mathrm{inj} \left(\tilde{R}_\mathrm{inj}^\frac{1}{2}-1\right)
\equiv
\left(-\frac{\partial M}{\partial t}\right)_\mathrm{typ}
\,,
\label{typdecrate}
\end{equation}
where $\left({\partial M}/{\partial t}\right)_\mathrm{typ}$ is defined below.

We may extend Eq.~(\ref{typdecrate}) to the general case of a time-dependent  $\dot M_\mathrm{inj}(\tilde{\tau})$, which would define, by the same equation, a time dependent asymptotic surface density, $\Sigma_0(\tilde{\tau})$. The function $\Sigma_0(\tilde{\tau})$, therefore, is just another way of specifying the history of mass injection from the star into the disc. 
It has, however, the advantage of being a surface density, which is a quantity that may be determined from, e.g, SED analyses, in contrast to the mass injection rate and the radius of mass injection, which are parameters that cannot be observationally determined.

The steady-state solution (Eq.~\ref{sigma_stat}) shows that, in the wider domain $\tilde{R}_\mathrm{inj} \leq \tilde{R}\leq \tilde{R}_\mathrm{out}$, the density profile of the disc is not altered if  $\tilde{R}_\mathrm{inj}$ is changed, provided that $\dot M_\mathrm{inj}$ is also changed in order to maintain $\Sigma_0$ fixed, according to Eq.~(\ref{typdecrate}). In fact, we verified that the time-dependent solutions of Eq.~(\ref{singlebe_eq1}) in the domain $\tilde{R}_\mathrm{inj} \leq \tilde{R}\leq \tilde{R}_\mathrm{out}$ are negligibly affected by the particular choice of $\tilde{R}_\mathrm{inj}$ or $\dot M_\mathrm{inj}(\tilde{\tau})$, as long as the quantity $\Sigma_0(\tilde{\tau})$ is kept fixed. This is a consequence of the fact that the dynamical solutions reach a near steady-state very quickly in the vicinity of the injection radius \citep{2012ApJ...756..156H}.
Furthermore, provided that mass is injected not too far from the stellar photosphere (i.e.,  assuming $\tilde{R}_\mathrm{inj} \gtrapprox
 1$), the domain $1\leq \tilde{R}\leq \tilde{R}_\mathrm{inj}$ is much narrower than the region where the continuum visual flux of Be stars is generated \citep{2011IAUS..272..325C}, which means that the emission from this region can be ignored. 
Consequently, we conclude that $\Sigma_0(\tilde{\tau})$ (with the assumption that $\tilde{R}_\mathrm{inj}\gtrapprox
 1$) is a much better parameter for describing the mass injection history of the disc than the pair of parameters $\dot{M}_\mathrm{inj}(\tilde{\tau})$ and $\tilde{R}_\mathrm{inj}$. 


The time-dependent solutions of Eq.~(\ref{singlebe_eq1}) generally show that, for Be stars dynamically feeding the disc but still far from steady-state, the mass flux close to $\tilde{R}_\mathrm{inj}$ has absolute values of the order of $\left(-\partial M/\partial t\right)_\mathrm{typ}$, defined by Eq.~(\ref{typdecrate}). Therefore, we refer to this quantity as the \emph{typical decretion rate}, which depends on parameters relatively easy to estimate from SEDs of Be stars. 


In our simulations, since the values of $\dot M_\mathrm{inj}$ and $\tilde{R}_\mathrm{inj}$ are of no interest, and the value of $\tilde{R}_\mathrm{out}$ is quite uncertain, we arbitrarily chose $\tilde{R}_\mathrm{inj}=1.017$ and $\tilde{R}_\mathrm{out}=1000$
(we discuss below how this choice of $\tilde{R}_\mathrm{out}$ might affect our results).
Eq.~\ref{typdecrate}, therefore, shows that the typical decretion rate is much smaller than the mass injection rate $\dot M_\mathrm{inj}(\tilde{\tau})$. In our case, 
the typical decretion rate is only $8.46\times 10^{-3} \dot M_\mathrm{inj}$. 
This means that the majority of the injected mass flows inwards and is absorbed by the inner boundary at the stellar equator, and only a small remaining fraction of the injected mass is responsible for the growth of the disc. 
These results were first obtained from SPH simulations of Be discs by \citet{2002MNRAS.337..967O}, 
who found that only about 0.1\% of the injected material flows outward, as a direct result of their choice for $R_\mathrm{inj}$.

It can be shown, by substitution of Eqs.~(\ref{sigma_stat}) and (\ref{typdecrate}) in Eq.~(\ref{fluxofmass}) that, in steady-state, only the fraction of the injected mass given by 
$\dot M_\mathrm{inj}(1-\Upsilon)$ is flowing outwards through the disc and crossing the outer radius $R_\mathrm{out}$, thus leaving the system.
For our assumed values for $\tilde{R}_\mathrm{inj}$ and $\tilde{R}_\mathrm{out}$, $1-\Upsilon=2.84\times 10^{-4}$.
Since the mass of the disc is not varying in steady-state, the mass flux given by $\dot M_\mathrm{inj}\left(1-\Upsilon\right)$ is actually the rate of mass being lost by the star, which we will indicate by $(-\partial M/\partial t)_\mathrm{steady}$. 
It is easily seen that it is related to the typical decretion rate by the following equation
\begin{equation}
\left(-\frac{\partial M}{\partial t}\right)_\mathrm{steady}=\frac{\Lambda}{\tilde{R}_\mathrm{out}^\frac{1}{2}}
\left(-\frac{\partial M}{\partial t}\right)_\mathrm{typ}
\,,
\label{statdecrate}
\end{equation}
where $\Lambda=1/(1-\tilde{R}_\mathrm{out}^{-1/2})$ is a number very close to 1 for any Be disc in general. 

In steady-state, the angular momentum escaping the system at the outer boundary (and also being lost by the star) is $(GMR_\mathrm{out})^{1/2} \dot M_\mathrm{inj}(1-\Upsilon)$, and is written as
\begin{equation}
\left(-\frac{\partial J}{\partial t}\right)_\mathrm{steady}=\Lambda \left(GMR_\mathrm{eq}\right)^\frac{1}{2}
\left(-\frac{\partial M}{\partial t}\right)_\mathrm{typ}
\,.
\label{statangmomlossrate}
\end{equation}
From Eq.~(\ref{statdecrate}) we see that knowing $\tilde{R}_\mathrm{out}$ is essential for estimating the rate of mass being lost by the star. Interestingly, this is not the case for the rate of angular momentum being lost by the star, given by Eq.~(\ref{statangmomlossrate}).

\subsection{The Mass Reservoir Effect}
\label{reservoir_effect}


It is important to stress that the solution $\Sigma(\tilde{R},\tilde{\tau})$ is shaped not just by the mass injection rate $\Sigma_0(\tilde{\tau})$ at the specific instant $\tilde{\tau}$, but by the whole mass injection history before the instant $\tilde{\tau}$. Therefore, the advantage of studying the relatively isolated bumps like the ones exemplified in Fig.~\ref{example_a}, which started after a clear inactive phase, is that there is no disc present when the bump starts developing; thus, no previous history of mass injection has to be taken into account in the beginning of the modelling. 

The light curves of several Be stars show that the duration of the build-up phase, which we refer to as the \emph{build-up time}, is variable between Be stars and even between different bumps from the same star, ranging from a few days to years. 
The following phase of disc dissipation, however, contrary to the build-up phase, depends of the previous history of mass injection. For this reason, the modelling of the dissipation phase must not be disconnected from the modelling of the build-up phase that happened before it.

One of the main consequences of this fact is the \emph{mass reservoir effect} \citep[see also][]{2017ASPC..508..323G}.
Basically,
discs that had a longer build-up phase necessarily transported more matter and angular momentum outwards and created a larger external reservoir of mass and angular momentum in its outer regions, which usually extend far beyond the first few stellar radii where the visible photometric observables are formed. It is common, for instance, that some bumps reach plateaus during the build-up phase. The plateau indicates that the density in the inner disc has reached near-steady-state values and, consequently, there is little photometric variation in the visible wavelengths. The outer disc, however, will likely be far from steady-state and thus will continue to increase in density and mass.
When mass injection ceases, the dissipation phase begins. Re-accretion occurs and, due to the more massive outer disc,  the inner disc remains relatively denser for a longer time. This makes the dissipation of the disc appear slower in the observed light curves. Conversely, a disc that had a small build-up time would dissipate much faster.


The importance of the mass reservoir effect can be assessed by the reevaluation of the $\alpha$ parameter in 28 CMa by \citet{2017ASPC..508..323G}.
\citet{2012ApJ...744L..15C} modelled the 2003 dissipation phase of 28 CMa by considering a very long previous build-up time, and found that a high value of $\alpha$ was necessary ($1.0 \pm 0.2$) to match the observed dissipation rate. \citeauthor{2017ASPC..508..323G} have shown that when the previous build-up phase is properly accounted for in the modelling, the value of $\alpha$ required to match the dissipation rate is much smaller ($0.21 \pm 0.05$).


\section{A model grid of disc formation and dissipation events}
\label{models_bumps}

In this section we describe the method we developed for fitting the light curves associated with events of disc formation and dissipation.
The method consists of precomputing a large grid of dynamical models of the time-dependent disc structure, covering the entire range of observed scenarios (Sect.~\ref{model_grid}) and  performing the radiative transfer in these models to produce synthetic light curves (Sect.~\ref{radiative_transfer}). The observed light curves are then fitted by the synthetic one using the procedure described in Sects.~\ref{emplaw_sect} and Sect.~\ref{pipeline_sect}.

\subsection{Dynamical model grid}
\label{model_grid}


For building a comprehensive grid of dynamical models that are solutions of Eq.~\ref{singlebe_eq1}, we used the definitions of Sect.~\ref{vdd_sect} that allow us to write the solution $\Sigma(\tilde{R},\tilde{\tau})$ in terms of $\Sigma_0(\tilde{\tau})$ and the dimensionless parameter $\tilde{\tau}$. 

As discussed in Sect.~\ref{reservoir_effect}, the advantage of studying relatively isolated bumps like the ones exemplified in Fig.~\ref{example_a}, which started after a clear inactive phase, is that there is no previous history of mass injection to be taken into account for the modelling, so that during build-up the shape of the curve is controlled solely by $\Sigma_0(\tilde{\tau})$ and $\tilde{\tau}$, while for dissipation the previous disc build-up time should also be considered (Sect.~\ref{reservoir_effect}).
By using the time parameter (Eq.~\ref{timepar1}) instead of the physical time, our dynamical models are independent of the specific physical parameters $M$, $T_\mathrm{eff}$, $R_\mathrm{eq}$ and $\alpha(t)$ of the Be star under consideration (Eq.~\ref{tau}).
Also, from the linearity of Eq.~\ref{singlebe_eq1}, it follows that multiplying $\Sigma_0(\tilde{\tau})$ by some constant results in the solution $\Sigma(\tilde{R},\tilde{\tau})$ multiplied by the same constant. Consequently, only one value of $\Sigma_0$ during the build-up phase is necessary. 

For our grid of dynamical models, we therefore assume that our Be stars start discless. At instant $\tilde{\tau}=0$, mass injection into the disc begins at an arbitrary constant rate ($\Sigma_0 > 0$) that lasts until $\tilde{\tau}=\tilde{\tau}_\mathrm{bu}$, which we refer to as the {\emph{scaled build-up time}}, since it is related to the above mentioned build-up time, but scaled by the timescale parameter. After that ($\tilde{\tau}>\tilde{\tau}_\mathrm{bu}$), mass injection no longer occurs ($\Sigma_0 = 0$) and the disc dissipates.
In Appendix~\ref{Appendix1}, we further discuss the properties of these dynamical models. 

We chose 11 values of $\tilde{\tau}_\mathrm{bu}$, listed in Table~\ref{star_and_disc}. 
Since the timescale parameter (Eq.~\ref{tau}) is roughly given by $\sim (100$--$200)/\alpha$ days for early Be stars in the main sequence with $\alpha\lesssim 1$, these values correspond to real build-up times of at least 15 days, which brackets the observed build-up times of the sample described below (Sect.~\ref{ogle_sect}, Table~\ref{my_selection}).
In this study, we decided not to model the bumps with observed build-up times lower than about 15 days, usually referred to as flickers \citep{2002AJ....124.2039K}. 

\subsection{Radiative transfer models}
\label{radiative_transfer}

%
%

Having selected a set of suitable hydrodynamic bump models, the next step is to produce photometric light curves of these models. The radiative transfer part of the problem requires a stellar model, which will be the primary source of radiation. 
The stellar model depends on the physical parameters $M$, $R_\mathrm{eq}$ and $T_\mathrm{eff}$, which were left unspecified in the dynamical model grid. In addition, three other parameters must be specified: the viewing angle, $i$ ($i=0$ means pole-on orientation), the distance to the star, $d$, and the interstellar reddening.

One important feature of the central stars of Be stars is that they are fast rotators. Fast rotation causes the star to be oblate, with hotter poles and colder equatorial regions. Rotation is specified by the ratio of the rotation velocity at the equator to the Keplerian velocity at the equator, $W=v_\mathrm{eq}/v_\mathrm{orb}$. 
The ratio between the equatorial radius to the polar radius is given by $R_\mathrm{eq}/R_\mathrm{pole}=1+W^2/2$ for a Roche-shaped star.
All these parameters evolve in time as a consequence of stellar evolution, and Be stars can be found in luminosity classes from V to III \citep{2013A&ARv..21...69R}. We adopt the Geneva evolutionary tracks  \citep{2013A&A...553A..24G} to determine $R_\mathrm{eq}$ and $T_\mathrm{eff}$ given $M$ and the age in the main sequence, $t_{\rm MS}$.


The current version of {\tt HDUST} allows for a spheroidal rotationally oblate star, with the latitude-dependent surface temperature being given by $T_\mathrm{surf}\propto g_\mathrm{eff}^\beta$ \citep{2008ApJ...676L..41C}. Here, the coefficient $\beta(W)$ is calculated by fitting a straight line to the gradient $\partial\ln T_\mathrm{surf}/\partial\ln g_\mathrm{eff}$ given by the flux theory of \citet{2011A&A...533A..43E}. 
For the disc scale height  (Eq.~\ref{scaleheight}), we assume  an isothermal disc with $T_\mathrm{disc}=0.6T_\mathrm{eff}$, where $T_\mathrm{eff}$ is the effective temperature of the star, defined by  $T_\mathrm{eff}={(L_*)^{1/4}(\sigma S_*)^{-1/4}}$, with $S_*$ being the surface area of the star. 

In order to generate synthetic absolute magnitudes from the computed SEDs, we used the standard $BVRI$ Johnson-Cousins passbands from \citet{1990PASP..102.1181B}
%
%
%
%
%
%
%
and the Vega flux from 
\citet{1994A&A...281..817C} as standard of calibration.


\begin{table}
\centering
\caption{Parameters of the grid of photometric models of bumps}
\label{star_and_disc}
\begin{tabular}{rrrr}
\hline
Star & $i\, [\mathrm{deg}]$ & $\Sigma_0\, [\mathrm{g\, cm^{-2}}]$ & $\tilde{\tau}_\mathrm{bu}$ 
\\
\hline
Star 1 & 00.0 & 0.30 & 00.15 \\
Star 2 & 21.8 & 0.41 & 00.45 \\
Star 3 & 31.0 & 0.56 & 00.75 \\
 & 38.2 & 0.75 & 01.50 \\
 & 44.4 & 1.01 & 02.25 \\
 & 50.0 & 1.37 & 03.00 \\
 & 55.2 & 1.85 & 04.50 \\
 & 60.0 & 2.50 & 06.00 \\
 & 64.6 &  & 09.00 \\
 & 69.1 &  & 15.00 \\
 & 73.4 &  & 30.00 \\
 & 77.6 &  &  \\
 & 81.8 &  &  \\
 & 85.9 &  &  \\
 & 90.0 &  &  \\
\hline
\end{tabular}
\end{table}

\begin{table}
\centering
\caption{Parameters of the stellar models of Table~\ref{star_and_disc}}
\label{stars_instardisc}
\begin{tabular}{rrrrrr}
\hline
Star & $Z$ & $M$\,[$M_\odot$] & $W$ & $t/t_\mathrm{MS}$ & $\alpha\tau\, [\mathrm{d}]$\\
\hline
Star 1 & 0.002 & 7 & 0.81 & 0.5 & 90.4  \\
Star 2 & 0.002 & 11 & 0.81 & 0.5 & 103.3 \\
Star 3 & 0.002 & 15 & 0.81 & 0.5 & 118.9 \\
\hline
\end{tabular}
\end{table}

A grid of model light curves was computed using the 11 dynamical models described in Sect.~\ref{model_grid}, with 8 different values of $\Sigma_0$ (third column of Table~\ref{star_and_disc}). For each of these disc models, radiative transfer models were calculated with {\tt HDUST} at 17 different time parameters (not shown in the table) and 15 equally-spaced values of $\cos i$ (second column). This whole process was done for 3 different stellar models \citep[``Star 1'', ``Star 2'' and ``Star 3'', first column of Table~\ref{star_and_disc}, according to the stellar models of][]{2013A&A...553A..24G}. Details on the stellar models are given in Table~\ref{stars_instardisc}. They were chosen to represent early B-type stars from the SMC ($Z=0.002$), in the middle of their life in the main sequence, with the rotation parameter given by the mean value obtained for Be stars \citep[$W=0.81$, ][]{2006A&A...459..137R}.
 In the sixth column of Table~\ref{stars_instardisc}, we present the values of $\alpha\tau$ (Eq.~\ref{tau}) for the discs of these stars (with the assumption that $T_\mathrm{disc}=0.6T_\mathrm{eff}$).
In short, a single light curve is specified by taking one element of each column of Table~\ref{star_and_disc}. The end result was a grid of $3\times15\times8\times11=3960$ light curves, for each of the $BVRI$ bands.


\begin{table}
\centering
\caption{Parameters of the grid of photometric models of discless stars \label{star_only}}
\begin{tabular}{rrrrr}
\hline
$Z$ & $M/M_\odot$ & $W$ & $t/t_\mathrm{MS}$ & $i\, [\mathrm{deg}]$
\\
\hline
0.002 & 2.50 & 0.447 & 0.00 & 00.0 \\
 & 3.68 & 0.633 & 0.20 & 27.3 \\
 & 4.85 & 0.775 & 0.40 & 38.9 \\
 & 6.00 & 0.894 & 0.60 & 48.2 \\
 & 7.15 & 0.949 & 0.80 & 56.3 \\
 & 8.29 &  & 1.00 & 63.6 \\
 & 9.42 &  & & 70.5 \\
 & 10.54 &  & & 77.2 \\
 & 11.66 &  & & 83.6 \\
 & 12.78 &  & & 90.0 \\
 & 13.89 &  & &  \\
 & 15.00 &  & &  \\
 & 20.00 &  & &  \\
\hline
\end{tabular}
\end{table}

A grid of inactive (discless) stellar models was also calculated. Because these models can be computed much faster than the bump models, we were able to cover a much finer grid of stellar parameters (Table~\ref{star_only}), aiming at a better determination of the stellar parameters.
The grid is composed by  models for 13 different masses (second column), 5 different rotation rates (third column), 6 equally spaced values for the age in the main sequence (forth column) and 10 equally spaced values of $\cos i$ (fifth column), resulting in a total of $13\times 5\times 6\times 10 = 3900$ photometric models for each of the $BVRI$ bands.

\subsection{Empirical law}
\label{emplaw_sect}


In order to facilitate the comparison of the synthetic light curves (Sect.~\ref{radiative_transfer}) with the observed ones (Sect.~\ref{ogle_sect}), we developed two empirical laws that match quite closely the synthetic light curves for build-up and dissipation. 
The usefulness of these formulae will become clear in the next section.


In our discussion of the features of the light curves, it is useful to separate them in three groups: 
(i) {\it pole-on light curves}, of stars seen at small inclination angles ($0\leq i \lesssim 70\deg$), which should statistically correspond to the majority of the observed light curves;
(ii) {\it edge-on light curves}, of shell stars ($i \approx 90\deg$); and,
(iii) {\it intermediate light curves}, of stars seen at intermediate angles ($70 \lesssim i \lesssim 85\deg$ -- the extension of this intermediate region varies depending the photometric band under consideration and will be defined below).
Pole-on light curves show an increase in apparent brightness, due to the additional flux coming from the disc. Conversely, edge-on light curves show a decrease in apparent brightness, due to obscuration of the star by the disc. 
The intermediate case shows the smallest variations in apparent brightness, and frequently the light curve has a more complicated shape, as it is influenced by variable amounts of disc emission/absorption.


{A computed light curve is a sequence of absolute magnitudes for a set of time parameters, in a given photometric band $X$, given by $$M_X(\tilde{\tau})=M_{X*}+\Delta X(\tilde{\tau})\,,$$ where $M_{X*}$ is the absolute magnitude of the inactive Be star at band $X$, and $\Delta X(\tilde{\tau})$ is the magnitude difference caused by the disc ($\Delta X(\tilde{\tau})$ can be either positive or negative). $M_{X*}$ can be estimated from the light curve during the inactive phase (e.g., the purple points of Fig.~\ref{example_a}).} 
A build-up light curve 
(for $\tilde{\tau}-\tilde{\tau}_\mathrm{bu} < 0$)
 is denoted  by 
$\Delta X_\mathrm{bu}(\tilde{\tau})$. 
Its limiting value, if the build-up phase were to have an infinite duration, is denoted by $\Delta X_\mathrm{bu}^\infty$ 
 -- the photometric excess of the disc when in steady-state. 
A dissipation light curve (for 
$\tilde{\tau}-\tilde{\tau}_\mathrm{bu}\geq 0$)
 is denoted by $\Delta X_\mathrm{d}(\tilde{\tau})$. Its value at the beginning of the dissipation ($\tilde{\tau}=\tilde{\tau}_\mathrm{bu}$) is given by $\Delta X_\mathrm{d}^0$. 
Since every dynamical model starts from a discless state and asymptotically ends at a discless state, it follows that, regardless of the viewing angle, 
for every build-up light curve the quantity 
$\Delta X_\mathrm{bu}(\tilde{\tau})/\Delta X_\mathrm{bu}^\infty$
 is a function that goes from 0 to 1 as $\tilde{\tau}$ goes from 0 to $\infty$, and for every dissipation light curve, the quantity 
$\Delta X_\mathrm{d}(\tilde{\tau})/\Delta X_\mathrm{d}^0$ 
 is a function that goes from 1 to 0 as $\tilde{\tau}-\tilde{\tau}_\mathrm{bu}$ goes from 0 to $\infty$. 
In Appendix~\ref{Appendix2}, we show examples of light curves that accompany the conclusions drawn on this section.

\begin{figure}
\centering{
\includegraphics[width=1.0\linewidth]{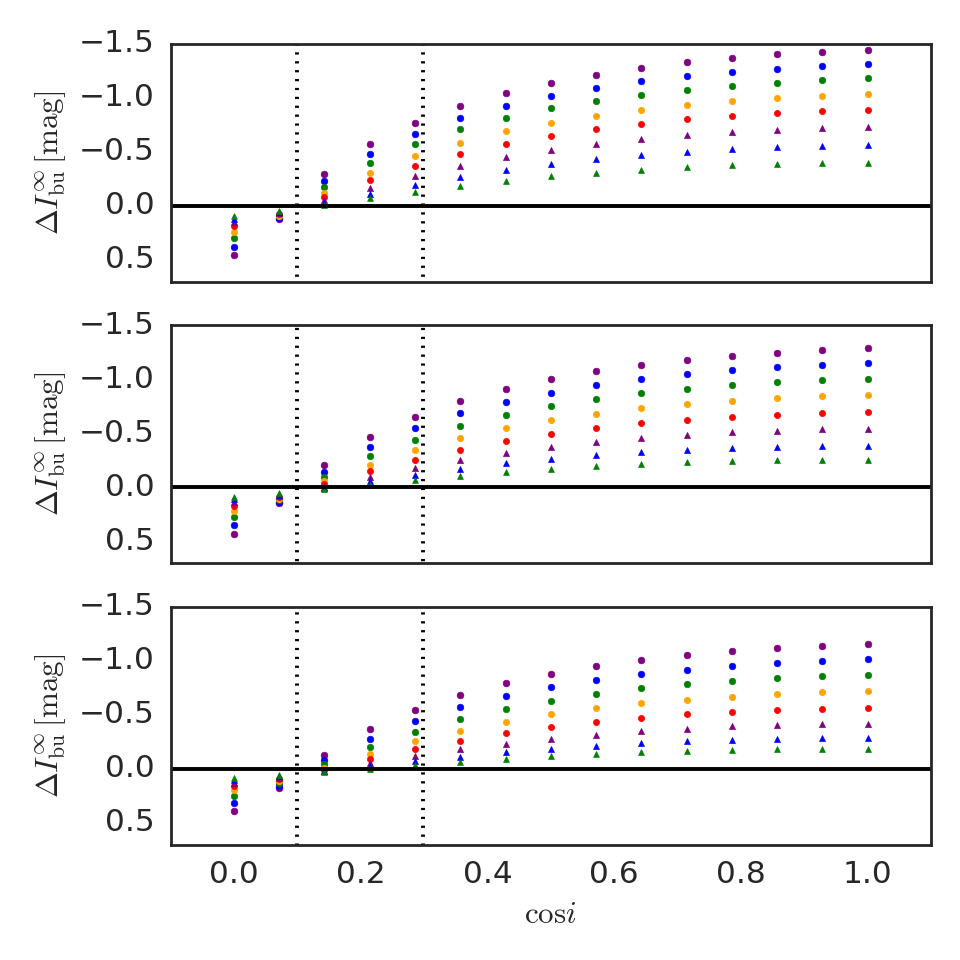}
}
\caption[]
{
The values of 
$\Delta I_\mathrm{bu}^\infty$
versus $\cos i$ for our grid. From top to bottom, the results are for Star 1, Star 2 and Star 3, respectively. 
Purple, blue, green, orange and red circles correspond to $\Sigma_0 = 2.50$, $1.85$, $1.37$, $1.01$, $0.75~\mathrm{g~cm^{-2}}$. Purple, blue and green triangles correspond to $\Sigma_0 = 0.56$, $0.41$, $0.30~\mathrm{g~cm^{-2}}$.
Vertical dotted lines define the region of intermediate angles for the $I$-band ($73 \lesssim i \lesssim 84\deg$).
}
\label{coeffs1}
\end{figure}
%


The values of 
$\Delta I_\mathrm{bu}^\infty$
for our grid are shown in Fig.~\ref{coeffs1}, plotted against $\cos i$. The values for the $BVR$ bands show qualitatively similar patterns to the ones presented in this figure. 
Each panel shows the results for a different star, and all 11 values of $\Sigma_0$ (Table~\ref{star_and_disc}) are represented in the figure by different colours and symbols.
The curves monotonically increase with $\cos i$, starting with negative values at edge-on orientation and reaching a maximum for pole-on viewing. The angle for which $\Delta X_\mathrm{bu}^\infty = 0$, where the disc excess emission is exactly matched by the absorption of photospheric light by the disc, depends both on the density scale (as shown in the figure) and (most importantly) on the band pass. 

An analysis of our model grid allowed us to determine the ranges in inclination angle for which the light curve displays the intermediate behaviour described above. They were determined by visual inspection of our model grid, as the angles for which the light curves present more complex shapes (see, e.g., the $I$-band light curves seen at $i=81.8\deg$ and $i=77.6\deg$ in Fig.~\ref{visual_events1} of Appendix~\ref{Appendix2}).
Their adopted values are 53--$78\deg$, 60--$78\deg$, 66--$84\deg$ and 73--$84\deg$ for the $BVRI$ bands, respectively.


For pole-on orientations, the observed excess is given by $\Delta X\approx -2.5\log(1+F_{\mathrm{disc}}/F_*)\approx -F_{\mathrm{disc}}/F_*$. \citet{2015MNRAS.454.2107V} studied the continuum emission from gaseous discs, and showed that it can be approximated by the sum of the flux coming from an optically thick inner part (the so-called pseudophotosphere) with the contribution from an optically thin outer part, i.e.,
\begin{equation}
F_{\mathrm{disc}}\propto F_\mathrm{thick}\cos i +F_\mathrm{thin}\,.
\end{equation}
If the contribution of the optically thin part of the disc were negligible {and the stellar flux, $F_*$, did not depend on $\cos i$}, $\Delta X$ would be a linear function of $\cos i$. Clearly this is not the case for the entire range of $\cos i$, and both the optically thin and thick parts of the disc contributes to observed behaviour of $\Delta X$  vs. $\cos i$. This pseudo-photosphere concept will be important to understand the growth and decay rates of the light curves, discussed below.

Figure~\ref{coeffs1} also shows that the excesses increase a little when moving from a low- to a high-mass star, for discs with the same other features. This is a consequence of the fact that the stellar flux relative to the disc flux increases with the luminosity of the star.

\begin{figure}
\centering{
\includegraphics[width=1.0\linewidth]{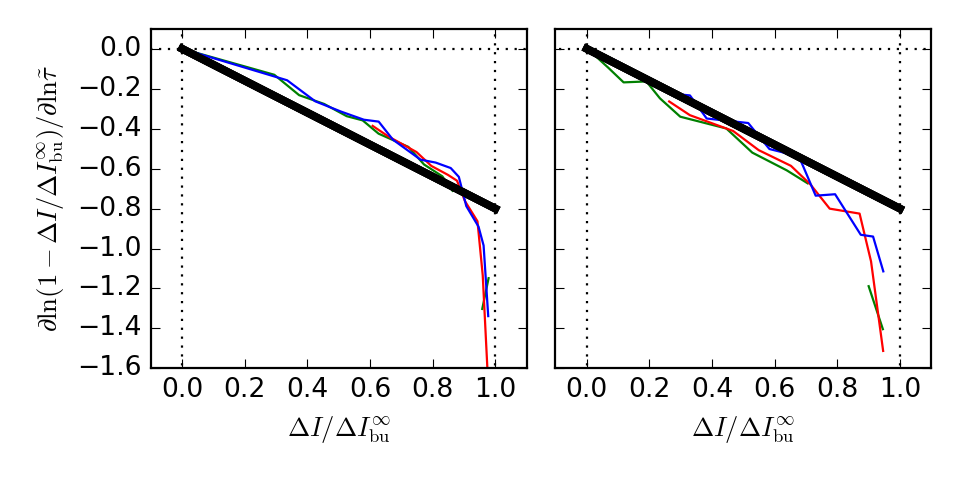}
\includegraphics[width=1.0\linewidth]{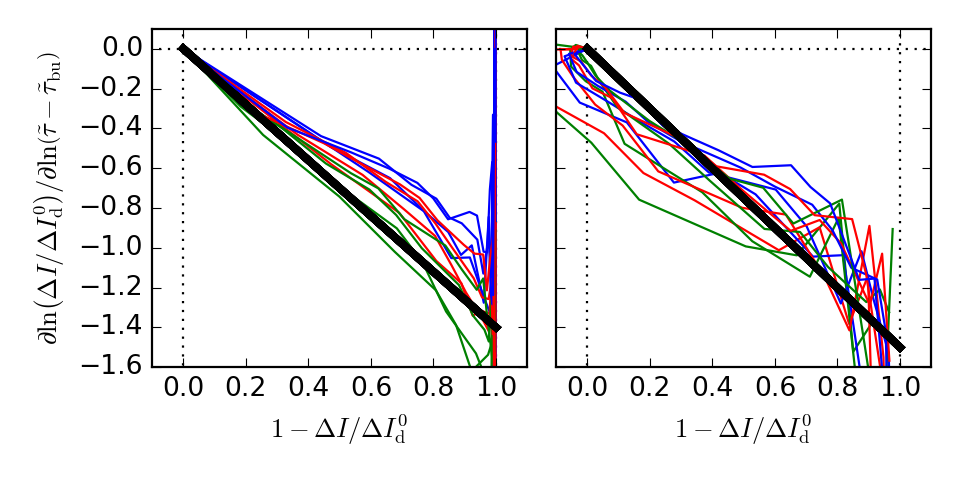}
}
\caption[]
{
Comparison of the empirical law (black straight lines) with the computed light curves for build-up and dissipation.
{\it Top panels}: $\log-\log$ derivative of $1-\Delta I_\mathrm{bu}(\tilde{\tau})/\Delta I_\mathrm{bu}^\infty$ vs. 
$\Delta I_\mathrm{bu}(\tilde{\tau})/\Delta I_\mathrm{bu}^\infty$.
{\it Bottom panels}: $\log-\log$ derivative of $\Delta {I}_\mathrm{d}/\Delta {I}_\mathrm{d}^0$ vs. 
$1-\Delta {I}_\mathrm{d}/\Delta {I}_\mathrm{d}^0$, for values of 
$\tilde{\tau}_\mathrm{bu}$ equal to $0.45$, $1.5$, $6$ and $30$.
 The results are shown for Star 2 at two inclination angles: $i=0\deg$ ({\it left}), and $i=90\deg$ ({\it right}). The green, red and blue curves correspond to $\Sigma_0$ equal to $1.37$, $0.75$ and $0.41\, \mathrm{g\, cm^{-2}}$.  
}
\label{obtlaw1}
\end{figure}
%

Both the pole-on and edge-on light curves have functions 
$\Delta X_\mathrm{bu}(\tilde{\tau})/\Delta X_\mathrm{bu}^\infty$
 and 
$\Delta X_\mathrm{d}(\tilde{\tau})/\Delta X_\mathrm{d}^0$ 
 that are 
 qualitatively similar to each other, suggesting that they could be approximated by simple and general formulas of $\tilde{\tau}$. 
This is illustrated in Fig.~\ref{obtlaw1}, 
where we compare the values of 
$\Delta I_\mathrm{bu}(\tilde{\tau})/\Delta I_\mathrm{bu}^\infty$
 and 
$\Delta I_\mathrm{d}(\tilde{\tau})/\Delta I_\mathrm{d}^0$ 
(in the horizontal axis) with their $\log-\log$ derivatives (in the vertical axis). 
As the panels exemplify, the curves 
are similar to each other in a wide range of parameters ($\cos i$, $\Sigma_0$ and $\tilde{\tau}_\mathrm{bu}$), and they can be roughly approximated by straight lines (shown in black). Therefore, the build-up and dissipation light curves (for inclinations not in the intermediate region) should roughly obey the following differential equations:
\begin{equation}
\frac{\partial\ln\phantom{\tilde{\tau}}}{\partial\ln\tilde{\tau}}\left(1-\frac{\Delta {X}_\mathrm{bu}}
{\Delta X_\mathrm{bu}^\infty}\right)
\approx -\eta_\mathrm{bu}\frac{\Delta {X}_\mathrm{bu}}
{\Delta X_\mathrm{bu}^\infty}
\,,
\label{empp1}
\end{equation}
and
\begin{equation}
\frac{\partial\ln\phantom{\left(\tilde{\tau}-\tilde{\tau}_\mathrm{bu}\right)}}{\partial\ln\left(\tilde{\tau}-\tilde{\tau}_\mathrm{bu}\right)}
\frac{\Delta {X}_\mathrm{d}}
{\Delta {X}_\mathrm{d}^0}
\approx -\eta_\mathrm{d}\left(1-\frac{\Delta {X}_\mathrm{d}}
{\Delta {X}_\mathrm{d}^0}\right)
\,,
\label{empp2}
\end{equation}
%
whose solutions are, respectively,
\begin{equation}
\Delta {X}_\mathrm{bu} = 
{\Delta X_\mathrm{bu}^\infty}
\left[
1-\frac{1}{1+\left(\xi_\mathrm{bu} \tilde{\tau}\right)^{\eta_\mathrm{bu}}}
\right]
\,,
\label{genpower_bu}
\end{equation}
and
\begin{equation}
\Delta {X}_\mathrm{d} = 
\Delta {X}_\mathrm{d}^0
\left[
\frac{1}{1+\left(\xi_\mathrm{d} \left(\tilde{\tau}-\tilde{\tau}_\mathrm{bu}\right)\right)^{\eta_\mathrm{d}}}
\right]
\,.
\label{genpower_diss}
\end{equation}
The continuity condition requires 
\begin{equation}
\Delta {X}_\mathrm{d}^0
= 
\Delta X_\mathrm{bu}^\infty
\left[
1-\frac{1}{1+\left(\xi_\mathrm{bu} \tilde{\tau}_\mathrm{bu}\right)^{\eta_\mathrm{bu}}}
\right]
\,.
\label{psilinha}
\end{equation}
The parameters $\xi_\mathrm{bu}$ and $\xi_\mathrm{d}$ are constants of integration whose values must be determined by fitting the above formulae to the computed light curves. 
The values of the exponents $\eta_\mathrm{bu}$ and $\eta_\mathrm{d}$ were empirically determined to best match the model light curves (Table~\ref{etas}). It was found that good fits for the whole set of parameters ($\cos i$, $\Sigma_0$ and $\tilde{\tau}_\mathrm{bu}$) could be obtained for certain fixed values of $\eta_\mathrm{bu}$ and $\eta_\mathrm{d}$ for each photometric band. 

\begin{table}
\centering
\caption{\label{etas} Adopted values for the $\eta$ exponent of Eqs.~\ref{genpower_bu} and \ref{genpower_diss}}
\begin{tabular}{rr|rrrr}
\hline
  &   & $B$ & $V$ & $R$ & $I$\\
\hline
$\eta_\mathrm{bu}$ & (edge-on) & 0.8 & 0.8 & 0.8 & 0.8\\
 & (pole-on) & 0.8 & 0.8 & 0.8 & 0.8\\
 \hline
$\eta_\mathrm{d}$ & (edge-on) & 1.5 & 1.5 & 1.5 & 1.5\\
 & (pole-on) & 1.1 & 1.2 & 1.3 & 1.4\\
\hline
\end{tabular}
\end{table}

\begin{figure}
\centering{
\includegraphics[width=1.0\linewidth]{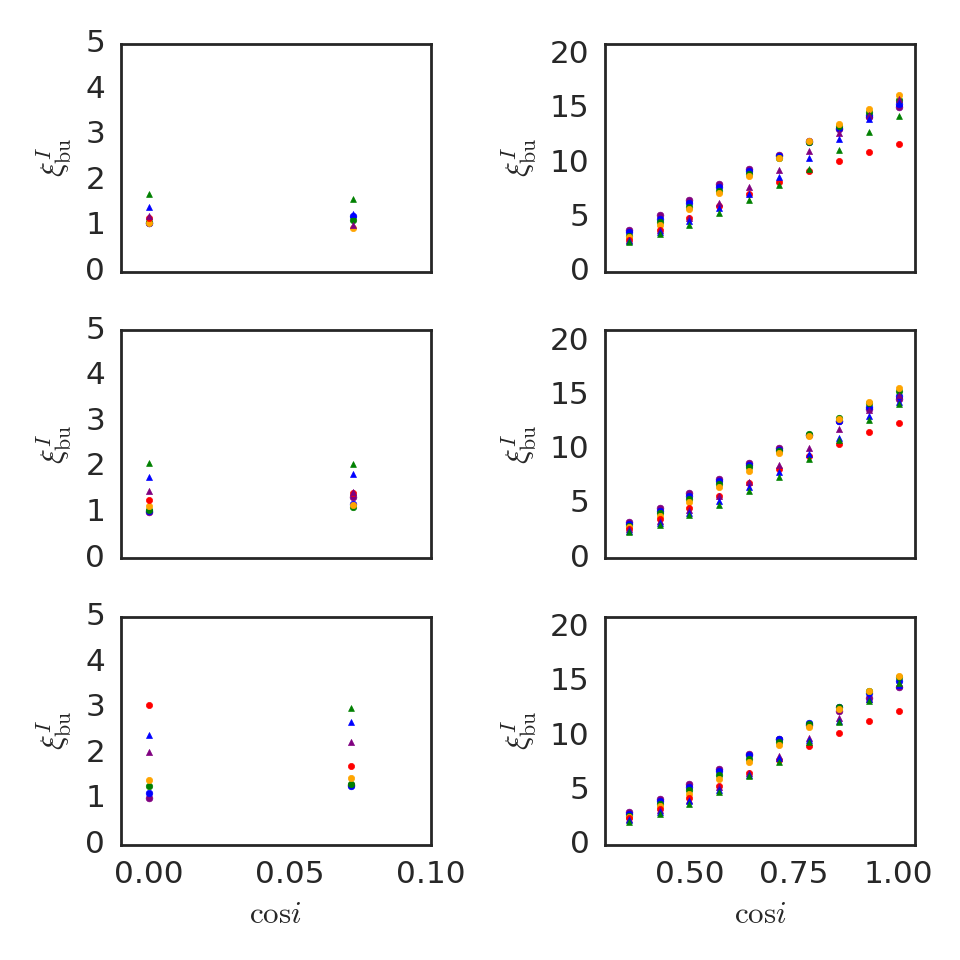}
}
\caption[]
{
The $I$-band values of $\xi_\mathrm{bu}$ 
for our grid vs. $\cos i$. {\it Left:} Edge-on models.  {\it Right:} Pole-on models. From top to bottom, the results for Star 1, 2, and 3, respectively. The markers are the same as in Fig.~\ref{coeffs1}.
}
\label{coeffs2}
\end{figure}
\begin{figure}
\centering{
\includegraphics[width=1.0\linewidth]{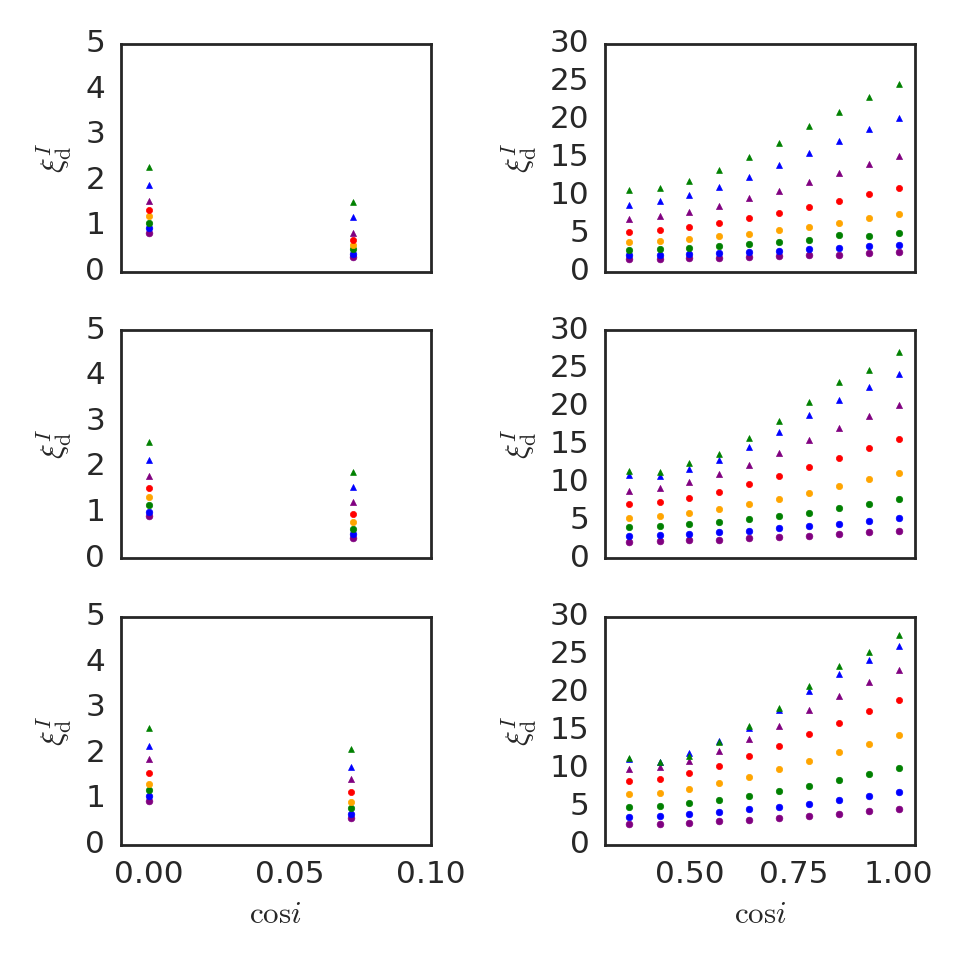}
}
\caption[]
{
Selected $I$-band values of $\xi_\mathrm{d}$ 
for our grid vs. $\cos i$. The scaled build-up time was fixed to $\tilde{\tau}_\mathrm{bu}=2.25$.
 {\it Left:} Edge-on models.  {\it Right:} Pole-on models. From top to bottom, the results for Star 1, 2, and 3, respectively. 
The markers are the same as in Fig.~\ref{coeffs1}.
}
\label{coeffs3b}
\end{figure}
\begin{figure}
\centering{
\includegraphics[width=1.0\linewidth]{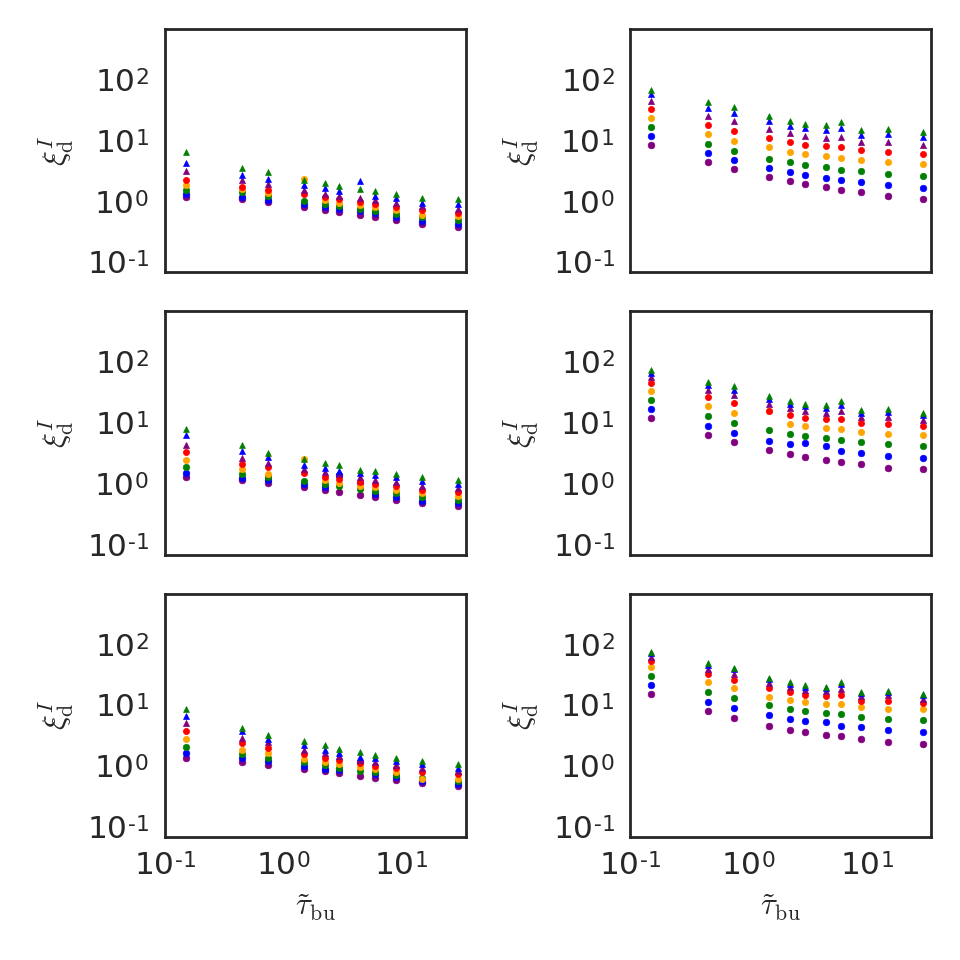}
}
\caption[]
{
Selected $I$-band values of $\xi_\mathrm{d}$ 
for our grid vs. $\tilde{\tau}_\mathrm{bu}$. 
{\it Left:} Edge-on models with $i=90\deg$.  {\it Right:} Pole-on models with $i=0\deg$. From top to bottom, the results for Star 1, 2, and 3, respectively. 
The markers are the same as in Fig.~\ref{coeffs1}.
}
\label{coeffs3}
\end{figure}
%


Representative $I$-band values of $\xi_\mathrm{bu}$ and $\xi_\mathrm{d}$ for our grid are shown in Figs.~\ref{coeffs2}, \ref{coeffs3b} and \ref{coeffs3} (the values of $\xi_\mathrm{bu}$ and $\xi_\mathrm{d}$ for the $BVR$ bands show qualitatively similar patterns to the ones presented in these figures).
Each row shows the results for a different star. The left (right) panels are for edge-on (pole-on) models. 
The values of $\xi$ are directly related to the rate of photometric variations: the smaller the $\xi$, the slower the variation (see Eqs.~\ref{genpower_bu} and \ref{genpower_diss}).

Figure~\ref{coeffs2} exemplifies the strong variation of $\xi_{\rm bu}$ with $i$, for the pole-on case.
This is probably due to the fact that, 
{in the build-up process, the density grows from the inside out (see Appendix~\ref{Appendix1}), which means that the optically thick part of the disc (an expanding pseudo-photosphere) forms first, and the optically thin part takes longer to be built.} 
{Since the 
optical excess of the disc is given by $\Delta X \propto  F_\mathrm{thick}\cos i+F_\mathrm{thin}$, it follows that, as we move from pole-on to edge-on angles, only the optically thick contribution (proportional to $\cos i$) varies. As a consequence, the observed rate of increase in flux moves from being more to less optically thick dominated.}

In the dissipation process, the density rapidly adjusts to a self-similar dissipation pattern in the inner disc (Appendix~\ref{Appendix1}). Therefore, the flux from the disc is the result of the decrease and disappearance of the optically thick region - transformed into an optically thin region - and the diminishing of the whole optically thin region. 
The pole-on values of $\xi_\mathrm{d}$ (right panels of Fig.~\ref{coeffs3b}) are affected by $\cos i$ to a less extent, when compared to the values of $\xi_\mathrm{bu}$. 
In the dissipation process, by the same reasoning applied to the build-up process, since the optically thick emission is attenuated by the effect of $\cos i$, its disappearance has a reduced effect for more inclined discs and, therefore, the disc should apparently disappear at a slower rate. 

The values of $\xi_\mathrm{d}$ also show great variation with the asymptotic surface density (Figs.~\ref{coeffs3b} and \ref{coeffs3}). More specifically,  increasing $\Sigma_0$  results in a light curve with a slower decay rate. This is probably due to different levels of saturation in the optically thick region. The denser the optically thick region, the bigger its optical depth and the greater the amount of time for it to turn into an optically thin region. 

In addition, Fig.~\ref{coeffs3} shows that $\xi_\mathrm{d}$ strongly depends on the scaled build-up time. As expected from the mass reservoir effect (Sect.~\ref{reservoir_effect}), increasing $\tilde{\tau}_\mathrm{bu}$ results in smaller values of $\xi_\mathrm{d}$, which implies slower decay rates. 

From the above an important conclusion can be drawn: viscosity  is not the only parameter affecting the rate of photometric variations in a Be light curve. The stellar parameters, the asymptotic surface density, as well as the inclination angle, all affect the observed shape of the light curve. Thus, extracting $\alpha$ from light curves, one of the main goals of this paper, cannot be done without some knowledge about these parameters.

%

\subsection{Fitting pipeline}
\label{pipeline_sect}

So far, our model light curves were given in terms of the adimensional time parameter, $\tilde{\tau}$. Thus, an equation is necessary to transform from the physical time $t$ to $\tilde{\tau}$, in order to connect the real light curves to our simulated ones.

A variation in the time parameter, $\mathrm{d}\tilde{\tau}$, is related to a variation in physical time by $\mathrm{d}\tilde{\tau}=\mathrm{d}t/\tau(t)$, where $\tau(t)$ depends on 3 stellar parameters ($M$, $R_\mathrm{eq}$, $T_\mathrm{eff}$) and the viscous parameter $\alpha(t)$ (Eq.~\ref{tau}). 
For a given Be star, the build-up phase starts at $t_1$,
and ends at $t_2$, when dissipation begins. Thus, the build-up time is given by $t_2-t_1$.
In this work, following the results of \citet{2017ASPC..508..323G}, we explore the possibility that the viscosity parameter may be different at build-up ($\alpha(t)=\alpha_\mathrm{bu}$, for $t_1\leq t < t_2$) and dissipation ($\alpha(t)=\alpha_\mathrm{d}$, for $t \geq t_2$). 
Therefore, the transformation equation from $t$ to $\tilde{\tau}$ is 
\begin{equation}
\tilde{\tau}=
\left\{
\begin{array}{ll}
\alpha_\mathrm{bu}\frac{t-t_1}{\alpha\tau}\,, &t_1\leq t < t_2\\
\alpha_\mathrm{bu}\frac{t_2-t_1}{\alpha\tau}+\alpha_\mathrm{d}\frac{t-t_2}{\alpha\tau}\,, &t\geq t_2
\end{array}
\right.\,,
\label{time_par_transfeq}
\end{equation}
which is such that, as $t$ goes from $t_1$ to $t_2$, $\tilde{\tau}$ goes from 0 to $\tilde{\tau}_\mathrm{bu}=\alpha_\mathrm{bu}(t_2-t_1)/\alpha\tau$, and 
for $t$ larger than $t_2$, we see that $\tilde{\tau} > \tilde{\tau}_\mathrm{bu}$.
Recall that $\alpha\tau$, defined in Eq.~\ref{tau}, is a quantity dependent only on the stellar parameters and the disc temperature.


Substitution of Eq.~(\ref{time_par_transfeq}) into Eqs.~(\ref{genpower_bu}), (\ref{genpower_diss}) and (\ref{psilinha}), gives the following equation for fitting an observed bump
\begin{eqnarray}
\Delta {X}(t) 
=
\left\{
\begin{array}{ll}
\Delta X_\mathrm{bu}^\infty\left(1-\frac{1}{1+\left[C_\mathrm{bu} (t-t_1)\right]^{\eta_\mathrm{bu}}}\right), & t_1\leq t < t_2
\\
\Delta X_\mathrm{bu}^\infty\left(1-\frac{1}{1+\left[C_\mathrm{bu} (t_2-t_1)\right]^{\eta_\mathrm{bu}}}\right)
\times & \\
\phantom{\log}\left(\frac{1}{1+\left[C_\mathrm{d} (t-t_2)\right]^{\eta_\mathrm{d}}}\right), & t \geq t_2
\end{array}
\right.\,
\label{genpowerfit_final}
\end{eqnarray}
%

%
where 
\begin{equation}
C_\mathrm{bu}=\alpha_\mathrm{bu}\frac{\xi_\mathrm{bu}
}{\alpha\tau
}
\,,
\label{c_bu}
\end{equation}
and
\begin{equation}
C_\mathrm{d}=\alpha_\mathrm{d}\frac{\xi_\mathrm{d}
}{\alpha\tau
}
\label{c_d}
\end{equation}
are coefficients related to the rate of photometric variations. 
The values of the parameters $\Delta X_\mathrm{bu}^\infty$, $\xi_\mathrm{bu}$ and $\xi_\mathrm{d}$ were tabulated in Sect.~\ref{emplaw_sect}, by fitting the respective empirical laws to the model grid. 



Our goal is to fit an observed light curve with Eq.~\ref{genpowerfit_final}, in order to obtain, in a self-consistent way, all the stellar and disc parameters of interest. For that, the following chain of procedures is adopted:

\begin{enumerate}
\setcounter{enumi}{0}

\item Find a light curve of a Be star that contains at least one clear inactive phase and one complete photometric bump. 

\item Obtain the magnitudes $X_*$ at the inactive phase. Subtract these magnitudes from the light curve and obtain the excesses $\Delta X(t)$.

\end{enumerate}


Without a clear inactive phase, it is not possible to obtain the pure photospheric level (e.g., the horizontal purple straight lines in Fig.~\ref{example_a}) and, consequently, it is not possible to know how much of the observed bumps represent the disc contribution to the total flux. 
In addition, the photometric bump must contain a completely identified build-up phase, from which the instants $t_1$ and $t_2$ can be extracted, and a considerable extension of the dissipation phase.

\begin{enumerate}
\setcounter{enumi}{2}

\item 
Fit Eq.~(\ref{genpowerfit_final}) to the selected bumps, obtaining the coefficients $\Delta X_\mathrm{bu}^\infty$, $C_\mathrm{bu}$ and $C_\mathrm{d}$, as well as the times $t_1$ and $t_2$ for the onsets of build-up and dissipation.

\end{enumerate}

\begin{enumerate}
\setcounter{enumi}{3}

\item Transform the magnitudes at the inactive phase, $X_*$, to absolute magnitudes, $M_{X*}$, by correcting for the distance to the star and reddening at each observed band.

\end{enumerate}

Given the theoretical dependence of the coefficients in Eq.~\ref{genpowerfit_final} on the stellar parameters, the absolute magnitudes are required to estimate the stellar parameters ($M$, $W$ and $t/t_\mathrm{MS}$). From them, the parameter $\alpha\tau$ (see Eq.~\ref{tau}) can be estimated. 
Clearly, if the stellar parameters are known from some other way (e.g., by spectroscopic analysis), this requirement is no longer necessary. Unfortunately, this is not the case for our sample.

\begin{enumerate}
\setcounter{enumi}{4}

\item Estimate the stellar parameters, the geometric parameter ($\cos i$) and the bump parameters ($\Sigma_0$, $\alpha_\mathrm{bu}$ and $\alpha_\mathrm{d}$, for each bump) that best reproduce the fitted stellar ($M_{X*}$) and bump ($\Delta X_\mathrm{bu}^\infty$, $C_\mathrm{bu}$ and $C_\mathrm{d}$) parameters (see Eqs.~\ref{c_bu} and \ref{c_d} and the parameters $\Delta X_\mathrm{bu}^\infty$, $\xi_\mathrm{bu}$ and $\xi_\mathrm{d}$ from Sect.~\ref{emplaw_sect}).

\end{enumerate}
In practice, the above process involves several complications (e.g., estimating the goodness of the fit) that are described in the next section.

\subsection{Fitting using a MCMC sampling}
\label{explicando_emcee}

The task of fitting the measured stellar absolute magnitude ($M_{X*}$) and bump parameters ($\Delta X_\mathrm{bu}^\infty$, $C_\mathrm{bu}$ and $C_\mathrm{d}$) for estimating the model parameters -- step 5, above -- was done using the Markov-Chain Monte Carlo (MCMC) sampling technique. We used the Python MCMC sampler {\tt emcee} \citep{2013PASP..125..306F}. 
The code samples a large collection of models by varying all model parameters within a pre-specified range.
The sampler provides a distribution of model parameters according to a posterior distribution $p\left(\mathrm{model}\mid \mathrm{data}\right)\propto L\left(\mathrm{data}\mid \mathrm{model}\right) \pi\left(\mathrm{model}\right)$, where $L$ and $\pi$ are the likelihood and the prior distributions, respectively.

In our fitting procedure, there are $4+5N_\mathrm{bumps}$ model parameters for each light curve containing $N_\mathrm{bumps}$ identified bumps. There are 3 stellar parameters ($M$, $t/t_\mathrm{MS}$, $W$) and one geometric parameter ($\cos i$), 
and, for each bump in the light curve, there are 5 parameters: the initial times of the build-up and dissipation phases ($t_1$ and $t_2$), the asymptotic surface density ($\Sigma_0$), and the viscosity parameters during the build-up and dissipation phases ($\alpha_\mathrm{bu}$ and $\alpha_\mathrm{d}$).

We assume that the errors of the observations follow a Gaussian distribution and, therefore, the likelihood of a Be star with certain model parameters, given the observed data, is given by 
%
\begin{equation}
L\left(\mathrm{data}\mid \mathrm{model}\right)\propto e^{-\frac{1}{2}\chi^2}\,,
\label{likelihood1}
\end{equation}
where 
\begin{equation}
\chi^2 = \chi_\mathrm{discless}^2+\chi^2_\mathrm{bump}
\,,
\label{chi2all}
\end{equation}
and
\begin{equation}
\chi_\mathrm{discless}^2=
\sum_{\rm bands}
\frac{
(M_{X*}^\mathrm{model}-M_{X*}^\mathrm{obs})^2
}{
\sigma^2({M_{X*}^\mathrm{obs}})
}
\,,
\label{chi2discless}
\end{equation}
and
\begin{equation}
\chi^2_\mathrm{bump}=
\sum_{\rm bands}
\sum_{\rm bumps}
\frac{1}{N_{t}}
\sum_{i=1}^{N_{t}}
\frac{
(\Delta X_{i}^\mathrm{model}-\Delta X_{i}^\mathrm{obs})^2
}{
\sigma^2({\Delta X_{i}^\mathrm{obs}})
}
\,,
\label{chi2bumps}
\end{equation}
where $N_{t}$ is the number of data points for a given bump at a given photometric band.



The prior distribution $\pi$ represents our prior knowledge of the distribution of Be stars. We assume it to be
\begin{equation}
\pi\left(\mathrm{model}\right) \propto M^{-2.3}f_\mathrm{Be}(M) e^{-\frac{(W-\langle W \rangle )^2}{2\sigma_W^2}}\,,
\label{prior1}
\end{equation}
where the factor $M^{-2.3}$ is the initial mass function (IMF) of \citet{2001MNRAS.322..231K}, and  $f_\mathrm{Be}(M)$ represents the fraction of Be stars relative to the number of B stars, estimated by \citet[][their Fig.~6]{2007A&A...462..683M}. Finally, the Gaussian factor 
comes from the distribution of rotational velocities in the sample of Be stars, here estimated from \citet{2006A&A...459..137R}, assuming $\langle W \rangle = 0.81$ and $\sigma_W=0.12$. 

For the parameter sampling, we have chosen hundreds of ``walkers''\footnote{Each walker can be viewed as a separated Markov Chain in the sample, although the walkers influence each another \citep{2013PASP..125..306F}.}, proportional to the number of $4+5N_\mathrm{bumps}$ model parameters. For a randomly chosen set of parameters sampled by {\tt emcee} in  the course of the simulation, the corresponding stellar and bump observables are calculated  by a multidimensional linear interpolation of the model grid. During the raffle of parameters the prior probability was set to zero if one of the values were sampled outside of the allowed range of a given parameter (Tables~\ref{star_and_disc} and \ref{star_only}). 
The simulation consists of two steps, the so-called ``burn-in phase'' and the sampling phase. We verified that 1000 iterations in the burn-in phase were sufficient for the convergence of all our models. 
For each parameter, the best-fitting values were chosen to be the median of distribution of the posterior probabilities, with upper (lower) uncertainties estimated from the differences between $84\%$($16\%$) of the sample and the median

In order to test our fitting routine, we applied it to synthetic light curves with levels of astrophysical noise and uncertainties similar to the observed light curves used in this work. In general, the parameters used to generate the synthetic light curves were fairly recovered, with the exception of $\cos i$ and $\alpha_\mathrm{bu}$, for which a strong correlation is expected on theoretical grounds (Fig.~\ref{coeffs2}).
As expected, when multi-band data was used the errors in the derived parameters were smaller. This test indicates that $\Sigma_0$ and $\alpha_\mathrm{d}$ can be reliably estimated from the lightcurves, while $\cos i$ and $\alpha_\mathrm{bu}$ less so.

In Sect.~\ref{ogle_sect} we present a selection of light curves of Be stars from the SMC, and measure their stellar and bump quantities (steps 1 to 4 of the pipeline). Later, in Sect.~\ref{sect_results}, we apply step 5 of the pipeline, as described in this section, in order to estimate the relevant parameters of the selected Be stars.



\section{OGLE light curves of Be star candidates}
\label{ogle_sect}


%
\begin{figure}
\centering{
\includegraphics[width=1.00\linewidth]{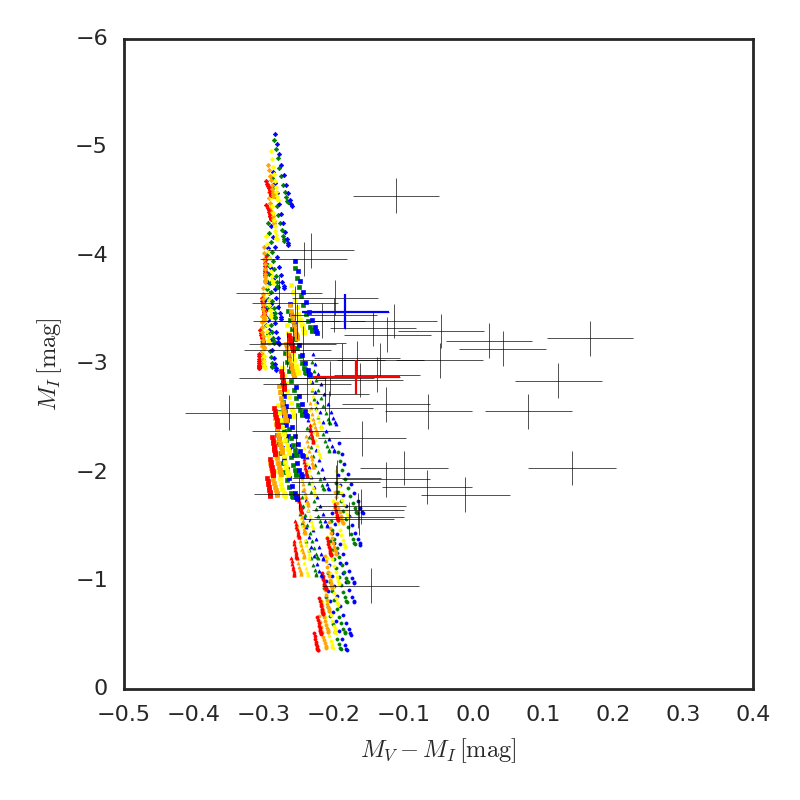}
}
\caption[]
{
Color-magnitude diagram of simulated discless stars (Table~\ref{star_only}). Circles, triangles, squares and diamonds correspond to stellar models with $M=6.0,\,8.3,\,11.7,\,20.0\, M_\odot$, respectively. The colours red, orange, yellow, green and blue correspond, respectively, to the 5 values of $W$ in increasing order. The 6 different stellar ages are seen as the groups of points move in the upper-right direction. For each star, 
the effect of going from $90\deg$ to $0\deg$ is to move in the upper-left direction. Also shown, as error bars, is the position of our selected stars (see eighth and ninth columns of Table~\ref{my_selection}). The blue and red error bars mark the positions of {SMC\_SC1 75701} and {SMC\_SC6 128831}, respectively.
}
\label{alphatau1}
\end{figure}
%

\citet{2002A&A...393..887M} selected roughly one thousand Be star candidates from the SMC, by studying light curve variations using the OGLE-II database \citep{1997AcA....47..319U}. 
They classified the morphologies found in the light curves into four categories. 
The majority of the light curves ($\sim 65\%$) belonged to their type-4 category, composed by the light curves showing irregular and non-periodic variations. These light curves should correspond to Be stars showing episodes of mass injection more complicated than the simple build-up followed by dissipation scenario described in Sect.~\ref{models_bumps}. 

Most interesting for us is the  type-1 group ($\sim$$13\%$ of the sample), composed by light curves that show single sharp or hump-like bumps, like the bumps of the light curve of {SMC\_SC1 75701} (Fig.~\ref{example_a}). These bumps should be the result of single nearly continuous episodes of mass injection followed by dissipation of the disc, like the theoretical scenario explored in Sect.~\ref{models_bumps}. 

The type-2 group ($\sim$$14\%$ of the sample), containing light curves showing high and low plateaus, also has some interesting cases for our purposes. High plateaus are usually the photometric result of a longer build-up process in which a near steady-state has been reached in the inner disc. The low plateaus are frequently the portions of the light curve during  inactive phases. 

\citet{2005MNRAS.361.1055S} selected roughly two thousand Be star candidates from the LMC and classified their light curves into the same four categories described by \citet{2002A&A...393..887M}. Previously, \citet{2002AJ....124.2039K} also studied light curves from the LMC using the MACHO survey. They spectroscopically analysed a subsample of their Be star candidates and found that $90\%$ of them were Be stars. They also classified morphologically their light curves in a slightly different manner. Their so called ``bumper events'' and ``flicker events'' more or less correspond to the bumps of type-1 light curves, but also to features of the more irregular type-4 light curves. The bumpers have duration of a few hundred days, while the flicker events are faster, with durations of a few dozens of days. 
Dips like the one exemplified by the light curve of {SMC\_SC6 128831} (Fig.~\ref{example_a}) were called ``fading events''. The frequency of these events was quite smaller than the bumpers, in accordance to the picture that fading events are associated with the less numerous shell stars.

\citet{2012MNRAS.421.3622P} studied the spectral properties of stars from the catalogues of \citet{2002A&A...393..887M} and \citet{2005MNRAS.361.1055S}. For the candidates from the SMC, they found that the majority of type-1 and type-2 light curves belong to early B type stars with emission features characteristic of circumstellar material \citep{2012MNRAS.421.3622P}. Therefore, these light curves are very likely to be from Be stars.

%
%
%
%
%
%
%
%
%

In this work, we selected light curves from the catalogue of Be star candidates from the SMC of \citet{2002A&A...393..887M}. 
In order to have light curves of a longer time baseline, we combined OGLE-II data with OGLE-III \citep{2008AcA....58...69U}. 
Due to a calibration issue between OGLE-II and OGLE-III, namely a shift in the zero points present in some of the light curves, it was necessary to find inactivity intervals in both the OGLE-II and OGLE-III portions of these light curves to measure and correct the problem.

The light curves were visually inspected according to the criteria of item 1, Sect.~\ref{pipeline_sect}, i.e., light curves with at least one clear inactive phase and one bump. In this initial work we focussed on well-behaved light curves with clear bumps. We also avoided the short events (flickers, with build-up times $\lesssim 15\,$days), due to the fact that most of them are poorly sampled. The end result was a sample of {54} stars, containing {81} selected bumps, shown in Table~\ref{my_selection}. In the table, horizontal lines separate the data for each of the {54} stars. Each row in the table contain the data for each of the {81} selected bumps. 
The fifth and sixth columns in the table contain the beginning and ending of the selected inactive interval for the light curve. The seventh, eighth and ninth columns contain the $B_*V_*I_*$ magnitudes obtained at the inactive phase for the light curve. Due to the nature of the OGLE survey, the $B_*V_*$ are not available for all sources. 
The eleventh column contains the bands that were considered in the fitting process of the specific bumps,  depending on the availability of measurements in each band. 
The last two columns are initial visual estimates of $t_1$ and $t_2$, which were used as input for {\tt emcee}. 

As explained in Sect.~\ref{pipeline_sect}, the magnitudes at the inactive phase are necessary to set the baseline level of the light curves and to provide an estimate of stellar parameters. In order to do the latter, these apparent magnitudes must be colour-corrected and converted to absolute magnitudes by the standard formula $M_{X*}=X_*-(5\log d-5)-A_X$.
We adopted as the distance to the SMC $d=60.3\pm 3.8\,$kpc from \citet{2005MNRAS.357..304H}.
The mean $E(V-I)$ of the RR Lyrae stars over the whole SMC is $E(V-I) = 0.07 \pm 0.06\,\mathrm{mag}$ \citep{2011AJ....141..158H}. The reddening in the $B$, $V$ and $I$ bands were obtained by the relations $A_X/A_V$ given by \citet{2003ApJ...594..279G}. 


The colour-magnitude diagram (CMD) of Fig.~\ref{alphatau1} compares the grid of discless models (Table~\ref{star_only}) to our sample of Be stars. The majority of our sample is comprised of early-type Be stars, as further discussed in Sect.~\ref{mass_distribution}. 

Our sample is also distributed in a broader range of $M_V-M_I$ than our grid of main sequence discless models, which is in accordance with the distribution Be star candidates of \citet{2002A&A...393..887M} (see their Fig.~7). It must be reminded also that our sample consists of type-1 and type-2 light curves from the SMC, meaning that they very likely belong to Be stars \citep{2012MNRAS.421.3622P}. There are two non-exclusive possibilities to explain these Be stars on the right of the main sequence discless models: 1) The usage of an average $E(V-I)$ for the entire SMS obviously does not take into consideration local variations of the reddening; therefore, the stars with large $M_V-M_I$ in our sample may be the ones for which the reddening is larger than the average value. 2) these stars are post-main sequence Be stars.

\section{Results}
\label{sect_results}

In this section, the results obtained by applying the pipeline  to the stars and bumps of our sample are described. Initially, the results for the two objects of Fig. 1 are examined in detail (Sect.~\ref{lc_example1}), followed by an analysis of the results obtained for the entire sample (Sect.~\ref{whole_sample}).

\begin{figure*}
\centering{
\includegraphics[width=1.0\linewidth]{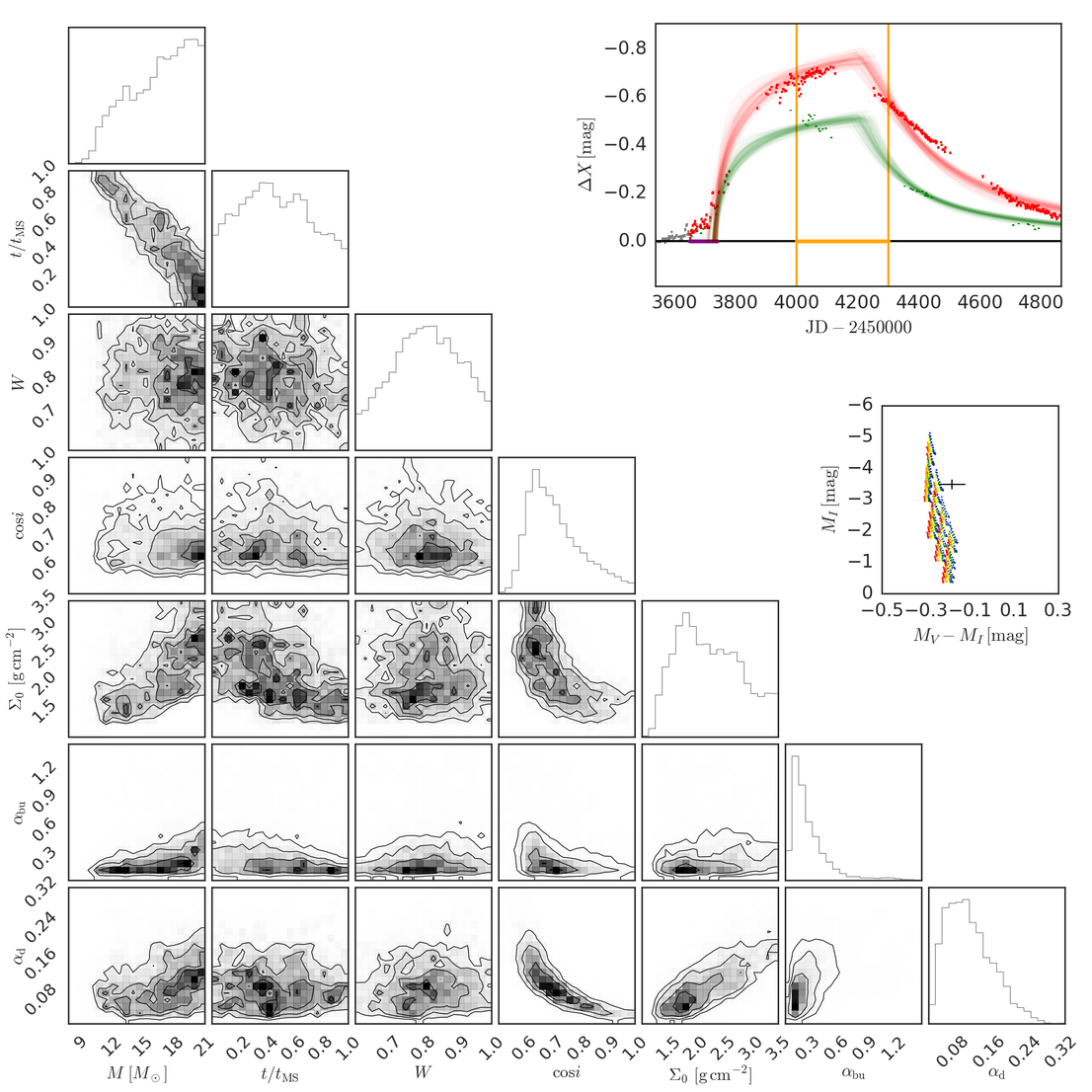}
}
\caption[]
{
{\it Upper-right}: Light curve of {SMC\_SC1 75701}. Thin lines: 100 model curves randomly selected from the stationary sample of the {\tt emcee} code.
The red (green) colour indicates the $I$ ($V$) band. The purple and orange time intervals marked in the horizontal straight lines are the allowed intervals for the model parameters $t_1$ and $t_2$, respectively.
{\it Middle-right}: CMD displaying the model grid of inactive Be stars and the position of {SMC\_SC1 75701} (see Fig.~\ref{alphatau1} for details).
{\it Below}: Results of the {\tt emcee} run for {SMC\_SC1 75701}. Histogram distributions of the posterior probabilities (top panels) and two-by-two correlations of the stellar ($M$, $t/t_\mathrm{MS}$ and $W$), geometrical ($\cos i$), and bump ($\Sigma_0$, $\alpha_\mathrm{bu}$ and $\alpha_\mathrm{d}$) parameters (off-diagonal panels). The parameters $t_1$ and $t_2$ were not shown for convenience. The normalised density levels shown in the off-diagonal panels are $12\%$, $39\%$, $68\%$, $87\%$ of the peak probability.
}
\label{example_bb1}
\end{figure*}
\begin{figure*}
\centering{
\includegraphics[width=1.0\linewidth]{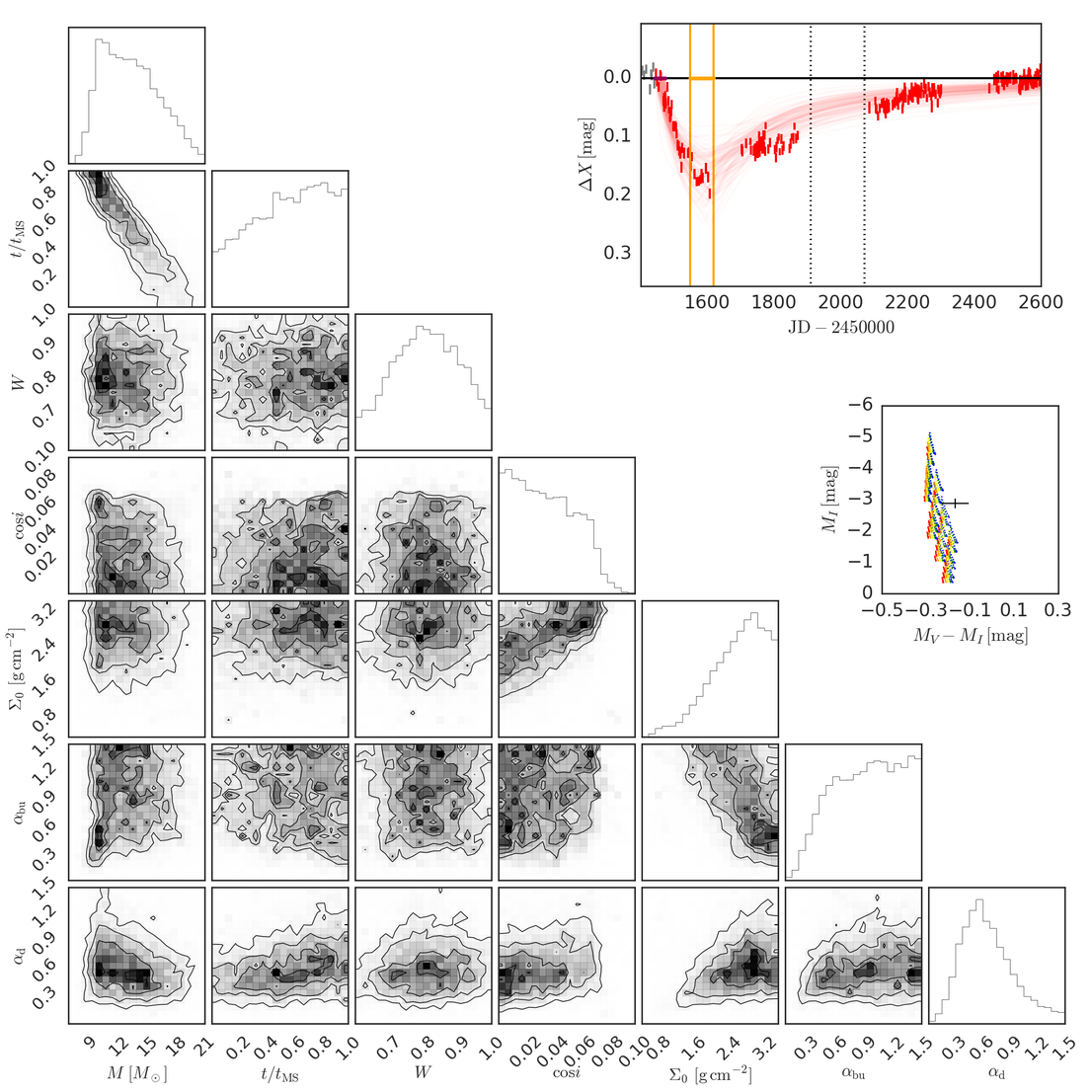}
}
\caption[]
{
Same as Fig.~\ref{example_bb1} for {SMC\_SC6 128831}.
}
\label{example_bb2}
\end{figure*}

\subsection{{SMC\_SC1 75701} and {SMC\_SC6 128831}}
\label{lc_example1}

The results for {SMC\_SC1 75701} and {SMC\_SC6 128831} are shown in Figs.~\ref{example_bb1} and \ref{example_bb2}, respectively. For {SMC\_SC1 75701} there was enough data for both the $I$ and $V$ bands to allow these two light curves to be fitted simultaneously. For {SMC\_SC6 128831}, however, only $I$-band data was available. 
The times for the beginning of the build-up ($t_1$) and dissipation ($t_2$) are fitted quantities in the pipeline, but an initial estimate for them is provided to {\tt emcee} by graphically analysing each light curve. These estimates are shown as the purple and orange segments in the horizontal straight lines in the figure. 
In the MCMC sampling, after a sufficient number of iterations, 
a stationary sample is obtained, for which the model parameters are more concentrated in the regions of higher posterior probability. In the plots we show 100 sets of randomly selected model curves obtained after the stationary sample was reached. The dispersion of the curves gives a visual measure of the goodness of the fits.




The goodness of the fit can be quantitatively assessed from the distribution of the posterior probabilities of each fitted parameter shown in Figs.~\ref{example_bb1} and \ref{example_bb2}. The main diagonal of the triangular diagram plots the distributions of the stellar ($M$, $t/t_\mathrm{MS}$ and $W$), geometrical ($\cos i$), and bump ($\Sigma_0$, $\alpha_\mathrm{bu}$ and $\alpha_\mathrm{d}$) parameters, and they can be used to assess how well-constrained each parameter is. The images below the diagonal show how the parameters correlate with each other. 


The stellar parameters ($M$, $t/t_\mathrm{MS}$ and $W$) are mainly constrained by the magnitudes at the inactive phase.
In Fig.~\ref{example_bb1}, the three leftmost histograms along the diagonal have broad distributions, which means that these parameters are not well constrained. The first histogram shows that {SMC\_SC1 75701} is an early Be star, perhaps even more massive than the available stellar models (Table~\ref{star_only}). 
The mass is anti-correlated with the main sequence age (see $t/t_\mathrm{MS}\times M$ plane), as expected from the fact that 
a less massive but more evolved star can have a similar absolute magnitude of a younger, more massive star.


The bump parameters ($\Sigma_0$, $\alpha_\mathrm{bu}$ and $\alpha_\mathrm{d}$) are mainly constrained by the shape of the observed bump. 
Roughly, the amplitude of the bump depends mostly on $\Sigma_0$ (and $\cos i$, see Fig.~\ref{coeffs1}), while the value of viscosity parameter in each phase controls the rate of brightness variation.
For  {SMC\_SC1 75701}, $\Sigma_0$ has a broad distribution peaking around $\sim$$1.5\,\mathrm{g\, cm^{-2}}$, indicating a quite dense disc, close to the densest cases in the sample of \citet{2017MNRAS.464.3071V} for the same spectral type. This fact can also be inferred from Fig.~\ref{coeffs1}, given the large observed $\Delta I_\mathrm{bu}^\infty$. 
The best-fit viscosity parameters are $\alpha_\mathrm{bu} = 0.25^{+0.21}_{-0.09}$ and $\alpha_\mathrm{d} = 0.11^{+0.06}_{-0.04}$.


Of the three bump parameters derived for SMC\_SC1 75701, $\Sigma_0$ and $\alpha_\mathrm{d}$ clearly anti-correlate with $\cos i$, while $\alpha_\mathrm{bu}$ shows a weaker correlation. In fact, an anti-correlation of these three parameters with $\cos i$ is expected, as a consequence of the dependency of $\Delta X_\mathrm{bu}^\infty$, $\xi_\mathrm{bu}$ and $\xi_\mathrm{d}$, defined in Sect.~\ref{emplaw_sect}, on $\cos i$. Fig.~\ref{coeffs1} shows that, if the star is seen more pole-on (higher values of $\cos i$), smaller values of $\Sigma_0$ are required in order to obtain the fitted $\Delta X_\mathrm{bu}^\infty$, hence the strong anti-correlation seen in the $\Sigma_0\times \cos i$ plane. 
Eq.~(\ref{c_bu}) shows that the fitted coefficient $C_\mathrm{bu}$ is proportional to the product of $\alpha_\mathrm{bu}$ and $\xi_\mathrm{bu}$, and it was shown (Fig.~\ref{coeffs2}) that discs seen more pole-on (higher values of $\cos i$) appear to build-up faster (having higher values of $\xi_\mathrm{bu}$). Therefore, for higher values of $\cos i$, smaller values of $\alpha_\mathrm{bu}$ are required to obtain the fitted $C_\mathrm{bu}$, which explains the anti-correlation in the $\alpha_\mathrm{bu}\times \cos i$ plane. 
Finally, Eq.~(\ref{c_d}) shows that the $C_\mathrm{d}$ is proportional to the product of $\alpha_\mathrm{d}$ and $\xi_\mathrm{d}$, and it was shown in Fig.~\ref{coeffs3b} that the more pole-on and the less dense the disc, the faster the rate of brightness variation in the dissipation, thus the anti-correlation expected in the $\alpha_\mathrm{d}\times \cos i$ plane. 





The results for {SMC\_SC6 128831} (Fig.~\ref{example_bb2}) point to a less massive star ($M=12.9^{+3.6}_{-2.9}\,M_\odot$)
surrounded by a much more massive disc ($\Sigma_0=2.7^{+0.5}_{-0.8}\,\rm g\,cm^{-2}$).
{SMC\_SC6 128831} is an example of a dip, which means that this Be star is seen at a near edge-on angles. 
The very steep build-up phase of {SMC\_SC6 128831} hints to large mass injection rate 
and viscosity 
during build-up, as confirmed by the fifth and sixth histograms along the diagonal  of
Fig.~\ref{example_bb2}.
The viscosity parameter during dissipation was found to be $\alpha_\mathrm{d} = 0.62^{+0.33}_{-0.23}$. 
The plane $\alpha_\mathrm{d}\times \Sigma_0$ shows a correlation, just as for the case of {SMC\_SC1 75701}, which is a consequence of the decrescent relationship between $\xi_\mathrm{d}$ with $\Sigma_0$, also expected for near-edge-on inclinations (see Fig.~\ref{coeffs3b}).

For SMC\_SC6 128831 a positive correlation between $\Sigma_0$, $\alpha_\mathrm{bu}$ and $\alpha_\mathrm{d}$ with $\cos i$ was observed. Fig.~\ref{coeffs1} shows that, if the star moves away from the edge-on case ($\cos i=0$), bigger values of $\Sigma_0$ are required in order to obtain the fitted $\Delta X_\mathrm{bu}^\infty$, hence the correlation seen in the $\Sigma_0\times \cos i$ plane. 
Fig.~\ref{coeffs2} shows that, for the edge-on case there is no strong variation of $\xi_\mathrm{bu}$ with $\cos i$. However, there is the trend that a more tenuous discs appear to build-up faster, specially for a hotter star. Therefore, since Eq.~\ref{c_bu} shows that $C_\mathrm{bu}\propto \alpha_\mathrm{bu}\xi_\mathrm{bu}$, it follows that, with the increase of $\Sigma_0$ with $\cos i$, the function $\xi_\mathrm{bu}$ decreases and, hence, $\alpha_\mathrm{bu}$ increases. 
Finally, Eq.~\ref{c_d} shows that $C_\mathrm{d}\propto \alpha_\mathrm{d}\xi_\mathrm{d}$ and Fig.~\ref{coeffs3} shows that $\xi_\mathrm{d}$ decreases with $\cos i$ and $\Sigma_0$. Therefore, as $\Sigma_0$ increases with $\cos i$, it follows that $\xi_\mathrm{d}$ decreases and, hence, $\alpha_\mathrm{d}$ must increase.
Similar trends were found for the other two edge-on stars in our sample (SMC\_SC1 92262, Figs.~\ref{smc_sc1_92262_01} and \ref{smc_sc1_92262_02}, and SMC\_SC4 179053 (Fig.~\ref{smc_sc4_179053_01}).


In general, the histograms of Fig.~\ref{example_bb2} are broader than the ones in Fig.~\ref{example_bb1}, indicating that parameters are worse constrained than for {SMC\_SC1 75701}. There are two main reasons for this. First, only $I$-band data was available for this star, which has a negative impact on the pipeline's ability to constrain the stellar parameters. The MCMC method ensures that the uncertainties in the stellar parameters are properly propagated into the other model parameters. Second,
the fact that the dips have smaller amplitudes than the bumps of pole-on stars, even for higher values of $\Sigma_0$, is a great disadvantage, because the bump amplitude is much closer to the noise level of the measurements. 

The results of the pipeline for all stars in Table~\ref{my_selection} are shown in Figs.~\ref{smc_sc1_7612_01} to \ref{smc_sc11_46587_01}, available electronically only. The best-fit model parameters for all stars and bumps are listed in Table~\ref{table_outputs}.

\subsection{Results for the whole sample}
\label{whole_sample}


As seen previously, the scarcity of information about the central star (one, two, or at most 3 photometric bands only) causes a poor determination of its fundamental stellar parameters, which, owing to the nature of the MCMC method, propagates onto the disc parameters. The main result of this work, therefore, does not lie on the individual determination of the bump parameters, but on the statistical properties of the sample as a whole.

\subsubsection{Mass distribution}
\label{mass_distribution}

\begin{figure}
\centering{
\includegraphics[width=1.00\linewidth]{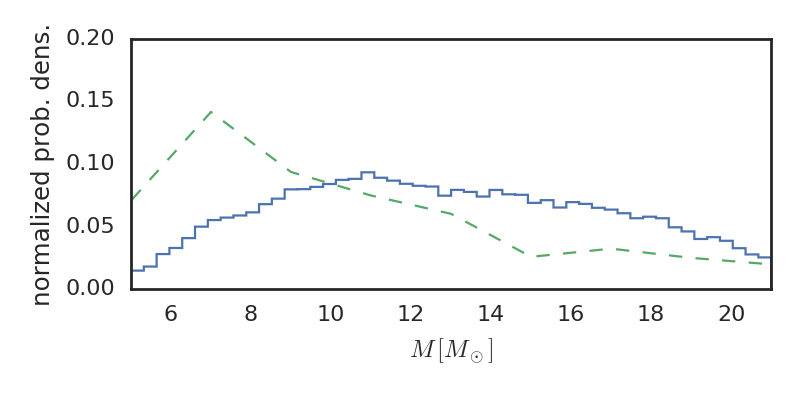}
}
\caption[]
{
{\it Solid line}: Histogram of the sum of the posterior probabilities of parameter $M$ for all  stars in our sample. {\it Dashed line}: IMF of \citet{2001MNRAS.322..231K} weighed by the fraction of Be stars over B stars of \citet{2007A&A...462..683M}, given by the factor $M^{-2.3}f_\mathrm{Be}(M)$ in Eq.~(\ref{prior1}).
}
\label{imfs1}
\end{figure}

Let us initially discuss the properties of our sample, in order to determine whether it represents a typical population of Be stars in SMC, or whether one or more selection biases where introduced.

Our selection of stars and bumps comes from the catalogue of visual photometric Be star candidates of \citet{2002A&A...393..887M}, where the candidate stars were selected according to the expected location of Be stars in colour-magnitude diagrams and according to the observed variability in the light curves.
In Fig.~\ref{imfs1}, we show
the sum of the posterior probabilities of parameter $M$ for all our sample of stars (solid line). Clearly, most of our stars are early-type Be stars, in agreement to the position of our sample in the CMD (Fig.~\ref{alphatau1}).
We also show the factor $\propto M^{-2.3}f_\mathrm{Be}(M)$ of Eq.~\ref{prior1} (dashed line).
We recall that this factor was assumed as a prior in the MCMC fitting, and it represents our current knowledge about the populations of Be stars in the SMC.
The green curve shows that, although the fraction of Be stars over B stars ($f_\mathrm{Be}(M)$, estimated by \citet{2007A&A...462..683M} from a cluster of the SMC) generally increases with $M$, the higher probability of the formation of less massive stars expressed in the IMF of \citet{2001MNRAS.322..231K} results in a bigger incidence of late type over early type Be stars. Our sample, therefore, is biased towards more massive stars.

This bias likely has several reasons:
\begin{enumerate}
\item The typical apparent $I$-band magnitudes of a B0 and B9 star in the SMC are $\sim$$15.5$ and $\sim$$19.5$, with rms uncertainties given by $\gtrsim 0.005$ and $\gtrsim 0.15$, respectively \citep{2009MNRAS.397.1228W}. Therefore, the threshold of detectability of a good bump increases for late type Be stars. 
\item Late-type Be stars tend to have more tenuous discs than early-type ones \citep[][see also next subsections]{2017MNRAS.464.3071V}, and therefore should develop bumps with much smaller amplitudes.
\item Late-type Be stars tend to show less variability \citep[][see also next subsections]{2013A&ARv..21...69R,2017arXiv170802594L}, which would make it less probable to identify bumps in their light curves.
\end{enumerate}

\subsubsection{Asymptotic surface density}

\begin{figure}
\centering{
\includegraphics[width=1.00\linewidth]{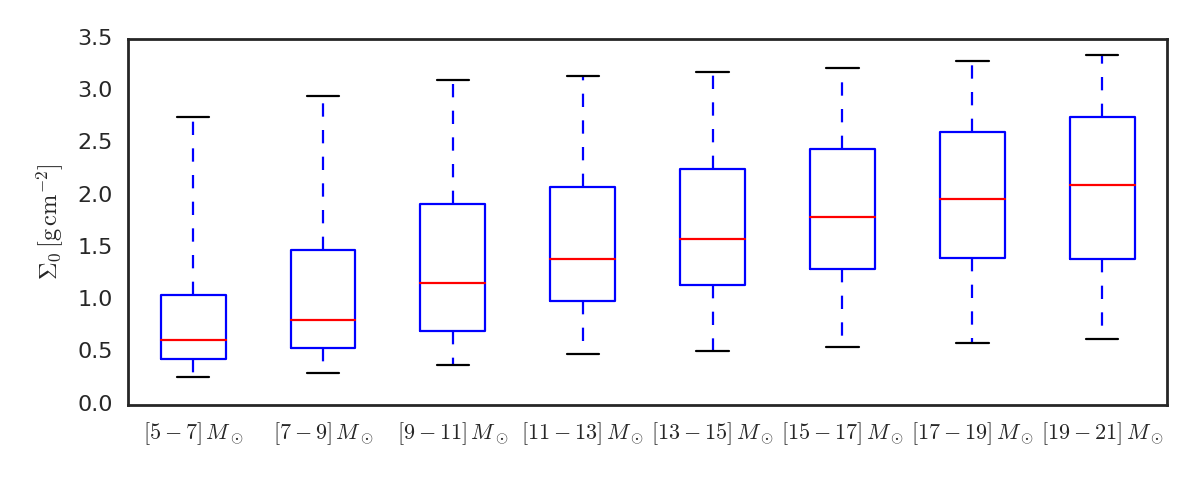}
\includegraphics[width=1.0\linewidth]{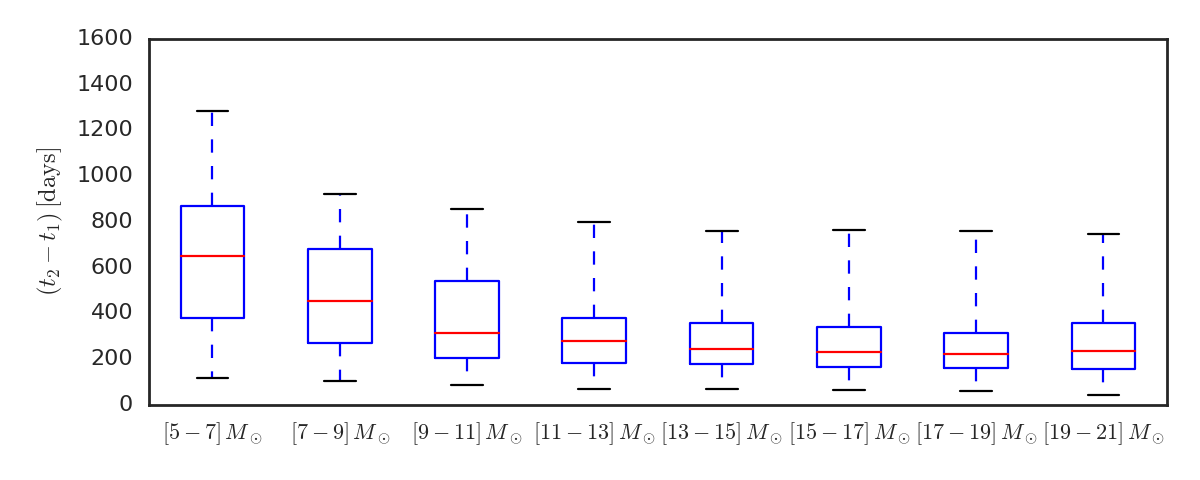}
}
\caption[]
{
Boxplots of $\Sigma_0$ (above) and $t_2-t_1$ (below) for the summed posterior probabilities of our sample of bumps, separated in eight equal intervals of mass, ranging from $5$ to $21$ 
solar masses. The middle line of the boxes mark the median ($50\%$) of the samples. The lower and upper ends of the boxes mark $25\%$ and $75\%$ of the samples. The lower and upper whiskers mark $5\%$ and $95\%$ of the samples.
}
\label{imfs2}
\end{figure}
\begin{figure}
\centering{
\includegraphics[width=1.0\linewidth]{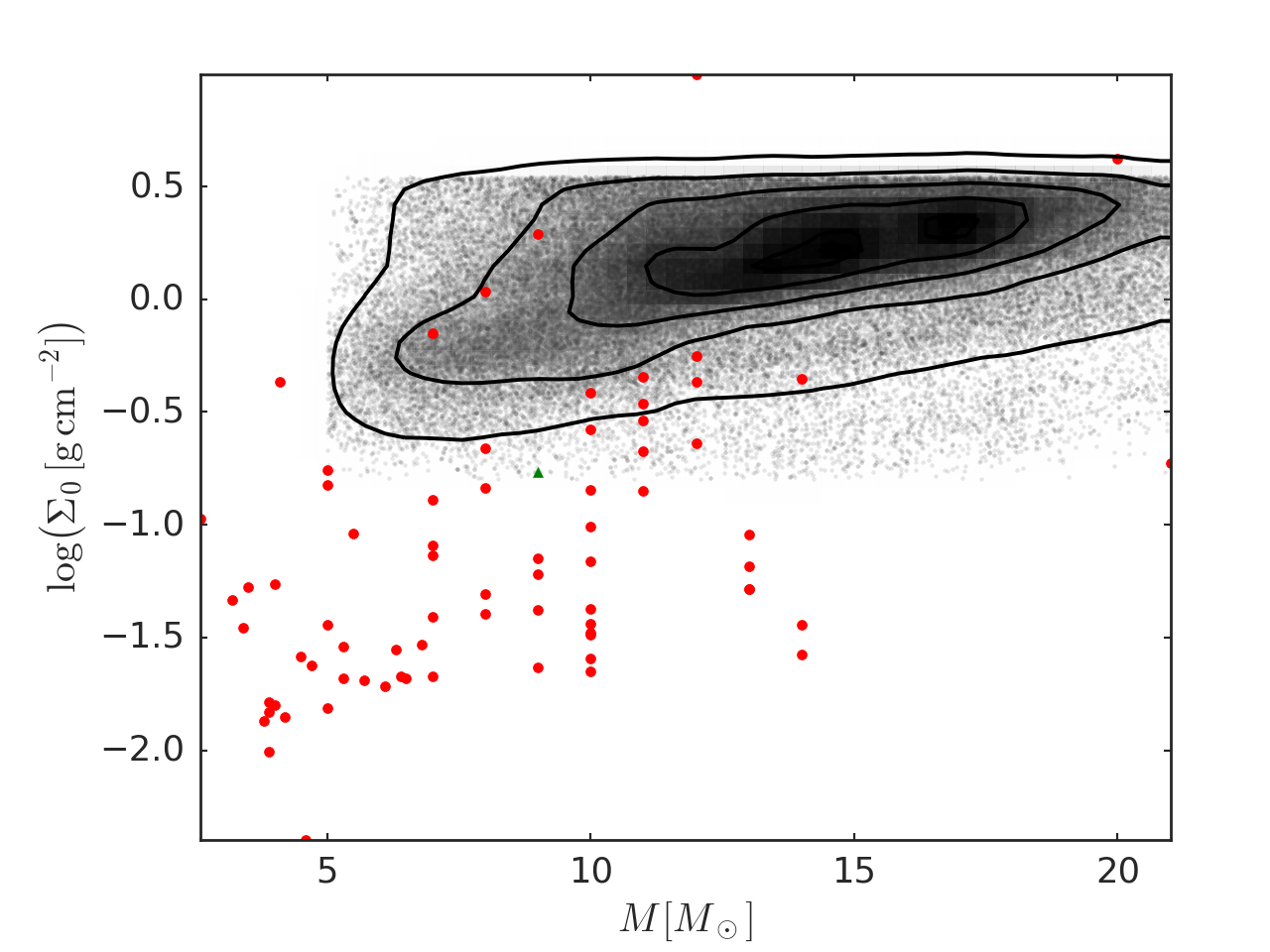}
}
\caption[]
{
Distribution of the parameters $M$ and $\Sigma_0$ for our sample. The contour levels are the same as in Fig.~\ref{example_bb1}. The red dots correspond to the surface densities at the base of the disc of Galactic Be stars, measured by \citet{2017MNRAS.464.3071V}. The green triangle corresponds to the initial state of the ablating disc model of \citet{2016MNRAS.458.2323K} for a B2e star. 
}
\label{asympfig}
\end{figure}

In the upper panel of Fig.~\ref{imfs2} we show how $\Sigma_0$ varies with stellar mass in our sample, demonstrating a clear tendency of denser discs around the more massive stars. 
\citet{2017MNRAS.464.3071V} have shown that, for the Be stars in the Galaxy, the incidence of denser discs increases with the mass of the stars. Comparison of our results with the ones of \citeauthor{2017MNRAS.464.3071V} is done in Fig.~\ref{asympfig}. 
While our sample is biased towards large masses, their sample is more evenly distributed in mass. Another difference is that our results are all concentrated in a region of high disc density, while theirs cover a much wider range of densities for all spectral types. The reason for this lies in fact that for this initial study we selected light curves with large and well-defined bumps, disregarding low-amplitude and short-duration ones.
In fact, the detection of tenuous discs by \citeauthor{2017MNRAS.464.3071V} was only possible because they studied the SED in the IR (typically between 9 and 60 $\mu\rm m$), where the disc emission is much stronger than in the visible range. 
Therefore, 
all but the most dense of their discs would be too tenuous to generate appreciable photometric excesses in visual photometric bands, suitable for our fitting procedure. 

We conclude that our sample of visual bumps should represent the upper limit for the densities found in the discs of SMC Be stars. In  the Galaxy, these large densities are only found in early type Be stars. The median of the $\Sigma_0$  for our sample is $\left\langle\Sigma_0\right\rangle = 1.50^{+1.12}_{-0.83}\,\mathrm{g\, cm^{-2}}$.
Furthermore, there may be some indication that the Be stars in the SMC may have more massive discs, on average, than their galactic counterparts, in line with results from the literature that report higher H$\alpha$ equivalent widths in the SMC Be stars than in the Galaxy \citep{2007A&A...472..577M}. This last point, however, should be viewed with some caution given the large biases present in our sample.

\subsubsection{Disc life cycles}

In the lower panel of Fig.~\ref{imfs2}, we plot the distribution of the build-up time, $t_2-t_1$, versus the stellar mass. 
We see that the duration of the bump is much shorter for massive stars, which indicates that these stars are much more variable than their late type siblings. Similar findings were reported in the Galaxy  \citep[e.g.][]{2013A&ARv..21...69R,2017arXiv170802594L}.
For a complete characterisation the disc life cycles a census of the number of bumps present during the timespan of OGLE-II and OGLE-III observations (roughly 12 years) would be required . Unfortunately, this cannot be done for our sample because in this study we focused only on the well-defined bumps.
The median of the build-up time for our sample is $\left\langle t_2-t_1\right\rangle=304^{+351}_{-168}$ days.

\subsubsection{Viscosity parameter}


%
\begin{figure}
\centering{
\includegraphics[width=1.0\linewidth]{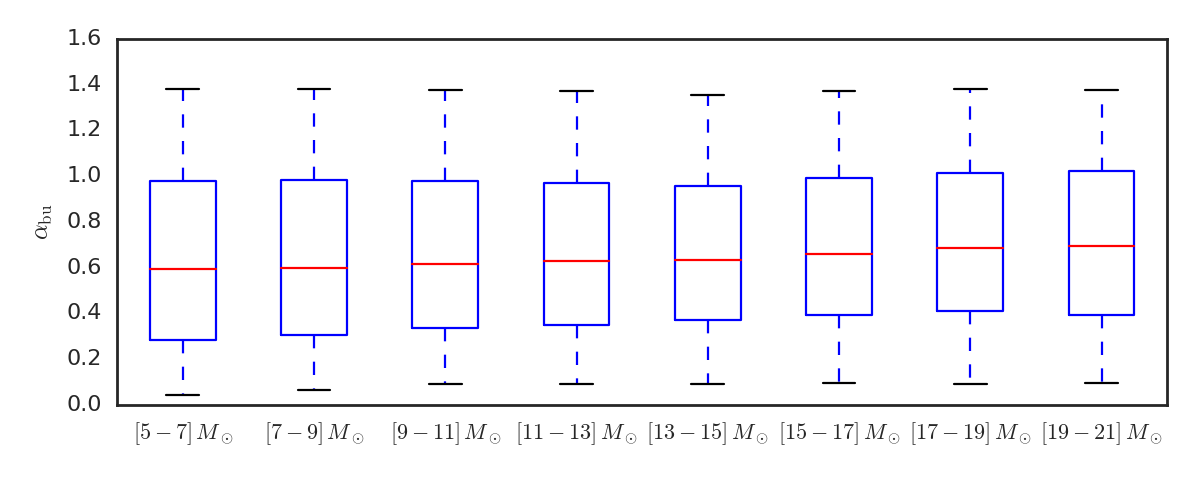}
\includegraphics[width=1.0\linewidth]{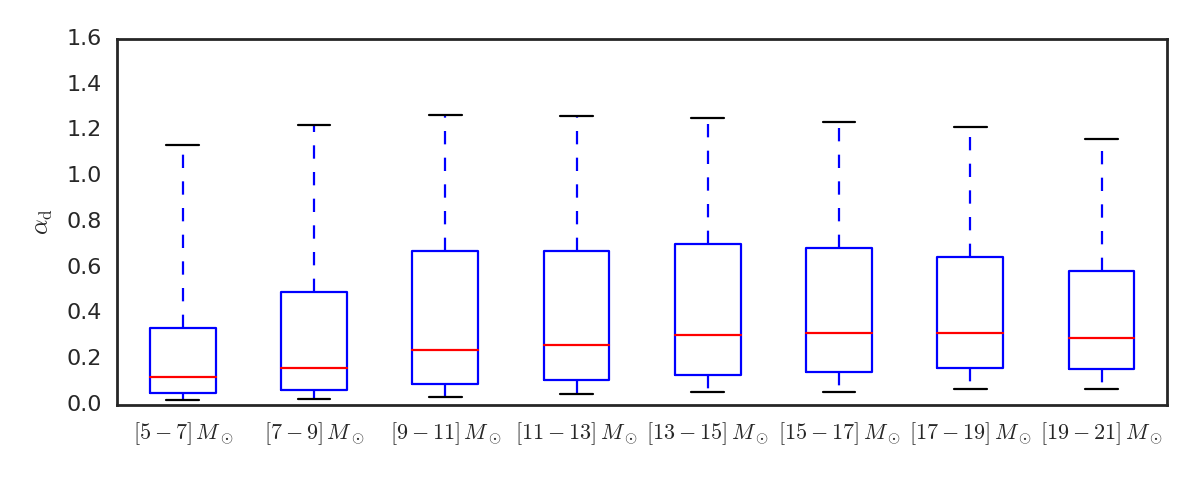}
}
\caption[]
{
Same as Fig.~\ref{imfs2} for $\alpha_\mathrm{bu}$ (above) and $\alpha_\mathrm{d}$ (below).
}
\label{corr_alphas2}
\end{figure}
\begin{figure}
\centering{
\includegraphics[width=1.00\linewidth]{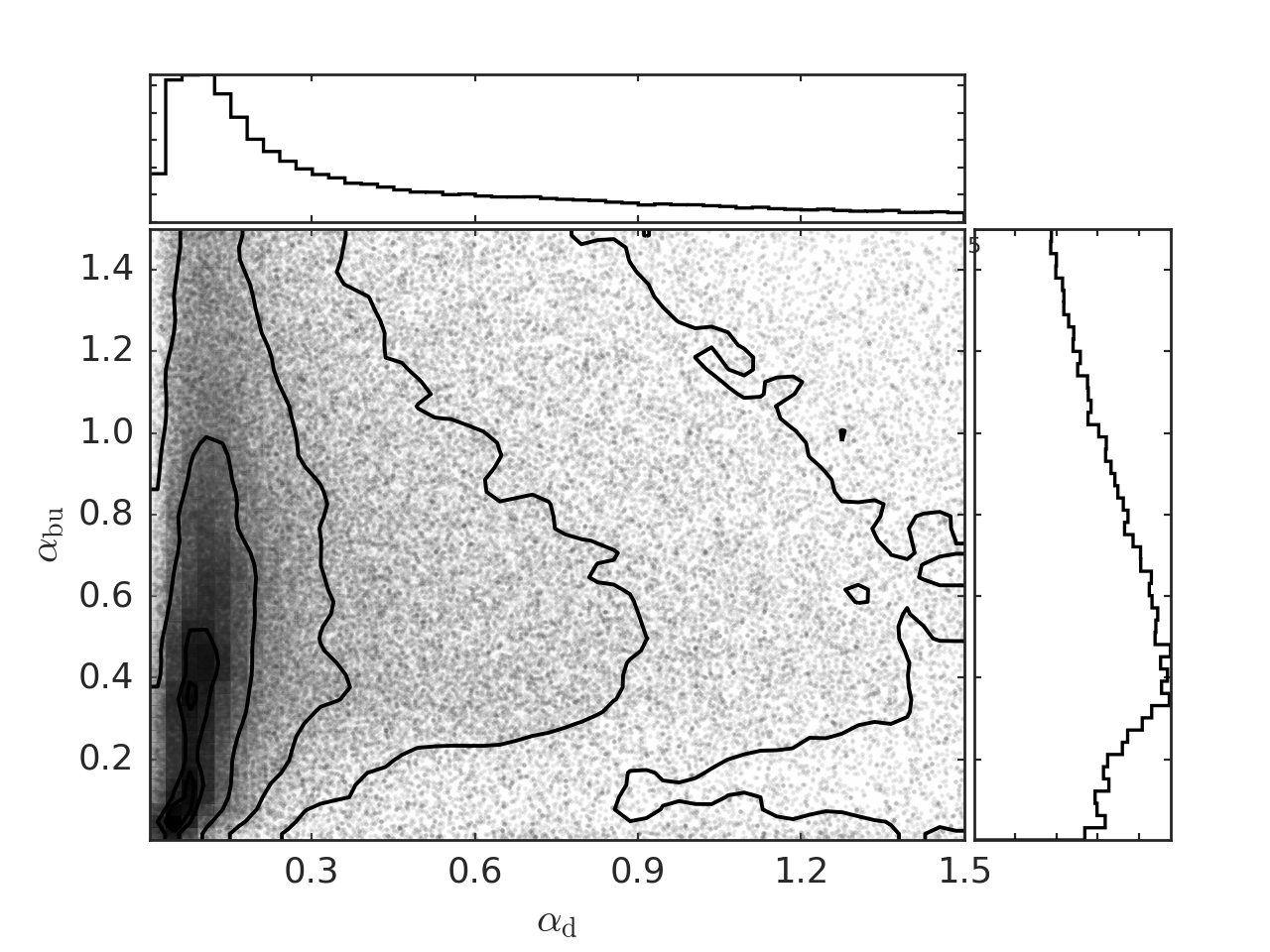}
}
\caption[]
{
Distribution of the two viscosity parameters $\alpha_\mathrm{bu}$ and $\alpha_\mathrm{d}$. The histograms above and on the right are the projections for $\alpha_\mathrm{d}$ and $\alpha_\mathrm{bu}$, respectively. 
The contour levels are the same as in Fig.~\ref{example_bb1}. 
}
\label{corr_alphas}
\end{figure}

The majority of our determinations of the viscosity parameter had broad uncertainty distributions, and it was found that the errors in the determinations of $\alpha_\mathrm{bu}$ were generally greater than those of $\alpha_\mathrm{d}$. 
This has to do with the fact that the inclination angle is a poorly constrained quantity in our analysis, and the rate of photometric variations during build-up depends much more on $\cos i$ (Fig.~\ref{coeffs2}) than the rate of photometric variations during dissipation (Fig.~\ref{coeffs3b}). 
Considering the sample as a whole, we find that there is no variation of $\alpha_\mathrm{bu}$ with the stellar mass (Fig.~\ref{corr_alphas2}, top), but there is a slight hint that $\alpha_\mathrm{d}$ may grow with the stellar mass  (Fig.~\ref{corr_alphas2}, bottom).
Furthermore, we find that on average the viscosity parameter is roughly two times larger at build-up than at dissipation 
($\left\langle\alpha_\mathrm{bu}\right\rangle = 0.64^{+0.50}_{-0.38}$ vs. $\left\langle\alpha_\mathrm{d}\right\rangle = 0.26^{+0.60}_{-0.18}$). 

The estimated values of $\alpha$ in the range of a few tenths to one is in agreement with the usual values obtained for the hot and variable discs of dwarf novae \citep{2007MNRAS.376.1740K, 2012A&A...545A.115K}, as well as with the values obtained for the Be star 28 CMa by \citet{2017ASPC..508..323G}. They are, however, an order of magnitude or more above the usual values obtained in magnetohydrodynamic (MHD) simulations, where the magnetorotational instability \citep[MRI,][]{1991ApJ...376..214B} is the main theoretical assumption for the mechanism that generates the necessary viscosity \citep{2007MNRAS.376.1740K}. 

In Fig.~\ref{corr_alphas}, we show the distributions of $\alpha_\mathrm{bu}$ (right) and $\alpha_\mathrm{d}$ (above) and the distribution in the $\alpha_\mathrm{bu}\times\alpha_\mathrm{d}$ plane. We found that, for most of the bumps, there was a correlation between $\alpha_\mathrm{bu}$ and $\alpha_\mathrm{d}$, with values of $\alpha_\mathrm{bu}$ greater than values of $\alpha_\mathrm{d}$ being more likely. This trend can be seen in the darker areas of the $\alpha_\mathrm{bu}\times\alpha_\mathrm{d}$ plane. \citet{2017ASPC..508..323G} found the same trend for 28 CMa.

It is unclear whether the higher likelihood of $\alpha_\mathrm{bu}>\alpha_\mathrm{d}$ is real a phenomenon or a result of the approximations employed in this work. One key approximation made in our model is that the hydrodynamical equations are solved assuming that the entire disc is isothermal. Earlier studies \citep[e.g.,][]{2004MNRAS.352..841J,2006ApJ...639.1081C} have shown the disc to be highly non-isothermal, which means that $c_s$ in Eq.~\ref{fluxofmass} is a complicated function both of $R$ and time. 
Another approximation is that the possible effects of line forces were neglected.
Recently, \citet{2016MNRAS.458.2323K} simulated the effects of a line-driven wind from early B-type stars over a non-viscous gaseous disc of solar metallicity and typical density. They showed that the line-driven wind was able to destroy the disc in timescales compatible with the observed large-amplitude photometric variations of Be stars. They argued that the presence of the line-driven wind might be the cause of the apparent abnormal value estimated for $\alpha$ by \citet{2012ApJ...744L..15C}, \citet{2017ASPC..508..323G}, and this work. Future work must explore if the viscous force and the line force working together can produce the variability of Be stars with smaller values of $\alpha$.

The line force of an outwardly diffusing near-Keplerian optically thin disc under irradiation by the hot star will have a negative component in the azimuthal direction, removing angular momentum from the gas \citep{2000ApJ...537..461G,2001ApJ...558..802G}. 
It is possible that this sink of angular momentum, during the build-up phase, may induce a clumping of material closer to the stellar equator, where the visual observables are generated, making the disc photometrically appear to build-up faster. In the dissipation phase, reaccretion occurs ($v_R$ becomes negative) and the line force would probably now point in the positive azimuthal direction, being a source of angular momentum to the disc and making the inner disc clear at a slower rate. If that is true, it would would contribute to the observed trend of $\alpha_\mathrm{bu}>\alpha_\mathrm{d}$. We expect, however, that for our low-metallicity SMC Be stars, the possible effect of the line force will be greatly diminished. It is also worth reminding that \citet{1996ApJ...472L.115O} showed that winds from rapidly rotating, oblate stars are weakest near the equatorial plane. Consequently, a wind alone is incapable of building the disc and, even at Solar metallicity, can only act to help viscosity in disc building, but not replace it.
Furthermore, so far all studies of the effects of line forces in gaseous Keplerian discs assumed that the gas is optically thin, 
which is not the case for our inner discs near the disc plane. {The green triangle in Fig.~\ref{asympfig} marks the mass and density at the stellar equator of the initial state of the ablating disc model of \citet{2016MNRAS.458.2323K} for a B2e star. Our calculations show that their initial state would generate only a modest excess $\Delta I = -0.1\,\mathrm{mag}$, if seen pole-on.}

On the other hand, it is possible that an opposite scenario might happen. The line force might operate ablating the tenuous material above the disc plane. These regions would receive radiation from the stellar surface and radiation reprocessed by the optically thick disc, behaving as a sink of mass and angular momentum of the disc \citep[e.g.][]{2011A&A...527A..84K}. In that case, the line-driven wind would actually slow down the build-up phase, because it would take a longer time for the disc to reach a near steady-state, and would speed up the dissipation phase. 
If the above were true, that would result in $\alpha_{\rm bu} < \alpha_{\rm d}$, contrary to the results of  \citet{2017ASPC..508..323G} and in this work.

{
We end this section by speculating about another possible cause for the observed trend $\alpha_\mathrm{bu}>\alpha_\mathrm{d}$. The disc formation is probably a mechanically violent process, with outbursts of matter injecting mechanical energy into the disc that is likely to disrupt its hydrostatic equilibrium and induce further turbulent motion. The dissipation, on the other hand, is expected to be a more gentle process, much less perturbed by stellar activity.
The mechanically-driven turbulence during outburst might account for the larger values of $\alpha$ at these phases.
}

\subsubsection{Mass and angular momentum loss}

\begin{figure}
\centering{
\includegraphics[width=1.0\linewidth]{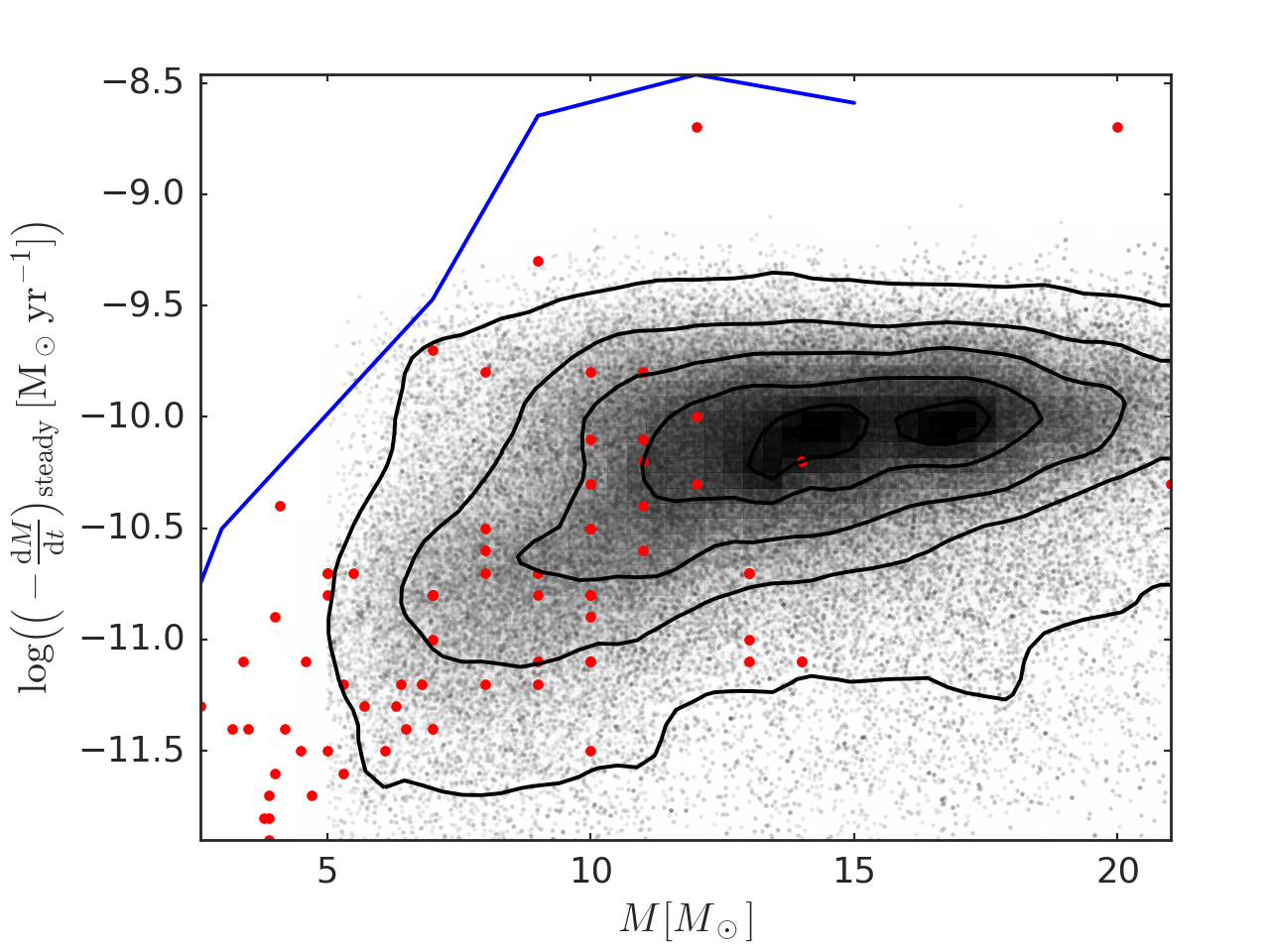}
\includegraphics[width=1.0\linewidth]{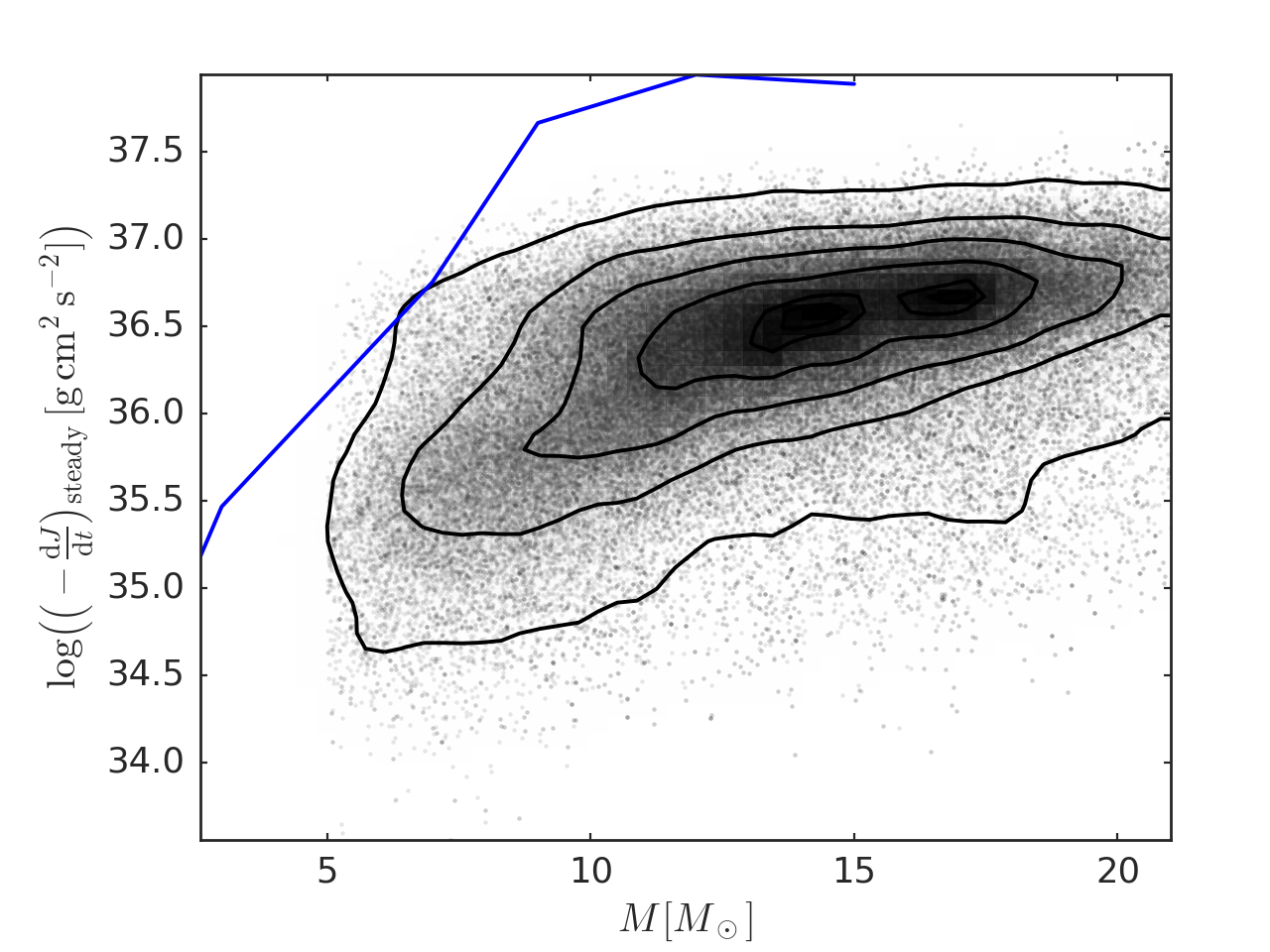}
}
\caption[]
{
Distributions of the steady-state mass (above) and angular momentum (below) loss rates for our sample. The contour levels are the same as in Fig.~\ref{example_bb1}.
The red dots (upper panel) are the values of $(-\partial M/\partial t)_\mathrm{steady}$, calculated from the results of \citet{2017MNRAS.464.3071V} for galactic Be stars. The blue curves are the estimations made by \citet{2013A&A...553A..25G} of the steady-state mass and angular momentum loss of their $Z=0.002$ stars during their episodes of disc formation. 
}
\label{corr_dynamical}
\end{figure}
%


Although the discs of Be stars in our sample are generally not in steady-state, their steady-state mass and angular momentum loss rates (Eqs.~\ref{statdecrate} and \ref{statangmomlossrate}) are
useful estimates of the actual quantities
that are lost by the star after the bump event ends (see Appendix~\ref{Appendix1}). 
The panels of Fig.~\ref{corr_dynamical} show our distributions of the steady-state mass (above) and angular momentum (below) loss rates. 
For the calculation of $(-\partial M/\partial t)_\mathrm{steady}$, we considered the radius of the outer boundary to be given by the radius at which the disc outflow becomes angular momentum conserving (which can be seen as an outer radius of the viscous disc), estimated as $\tilde{R}_\mathrm{out}=0.3(v_\mathrm{orb}/c_s)^2$ by \citet{2011A&A...527A..84K}. 
The red dots (upper panel) are the estimates of $(-\partial M/\partial t)_\mathrm{steady}$ made from the results of \citet{2017MNRAS.464.3071V} for Galactic Be stars, assuming that $\alpha=1$.

The steady-state mass and angular momentum loss rates for our densest bumps are of the order of $\sim$$10^{-10}\, M_\odot\,\mathrm{yr^{-1}}$ and $\sim$$5\times 10^{36}\, \mathrm{g\, cm^{2}\, s^{-2}}$, respectively. 
The typical decretion rate, which estimates the flux of mass in the disc near the stellar equator, is an order of magnitude higher than $(-\partial M/\partial t)_\mathrm{steady}$, being of the order of $\sim$$10^{-9}\, M_\odot\,\mathrm{yr^{-1}}$
{, which also corresponds to the upper limit of the observed wind mass loss rate of B stars \citep{1981ApJ...251..139S,2008A&ARv..16..209P}.}
Values of the typical decretion rate and steady-state angular momentum loss rate are given in the eleventh and twelfth columns of Table~\ref{table_outputs}, respectively. 
The total angular momentum lost by the star as a consequence of the bump, $-\Delta J_{*}$, is given by $(-\partial J/\partial t)_\mathrm{steady}$ times the build-up time (Eq.~\ref{angmom_lost2}), and the total mass lost is simply $-\Delta M_{*}=-\Delta J_{*}/(GMR_\mathrm{out})^{1/2}$, if we still approximate the VDD as a Keplerian disc at $R_\mathrm{out}$.
If we consider that a typical bump has a build-up time of roughly one year, then the mass and angular momentum lost by the star as a consequence of one dense complete bump are of the order of $\sim 10^{-10}\, M_\odot$ and $\sim 10^{44}\, \mathrm{g\, cm^{2}\, s^{-1}}$ (or $\sim 0.01$ Moon masses and $\sim 10^{-3}$ times Earth's orbital angular momentum around the Sun, respectively). 
The VDD is, thus, a physical mechanism capable of extracting a large quantity of angular momentum from the outer layers of the star, without requiring the loss of too much mass. It is, therefore, a breaking mechanism of the outer layers.
The angular momentum lost by the star for each individual bump is given in the thirteenth column of Table~\ref{table_outputs}.

It was proposed \citep[e.g.,][]{2011A&A...527A..84K} that, with the evolution of the star, the formation of the VDD might be a natural mechanism to extract angular momentum from the outer layers of the star, preventing it to exceed the break up velocity. \citet{2013A&A...553A..25G} assumed the appearance of a  steady-state VDD in the Geneva stellar evolution code every time the outer layers of the star reached $W>0.88$ (or $\omega>0.99$, in their notation, where $\omega=\Omega/(8GM/27R_\mathrm{pole}^3)^{1/2}$). The blue curves in Fig.~\ref{corr_dynamical} are the estimates made by \citet{2013A&A...553A..25G} of the steady-state mass and angular momentum loss of their $Z=0.002$ stars during their episodes of disc formation.
The fact that the curves of \citet{2013A&A...553A..25G} lie up to one order of magnitude above our results and the ones of \citeauthor{2017MNRAS.464.3071V} suggests that their assumed discs were made much too dense to be able to remove the needed angular momentum. 

Although the Be phenomenon is probably a powerful velocity breaking mechanism for the outer layers of the star, its effect on the star as a whole is expected to be modest. 
Our determinations of angular momentum loss show that, even if the Be phenomenon happened during $\sim$$30\%$ 
of the stellar main sequence lifetime,  and the integrated time of all the build-up phases was $\sim$$30\%$ 
of that time, that would lead to the removal of $\sim$$1\%$ 
of the initial angular momentum of the star, which, for fast rotating stars of masses from $7M_\odot$ to $15M_\odot$ is $8-30\times 10^{51}\, \mathrm{g\, cm^{2}\, s^{-1}}$ \citep{2013A&A...553A..25G}. 

{
In conclusion, light-curve modelling offers a reliable way to measure directly the amount of angular momentum lost by disc formation events in Be stars. Coupling this information with future studies about the fraction of the main-sequence lifetime a Be star spends in outburst will allow us to estimate the total amount of angular momentum that is lost by the disc during the main sequence. To obtain the total angular momentum lost we must sum the amount lost by the stellar wind.
This quantity, hitherto unknown, will provide an essential constraint on stellar evolution models, in effect allowing for calibration of the core-surface angular momentum coupling of a star.
}

\section{Conclusions}
\label{sect_conclusions}

We present a new method to model the light curves of Be stars with the goal of extracting quantitative information about the fundamental parameters of their discs, such as the viscosity parameter, $\alpha$, and the asymptotic surface density ($\Sigma_0$, which is related to the disc mass injection rate).
The method uses  a large pre-computed grid of synthetic Be light curves, calculated using detailed hydrodynamic calculations coupled with three-dimensional NLTE radiative transfer calculations. The comparison between the model grid and the observed light curves was made possible by the identification of two empirical laws, that consists of simple formulas that closely match the photometric behaviour during disc build-up and dissipation.

An initial analysis of our model grid allowed us to draw important conclusions about the properties of viscous decretion discs (VDD) around Be stars, and their effect on the stellar SED as they evolve in time:
\begin{itemize}
\item The viscosity parameter $\alpha$ is the most important parameter controlling the observed rate of photometric variations in Be light curves, but it is not the only one. Stellar parameters (mass, radius and effective temperature), as well as the disc viewing angle and density level, all concur to change the rate of brightness variations in complicated ways, which means that if meaningful information about $\alpha$ is to be extracted from the data, these parameters must be somehow estimated;
\item We identified a previously unknown effect, dubbed mass-reservoir effect, which also controls the rate of photometric variations during phases of disc dissipation. This effect is a consequence of the fact that VDDs build a mass reservoir at their outer regions, which is undetectable at short wavelengths (e.g., visible). The longer the build-up phase of a disc, the bigger its mass reservoir. When mass injection from the star stops and reaccretion occurs, the reservoir feeds the inner disc with mass. Clearly, the larger the reservoir, the longer it will be able to supply mass to the inner disc, and the slower the photometric dissipation will appear.
\end{itemize}

We applied our fitting pipeline to a sample of light curves of 54 Be star candidates from the SMC \citep{2002A&A...393..887M} containing 81 clearly identified events of disc formation/dissipation (here referred to as bumps). The light curves come from OGLE-II and OGLE-III data, covering roughly 12 years. A Markov Chain Monte Carlo technique was used to properly estimate the posterior probabilities of each fitted parameter.


It was found that our sample is biased towards early type Be stars, likely because these stars are more variable and their discs are denser, resulting in clearer bumps. Also, photometric uncertainties increase for late type Be stars. Since our sample was selected based on the appearance of their bumps, we conclude that our Be discs must be among the densest found for Be stars in the SMC. 
We verified an increase of $\Sigma_0$ with the stellar mass and the median for our whole sample is $\left\langle\Sigma_0\right\rangle = 1.50^{+1.12}_{-0.83}\,\mathrm{g\, cm^{-2}}$. Our results may suggest that Be discs in the SMC are denser than their siblings in the Galaxy, in line with H$\alpha$ surveys that found stronger line emission in the SMC than in our Galaxy. 

The durations of the build-up phases become shorter for more massive stars, which indicates that, as it was already observed in the Galaxy, late-type Be stars are less variable than early-type ones. The median of the build-up time for our sample is $\left\langle t_2-t_1\right\rangle=304^{+351}_{-168}$ days.

We obtained, for the first time, estimates of $\alpha$ for a statistically significant sample of Be stars. In our work, we explored the possibility that the viscosity parameter might be different at build-up ($\alpha_\mathrm{bu}$) and dissipation ($\alpha_\mathrm{d}$). We found no significant variation of $\alpha_\mathrm{bu}$ with the stellar mass (Fig.~\ref{corr_alphas2}, top),
but some evidence points to a correlation between $\alpha_\mathrm{d}$ with $M$ (Fig.~\ref{corr_alphas2}, bottom). 
Furthermore, we find that on average the viscosity parameter is sensibly larger at build-up than at dissipation. Our medians of the two viscosity parameters are $\left\langle\alpha_\mathrm{bu}\right\rangle = 0.63^{+0.50}_{-0.38}$ and $\left\langle\alpha_\mathrm{d}\right\rangle = 0.26^{+0.59}_{-0.18}$. These values are in agreement in magnitude with the determinations of \citet{2012ApJ...744L..15C} and \citet{2017ASPC..508..323G} for the Galactic Be star 28 CMa. They are also similar to the values of $\alpha$ usually found in cataclysmic variables \citep{2007MNRAS.376.1740K,2012A&A...545A.115K}. 

The trend that $\alpha_\mathrm{bu}>\alpha_\mathrm{d}$ was also seen by \citet{2017ASPC..508..323G} in the different cycles of activity of the Be star 28 CMa. Further work is necessary to establish whether this trend is real or simply a result of our model assumptions. In particular, two important physical effects were ignored in this work, namely the fact that Be discs are non-isothermal and the line forces known to act on the disc material. 
This last point, however, may be of little importance for Be stars in the SMC, given their low metallicity.

{
It must be further emphasised that in this work what is really measured are the timescales for disc build-up and dissipation. Under the assumption that viscosity is the only driving mechanism operating on the disc, these
timescales can in turn be converted to estimates of the viscosity parameter. 
The presence of other driving mechanisms (such as the aforementioned ablation) might affect the determination of $\alpha$ in unpredictable ways.
}

The steady-state mass and angular momentum loss rates for the studied bumps are of the order of $\sim$$10^{-10}\, M_\odot\,\mathrm{yr^{-1}}$ and $\sim$$5\times 10^{36}\, \mathrm{g\, cm^{2}\, s^{-2}}$, respectively. 
The typical decretion rate is of the order of $\sim 10^{-9}\, M_\odot\,\mathrm{yr^{-1}}$. These values are in agreement  with the upper limit of the observed wind mass loss rate of B stars \citep{1981ApJ...251..139S,2008A&ARv..16..209P}.
In addition, these values roughly agree in magnitude with the work of 
\citet{2017MNRAS.464.3071V}, who studied a sample of 80 Galactic Be stars.

{
Future perspectives for this work are threefold. First, an effort must be made to remove the biases of our current sample, by including late-type Be stars and smaller-amplitude bumps. 
Second, the availability of a great number of past and current automated surveys (e.g., MACHO -- \citealt{1997ApJ...486..697A}, EROS -- \citealt{1993Natur.365..623A}, ASAS -- \citealt{1997AcA....47..467P}, VISTA-VVV -- \citealt{2010NewA...15..433M}, KELT -- \citealt{2007PASP..119..923P}) will allow us not only to greatly increase the number of Be stars studied, but also to explore the Be phenomenon and associated disc properties for other metallicities (Galaxy, LMC, etc.). Finally, also important is to obtain a better estimate of the central star properties, e.g., via spectroscopic modelling or using stars belonging to clusters with known age. Our current analysis was bound by the limited amount of information available on the central stars.
}

The $\alpha$ determinations made in this work should help in  investigating the physical mechanisms originating the anomalous viscosity in circumstellar discs environments. In addition, the estimates of the net mass and angular momentum loss rates are important for understanding the conditions in which the Be phenomenon appears, and its consequences for the evolution of B-type stars.



\begin{landscape}
\begin{table}
\caption{List of Be stars and their respective bumps selected for this study \label{my_selection}}
\begin{tabular}{rrrr|cc|rrr|r|r|cc}
\hline
\multicolumn{2}{r}{OGLE-II} & \multicolumn{2}{r}{OGLE-III} & \multicolumn{2}{|r|}{diskless interval} & $B_*$ & $V_*$ & $I_*$ & Bump & Bands & \multicolumn{2}{r}{Bump interval} \\
Field & ID & Field & ID & \multicolumn{2}{|r|}{(JD-2450000)} &  &  & & ID & & \multicolumn{2}{r}{(JD-2450000)} \\
\hline

SMC\_SC1 & 7612 & SMC133.4 & 8877 & 600 & 1300 & 15.926$\pm$0.01 & 16.164$\pm$0.009 & 16.443$\pm$0.009 & 01 & $I$  & 1420 & 2000 \\
 &  &  &  &  &  &  &  &  & 02 & $I$  & 3250 & 3800 \\
\hline 
SMC\_SC1 & 60553 & SMC128.6 & 57 & 3000 & 3500 & --- & 15.418$\pm$0.003 & 15.601$\pm$0.007 & 01 & $V$ $I$  & 3500 & 5000 \\
\hline 
SMC\_SC1 & 75701 & SMC125.7 & 20383 & 3000 & 3500 & --- & 15.397$\pm$0.003 & 15.51$\pm$0.006 & 01 & $V$ $I$  & 3650 & 5000 \\
\hline 
SMC\_SC1 & 92262 & SMC128.6 & 147 & 3500 & 3800 & --- & 15.623$\pm$0.003 & 15.811$\pm$0.007 & 01 & $I$  & 2600 & 3100 \\
 &  &  &  &  &  &  &  &  & 02 & $V$ $I$  & 3900 & 5000 \\
\hline 
SMC\_SC2 & 94939 & SMC125.3 & 52 & 1000 & 1100 & 15.832$\pm$0.008 & 15.991$\pm$0.013 & 16.126$\pm$0.007 & 01 & $I$  & 1100 & 2000 \\
\hline 
SMC\_SC3 & 5719 & SMC125.1 & 20231 & 2980 & 3020 & --- & --- & 16.233$\pm$0.01 & 01 & $I$  & 1200 & 4500 \\
\hline 
SMC\_SC3 & 15970 & SMC125.2 & 28056 & 700 & 750 & 15.282$\pm$0.01 & 15.412$\pm$0.006 & 15.541$\pm$0.006 & 01 & $B$ $V$ $I$  & 750 & 2000 \\
 &  &  &  &  &  &  &  &  & 02 & $I$  & 2200 & 3500 \\
 &  &  &  &  &  &  &  &  & 03 & $V$ $I$  & 4000 & 5000 \\
\hline 
SMC\_SC3 & 71445 & SMC125.2 & 34818 & 700 & 800 & 16.227$\pm$0.01 & 16.425$\pm$0.009 & 16.608$\pm$0.012 & 01 & $I$  & 1450 & 2400 \\
 &  &  &  &  &  &  &  &  & 02 & $I$  & 4400 & 4700 \\
\hline 
SMC\_SC3 & 125899 & SMC125.2 & 6200 & 3500 & 3750 & --- & 15.837$\pm$0.004 & 15.956$\pm$0.008 & 01 & $I$  & 2320 & 3400 \\
\hline 
SMC\_SC3 & 197941 & SMC125.3 & 25034 & 4000 & 4060 & --- & 15.671$\pm$0.003 & 15.724$\pm$0.006 & 01 & $V$ $I$  & 2350 & 4000 \\
\hline 
SMC\_SC4 & 22859 & SMC125.4 & 22723 & 700 & 1200 & 17.068$\pm$0.013 & 17.133$\pm$0.012 & 17.129$\pm$0.015 & 01 & $V$ $I$  & 1200 & 3500 \\
\hline 
SMC\_SC4 & 71499 & SMC100.7 & 34896 & 1000 & 1450 & 15.414$\pm$0.011 & 15.553$\pm$0.007 & 15.597$\pm$0.006 & 01 & $V$ $I$  & 1430 & 2400 \\
 &  &  &  &  &  &  &  &  & 02 & $V$ $I$  & 2900 & 4400 \\
\hline 
SMC\_SC4 & 120783 & SMC100.6 & 7129 & 4250 & 4500 & --- & 14.401$\pm$0.003 & 14.442$\pm$0.005 & 01 & $V$ $I$  & 2750 & 4500 \\
\hline 
SMC\_SC4 & 127840 & SMC100.6 & 38372 & 3500 & 4500 & --- & 14.853$\pm$0.003 & 15.026$\pm$0.006 & 01 & $I$  & 2800 & 3500 \\
\hline 
SMC\_SC4 & 156248 & SMC100.8 & 14683 & 1000 & 1200 & 15.91$\pm$0.01 & 15.964$\pm$0.007 & 15.852$\pm$0.006 & 01 & $I$  & 1320 & 2300 \\
\hline 
SMC\_SC4 & 156251 & SMC100.8 & 14642 & 600 & 680 & 14.881$\pm$0.005 & 15.13$\pm$0.007 & 15.338$\pm$0.005 & 01 & $B$ $V$ $I$  & 650 & 1200 \\
 &  &  &  &  &  &  &  &  & 02 & $V$ $I$  & 3170 & 3600 \\
 &  &  &  &  &  &  &  &  & 03 & $V$ $I$  & 3770 & 4600 \\
\hline 
SMC\_SC4 & 159829 & SMC100.8 & 37214 & 4450 & 4500 & --- & 15.895$\pm$0.005 & 15.959$\pm$0.009 & 01 & $I$  & 2600 & 3000 \\
 &  &  &  &  &  &  &  &  & 02 & $V$ $I$  & 3430 & 4500 \\
\hline 
SMC\_SC4 & 159857 & SMC100.8 & 45127 & 3000 & 3500 & --- & 15.626$\pm$0.004 & 15.8$\pm$0.008 & 01 & $I$  & 780 & 1400 \\
 &  &  &  &  &  &  &  &  & 02 & $V$ $I$  & 3600 & 4300 \\
\hline 
SMC\_SC4 & 163828 & SMC100.7 & 8813 & 600 & 700 & 17.171$\pm$0.009 & 17.164$\pm$0.009 & 16.953$\pm$0.013 & 01 & $B$ $V$ $I$  & 700 & 2400 \\
\hline 
SMC\_SC4 & 167554 & SMC100.7 & 51098 & 3000 & 4500 & --- & 17.258$\pm$0.009 & 17.2$\pm$0.016 & 01 & $I$  & 1620 & 1900 \\
 &  &  &  &  &  &  &  &  & 02 & $V$ $I$  & 4700 & 5000 \\
\hline 
\end{tabular}
\end{table}
\end{landscape}

\begin{landscape}
\begin{table}
\contcaption{List of Be stars and their respective bumps selected for this study}
 \label{tab:continued}
\begin{tabular}{rrrr|cc|rrr|r|r|cc}
\hline
\multicolumn{2}{r}{OGLE-II} & \multicolumn{2}{r}{OGLE-III} & \multicolumn{2}{|r|}{diskless interval} & $B_*$ & $V_*$ & $I_*$ & Bump & Bands & \multicolumn{2}{r}{Bump interval} \\
Field & ID & Field & ID & \multicolumn{2}{|r|}{(JD-2450000)} &  &  & & ID & & \multicolumn{2}{r}{(JD-2450000)} \\
\hline

SMC\_SC4 & 171253 & SMC100.7 & 42620 & 700 & 1200 & 15.674$\pm$0.01 & 15.714$\pm$0.009 & 15.691$\pm$0.006 & 01 & $I$  & 1410 & 1900 \\
 &  &  &  &  &  &  &  &  & 02 & $V$ $I$  & 3245 & 4000 \\
\hline 
SMC\_SC4 & 175272 & SMC100.6 & 7362 & 3950 & 4000 & --- & 16.579$\pm$0.005 & 16.431$\pm$0.009 & 01 & $I$  & 2600 & 3800 \\
\hline 
SMC\_SC4 & 179053 & SMC100.6 & 38443 & 0 & 1000 & 16.304$\pm$0.009 & 16.339$\pm$0.009 & 16.148$\pm$0.008 & 01 & $I$  & 1300 & 2400 \\
\hline 
SMC\_SC5 & 11453 & SMC100.8 & 14734 & 3300 & 3400 & --- & 15.871$\pm$0.004 & 15.779$\pm$0.008 & 01 & $I$  & 3400 & 5000 \\
\hline 
SMC\_SC5 & 21117 & SMC100.8 & 52883 & 3650 & 3750 & --- & 16.051$\pm$0.004 & 16.143$\pm$0.009 & 01 & $V$ $I$  & 970 & 2000 \\
 &  &  &  &  &  &  &  &  & 02 & $I$  & 2200 & 2800 \\
 &  &  &  &  &  &  &  &  & 03 & $I$  & 2900 & 3400 \\
 &  &  &  &  &  &  &  &  & 04 & $V$ $I$  & 3850 & 5000 \\
\hline 
SMC\_SC5 & 21134 & SMC100.8 & 45175 & 1000 & 1500 & 15.994$\pm$0.013 & 16.023$\pm$0.006 & 16.091$\pm$0.007 & 01 & $I$  & 1600 & 2000 \\
\hline 
SMC\_SC5 & 32377 & SMC100.7 & 50838 & 4050 & 4100 & --- & 15.844$\pm$0.003 & 15.941$\pm$0.008 & 01 & $I$  & 3030 & 3400 \\
 &  &  &  &  &  &  &  &  & 02 & $I$  & 4350 & 5000 \\
\hline 
SMC\_SC5 & 43650 & SMC100.6 & 15248 & 1400 & 2000 & 17.177$\pm$0.013 & 17.214$\pm$0.01 & 17.305$\pm$0.017 & 01 & $V$ $I$  & 750 & 2000 \\
 &  &  &  &  &  &  &  &  & 02 & $V$ $I$  & 2550 & 5000 \\
\hline 
SMC\_SC5 & 54851 & SMC100.5 & 14725 & 600 & 850 & 16.264$\pm$0.01 & 16.311$\pm$0.009 & 16.366$\pm$0.008 & 01 & $B$ $V$ $I$  & 850 & 2000 \\
 &  &  &  &  &  &  &  &  & 02 & $I$  & 2120 & 3600 \\
 &  &  &  &  &  &  &  &  & 03 & $V$ $I$  & 3650 & 5000 \\
\hline 
SMC\_SC5 & 65500 & SMC101.8 & 21127 & 1000 & 1500 & 16.034$\pm$0.011 & 15.981$\pm$0.007 & 15.959$\pm$0.006 & 01 & $B$ $V$ $I$  & 600 & 1000 \\
\hline 
SMC\_SC5 & 129535 & SMC100.6 & 53957 & 4700 & 4800 & --- & 16.923$\pm$0.008 & 16.953$\pm$0.013 & 01 & $V$ $I$  & 3300 & 5000 \\
\hline 
SMC\_SC5 & 145724 & SMC101.8 & 21370 & 3000 & 3200 & --- & --- & 17.116$\pm$0.019 & 01 & $I$  & 3230 & 5000 \\
\hline 
SMC\_SC5 & 180034 & SMC100.1 & 27826 & 3900 & 4100 & --- & 16.436$\pm$0.005 & 16.431$\pm$0.009 & 01 & $V$ $I$  & 4120 & 5000 \\
\hline 
SMC\_SC5 & 260841 & SMC100.1 & 36050 & 800 & 900 & 15.858$\pm$0.011 & 16.013$\pm$0.008 & 16.182$\pm$0.008 & 01 & $I$  & 1500 & 2200 \\
 &  &  &  &  &  &  &  &  & 02 & $V$ $I$  & 3800 & 4800 \\
\hline 
SMC\_SC5 & 260957 & SMC100.1 & 36101 & 1200 & 1700 & 16.747$\pm$0.009 & 16.917$\pm$0.009 & 17.043$\pm$0.015 & 01 & $V$ $I$  & 3620 & 5000 \\
\hline 
SMC\_SC5 & 266088 & SMC100.2 & 9240 & 700 & 750 & 17.227$\pm$0.012 & 17.315$\pm$0.01 & 17.408$\pm$0.016 & 01 & $B$ $V$ $I$  & 750 & 2000 \\
\hline 
SMC\_SC5 & 276982 & SMC100.3 & 9403 & 4400 & 4500 & --- & 15.993$\pm$0.005 & 15.757$\pm$0.008 & 01 & $V$ $I$  & 3030 & 4500 \\
\hline 
SMC\_SC5 & 282963 & SMC100.3 & 9408 & 1000 & 1500 & 15.431$\pm$0.009 & 15.591$\pm$0.006 & 15.665$\pm$0.005 & 01 & $I$  & 1600 & 4300 \\
\hline 
SMC\_SC6 & 11085 & SMC100.1 & 36096 & 700 & 1100 & 15.52$\pm$0.01 & 15.667$\pm$0.007 & 15.863$\pm$0.006 & 01 & $V$ $I$  & 1400 & 2000 \\
 &  &  &  &  &  &  &  &  & 02 & $V$ $I$  & 3900 & 5000 \\
\hline 
SMC\_SC6 & 17538 & SMC100.2 & 9240 & 0 & 750 & 17.213$\pm$0.012 & 17.315$\pm$0.013 & 17.424$\pm$0.015 & 01 & $V$ $I$  & 800 & 3000 \\
\hline 
SMC\_SC6 & 42440 & SMC100.3 & 56046 & 700 & 1100 & 16.867$\pm$0.012 & 17.01$\pm$0.011 & 17.19$\pm$0.017 & 01 & $I$  & 1400 & 3500 \\
\hline 

\end{tabular}
\end{table}
\end{landscape}

\begin{landscape}
\begin{table}
\contcaption{List of Be stars and their respective bumps selected for this study}
 \label{tab:continued}
\begin{tabular}{rrrr|cc|rrr|r|r|cc}
\hline
\multicolumn{2}{r}{OGLE-II} & \multicolumn{2}{r}{OGLE-III} & \multicolumn{2}{|r|}{diskless interval} & $B_*$ & $V_*$ & $I_*$ & Bump & Bands & \multicolumn{2}{r}{Bump interval} \\
Field & ID & Field & ID & \multicolumn{2}{|r|}{(JD-2450000)} &  &  & & ID & & \multicolumn{2}{r}{(JD-2450000)} \\
\hline

SMC\_SC6 & 99991 & SMC100.1 & 43700 & 1300 & 1800 & 15.755$\pm$0.007 & 15.923$\pm$0.007 & 16.126$\pm$0.009 & 01 & $I$  & 650 & 1400 \\
 &  &  &  &  &  &  &  &  & 02 & $I$  & 1800 & 2500 \\
 &  &  &  &  &  &  &  &  & 03 & $I$  & 3150 & 4000 \\
 &  &  &  &  &  &  &  &  & 04 & $I$  & 4200 & 5000 \\
\hline 
SMC\_SC6 & 105368 & SMC100.2 & 17645 & 600 & 1200 & 16.462$\pm$0.011 & 16.592$\pm$0.008 & 16.681$\pm$0.011 & 01 & $V$ $I$  & 1150 & 2000 \\
 &  &  &  &  &  &  &  &  & 02 & $I$  & 2700 & 4000 \\
\hline 
SMC\_SC6 & 116294 & SMC100.2 & 49901 & 0 & 900 & 16.775$\pm$0.012 & 16.958$\pm$0.01 & 17.085$\pm$0.013 & 01 & $I$  & 985 & 2000 \\
\hline 
SMC\_SC6 & 128831 & SMC100.3 & 55954 & 600 & 1200 & 15.849$\pm$0.009 & 16.018$\pm$0.007 & 16.116$\pm$0.007 & 01 & $I$  & 1445 & 2600 \\
\hline 
SMC\_SC6 & 199611 & SMC100.3 & 29080 & 600 & 1300 & 15.265$\pm$0.011 & 15.447$\pm$0.008 & 15.594$\pm$0.006 & 01 & $I$  & 1500 & 2000 \\
\hline 
SMC\_SC6 & 272665 & SMC106.6 & 26640 & 1000 & 1500 & 17.784$\pm$0.016 & 17.962$\pm$0.016 & 18.04$\pm$0.029 & 01 & $I$  & 1620 & 4500 \\
\hline 
SMC\_SC7 & 57131 & SMC105.6 & 33029 & 1200 & 2000 & 16.037$\pm$0.015 & 16.127$\pm$0.008 & 16.269$\pm$0.008 & 01 & $I$  & 2780 & 3500 \\
\hline 
SMC\_SC8 & 183240 & SMC105.2 & 32029 & 3000 & 4000 & --- & 14.783$\pm$0.003 & 14.946$\pm$0.005 & 01 & $V$ $I$  & 4150 & 4800 \\
\hline 
SMC\_SC9 & 105383 & SMC110.6 & 114 & 1000 & 1300 & 16.115$\pm$0.01 & 16.264$\pm$0.007 & 16.4$\pm$0.009 & 01 & $V$ $I$  & 1240 & 3500 \\
 &  &  &  &  &  &  &  &  & 02 & $V$ $I$  & 3780 & 4800 \\
\hline 
SMC\_SC9 & 168422 & SMC113.7 & 6330 & 4700 & 4850 & --- & 17.002$\pm$0.009 & 17.057$\pm$0.014 & 01 & $V$ $I$  & 2700 & 4850 \\
\hline 
SMC\_SC10 & 8906 & SMC110.6 & 22338 & 4500 & 5000 & --- & 15.253$\pm$0.003 & 15.382$\pm$0.006 & 01 & $I$  & 2935 & 3300 \\
 &  &  &  &  &  &  &  &  & 02 & $V$ $I$  & 3650 & 3710 \\
\hline 
SMC\_SC11 & 28090 & SMC113.2 & 4458 & 4500 & 5000 & --- & 15.248$\pm$0.004 & 15.433$\pm$0.006 & 01 & $V$ $I$  & 2300 & 4500 \\
\hline 
SMC\_SC11 & 46587 & SMC110.3 & 16096 & 600 & 1000 & 17.087$\pm$0.013 & 17.248$\pm$0.01 & 17.343$\pm$0.021 & 01 & $I$  & 1110 & 4500 
\end{tabular}
\end{table}
\end{landscape}


\begin{landscape}
\begin{table}
\caption{Results of the pipeline for each star and bump of the sample}
\label{table_outputs}
\begin{tabular}{lr|rrrr|rrr|rrrr}
\hline 
OGLE-II ID& Bump& $M\,[M_\odot$]& $t/t_\mathrm{MS}$& $W$& $\cos i$& $\Sigma_0$& $\alpha_\mathrm{bu}$& $\alpha_\mathrm{d}$& $\tilde{\tau}_\mathrm{bu}$& $\left(-\frac{\partial M}{\partial t}\right)_\mathrm{typ}$ & $\left(-\frac{\partial J}{\partial t}\right)_\mathrm{std}$ & $-\Delta J_*$ \\ 
& ID & & & & & $[\mathrm{g\, cm^{-2}}]$ & & & & $[10^{-9}\times$ & $[10^{36}\times$ & $[10^{44}\times$ \\
& & & & & & & & & & $ M_\odot\, \mathrm{yr}^{-1}]$ & $ \mathrm{g\, cm^{2}\, s^{-2}}]$ & $ \mathrm{g\, cm^{2}\, s^{-1}}]$\\
\hline 

SMC\_SC1 7612 & 01 & $13.7^{+3.0}_{-3.0}$ & $0.4^{+0.3}_{-0.2}$ & $0.82^{+0.1}_{-0.1}$ & $0.37^{+0.16}_{-0.05}$ & $0.9^{+1.4}_{-0.5}$ & $0.44^{+0.68}_{-0.34}$ & $0.62^{+0.52}_{-0.4}$ & $0.23^{+0.35}_{-0.18}$ & $0.55^{+0.66}_{-0.34}$ & $0.94^{+1.31}_{-0.6}$ & $0.06^{+0.08}_{-0.04}$ \\ 
  & 02 &   &   &   &   & $1.1^{+1.5}_{-0.6}$ & $0.47^{+0.64}_{-0.34}$ & $0.56^{+0.62}_{-0.38}$ & $0.2^{+0.3}_{-0.14}$ & $0.7^{+0.76}_{-0.44}$ & $1.18^{+1.51}_{-0.76}$ & $0.06^{+0.08}_{-0.04}$ \\ 
\hline 
SMC\_SC1 60553 & 01 & $13.8^{+3.5}_{-2.7}$ & $0.7^{+0.2}_{-0.3}$ & $0.8^{+0.1}_{-0.1}$ & $0.62^{+0.1}_{-0.05}$ & $1.9^{+1.0}_{-0.5}$ & $1.26^{+0.17}_{-0.28}$ & $0.1^{+0.04}_{-0.03}$ & $3.07^{+0.57}_{-0.72}$ & $4.72^{+1.94}_{-1.59}$ & $9.12^{+4.64}_{-3.45}$ & $2.93^{+1.45}_{-1.11}$ \\ 
\hline 
SMC\_SC1 75701 & 01 & $17.0^{+2.8}_{-3.9}$ & $0.5^{+0.3}_{-0.3}$ & $0.81^{+0.11}_{-0.11}$ & $0.7^{+0.13}_{-0.07}$ & $2.2^{+0.7}_{-0.5}$ & $0.25^{+0.21}_{-0.09}$ & $0.11^{+0.06}_{-0.04}$ & $0.81^{+0.65}_{-0.3}$ & $1.11^{+1.24}_{-0.51}$ & $2.33^{+2.92}_{-1.17}$ & $0.97^{+1.22}_{-0.49}$ \\ 
\hline 
SMC\_SC1 92262 & 01 & $14.7^{+3.3}_{-2.9}$ & $0.6^{+0.2}_{-0.3}$ & $0.8^{+0.1}_{-0.1}$ & $0.04^{+0.02}_{-0.02}$ & $1.3^{+1.1}_{-0.7}$ & $0.49^{+0.53}_{-0.3}$ & $0.89^{+0.41}_{-0.47}$ & $0.53^{+0.58}_{-0.31}$ & $1.11^{+1.2}_{-0.6}$ & $2.2^{+2.42}_{-1.21}$ & $0.32^{+0.33}_{-0.17}$ \\ 
  & 02 &   &   &   &   & $2.6^{+0.5}_{-0.5}$ & $0.99^{+0.33}_{-0.36}$ & $0.94^{+0.34}_{-0.32}$ & $1.14^{+0.64}_{-0.44}$ & $4.86^{+1.81}_{-1.35}$ & $9.61^{+4.13}_{-2.88}$ & $1.52^{+0.72}_{-0.48}$ \\ 
\hline 
SMC\_SC2 94939 & 01 & $12.0^{+3.0}_{-2.4}$ & $0.6^{+0.3}_{-0.4}$ & $0.81^{+0.1}_{-0.1}$ & $0.46^{+0.14}_{-0.07}$ & $2.4^{+0.8}_{-0.8}$ & $0.9^{+0.42}_{-0.43}$ & $0.63^{+0.34}_{-0.32}$ & $1.31^{+0.59}_{-0.61}$ & $3.26^{+2.5}_{-1.74}$ & $5.47^{+4.52}_{-2.98}$ & $0.96^{+0.77}_{-0.53}$ \\ 
\hline 
SMC\_SC3 5719 & 01 & $12.1^{+2.9}_{-2.6}$ & $0.6^{+0.3}_{-0.3}$ & $0.81^{+0.1}_{-0.11}$ & $0.41^{+0.1}_{-0.06}$ & $2.6^{+0.6}_{-0.8}$ & $0.68^{+0.46}_{-0.33}$ & $0.38^{+0.21}_{-0.18}$ & $4.15^{+2.74}_{-1.98}$ & $2.74^{+2.49}_{-1.6}$ & $4.59^{+4.31}_{-2.72}$ & $3.28^{+3.06}_{-1.95}$ \\ 
\hline 
SMC\_SC3 15970 & 01 & $15.0^{+2.7}_{-2.4}$ & $0.7^{+0.2}_{-0.2}$ & $0.82^{+0.09}_{-0.1}$ & $0.62^{+0.03}_{-0.02}$ & $1.3^{+0.4}_{-0.2}$ & $0.72^{+0.3}_{-0.29}$ & $0.12^{+0.05}_{-0.03}$ & $1.52^{+0.69}_{-0.67}$ & $2.08^{+1.17}_{-0.76}$ & $4.3^{+3.07}_{-1.71}$ & $1.26^{+0.81}_{-0.5}$ \\ 
  & 02 &   &   &   &   & $1.5^{+0.4}_{-0.3}$ & $0.78^{+0.37}_{-0.3}$ & $0.17^{+0.05}_{-0.03}$ & $2.05^{+1.01}_{-0.81}$ & $2.68^{+1.64}_{-1.01}$ & $5.53^{+4.21}_{-2.24}$ & $2.02^{+1.58}_{-0.82}$ \\ 
  & 03 &   &   &   &   & $1.3^{+0.3}_{-0.2}$ & $1.08^{+0.26}_{-0.27}$ & $0.13^{+0.05}_{-0.05}$ & $3.17^{+0.93}_{-0.96}$ & $3.13^{+1.26}_{-0.99}$ & $6.5^{+3.54}_{-2.39}$ & $2.62^{+1.39}_{-0.96}$ \\ 
\hline 
SMC\_SC3 71445 & 01 & $10.5^{+2.4}_{-2.1}$ & $0.6^{+0.3}_{-0.3}$ & $0.81^{+0.1}_{-0.1}$ & $0.43^{+0.09}_{-0.06}$ & $2.5^{+0.6}_{-0.7}$ & $0.58^{+0.45}_{-0.28}$ & $0.67^{+0.38}_{-0.29}$ & $1.3^{+1.02}_{-0.6}$ & $1.96^{+1.67}_{-1.05}$ & $2.88^{+2.58}_{-1.6}$ & $0.7^{+0.64}_{-0.39}$ \\ 
  & 02 &   &   &   &   & $1.2^{+0.6}_{-0.4}$ & $0.67^{+0.52}_{-0.35}$ & $1.0^{+0.34}_{-0.37}$ & $0.59^{+0.5}_{-0.31}$ & $1.12^{+0.71}_{-0.49}$ & $1.64^{+1.12}_{-0.74}$ & $0.16^{+0.12}_{-0.07}$ \\ 
\hline 
SMC\_SC3 125899 & 01 & $13.9^{+3.3}_{-3.2}$ & $0.5^{+0.3}_{-0.3}$ & $0.81^{+0.1}_{-0.1}$ & $0.39^{+0.2}_{-0.07}$ & $1.6^{+1.2}_{-0.8}$ & $0.73^{+0.48}_{-0.4}$ & $0.35^{+0.29}_{-0.19}$ & $1.09^{+0.73}_{-0.61}$ & $1.88^{+1.82}_{-1.13}$ & $3.49^{+3.65}_{-2.16}$ & $0.64^{+0.68}_{-0.39}$ \\ 
\hline 
SMC\_SC3 197941 & 01 & $12.1^{+4.5}_{-2.0}$ & $0.7^{+0.2}_{-0.4}$ & $0.83^{+0.09}_{-0.11}$ & $0.56^{+0.14}_{-0.04}$ & $0.8^{+0.4}_{-0.2}$ & $0.36^{+0.57}_{-0.21}$ & $0.1^{+0.04}_{-0.03}$ & $1.38^{+2.24}_{-0.82}$ & $0.62^{+0.77}_{-0.34}$ & $1.14^{+1.5}_{-0.66}$ & $0.58^{+0.76}_{-0.33}$ \\ 
\hline 
SMC\_SC4 22859 & 01 & $8.2^{+1.7}_{-1.6}$ & $0.4^{+0.3}_{-0.3}$ & $0.8^{+0.11}_{-0.1}$ & $0.64^{+0.19}_{-0.11}$ & $0.3^{+0.2}_{-0.1}$ & $0.68^{+0.56}_{-0.55}$ & $0.03^{+0.66}_{-0.02}$ & $4.16^{+4.08}_{-3.37}$ & $0.2^{+0.18}_{-0.15}$ & $0.23^{+0.23}_{-0.17}$ & $0.13^{+0.15}_{-0.1}$ \\ 
\hline 
SMC\_SC4 71499 & 01 & $17.3^{+2.4}_{-3.7}$ & $0.5^{+0.3}_{-0.3}$ & $0.81^{+0.1}_{-0.11}$ & $0.53^{+0.05}_{-0.03}$ & $1.5^{+0.6}_{-0.4}$ & $0.7^{+0.46}_{-0.32}$ & $0.25^{+0.1}_{-0.07}$ & $1.1^{+0.7}_{-0.51}$ & $2.09^{+1.16}_{-0.86}$ & $4.41^{+2.86}_{-2.0}$ & $0.9^{+0.59}_{-0.4}$ \\ 
  & 02 &   &   &   &   & $1.8^{+0.7}_{-0.5}$ & $0.68^{+0.44}_{-0.31}$ & $0.19^{+0.07}_{-0.06}$ & $1.37^{+0.93}_{-0.62}$ & $2.32^{+1.49}_{-1.0}$ & $4.9^{+3.62}_{-2.28}$ & $1.34^{+0.97}_{-0.62}$ \\ 
\hline 
SMC\_SC4 120783 & 01 & $18.2^{+1.9}_{-2.4}$ & $0.9^{+0.1}_{-0.1}$ & $0.82^{+0.1}_{-0.11}$ & $0.6^{+0.18}_{-0.08}$ & $0.9^{+0.3}_{-0.2}$ & $0.73^{+0.5}_{-0.47}$ & $0.13^{+0.07}_{-0.05}$ & $1.72^{+1.11}_{-1.1}$ & $2.01^{+1.61}_{-1.05}$ & $5.47^{+4.82}_{-2.84}$ & $2.25^{+1.97}_{-1.16}$ \\ 
\hline 
SMC\_SC4 127840 & 01 & $15.2^{+3.0}_{-2.3}$ & $0.8^{+0.1}_{-0.2}$ & $0.82^{+0.1}_{-0.1}$ & $0.65^{+0.12}_{-0.09}$ & $2.8^{+0.5}_{-0.6}$ & $0.39^{+0.21}_{-0.13}$ & $0.76^{+0.4}_{-0.32}$ & $0.57^{+0.28}_{-0.18}$ & $2.95^{+2.15}_{-1.19}$ & $6.77^{+5.27}_{-2.88}$ & $1.52^{+1.16}_{-0.66}$ \\ 
\hline 
SMC\_SC4 156248 & 01 & $12.9^{+3.2}_{-2.4}$ & $0.7^{+0.2}_{-0.4}$ & $0.82^{+0.1}_{-0.11}$ & $0.07^{+0.02}_{-0.04}$ & $1.2^{+1.3}_{-0.8}$ & $0.46^{+0.63}_{-0.32}$ & $0.89^{+0.42}_{-0.48}$ & $1.5^{+2.08}_{-1.05}$ & $0.69^{+1.94}_{-0.45}$ & $1.28^{+3.51}_{-0.83}$ & $0.53^{+1.51}_{-0.35}$ \\ 
\hline 
SMC\_SC4 156251 & 01 & $17.3^{+2.0}_{-2.4}$ & $0.6^{+0.2}_{-0.2}$ & $0.83^{+0.09}_{-0.1}$ & $0.78^{+0.06}_{-0.05}$ & $2.3^{+0.3}_{-0.4}$ & $0.41^{+0.16}_{-0.09}$ & $0.67^{+0.36}_{-0.27}$ & $0.43^{+0.17}_{-0.1}$ & $2.14^{+0.95}_{-0.63}$ & $4.84^{+2.55}_{-1.58}$ & $0.73^{+0.38}_{-0.25}$ \\ 
  & 02 &   &   &   &   & $1.6^{+0.2}_{-0.2}$ & $0.6^{+0.31}_{-0.2}$ & $0.49^{+0.36}_{-0.16}$ & $0.64^{+0.29}_{-0.2}$ & $2.24^{+1.19}_{-0.7}$ & $5.08^{+3.01}_{-1.77}$ & $0.76^{+0.43}_{-0.26}$ \\ 
  & 03 &   &   &   &   & $2.8^{+0.4}_{-0.4}$ & $0.93^{+0.36}_{-0.34}$ & $0.49^{+0.16}_{-0.11}$ & $1.25^{+0.48}_{-0.42}$ & $5.88^{+2.75}_{-2.18}$ & $13.27^{+7.36}_{-5.32}$ & $2.52^{+1.42}_{-0.96}$ \\ 
\hline 
SMC\_SC4 159829 & 01 & $14.5^{+2.8}_{-3.0}$ & $0.4^{+0.3}_{-0.3}$ & $0.81^{+0.1}_{-0.1}$ & $0.59^{+0.07}_{-0.05}$ & $2.5^{+0.7}_{-0.6}$ & $1.05^{+0.29}_{-0.31}$ & $0.26^{+0.09}_{-0.07}$ & $1.64^{+0.5}_{-0.48}$ & $4.16^{+1.75}_{-1.48}$ & $7.59^{+3.8}_{-2.93}$ & $1.44^{+0.68}_{-0.55}$ \\ 
  & 02 &   &   &   &   & $2.5^{+0.6}_{-0.6}$ & $0.96^{+0.34}_{-0.37}$ & $0.72^{+0.28}_{-0.23}$ & $2.6^{+0.91}_{-0.95}$ & $3.82^{+1.64}_{-1.47}$ & $6.92^{+3.45}_{-2.79}$ & $2.25^{+1.11}_{-0.91}$ \\ 
\hline 
SMC\_SC4 159857 & 01 & $15.2^{+3.4}_{-3.3}$ & $0.5^{+0.3}_{-0.3}$ & $0.82^{+0.1}_{-0.1}$ & $0.54^{+0.08}_{-0.03}$ & $0.7^{+1.2}_{-0.3}$ & $0.45^{+0.63}_{-0.35}$ & $0.67^{+0.49}_{-0.37}$ & $0.33^{+0.5}_{-0.26}$ & $0.59^{+0.48}_{-0.3}$ & $1.15^{+1.08}_{-0.61}$ & $0.11^{+0.11}_{-0.06}$ \\ 
  & 02 &   &   &   &   & $1.1^{+0.7}_{-0.3}$ & $0.62^{+0.48}_{-0.37}$ & $0.32^{+0.62}_{-0.15}$ & $0.43^{+0.3}_{-0.27}$ & $1.33^{+0.77}_{-0.61}$ & $2.63^{+1.79}_{-1.29}$ & $0.22^{+0.15}_{-0.11}$ \\ 
\hline 
SMC\_SC4 163828 & 01 & $8.7^{+1.8}_{-1.8}$ & $0.4^{+0.4}_{-0.3}$ & $0.8^{+0.1}_{-0.1}$ & $0.72^{+0.14}_{-0.08}$ & $0.6^{+0.2}_{-0.1}$ & $0.48^{+0.58}_{-0.31}$ & $0.04^{+0.04}_{-0.02}$ & $3.76^{+4.54}_{-2.4}$ & $0.29^{+0.33}_{-0.18}$ & $0.34^{+0.42}_{-0.21}$ & $0.25^{+0.31}_{-0.15}$ \\ 
\hline 
SMC\_SC4 167554 & 01 & $9.2^{+1.9}_{-2.1}$ & $0.4^{+0.3}_{-0.3}$ & $0.8^{+0.1}_{-0.1}$ & $0.59^{+0.08}_{-0.05}$ & $0.8^{+0.5}_{-0.2}$ & $0.56^{+0.46}_{-0.34}$ & $0.54^{+0.45}_{-0.24}$ & $0.54^{+0.44}_{-0.34}$ & $0.47^{+0.32}_{-0.2}$ & $0.57^{+0.44}_{-0.25}$ & $0.05^{+0.04}_{-0.02}$ \\ 
  & 02 &   &   &   &   & $1.5^{+0.6}_{-0.4}$ & $0.42^{+0.36}_{-0.18}$ & $0.84^{+0.42}_{-0.34}$ & $0.51^{+0.47}_{-0.21}$ & $0.7^{+0.5}_{-0.29}$ & $0.86^{+0.71}_{-0.39}$ & $0.1^{+0.08}_{-0.05}$ \\ 
\hline 
\end{tabular}
\end{table}
\end{landscape}

\begin{landscape}
\begin{table}
\contcaption{Results of the pipeline for each star and bump of the sample}
\label{tab:continued}
\begin{tabular}{lr|rrrr|rrr|rrrr}
\hline 
OGLE-II ID& Bump& $M\,[M_\odot$]& $t/t_\mathrm{MS}$& $W$& $\cos i$& $\Sigma_0$& $\alpha_\mathrm{bu}$& $\alpha_\mathrm{d}$& $\tilde{\tau}_\mathrm{bu}$& $\left(-\frac{\partial M}{\partial t}\right)_\mathrm{typ}$ & $\left(-\frac{\partial J}{\partial t}\right)_\mathrm{std}$ & $-\Delta J_*$ \\ 
& ID & & & & & $[\mathrm{g\, cm^{-2}}]$ & & & & $[10^{-9}\times$ & $[10^{36}\times$ & $[10^{44}\times$ \\
& & & & & & & & & & $ M_\odot\, \mathrm{yr}^{-1}]$ & $ \mathrm{g\, cm^{2}\, s^{-2}}]$ & $ \mathrm{g\, cm^{2}\, s^{-1}}]$\\
\hline

SMC\_SC4 171253 & 01 & $16.5^{+2.5}_{-3.4}$ & $0.4^{+0.3}_{-0.3}$ & $0.81^{+0.11}_{-0.1}$ & $0.59^{+0.14}_{-0.07}$ & $0.3^{+1.2}_{-0.1}$ & $0.41^{+0.73}_{-0.39}$ & $0.78^{+0.5}_{-0.47}$ & $0.38^{+0.65}_{-0.36}$ & $0.2^{+0.28}_{-0.15}$ & $0.41^{+0.58}_{-0.31}$ & $0.05^{+0.07}_{-0.03}$ \\ 
  & 02 &   &   &   &   & $0.4^{+0.5}_{-0.1}$ & $0.54^{+0.6}_{-0.47}$ & $0.67^{+0.5}_{-0.46}$ & $0.4^{+0.44}_{-0.36}$ & $0.3^{+0.32}_{-0.2}$ & $0.59^{+0.69}_{-0.4}$ & $0.05^{+0.07}_{-0.04}$ \\ 
\hline 
SMC\_SC4 175272 & 01 & $12.1^{+2.6}_{-2.8}$ & $0.5^{+0.3}_{-0.3}$ & $0.82^{+0.1}_{-0.12}$ & $0.37^{+0.15}_{-0.05}$ & $1.8^{+1.1}_{-0.7}$ & $0.72^{+0.46}_{-0.39}$ & $0.46^{+0.34}_{-0.23}$ & $2.54^{+1.6}_{-1.38}$ & $1.87^{+1.68}_{-1.13}$ & $2.96^{+2.97}_{-1.82}$ & $1.19^{+1.17}_{-0.74}$ \\ 
\hline 
SMC\_SC4 179053 & 01 & $12.5^{+3.0}_{-2.7}$ & $0.5^{+0.3}_{-0.3}$ & $0.82^{+0.1}_{-0.11}$ & $0.03^{+0.03}_{-0.02}$ & $2.1^{+0.9}_{-0.7}$ & $0.78^{+0.46}_{-0.43}$ & $1.08^{+0.3}_{-0.35}$ & $2.12^{+1.4}_{-1.18}$ & $2.28^{+1.67}_{-1.02}$ & $3.81^{+3.09}_{-1.77}$ & $1.2^{+0.98}_{-0.54}$ \\ 
\hline 
SMC\_SC5 11453 & 01 & $14.6^{+3.3}_{-3.4}$ & $0.5^{+0.3}_{-0.3}$ & $0.81^{+0.11}_{-0.11}$ & $0.39^{+0.14}_{-0.06}$ & $1.9^{+1.0}_{-0.8}$ & $0.89^{+0.38}_{-0.39}$ & $0.24^{+0.15}_{-0.11}$ & $4.72^{+2.18}_{-2.0}$ & $2.96^{+2.47}_{-1.71}$ & $5.66^{+5.16}_{-3.43}$ & $3.85^{+3.47}_{-2.35}$ \\ 
\hline 
SMC\_SC5 21117 & 01 & $16.0^{+2.3}_{-2.6}$ & $0.5^{+0.2}_{-0.2}$ & $0.8^{+0.1}_{-0.1}$ & $0.52^{+0.03}_{-0.01}$ & $1.4^{+0.4}_{-0.3}$ & $0.55^{+0.37}_{-0.21}$ & $0.55^{+0.42}_{-0.25}$ & $0.83^{+0.57}_{-0.31}$ & $1.48^{+0.84}_{-0.55}$ & $3.0^{+2.0}_{-1.25}$ & $0.58^{+0.39}_{-0.24}$ \\ 
  & 02 &   &   &   &   & $2.1^{+0.8}_{-0.4}$ & $0.75^{+0.38}_{-0.29}$ & $0.15^{+0.08}_{-0.04}$ & $1.15^{+0.55}_{-0.41}$ & $2.89^{+2.02}_{-1.16}$ & $5.77^{+4.94}_{-2.51}$ & $1.16^{+0.94}_{-0.5}$ \\ 
  & 03 &   &   &   &   & $2.1^{+0.6}_{-0.5}$ & $0.62^{+0.28}_{-0.21}$ & $0.33^{+0.21}_{-0.11}$ & $0.74^{+0.37}_{-0.25}$ & $2.42^{+1.36}_{-0.86}$ & $4.86^{+3.4}_{-1.97}$ & $0.75^{+0.51}_{-0.3}$ \\ 
  & 04 &   &   &   &   & $1.6^{+0.3}_{-0.3}$ & $0.72^{+0.31}_{-0.23}$ & $0.13^{+0.07}_{-0.04}$ & $2.96^{+1.15}_{-1.03}$ & $2.23^{+1.11}_{-0.78}$ & $4.5^{+2.69}_{-1.75}$ & $2.28^{+1.48}_{-0.92}$ \\ 
\hline 
SMC\_SC5 21134 & 01 & $12.9^{+3.7}_{-2.8}$ & $0.5^{+0.3}_{-0.4}$ & $0.81^{+0.1}_{-0.1}$ & $0.41^{+0.22}_{-0.08}$ & $0.7^{+1.6}_{-0.4}$ & $0.43^{+0.66}_{-0.37}$ & $0.15^{+0.3}_{-0.09}$ & $0.44^{+0.74}_{-0.38}$ & $0.42^{+0.56}_{-0.3}$ & $0.72^{+1.06}_{-0.53}$ & $0.09^{+0.13}_{-0.06}$ \\ 
\hline 
SMC\_SC5 32377 & 01 & $14.3^{+3.1}_{-3.1}$ & $0.6^{+0.3}_{-0.3}$ & $0.82^{+0.1}_{-0.1}$ & $0.35^{+0.08}_{-0.04}$ & $1.8^{+1.0}_{-0.7}$ & $0.75^{+0.48}_{-0.41}$ & $0.57^{+0.4}_{-0.26}$ & $0.37^{+0.33}_{-0.2}$ & $2.25^{+1.85}_{-1.11}$ & $4.31^{+3.82}_{-2.24}$ & $0.28^{+0.26}_{-0.15}$ \\ 
  & 02 &   &   &   &   & $2.1^{+0.8}_{-0.7}$ & $0.6^{+0.37}_{-0.27}$ & $0.27^{+0.12}_{-0.11}$ & $1.13^{+0.73}_{-0.49}$ & $2.26^{+1.73}_{-1.11}$ & $4.23^{+3.75}_{-2.13}$ & $1.02^{+0.89}_{-0.52}$ \\ 
\hline 
SMC\_SC5 43650 & 01 & $8.9^{+2.0}_{-1.9}$ & $0.4^{+0.3}_{-0.3}$ & $0.79^{+0.11}_{-0.1}$ & $0.55^{+0.1}_{-0.04}$ & $0.4^{+0.7}_{-0.1}$ & $0.5^{+0.64}_{-0.45}$ & $0.34^{+0.73}_{-0.26}$ & $1.07^{+1.36}_{-0.95}$ & $0.18^{+0.18}_{-0.12}$ & $0.21^{+0.23}_{-0.14}$ & $0.04^{+0.05}_{-0.03}$ \\ 
  & 02 &   &   &   &   & $0.5^{+0.2}_{-0.1}$ & $0.58^{+0.59}_{-0.43}$ & $0.62^{+0.59}_{-0.51}$ & $6.62^{+6.56}_{-4.92}$ & $0.29^{+0.27}_{-0.19}$ & $0.34^{+0.35}_{-0.23}$ & $0.36^{+0.41}_{-0.24}$ \\ 
\hline 
SMC\_SC5 54851 & 01 & $12.5^{+2.4}_{-2.0}$ & $0.6^{+0.2}_{-0.2}$ & $0.82^{+0.09}_{-0.09}$ & $0.62^{+0.03}_{-0.01}$ & $1.1^{+0.3}_{-0.2}$ & $0.74^{+0.35}_{-0.33}$ & $0.1^{+0.04}_{-0.03}$ & $1.85^{+0.85}_{-0.85}$ & $1.41^{+0.74}_{-0.56}$ & $2.44^{+1.54}_{-1.04}$ & $0.73^{+0.48}_{-0.32}$ \\ 
  & 02 &   &   &   &   & $1.2^{+0.3}_{-0.2}$ & $0.76^{+0.35}_{-0.34}$ & $0.11^{+0.03}_{-0.03}$ & $2.08^{+0.89}_{-0.95}$ & $1.55^{+0.8}_{-0.61}$ & $2.67^{+1.67}_{-1.13}$ & $0.88^{+0.52}_{-0.37}$ \\ 
  & 03 &   &   &   &   & $1.1^{+0.2}_{-0.1}$ & $0.5^{+0.31}_{-0.18}$ & $0.05^{+0.01}_{-0.01}$ & $2.42^{+1.6}_{-0.86}$ & $0.97^{+0.63}_{-0.39}$ & $1.69^{+1.3}_{-0.76}$ & $0.99^{+0.81}_{-0.45}$ \\ 
\hline 
SMC\_SC5 65500 & 01 & $12.8^{+3.3}_{-2.7}$ & $0.5^{+0.3}_{-0.3}$ & $0.81^{+0.1}_{-0.11}$ & $0.66^{+0.11}_{-0.05}$ & $2.0^{+1.0}_{-0.6}$ & $0.61^{+0.38}_{-0.27}$ & $0.29^{+0.13}_{-0.11}$ & $0.46^{+0.27}_{-0.19}$ & $1.91^{+1.23}_{-0.88}$ & $3.21^{+2.57}_{-1.62}$ & $0.28^{+0.22}_{-0.14}$ \\ 
\hline 
SMC\_SC5 129535 & 01 & $9.1^{+2.1}_{-2.0}$ & $0.4^{+0.3}_{-0.3}$ & $0.81^{+0.1}_{-0.1}$ & $0.57^{+0.13}_{-0.05}$ & $0.7^{+0.2}_{-0.2}$ & $0.63^{+0.53}_{-0.36}$ & $0.22^{+0.27}_{-0.11}$ & $4.4^{+3.69}_{-2.6}$ & $0.49^{+0.39}_{-0.27}$ & $0.61^{+0.52}_{-0.34}$ & $0.42^{+0.37}_{-0.23}$ \\ 
\hline 
SMC\_SC5 145724 & 01 & $8.6^{+2.1}_{-1.8}$ & $0.5^{+0.3}_{-0.3}$ & $0.81^{+0.11}_{-0.11}$ & $0.42^{+0.1}_{-0.05}$ & $2.4^{+0.7}_{-0.8}$ & $0.81^{+0.44}_{-0.41}$ & $0.13^{+0.07}_{-0.06}$ & $4.2^{+2.39}_{-2.09}$ & $2.05^{+1.73}_{-1.15}$ & $2.51^{+2.24}_{-1.4}$ & $1.31^{+1.21}_{-0.75}$ \\ 
\hline 
SMC\_SC5 180034 & 01 & $12.0^{+2.5}_{-2.5}$ & $0.4^{+0.3}_{-0.3}$ & $0.8^{+0.1}_{-0.11}$ & $0.63^{+0.15}_{-0.1}$ & $1.0^{+0.5}_{-0.3}$ & $0.32^{+0.55}_{-0.19}$ & $0.19^{+0.15}_{-0.08}$ & $0.91^{+1.56}_{-0.54}$ & $0.45^{+0.57}_{-0.25}$ & $0.71^{+0.95}_{-0.42}$ & $0.22^{+0.27}_{-0.13}$ \\ 
\hline 
SMC\_SC5 260841 & 01 & $13.4^{+4.0}_{-3.2}$ & $0.6^{+0.3}_{-0.3}$ & $0.81^{+0.1}_{-0.1}$ & $0.53^{+0.06}_{-0.02}$ & $0.6^{+0.5}_{-0.1}$ & $0.78^{+0.45}_{-0.52}$ & $0.06^{+0.26}_{-0.02}$ & $1.53^{+0.95}_{-1.05}$ & $0.77^{+0.52}_{-0.4}$ & $1.34^{+1.17}_{-0.69}$ & $0.33^{+0.29}_{-0.17}$ \\ 
  & 02 &   &   &   &   & $0.6^{+0.3}_{-0.1}$ & $0.75^{+0.44}_{-0.44}$ & $0.06^{+0.77}_{-0.03}$ & $2.74^{+1.74}_{-1.78}$ & $0.72^{+0.44}_{-0.35}$ & $1.27^{+0.92}_{-0.62}$ & $0.54^{+0.43}_{-0.28}$ \\ 
\hline 
SMC\_SC5 260957 & 01 & $8.9^{+2.1}_{-2.0}$ & $0.4^{+0.4}_{-0.3}$ & $0.82^{+0.09}_{-0.11}$ & $0.58^{+0.14}_{-0.06}$ & $1.2^{+0.4}_{-0.3}$ & $0.55^{+0.5}_{-0.24}$ & $0.04^{+0.02}_{-0.02}$ & $2.45^{+2.21}_{-1.1}$ & $0.78^{+0.63}_{-0.4}$ & $0.98^{+0.86}_{-0.54}$ & $0.43^{+0.4}_{-0.23}$ \\ 
\hline 
SMC\_SC5 266088 & 01 & $7.8^{+1.5}_{-1.6}$ & $0.3^{+0.3}_{-0.2}$ & $0.81^{+0.1}_{-0.11}$ & $0.7^{+0.14}_{-0.08}$ & $0.7^{+0.4}_{-0.1}$ & $0.3^{+0.69}_{-0.24}$ & $0.06^{+0.05}_{-0.02}$ & $1.92^{+4.42}_{-1.56}$ & $0.17^{+0.3}_{-0.11}$ & $0.18^{+0.34}_{-0.12}$ & $0.11^{+0.19}_{-0.07}$ \\ 
\hline 
SMC\_SC5 276982 & 01 & $17.1^{+2.2}_{-2.6}$ & $0.2^{+0.2}_{-0.2}$ & $0.8^{+0.1}_{-0.11}$ & $0.75^{+0.11}_{-0.1}$ & $1.2^{+0.4}_{-0.3}$ & $0.06^{+0.19}_{-0.03}$ & $0.05^{+0.04}_{-0.02}$ & $0.31^{+1.0}_{-0.16}$ & $0.12^{+0.26}_{-0.06}$ & $0.23^{+0.54}_{-0.13}$ & $0.15^{+0.33}_{-0.08}$ \\ 
\hline 
SMC\_SC5 282963 & 01 & $15.6^{+3.1}_{-3.6}$ & $0.6^{+0.3}_{-0.3}$ & $0.82^{+0.1}_{-0.11}$ & $0.38^{+0.19}_{-0.06}$ & $1.7^{+1.0}_{-0.8}$ & $0.62^{+0.51}_{-0.33}$ & $0.39^{+0.32}_{-0.21}$ & $5.62^{+4.36}_{-2.92}$ & $2.07^{+2.43}_{-1.35}$ & $4.14^{+5.39}_{-2.78}$ & $4.96^{+6.18}_{-3.36}$ \\ 
\hline 
SMC\_SC6 11085 & 01 & $17.8^{+1.9}_{-2.9}$ & $0.3^{+0.3}_{-0.2}$ & $0.81^{+0.1}_{-0.11}$ & $0.74^{+0.06}_{-0.05}$ & $2.4^{+0.6}_{-0.5}$ & $0.59^{+0.26}_{-0.18}$ & $0.13^{+0.05}_{-0.04}$ & $0.99^{+0.45}_{-0.3}$ & $2.53^{+1.13}_{-0.87}$ & $5.19^{+2.71}_{-1.99}$ & $1.06^{+0.57}_{-0.4}$ \\ 
  & 02 &   &   &   &   & $2.6^{+0.5}_{-0.4}$ & $0.78^{+0.34}_{-0.27}$ & $0.27^{+0.09}_{-0.07}$ & $1.49^{+0.63}_{-0.5}$ & $3.41^{+1.75}_{-1.29}$ & $7.07^{+4.11}_{-3.0}$ & $1.64^{+0.95}_{-0.68}$ \\ 
\hline 
SMC\_SC6 17538 & 01 & $7.8^{+1.6}_{-1.4}$ & $0.3^{+0.3}_{-0.2}$ & $0.8^{+0.1}_{-0.11}$ & $0.61^{+0.15}_{-0.08}$ & $0.7^{+0.3}_{-0.2}$ & $0.59^{+0.58}_{-0.41}$ & $0.08^{+0.09}_{-0.04}$ & $3.55^{+3.44}_{-2.47}$ & $0.37^{+0.34}_{-0.22}$ & $0.4^{+0.39}_{-0.24}$ & $0.22^{+0.21}_{-0.13}$ \\ 
\hline 
SMC\_SC6 42440 & 01 & $8.9^{+2.0}_{-2.0}$ & $0.4^{+0.4}_{-0.3}$ & $0.81^{+0.1}_{-0.11}$ & $0.36^{+0.15}_{-0.05}$ & $1.8^{+1.2}_{-0.8}$ & $0.78^{+0.45}_{-0.42}$ & $0.19^{+0.17}_{-0.1}$ & $2.82^{+1.81}_{-1.46}$ & $1.36^{+1.32}_{-0.79}$ & $1.66^{+1.84}_{-0.99}$ & $0.61^{+0.72}_{-0.37}$ \\ 
\hline 
\end{tabular}
\end{table}
\end{landscape}

\begin{landscape}
\begin{table}
\contcaption{Results of the pipeline for each star and bump of the sample}
\label{tab:continued}
\begin{tabular}{lr|rrrr|rrr|rrrr}

SMC\_SC6 99991 & 01 & $14.8^{+2.6}_{-2.8}$ & $0.5^{+0.2}_{-0.2}$ & $0.83^{+0.09}_{-0.09}$ & $0.34^{+0.04}_{-0.02}$ & $1.7^{+0.6}_{-0.5}$ & $0.62^{+0.43}_{-0.25}$ & $0.98^{+0.31}_{-0.29}$ & $0.88^{+0.64}_{-0.36}$ & $1.85^{+1.11}_{-0.75}$ & $3.54^{+2.45}_{-1.55}$ & $0.64^{+0.44}_{-0.27}$ \\ 
  & 02 &   &   &   &   & $2.2^{+0.7}_{-0.6}$ & $0.62^{+0.33}_{-0.25}$ & $0.74^{+0.33}_{-0.26}$ & $1.13^{+0.62}_{-0.44}$ & $2.37^{+1.58}_{-0.97}$ & $4.47^{+3.48}_{-1.93}$ & $1.02^{+0.8}_{-0.43}$ \\ 
  & 03 &   &   &   &   & $1.9^{+0.6}_{-0.5}$ & $0.55^{+0.45}_{-0.22}$ & $0.76^{+0.3}_{-0.24}$ & $1.03^{+0.83}_{-0.43}$ & $1.94^{+1.17}_{-0.77}$ & $3.72^{+2.63}_{-1.65}$ & $0.85^{+0.62}_{-0.37}$ \\ 
  & 04 &   &   &   &   & $1.7^{+0.5}_{-0.4}$ & $0.53^{+0.36}_{-0.22}$ & $0.92^{+0.3}_{-0.26}$ & $1.12^{+0.77}_{-0.47}$ & $1.56^{+0.97}_{-0.59}$ & $2.96^{+2.18}_{-1.24}$ & $0.79^{+0.59}_{-0.33}$ \\ 
\hline 
SMC\_SC6 105368 & 01 & $11.2^{+2.7}_{-2.9}$ & $0.5^{+0.3}_{-0.3}$ & $0.81^{+0.1}_{-0.1}$ & $0.53^{+0.04}_{-0.02}$ & $0.9^{+0.4}_{-0.2}$ & $0.51^{+0.57}_{-0.31}$ & $0.17^{+0.1}_{-0.06}$ & $1.12^{+1.23}_{-0.68}$ & $0.61^{+0.56}_{-0.3}$ & $0.92^{+0.91}_{-0.47}$ & $0.22^{+0.22}_{-0.11}$ \\ 
  & 02 &   &   &   &   & $1.2^{+0.5}_{-0.4}$ & $0.28^{+0.48}_{-0.13}$ & $0.19^{+0.13}_{-0.07}$ & $0.78^{+1.38}_{-0.37}$ & $0.5^{+0.48}_{-0.21}$ & $0.76^{+0.77}_{-0.34}$ & $0.23^{+0.25}_{-0.1}$ \\ 
\hline 
SMC\_SC6 116294 & 01 & $8.6^{+2.1}_{-1.7}$ & $0.5^{+0.3}_{-0.3}$ & $0.82^{+0.1}_{-0.11}$ & $0.41^{+0.1}_{-0.05}$ & $2.6^{+0.7}_{-1.0}$ & $0.87^{+0.4}_{-0.39}$ & $0.16^{+0.09}_{-0.07}$ & $2.71^{+1.35}_{-1.23}$ & $2.42^{+1.65}_{-1.34}$ & $2.99^{+2.11}_{-1.69}$ & $0.93^{+0.71}_{-0.53}$ \\ 
\hline 
SMC\_SC6 128831 & 01 & $12.9^{+3.6}_{-2.9}$ & $0.6^{+0.3}_{-0.3}$ & $0.81^{+0.1}_{-0.1}$ & $0.03^{+0.03}_{-0.02}$ & $2.7^{+0.5}_{-0.8}$ & $0.96^{+0.37}_{-0.42}$ & $0.62^{+0.33}_{-0.23}$ & $0.85^{+0.38}_{-0.37}$ & $3.93^{+2.09}_{-1.61}$ & $6.95^{+4.2}_{-2.98}$ & $0.79^{+0.45}_{-0.35}$ \\ 
\hline 
SMC\_SC6 199611 & 01 & $15.5^{+3.6}_{-3.3}$ & $0.6^{+0.3}_{-0.3}$ & $0.81^{+0.11}_{-0.11}$ & $0.39^{+0.17}_{-0.07}$ & $1.5^{+1.3}_{-0.8}$ & $0.51^{+0.58}_{-0.36}$ & $0.29^{+0.26}_{-0.16}$ & $0.44^{+0.52}_{-0.31}$ & $1.33^{+1.47}_{-0.84}$ & $2.75^{+3.38}_{-1.76}$ & $0.33^{+0.39}_{-0.21}$ \\ 
\hline 
SMC\_SC6 272665 & 01 & $6.3^{+1.2}_{-0.9}$ & $0.3^{+0.3}_{-0.2}$ & $0.81^{+0.12}_{-0.11}$ & $0.39^{+0.19}_{-0.07}$ & $0.6^{+0.3}_{-0.2}$ & $0.7^{+0.51}_{-0.47}$ & $0.07^{+0.41}_{-0.05}$ & $7.81^{+6.55}_{-5.28}$ & $0.26^{+0.25}_{-0.15}$ & $0.23^{+0.24}_{-0.14}$ & $0.22^{+0.23}_{-0.14}$ \\ 
\hline 
SMC\_SC7 57131 & 01 & $12.0^{+2.8}_{-2.4}$ & $0.5^{+0.3}_{-0.3}$ & $0.81^{+0.1}_{-0.11}$ & $0.39^{+0.14}_{-0.06}$ & $2.0^{+0.8}_{-0.7}$ & $0.73^{+0.43}_{-0.4}$ & $0.65^{+0.41}_{-0.36}$ & $1.81^{+1.04}_{-0.98}$ & $2.11^{+1.81}_{-1.3}$ & $3.4^{+3.27}_{-2.12}$ & $0.99^{+0.93}_{-0.62}$ \\ 
\hline 
SMC\_SC8 183240 & 01 & $17.6^{+2.3}_{-2.9}$ & $0.7^{+0.2}_{-0.2}$ & $0.82^{+0.1}_{-0.11}$ & $0.61^{+0.15}_{-0.08}$ & $0.9^{+0.4}_{-0.2}$ & $0.64^{+0.51}_{-0.44}$ & $0.28^{+0.23}_{-0.13}$ & $1.18^{+0.87}_{-0.77}$ & $1.45^{+1.3}_{-0.79}$ & $3.46^{+3.52}_{-1.93}$ & $0.98^{+1.0}_{-0.54}$ \\ 
\hline 
SMC\_SC9 105383 & 01 & $13.0^{+3.0}_{-2.6}$ & $0.4^{+0.3}_{-0.3}$ & $0.82^{+0.1}_{-0.11}$ & $0.56^{+0.06}_{-0.04}$ & $1.9^{+0.7}_{-0.5}$ & $0.53^{+0.44}_{-0.26}$ & $0.09^{+0.04}_{-0.02}$ & $1.43^{+1.21}_{-0.69}$ & $1.47^{+1.17}_{-0.68}$ & $2.42^{+2.29}_{-1.19}$ & $0.74^{+0.7}_{-0.37}$ \\ 
  & 02 &   &   &   &   & $1.7^{+0.6}_{-0.4}$ & $0.77^{+0.41}_{-0.32}$ & $0.09^{+0.03}_{-0.03}$ & $1.9^{+0.91}_{-0.78}$ & $1.94^{+1.22}_{-0.81}$ & $3.2^{+2.55}_{-1.44}$ & $0.87^{+0.71}_{-0.38}$ \\ 
\hline 
SMC\_SC9 168422 & 01 & $8.8^{+1.8}_{-1.8}$ & $0.4^{+0.4}_{-0.3}$ & $0.8^{+0.1}_{-0.1}$ & $0.61^{+0.18}_{-0.08}$ & $0.5^{+0.3}_{-0.1}$ & $0.56^{+0.6}_{-0.43}$ & $0.07^{+0.11}_{-0.03}$ & $2.55^{+2.96}_{-2.01}$ & $0.27^{+0.27}_{-0.18}$ & $0.33^{+0.34}_{-0.21}$ & $0.15^{+0.16}_{-0.1}$ \\ 
\hline 
SMC\_SC10 8906 & 01 & $16.1^{+3.3}_{-3.4}$ & $0.6^{+0.2}_{-0.3}$ & $0.82^{+0.1}_{-0.1}$ & $0.56^{+0.09}_{-0.04}$ & $2.2^{+0.9}_{-0.7}$ & $0.89^{+0.38}_{-0.31}$ & $0.44^{+0.23}_{-0.16}$ & $0.2^{+0.1}_{-0.07}$ & $4.23^{+2.33}_{-1.6}$ & $9.05^{+5.83}_{-3.69}$ & $0.3^{+0.19}_{-0.12}$ \\ 
  & 02 &   &   &   &   & $2.0^{+0.8}_{-0.7}$ & $0.82^{+0.36}_{-0.31}$ & $0.9^{+0.35}_{-0.31}$ & $0.07^{+0.03}_{-0.02}$ & $3.5^{+1.98}_{-1.35}$ & $7.5^{+4.86}_{-3.08}$ & $0.09^{+0.05}_{-0.04}$ \\ 
\hline 
SMC\_SC11 28090 & 01 & $14.5^{+3.8}_{-2.8}$ & $0.7^{+0.2}_{-0.3}$ & $0.82^{+0.1}_{-0.11}$ & $0.57^{+0.13}_{-0.05}$ & $1.8^{+0.8}_{-0.7}$ & $0.1^{+0.09}_{-0.05}$ & $0.12^{+0.05}_{-0.05}$ & $0.46^{+0.41}_{-0.2}$ & $0.45^{+0.38}_{-0.22}$ & $0.94^{+0.96}_{-0.5}$ & $0.59^{+0.59}_{-0.32}$ \\ 
\hline 
SMC\_SC11 46587 & 01 & $8.2^{+1.8}_{-1.6}$ & $0.4^{+0.4}_{-0.3}$ & $0.8^{+0.1}_{-0.11}$ & $0.38^{+0.21}_{-0.06}$ & $0.8^{+0.6}_{-0.3}$ & $0.61^{+0.57}_{-0.46}$ & $0.3^{+0.47}_{-0.18}$ & $4.93^{+4.65}_{-3.7}$ & $0.39^{+0.47}_{-0.25}$ & $0.44^{+0.57}_{-0.29}$ & $0.34^{+0.43}_{-0.22}$ \\ 
\hline 
\end{tabular}
\end{table}
\end{landscape}

\section*{Acknowledgements}

This work made use of the computing facilities of the Laboratory of Astroinformatics (IAG/USP, NAT/Unicsul), whose purchase was made possible by the Brazilian agency FAPESP (grant 2009/54006-4) and the \mbox{INCT-A}. 
L.~R.~R. acknowledges the support from FAPESP (grant 2012/21518-5) and from CNPq (grant 142411/2011-6). 
A.~C.~C. acknowledges the support from CNPq (grant 307594/2015-7) and FAPESP (grant 2015/17967-7).
D.~M.~F. acknowledges the support from FAPESP (grant 2016/16844-1).
R.~G.~V. acknowledges the support from FAPESP (grant 2012/20364-4).
M.~R.~G. acknowledges the support from CAPES PROEX Programa Astronomia.
J.~E.~B. was supported by NSF grant AST-1412135.
The OGLE project has received funding from the National Science Centre, Poland, grant MAESTRO 2014/14/A/ST9/00121.
We also thank Nathaniel Dylan Kee for the very fruitful discussions on the results of this work.

\bibliographystyle{mnras}
\bibliography{referencias1}

\begin{appendix}

\section{Properties of the hydrodynamical solutions of the VDD model}
\label{Appendix1}


\begin{figure}
        \includegraphics[width=\linewidth]{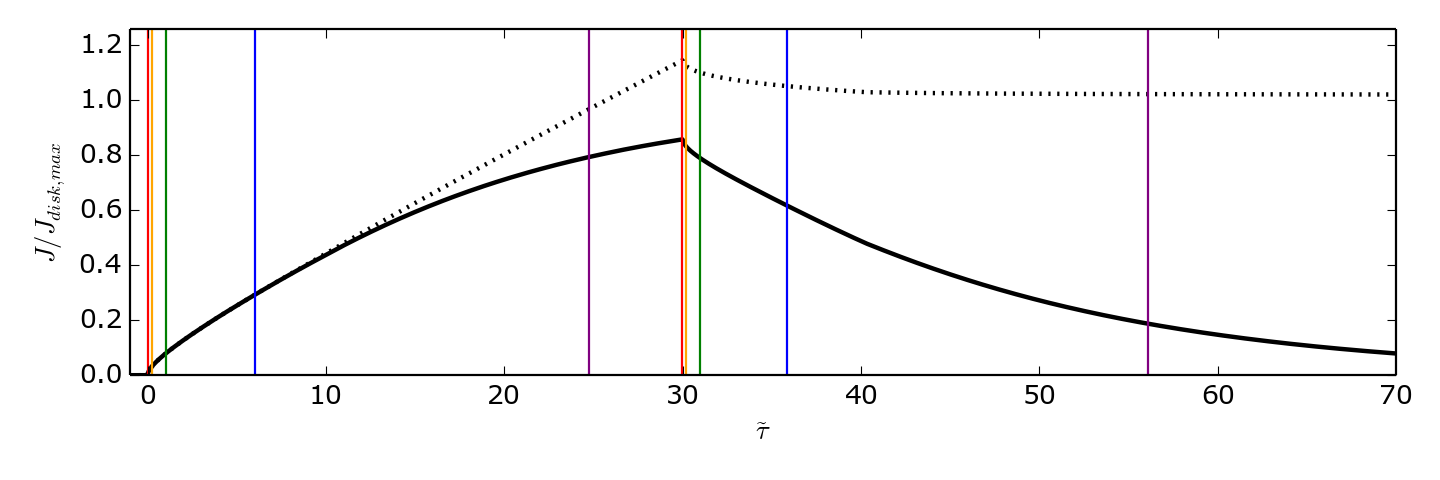}
        \includegraphics[width=\linewidth]{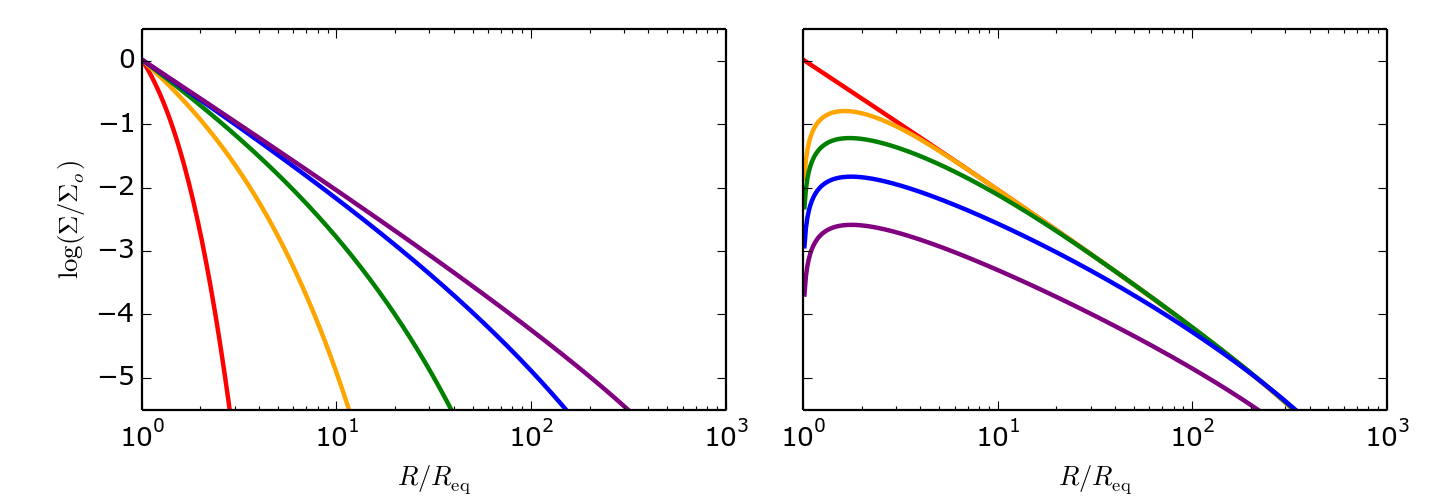}
        \caption{Dynamical bump model with $\tilde{\tau}_\mathrm{bu}=30$. {\it Top}: the amount of angular momentum in the disc (solid black curve) and the amount of angular momentum that is lost by the star (dotted black curve). Colored vertical straight lines mark the 5 instants $\tilde{\tau}=0,0.2,1,6,25$ (during the build-up phase), and the 5 instants $\tilde{\tau}=30,30.2,31,36,55$ (during the dissipation phase). 
Surface density profiles are shown at the first 5 instants ({\it bottom left}) and at the last 5 instants ({\it bottom right}).}

        \label{hydroplot1}
\end{figure}


A particular bump model  (Sect.~\ref{model_grid}) with arbitrary density $\Sigma_0$ and scaled build-up time $\tilde{\tau}_\mathrm{bu}=30$ is used to illustrate features of the bump models in Fig.~\ref{hydroplot1}.
The top panel shows the amount of angular momentum in the disc, given by 
\begin{equation}
J_\mathrm{disc}=\int (GMR)^\frac{1}{2}\Sigma 2\pi R\mathrm{d}R\,, 
\end{equation}
and the amount of angular momentum that is lost by the star, given by the angular momentum that is injected at the radius of mass injection minus the angular momentum that falls back to the stellar equator.
By the continuity of angular momentum, the difference between the dotted and the solid curves is the angular momentum that escapes the system through the outer boundary at $R_\mathrm{out}$. All values were scaled by the maximum angular momentum supported by the disc, which is 
\begin{equation}
J_\mathrm{disc,max}=\int (GMR)^\frac{1}{2}\Sigma_\mathrm{steady} 2\pi R\mathrm{d}R \,.
\end{equation}
The plot shows that, as the build-up process occurs, the disc mass and angular momentum content increase continuously. Eventually (in our example, roughly after $\tilde{\tau}=10$), a non-negligible amount of angular momentum starts to reach the outer radius $R_\mathrm{out}=1000R_\mathrm{eq}$, leaving the system through the outer boundary. After the end of the build-up phase (which, in our example, happens at $\tilde{\tau}=30$), the disc starts to dissipate: the black curve shows that the disc loses angular momentum until it reaches zero. However, as the dotted curve shows, only a fraction of the angular momentum of the disc returns to the star by re-accretion. The dotted curve tends to a non-zero value, which is the angular momentum that was lost by the star in the whole process. This non-zero total angular momentum lost by the star was verified in our simulations to be given exactly by
%
\begin{equation}
-\Delta J_*=\int_0^{\tilde{\tau}_\mathrm{bu}}\tau(t)\left(-\frac{\partial J}{\partial t}\right)_\mathrm{steady}\mathrm{d}\tilde{\tau}\,,
\label{angmom_lost}
\end{equation}
or, in the case of a constant $\alpha$  in time during build-up (Eq.~\ref{time_par_transfeq}), 
\begin{equation}
-\Delta J_*=\left(-\frac{\partial J}{\partial t}\right)_\mathrm{steady}(t_2-t_1)\,.
\label{angmom_lost2}
\end{equation}
This quantity, therefore, is nearly independent of $R_\mathrm{out}$.

The bottom panels of Fig.~\ref{hydroplot1} show surface density radial profiles at the specify instants marked in the top panel.
During build-up, the disc grows in an inside out pattern, with the inner regions reaching a near stationary regime earlier than the outer parts. During dissipation, however, the disc becomes less and less dense as a whole, more or less self-similar way, because the entire disc is coupled by viscous forces.

\section{Examples of model light curves}
\label{Appendix2}

\begin{figure*}
        \includegraphics[width=0.48\linewidth]{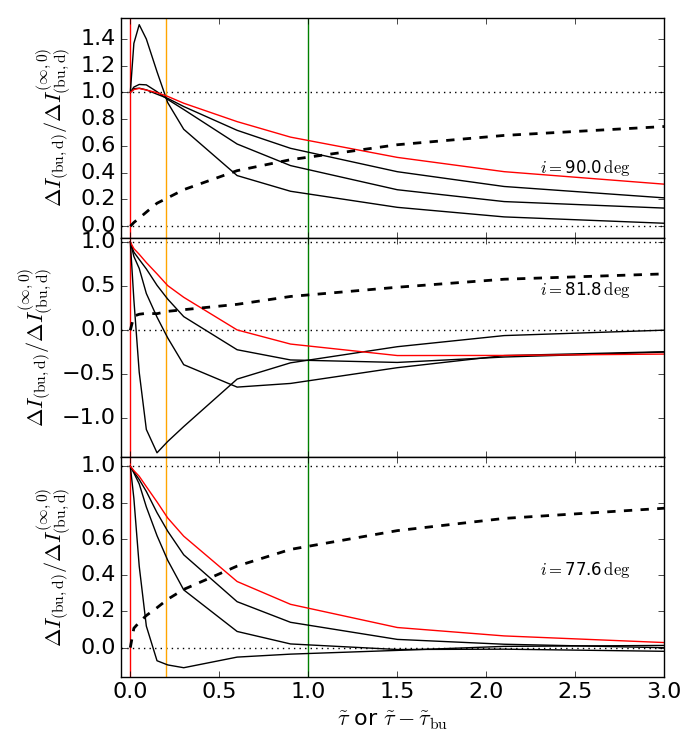}
        \includegraphics[width=0.48\linewidth]{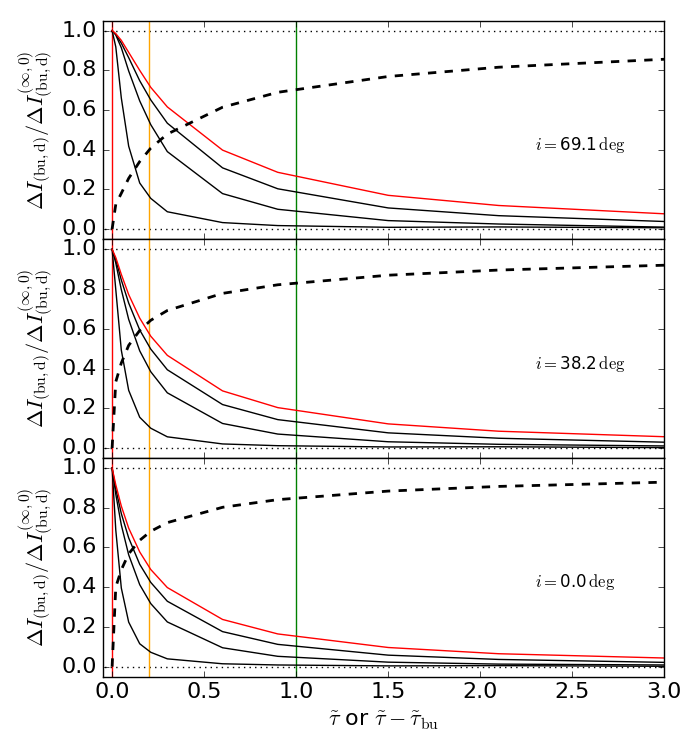}
        \caption{Examples of model $I$-band light curves. Each panel shows the results for different inclination angles, as indicated. The dashed black curves correspond to $\Delta I_\mathrm{bu}/\Delta I_\mathrm{bu}^\infty$ versus $\tilde{\tau}$, and the solid curves correspond to $\Delta I_\mathrm{d}/\Delta I_\mathrm{d}^0$ versus $\tilde{\tau}-\tilde{\tau}_\mathrm{bu}$, for four dissipating light curves with scaled build-up times given by $\tilde{\tau}_\mathrm{bu}=0.15,1.5,6$ (in black) and $\tilde{\tau}_\mathrm{bu}=30$ (in red). Vertical colored straight lines mark the instants $\tilde{\tau}~\mathrm{or}~\tilde{\tau}-\tilde{\tau}_\mathrm{bu}=0,0.2,1$ (same color-code as in Fig.~\ref{hydroplot1}). All light curves are from Star 2 and $\Sigma_0=1.37\,\mathrm{g\, cm^{-2}}$.}
        \label{visual_events1}
\end{figure*}

Figure~\ref{visual_events1} shows examples of $I$-band light curves from our grid (see Table~\ref{star_and_disc} and Sects.~\ref{radiative_transfer} and \ref{emplaw_sect}). The dashed black curves correspond to $\Delta I_\mathrm{bu}/\Delta I_\mathrm{bu}^\infty$ versus $\tilde{\tau}$. 
The solid curves correspond to $\Delta I_\mathrm{d}/\Delta I_\mathrm{d}^0$ versus $\tilde{\tau}-\tilde{\tau}_\mathrm{bu}$ 
for four dissipating light curves with increasing scaled build-up times. 
Since $\Delta I_\mathrm{bu}^\infty$ is the limiting magnitude of the build-up light curves and $\Delta I_\mathrm{d}^0$ is the magnitude at the instant of the beginning of dissipation, it follows that all $\Delta I_\mathrm{bu}/\Delta I_\mathrm{bu}^\infty$ curves go from 0 to 1 and all $\Delta I_\mathrm{d}/\Delta I_\mathrm{d}^0$ curves go from 1 to 0. 

The build-up and dissipation light curves of the edge-on (upper-left panel) and nearly-pole-on cases (right panels) can be approximated by the mathematical formulae given by Eqs.~\ref{genpower_bu} and \ref{genpower_diss}, respectively. The light curves at intermediate angles like the ones of the middle-left and lower-left panels show more complex forms that cannot be described by Eqs.~\ref{genpower_bu} and \ref{genpower_diss}.
The light curves show that, at $\tilde{\tau}\approx 1$, the simulated bump has reached a significant fraction of its limiting value, and, at $\tilde{\tau}-\tilde{\tau}_\mathrm{bu}\approx 1$, the bump has already fallen considerably from its previous magnitude before the beginning of the dissipation.

Furthermore, dissipating curves with larger scaled build-up times dissipate at slower rates, as a result of the mass reservoir effect (Sect.~\ref{reservoir_effect}). Also, as discussed in Sect.~\ref{emplaw_sect},  both the dissipation and growth rates depend on the viewing angle.

%


\section{Online figures}
\label{Appendix3}

\begin{figure*}
\centering{
\includegraphics[width=1.0\linewidth]{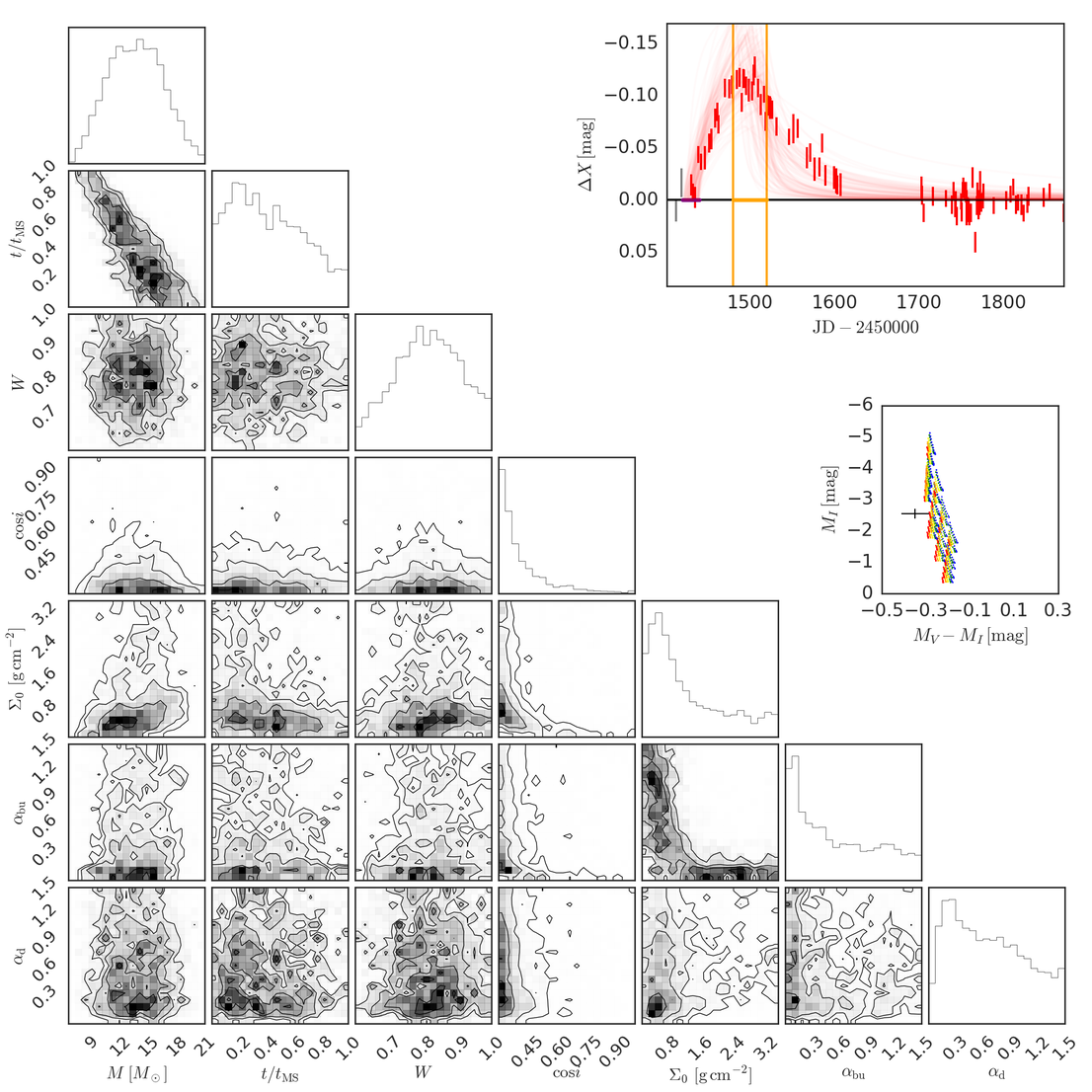}
}
\caption[]
{
Same as Fig.~\ref{example_bb1} for SMC\_SC1 7612 and bump ID 01. 
}
\label{smc_sc1_7612_01}
\end{figure*}
\clearpage

\begin{figure*}
\centering{
\includegraphics[width=1.0\linewidth]{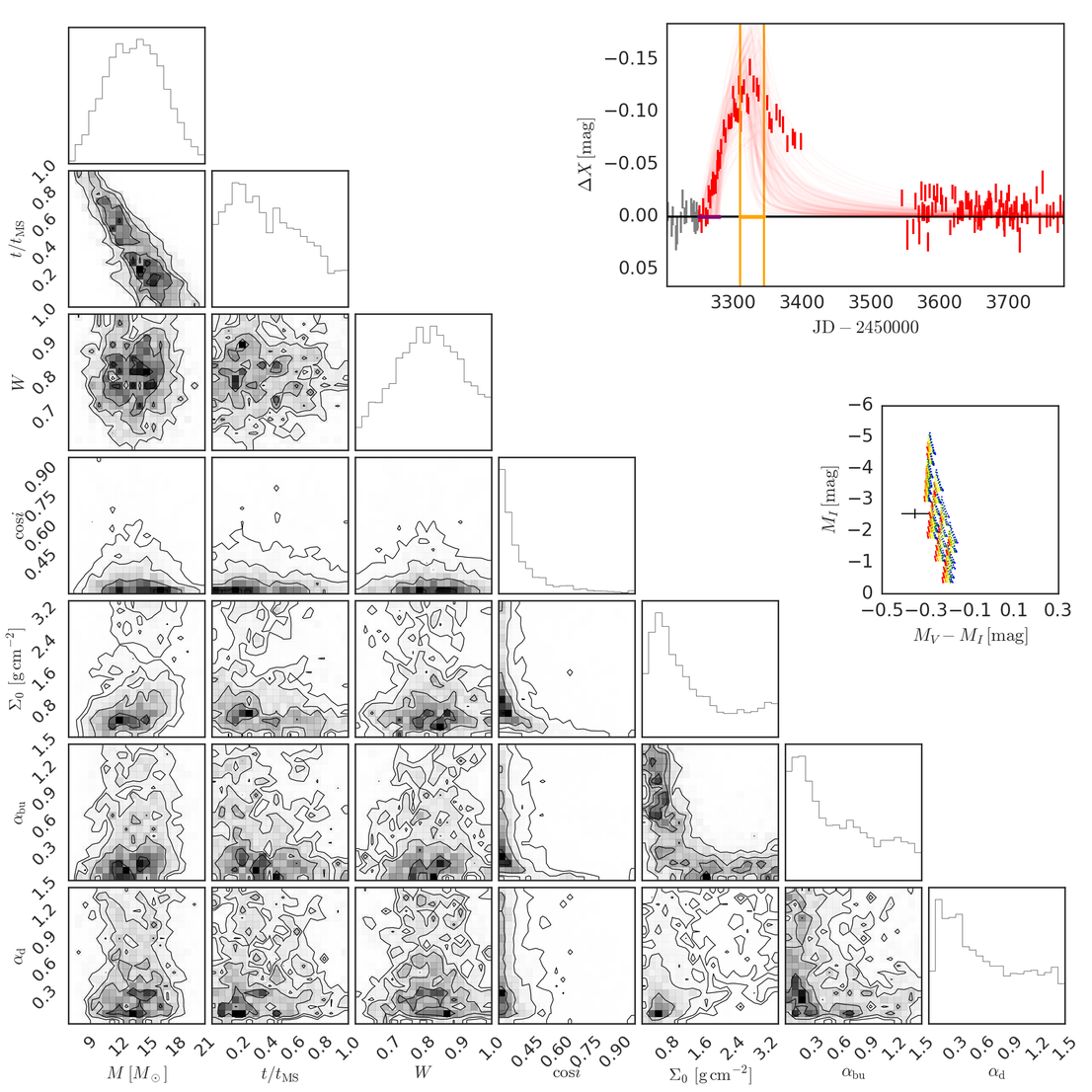}
}
\caption[]
{
Same as Fig.~\ref{example_bb1} for SMC\_SC1 7612 and bump ID 02. 
}
\label{smc_sc1_7612_02}
\end{figure*}
\clearpage

\begin{figure*}
\centering{
\includegraphics[width=1.0\linewidth]{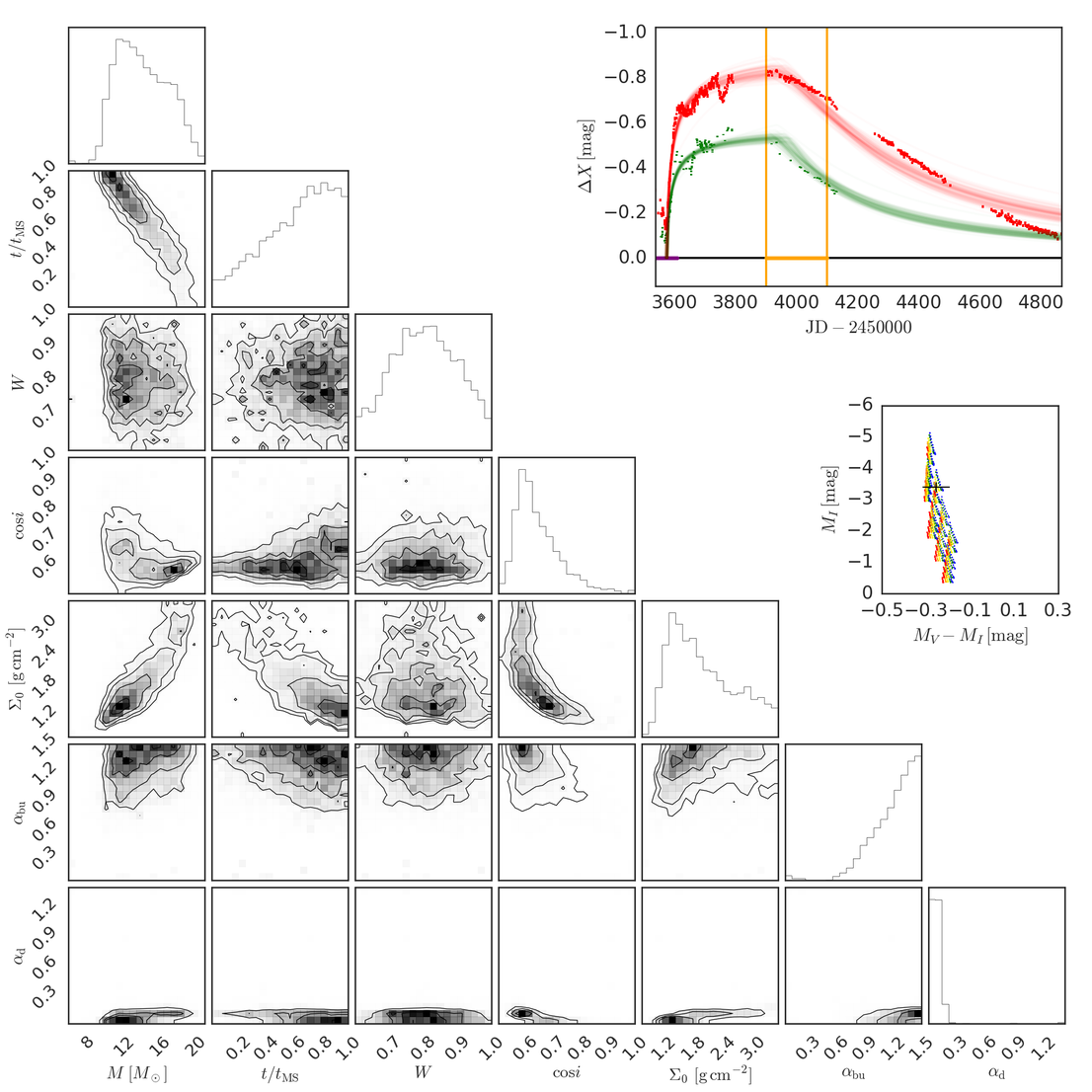}
}
\caption[]
{
Same as Fig.~\ref{example_bb1} for SMC\_SC1 60553 and bump ID 01. 
}
\label{smc_sc1_60553_01}
\end{figure*}
\clearpage


%
\begin{figure*}
\centering{
\includegraphics[width=1.0\linewidth]{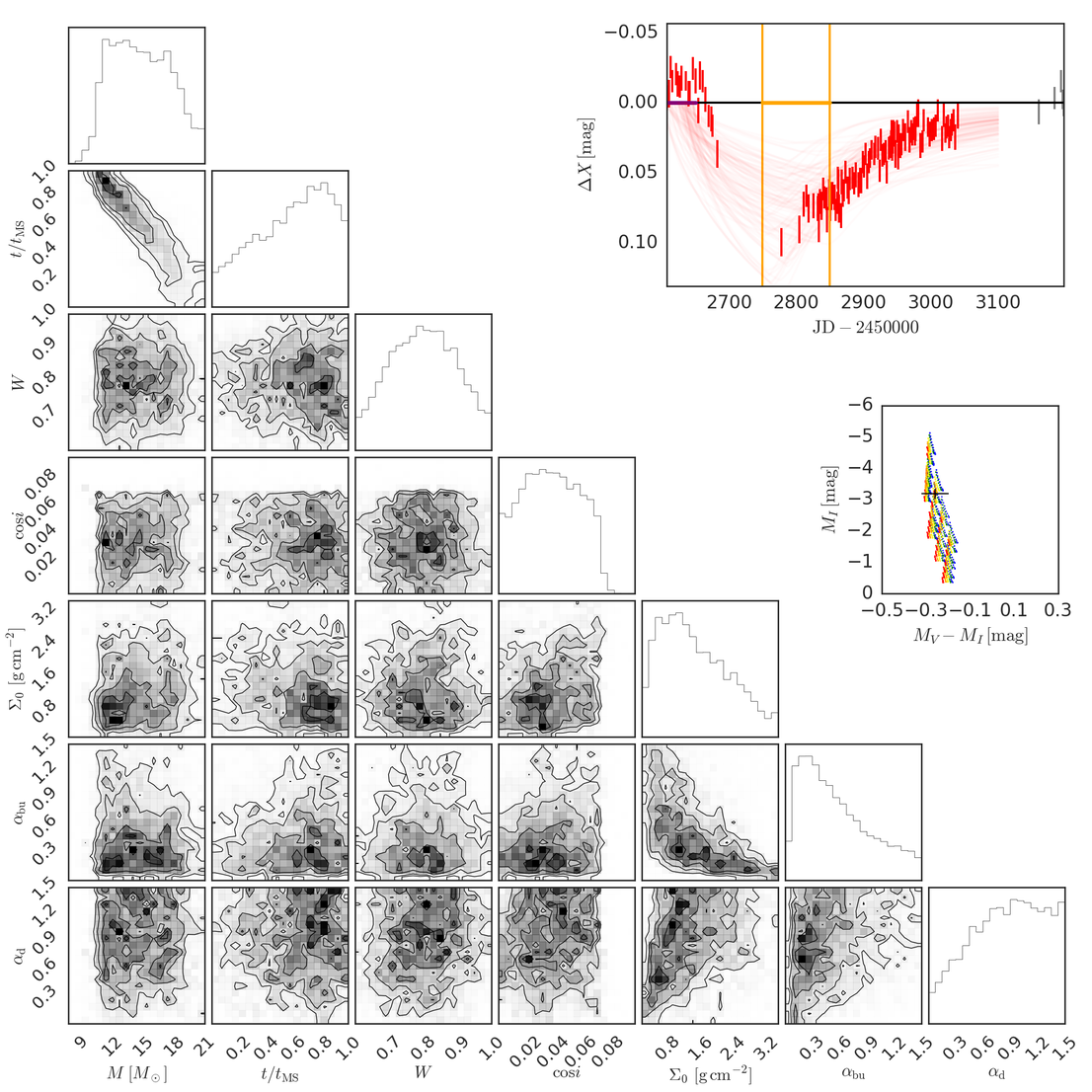}
}
\caption[]
{
Same as Fig.~\ref{example_bb1} for SMC\_SC1 92262 and bump ID 01. 
}
\label{smc_sc1_92262_01}
\end{figure*}
\clearpage

\begin{figure*}
\centering{
\includegraphics[width=1.0\linewidth]{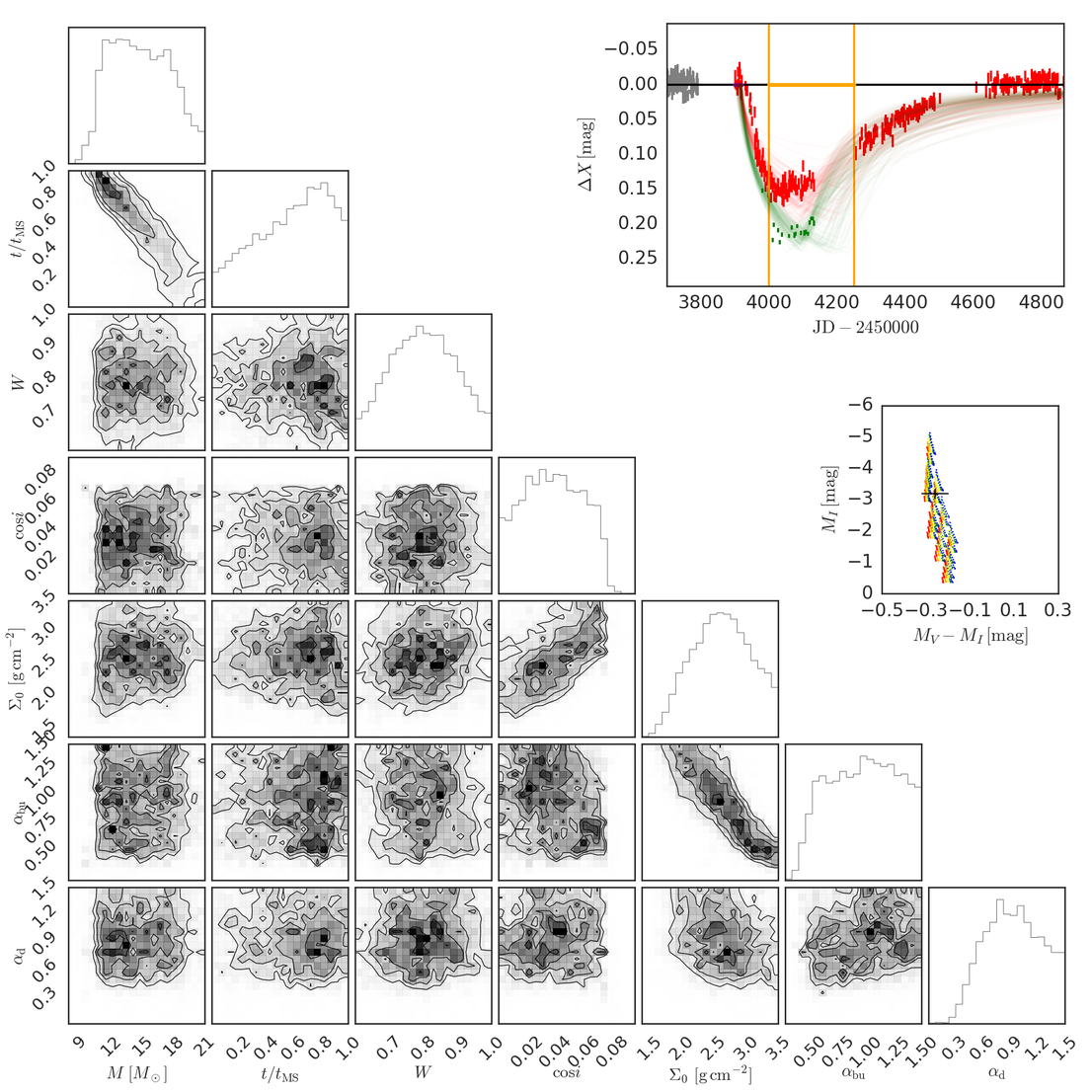}
}
\caption[]
{
Same as Fig.~\ref{example_bb1} for SMC\_SC1 92262 and bump ID 02. 
}
\label{smc_sc1_92262_02}
\end{figure*}
\clearpage

\begin{figure*}
\centering{
\includegraphics[width=1.0\linewidth]{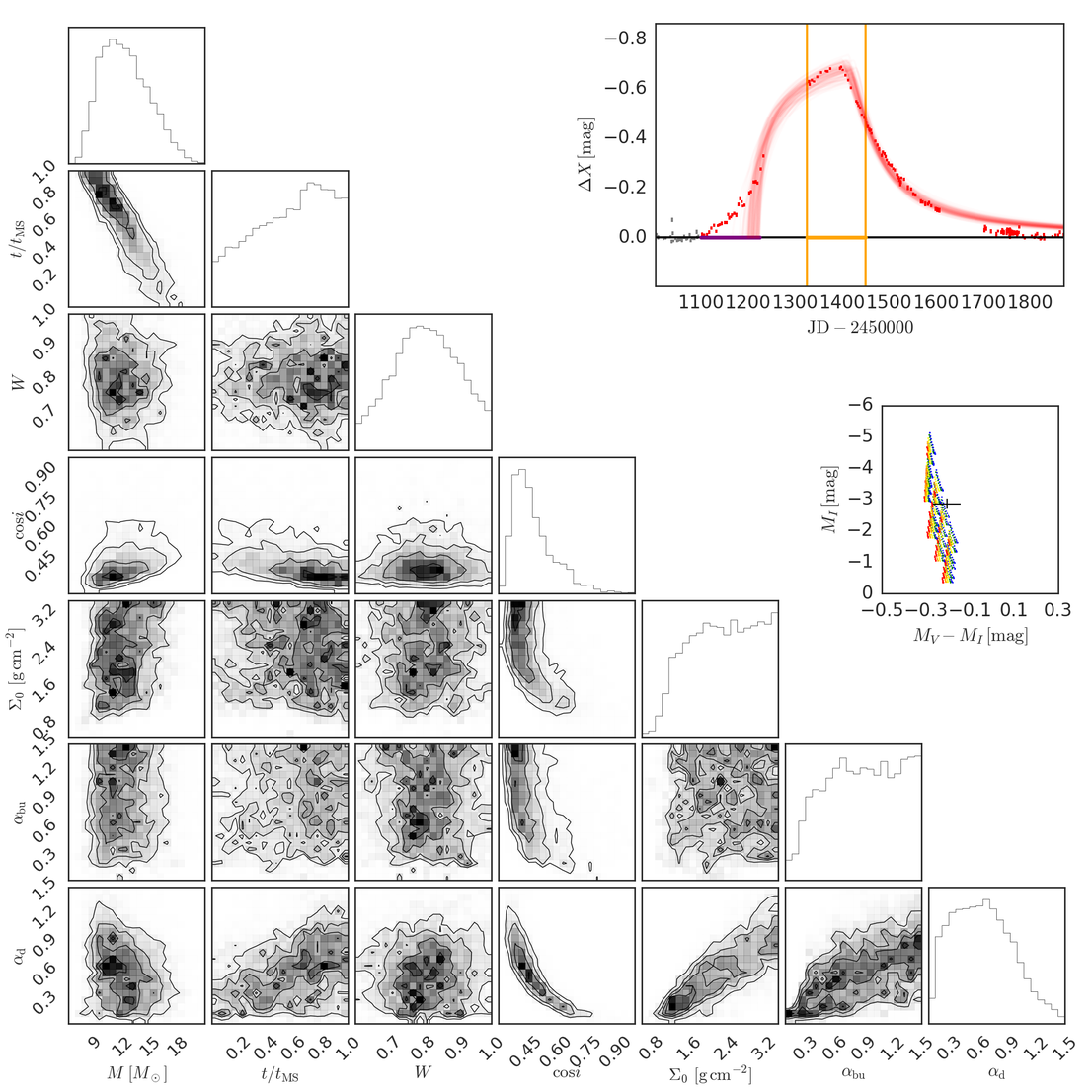}
}
\caption[]
{
Same as Fig.~\ref{example_bb1} for SMC\_SC2 94939 and bump ID 01. 
}
\label{smc_sc2_94939_01}
\end{figure*}
\clearpage

\begin{figure*}
\centering{
\includegraphics[width=1.0\linewidth]{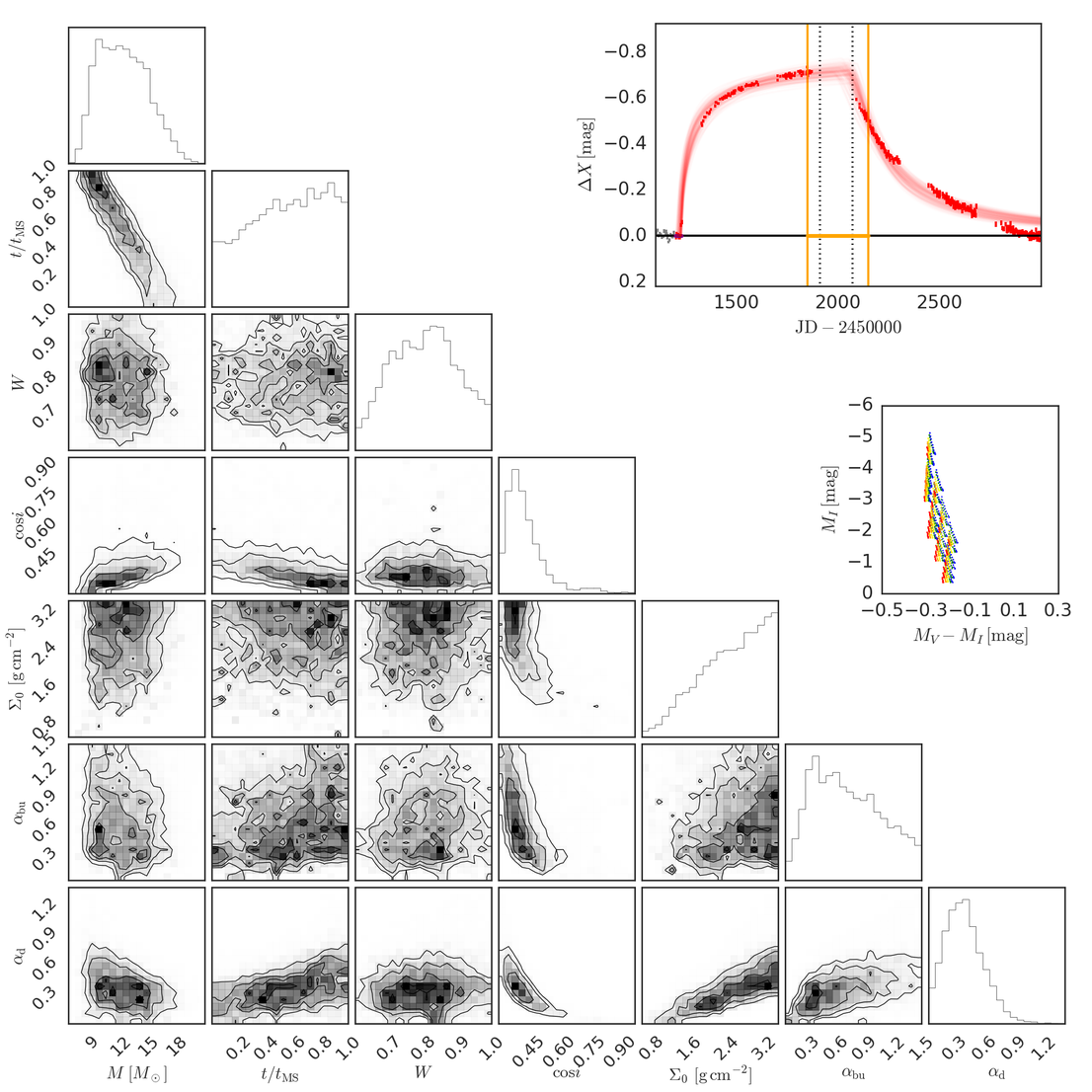}
}
\caption[]
{
Same as Fig.~\ref{example_bb1} for SMC\_SC3 5719 and bump ID 01. 
}
\label{smc_sc3_5719_01}
\end{figure*}
\clearpage

\begin{figure*}
\centering{
\includegraphics[width=1.0\linewidth]{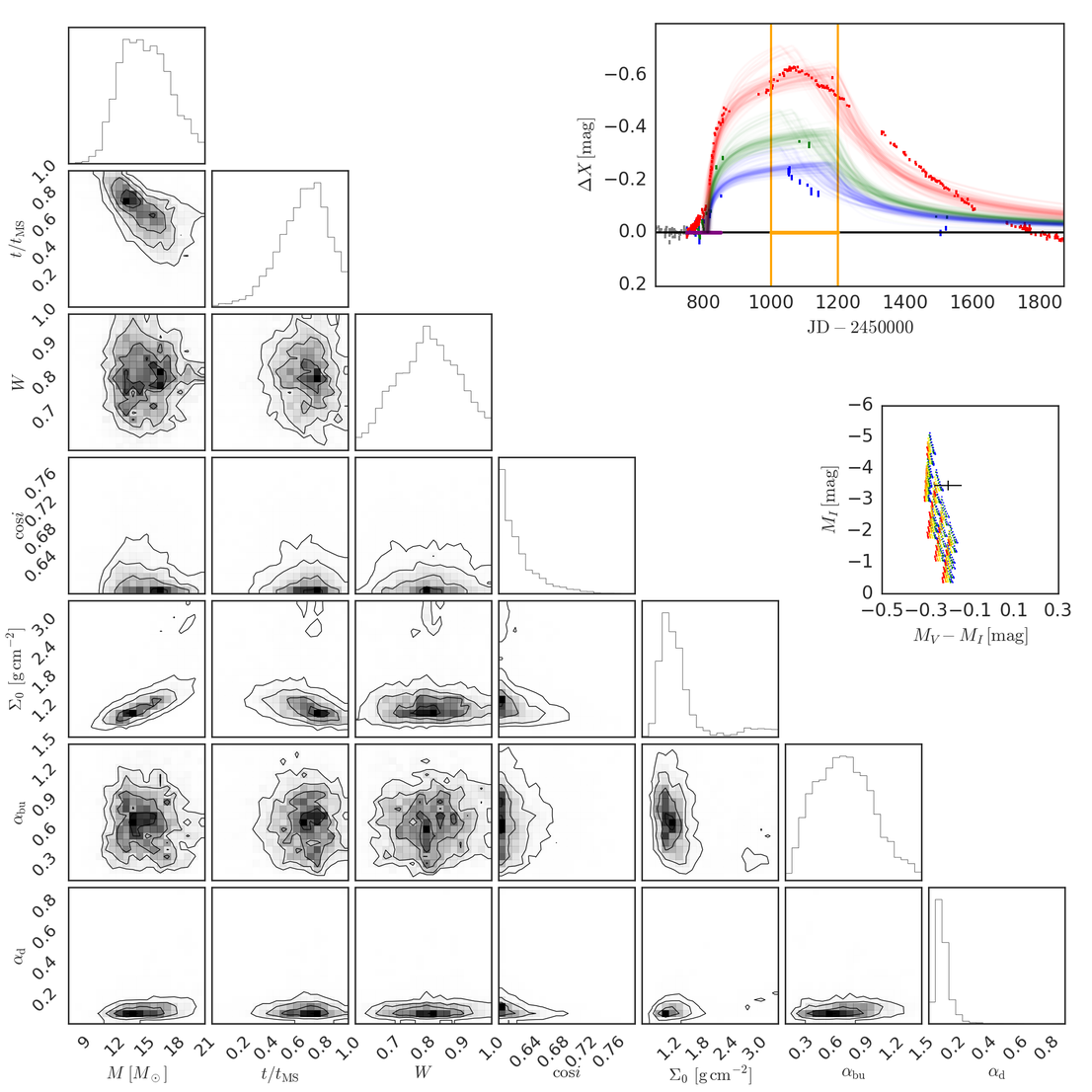}
}
\caption[]
{
Same as Fig.~\ref{example_bb1} for SMC\_SC3 15970 and bump ID 01. 
}
\label{smc_sc3_15970_01}
\end{figure*}
\clearpage

\begin{figure*}
\centering{
\includegraphics[width=1.0\linewidth]{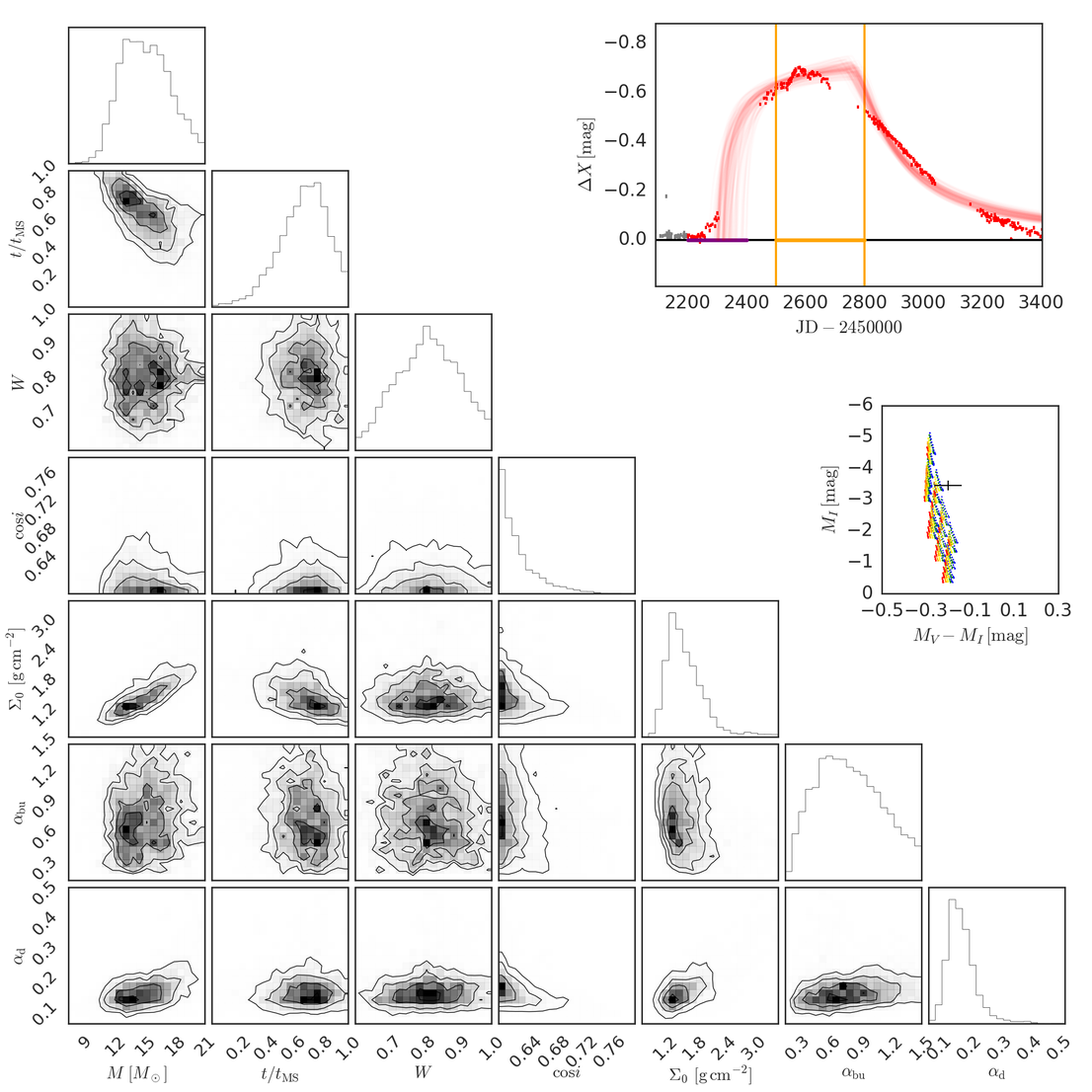}
}
\caption[]
{
Same as Fig.~\ref{example_bb1} for SMC\_SC3 15970 and bump ID 02. 
}
\label{smc_sc3_15970_02}
\end{figure*}
\clearpage

\begin{figure*}
\centering{
\includegraphics[width=1.0\linewidth]{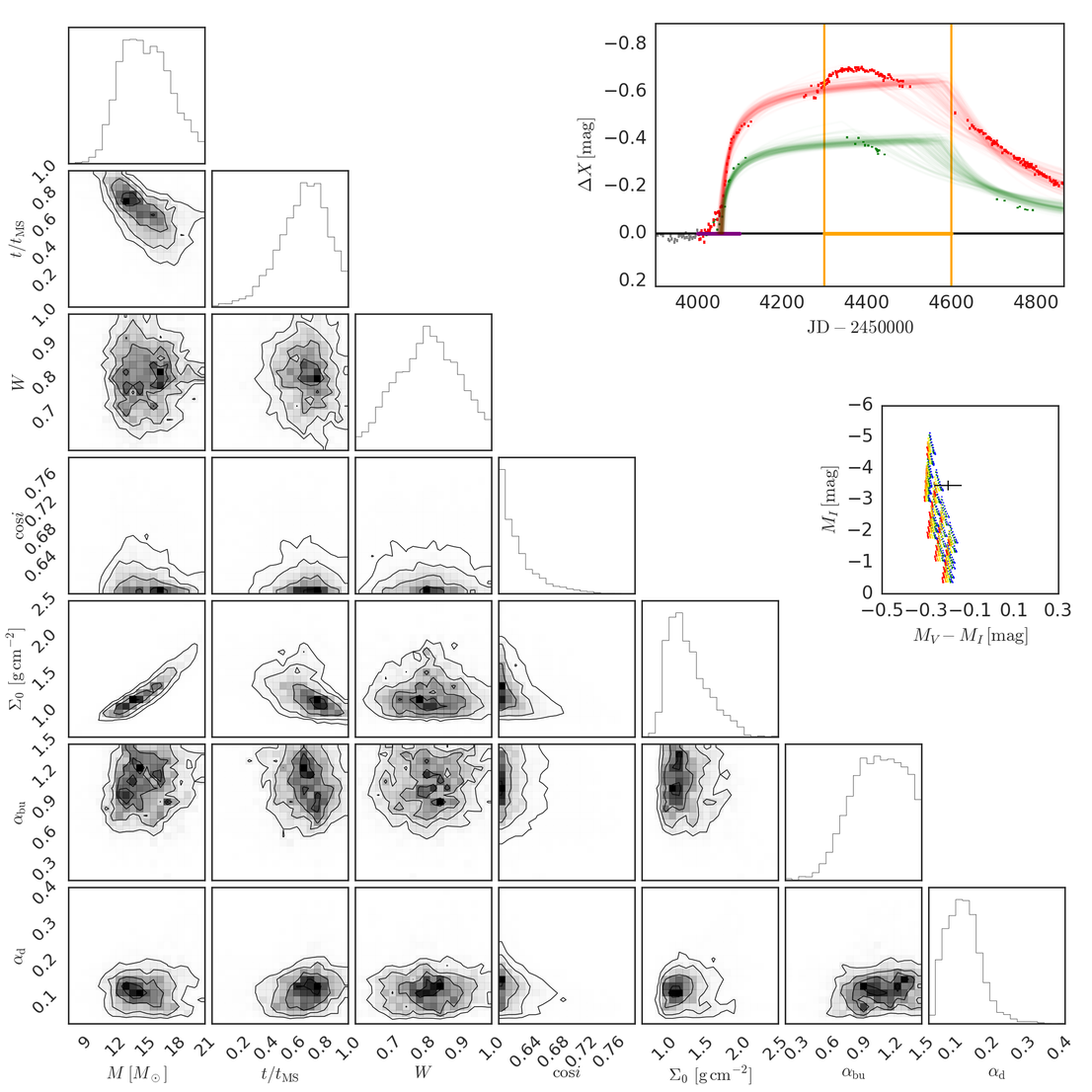}
}
\caption[]
{
Same as Fig.~\ref{example_bb1} for SMC\_SC3 15970 and bump ID 03. 
}
\label{smc_sc3_15970_03}
\end{figure*}
\clearpage

\begin{figure*}
\centering{
\includegraphics[width=1.0\linewidth]{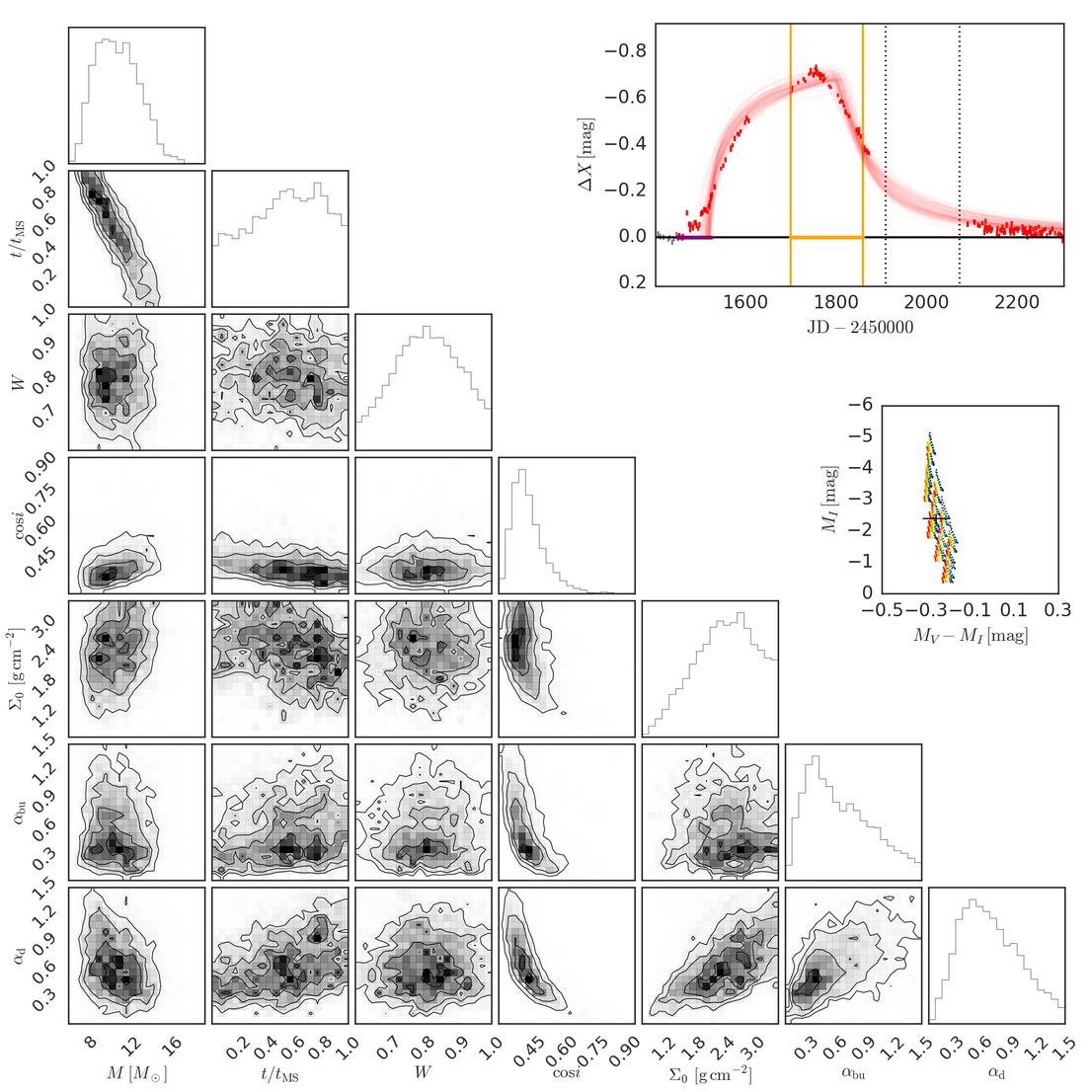}
}
\caption[]
{
Same as Fig.~\ref{example_bb1} for SMC\_SC3 71445 and bump ID 01. 
}
\label{smc_sc3_71445_01}
\end{figure*}
\clearpage

\begin{figure*}
\centering{
\includegraphics[width=1.0\linewidth]{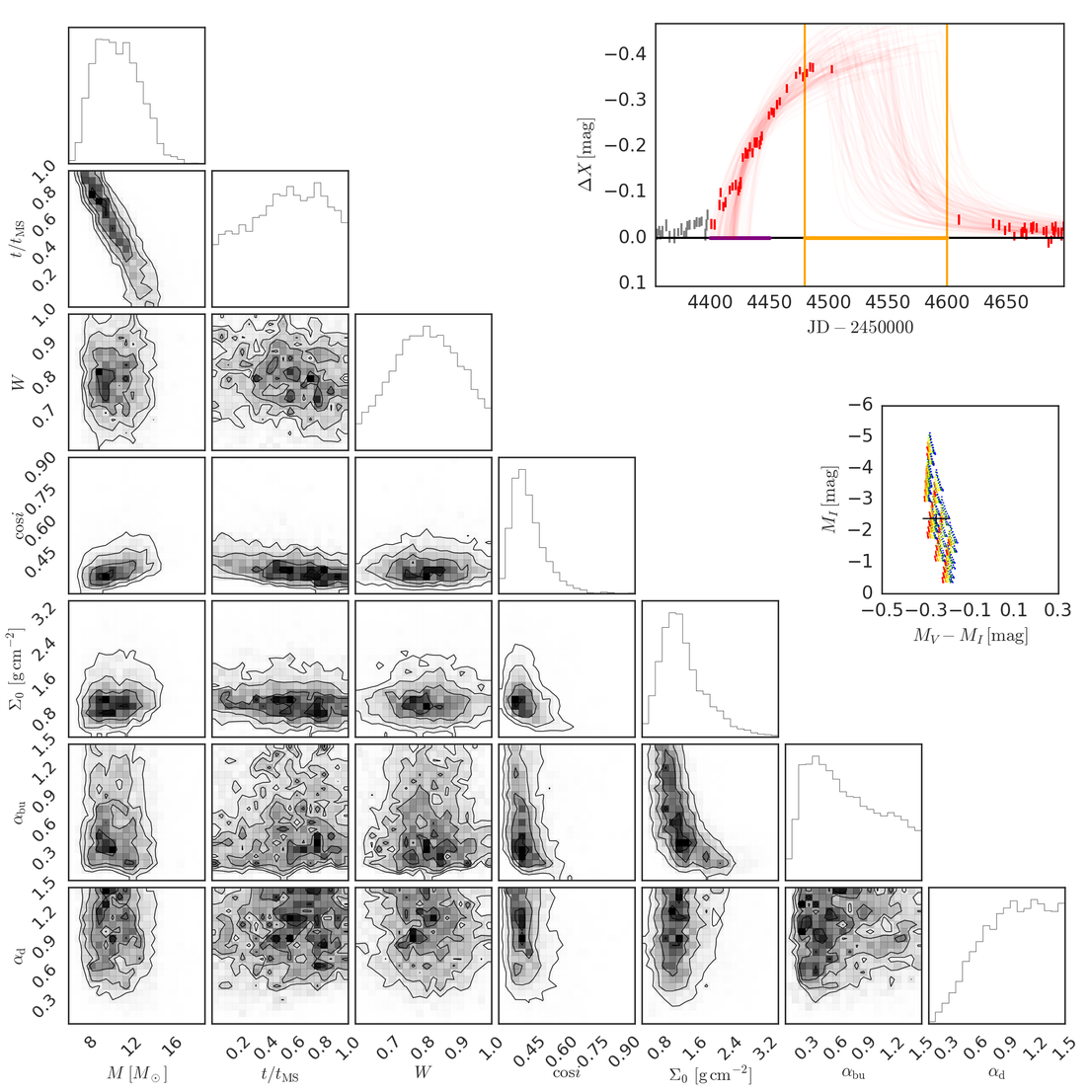}
}
\caption[]
{
Same as Fig.~\ref{example_bb1} for SMC\_SC3 71445 and bump ID 02. 
}
\label{smc_sc3_71445_02}
\end{figure*}
\clearpage

\begin{figure*}
\centering{
\includegraphics[width=1.0\linewidth]{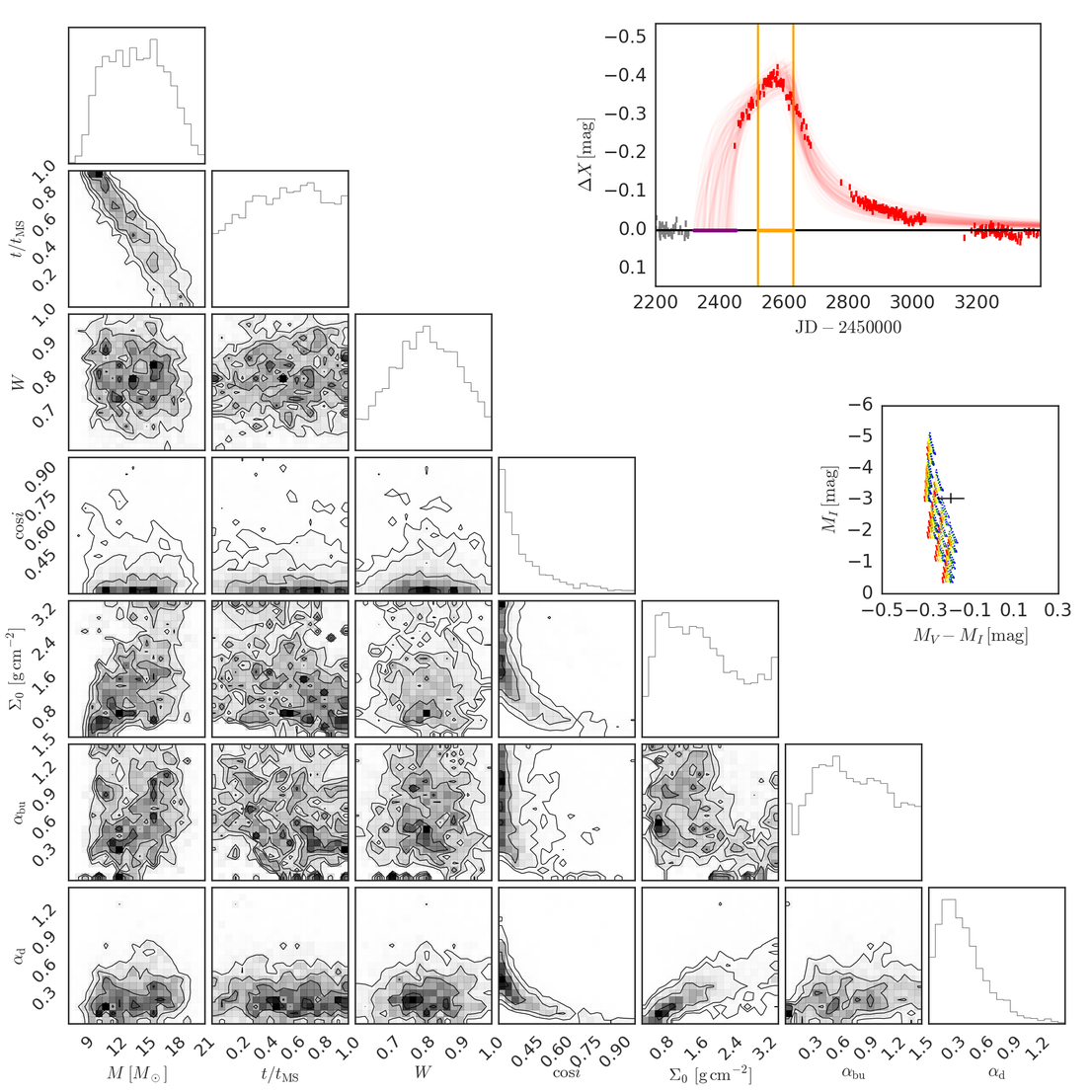}
}
\caption[]
{
Same as Fig.~\ref{example_bb1} for SMC\_SC3 125899 and bump ID 01. 
}
\label{smc_sc3_125899_01}
\end{figure*}
\clearpage

\begin{figure*}
\centering{
\includegraphics[width=1.0\linewidth]{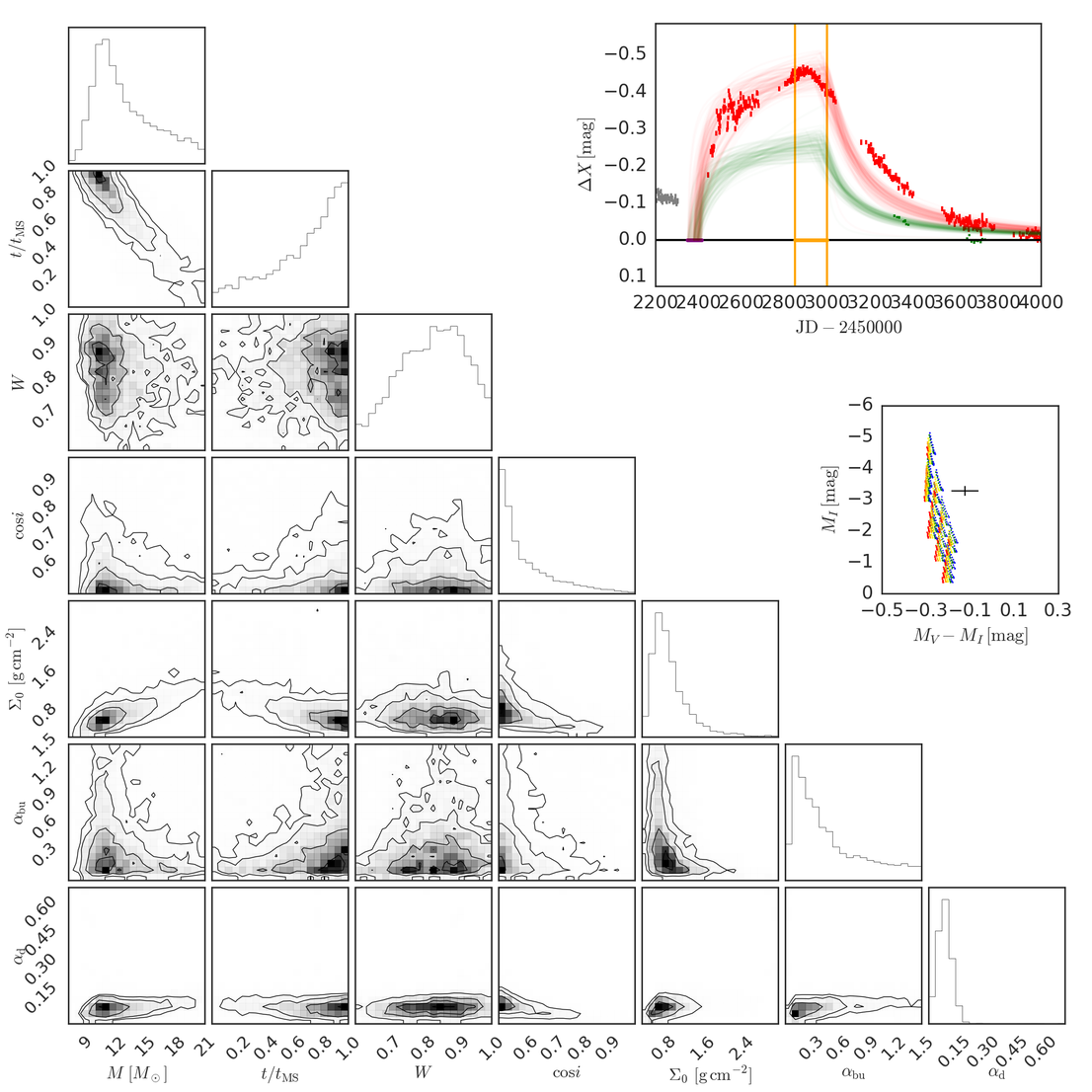}
}
\caption[]
{
Same as Fig.~\ref{example_bb1} for SMC\_SC3 197941 and bump ID 01. 
}
\label{smc_sc3_197941_01}
\end{figure*}
\clearpage

\begin{figure*}
\centering{
\includegraphics[width=1.0\linewidth]{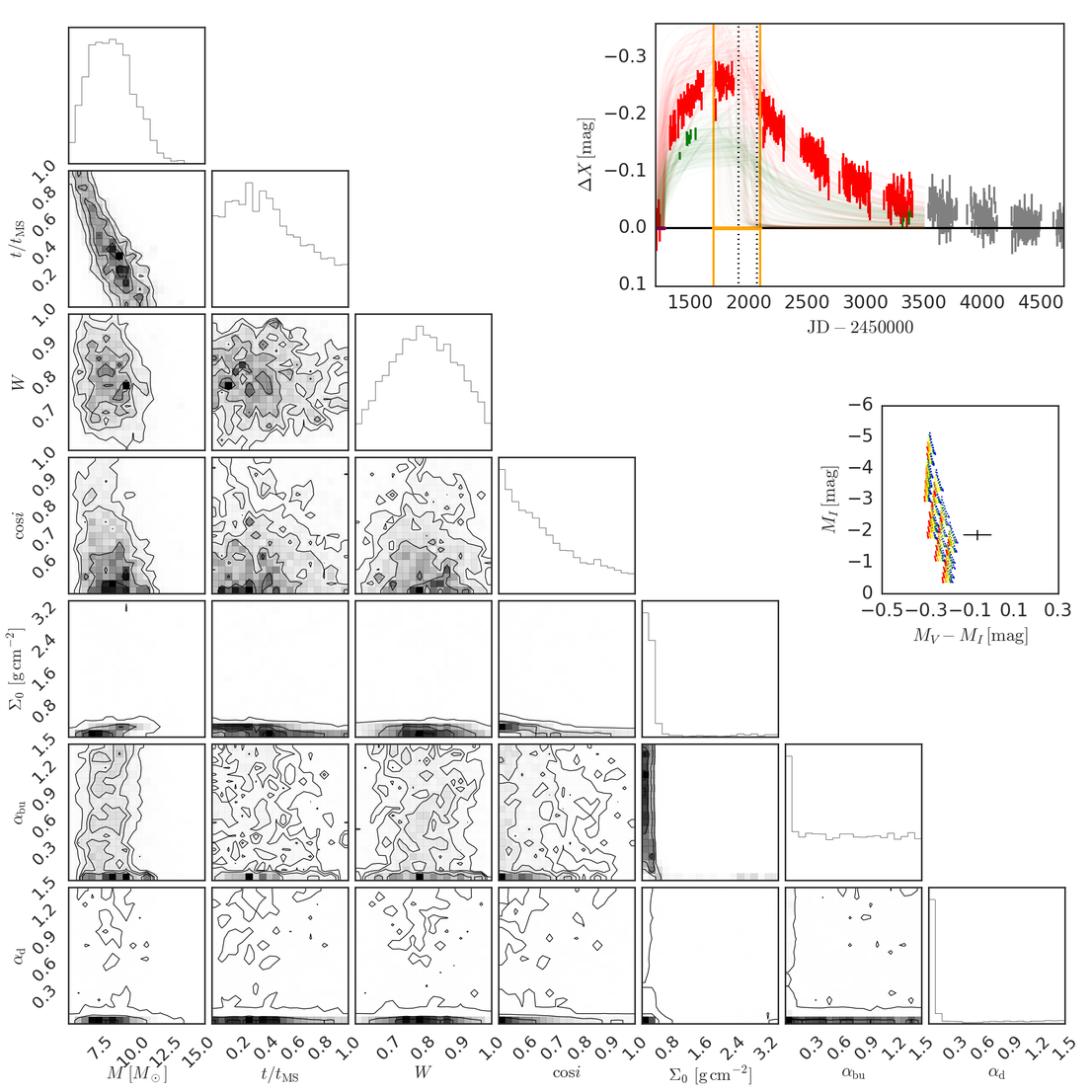}
}
\caption[]
{
Same as Fig.~\ref{example_bb1} for SMC\_SC4 22859 and bump ID 01. 
}
\label{smc_sc4_22859_01}
\end{figure*}
\clearpage

\begin{figure*}
\centering{
\includegraphics[width=1.0\linewidth]{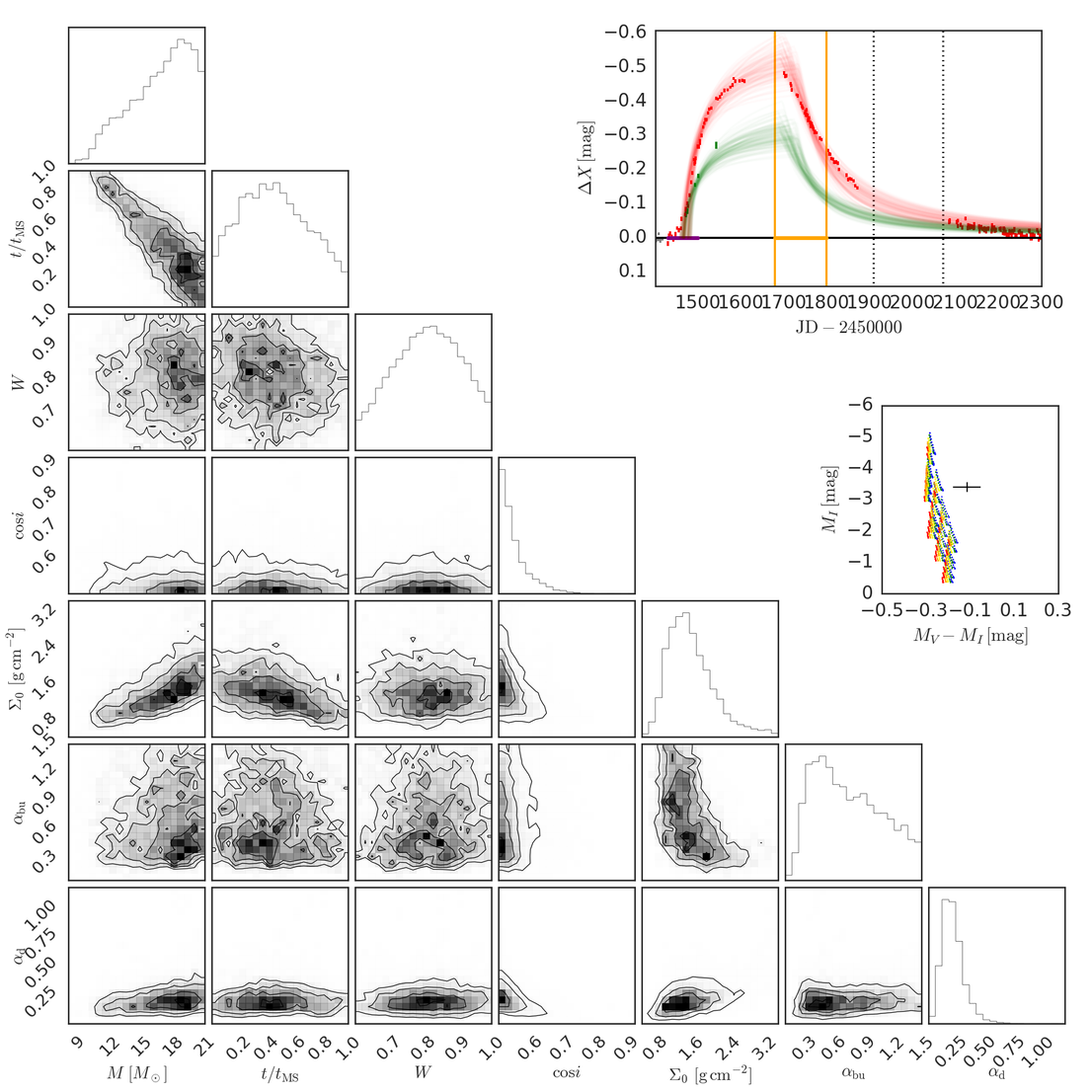}
}
\caption[]
{
Same as Fig.~\ref{example_bb1} for SMC\_SC4 71499 and bump ID 01. 
}
\label{smc_sc4_71499_01}
\end{figure*}
\clearpage

\begin{figure*}
\centering{
\includegraphics[width=1.0\linewidth]{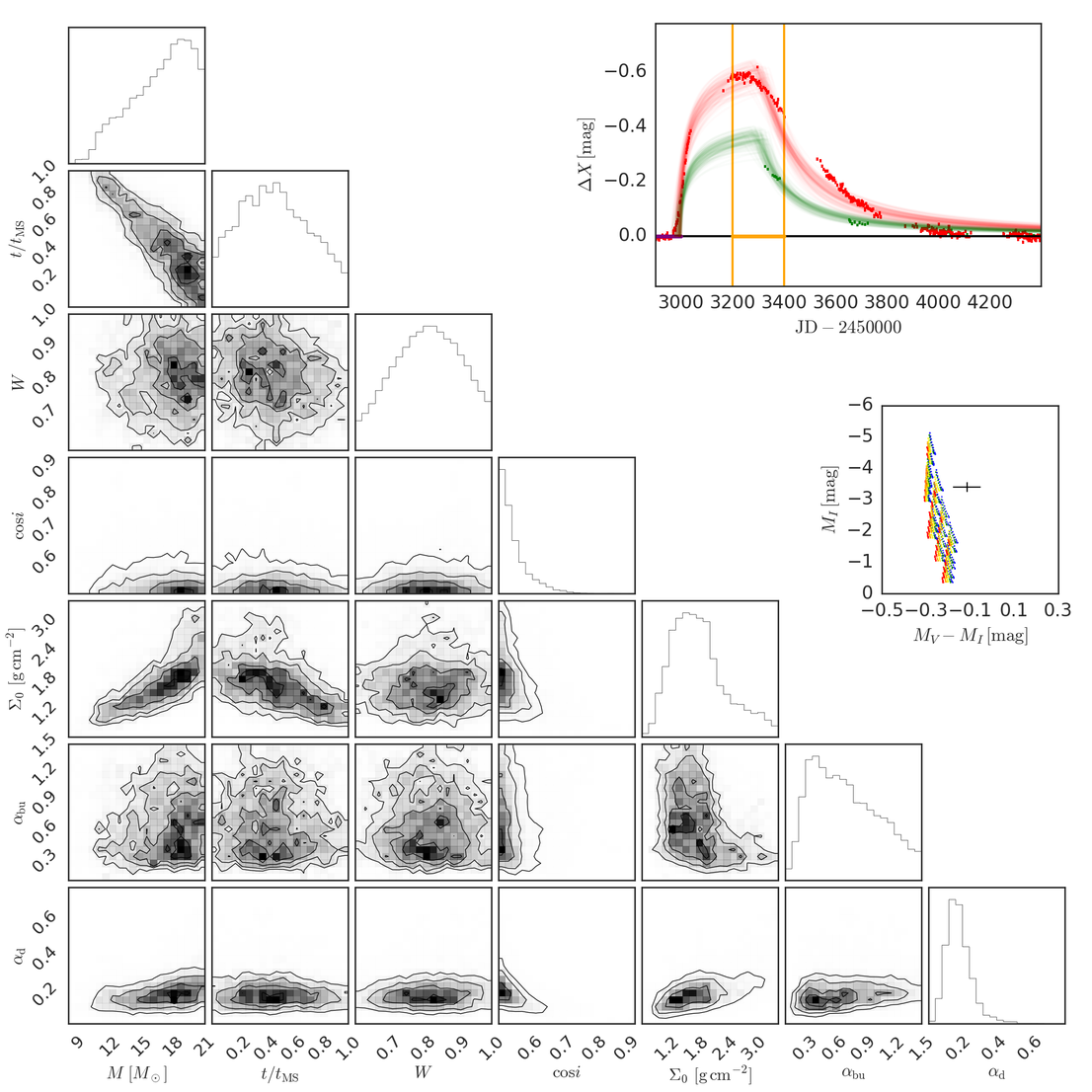}
}
\caption[]
{
Same as Fig.~\ref{example_bb1} for SMC\_SC4 71499 and bump ID 02. 
}
\label{smc_sc4_71499_02}
\end{figure*}
\clearpage

\begin{figure*}
\centering{
\includegraphics[width=1.0\linewidth]{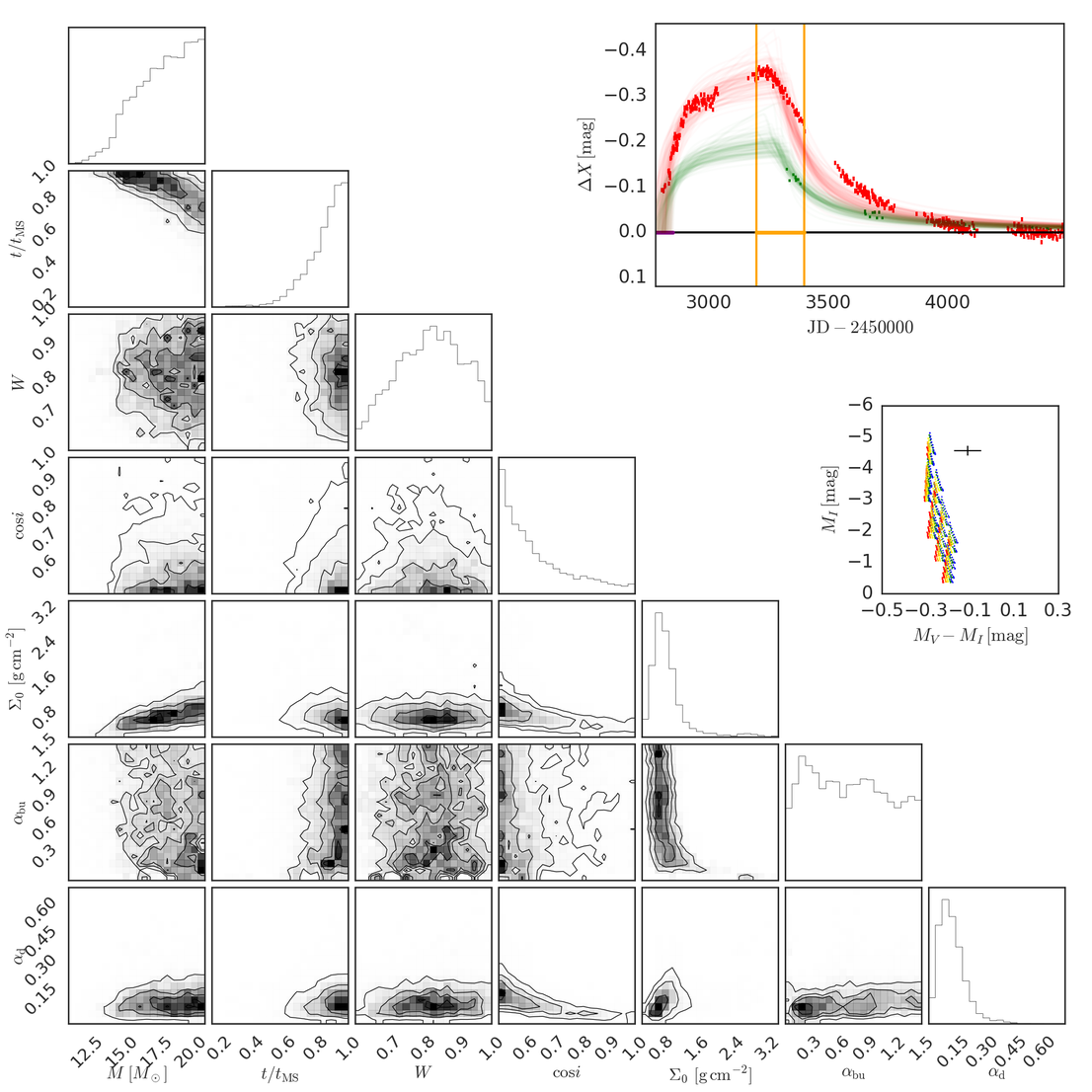}
}
\caption[]
{
Same as Fig.~\ref{example_bb1} for SMC\_SC4 120783 and bump ID 01. 
}
\label{smc_sc4_120783_01}
\end{figure*}
\clearpage

\begin{figure*}
\centering{
\includegraphics[width=1.0\linewidth]{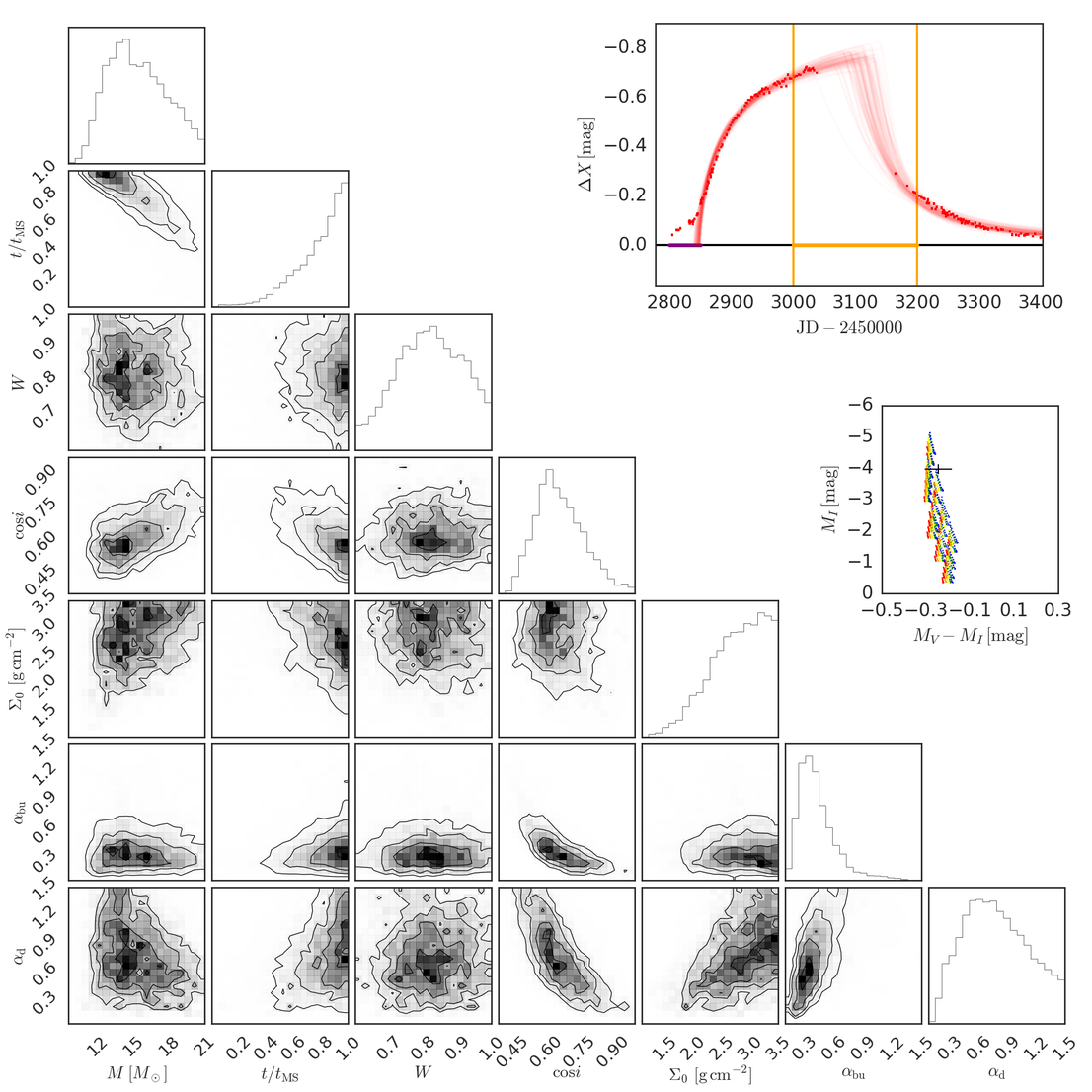}
}
\caption[]
{
Same as Fig.~\ref{example_bb1} for SMC\_SC4 127840 and bump ID 01. 
}
\label{smc_sc4_127840_01}
\end{figure*}
\clearpage

\begin{figure*}
\centering{
\includegraphics[width=1.0\linewidth]{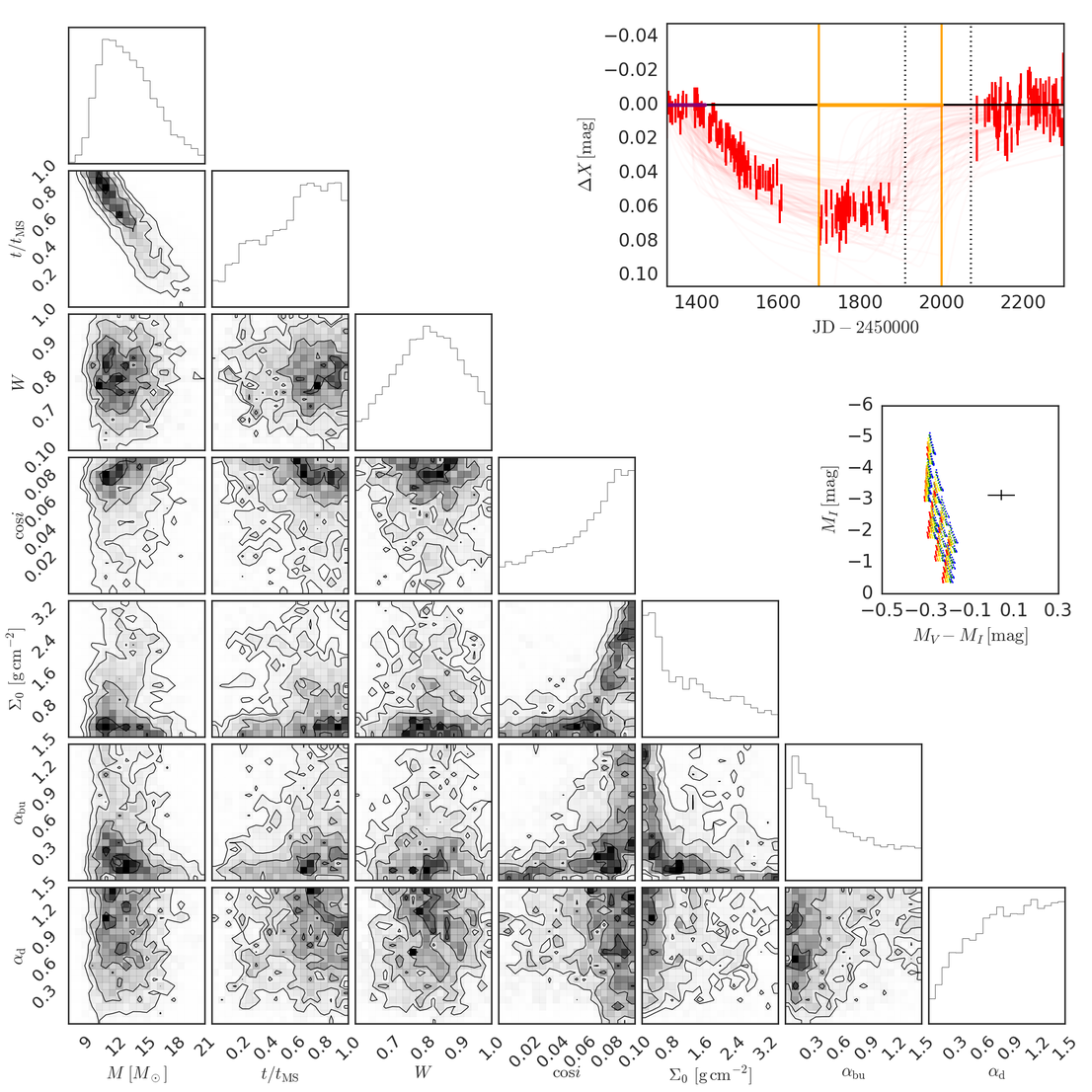}
}
\caption[]
{
Same as Fig.~\ref{example_bb1} for SMC\_SC4 156248 and bump ID 01. 
}
\label{smc_sc4_156248_01}
\end{figure*}
\clearpage

\begin{figure*}
\centering{
\includegraphics[width=1.0\linewidth]{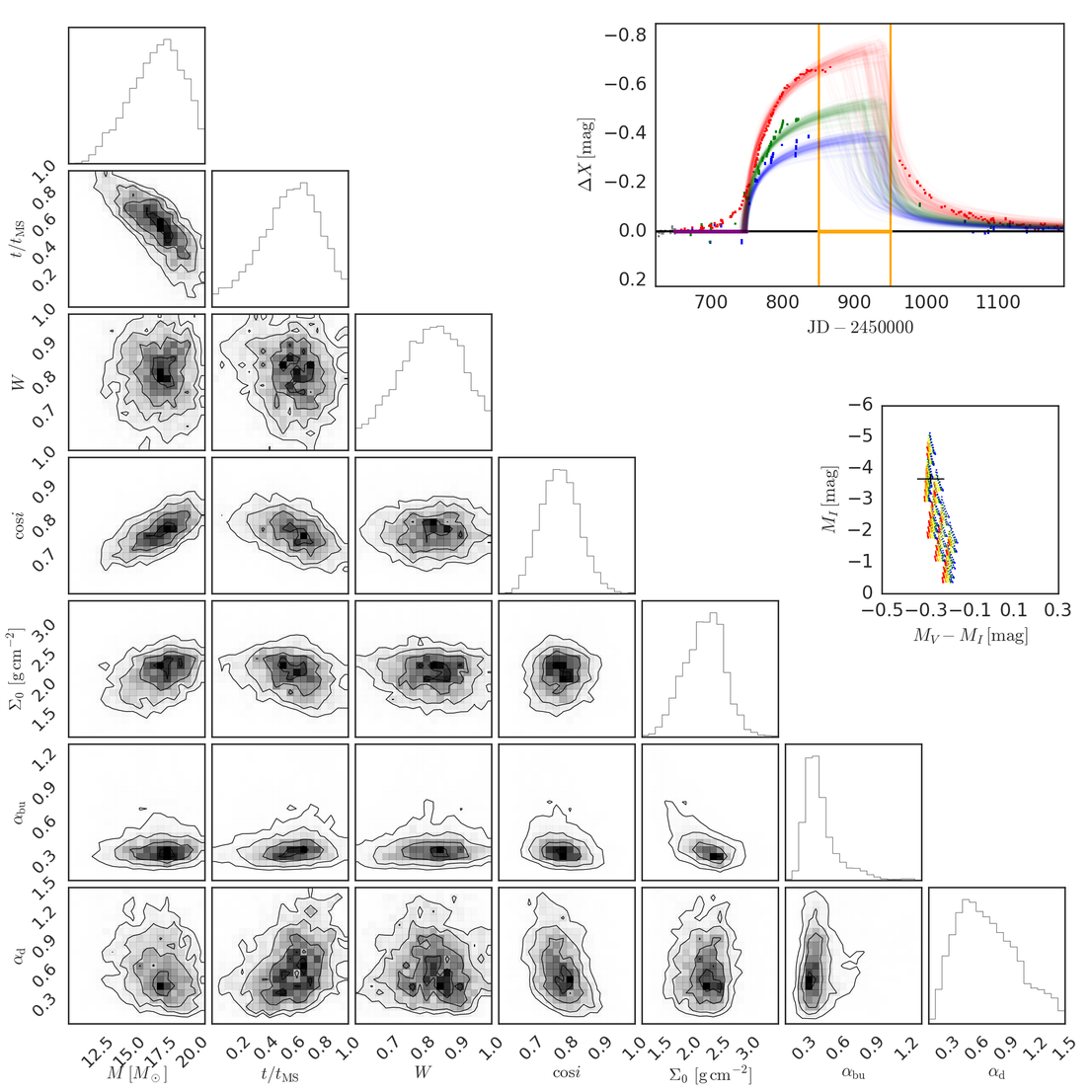}
}
\caption[]
{
Same as Fig.~\ref{example_bb1} for SMC\_SC4 156251 and bump ID 01. 
}
\label{smc_sc4_156251_01}
\end{figure*}
\clearpage

\begin{figure*}
\centering{
\includegraphics[width=1.0\linewidth]{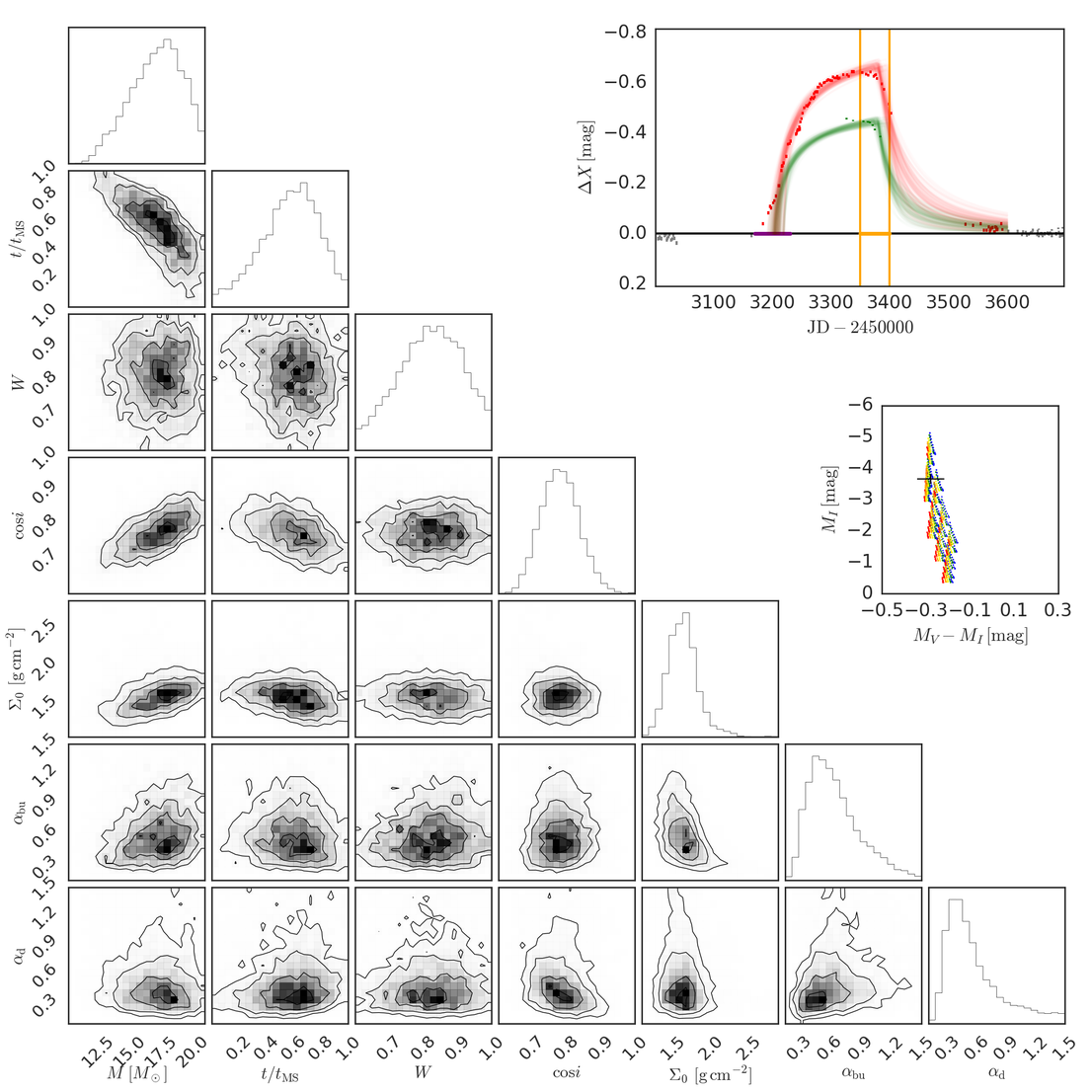}
}
\caption[]
{
Same as Fig.~\ref{example_bb1} for SMC\_SC4 156251 and bump ID 02. 
}
\label{smc_sc4_156251_02}
\end{figure*}
\clearpage

\begin{figure*}
\centering{
\includegraphics[width=1.0\linewidth]{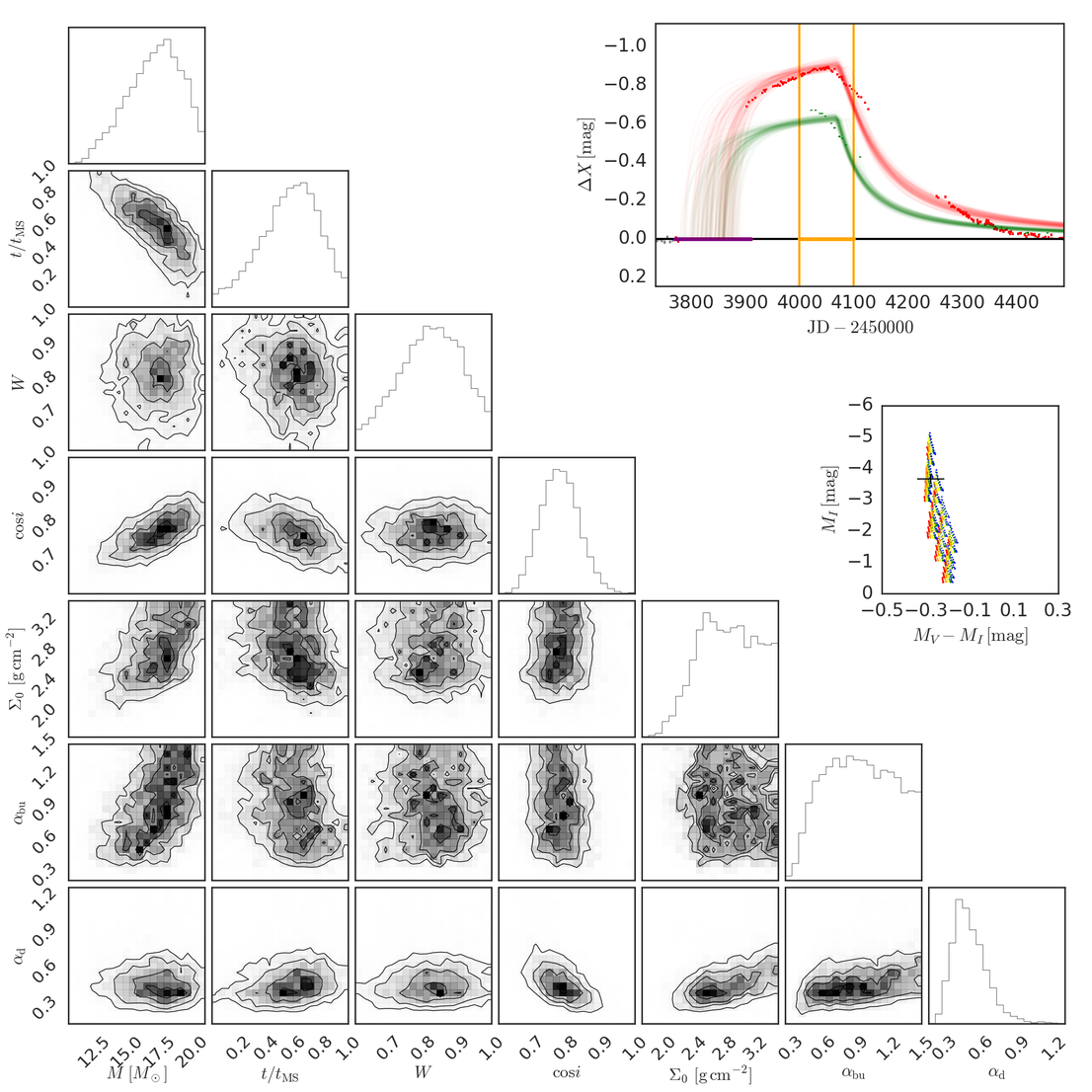}
}
\caption[]
{
Same as Fig.~\ref{example_bb1} for SMC\_SC4 156251 and bump ID 03. 
}
\label{smc_sc4_156251_03}
\end{figure*}
\clearpage

\begin{figure*}
\centering{
\includegraphics[width=1.0\linewidth]{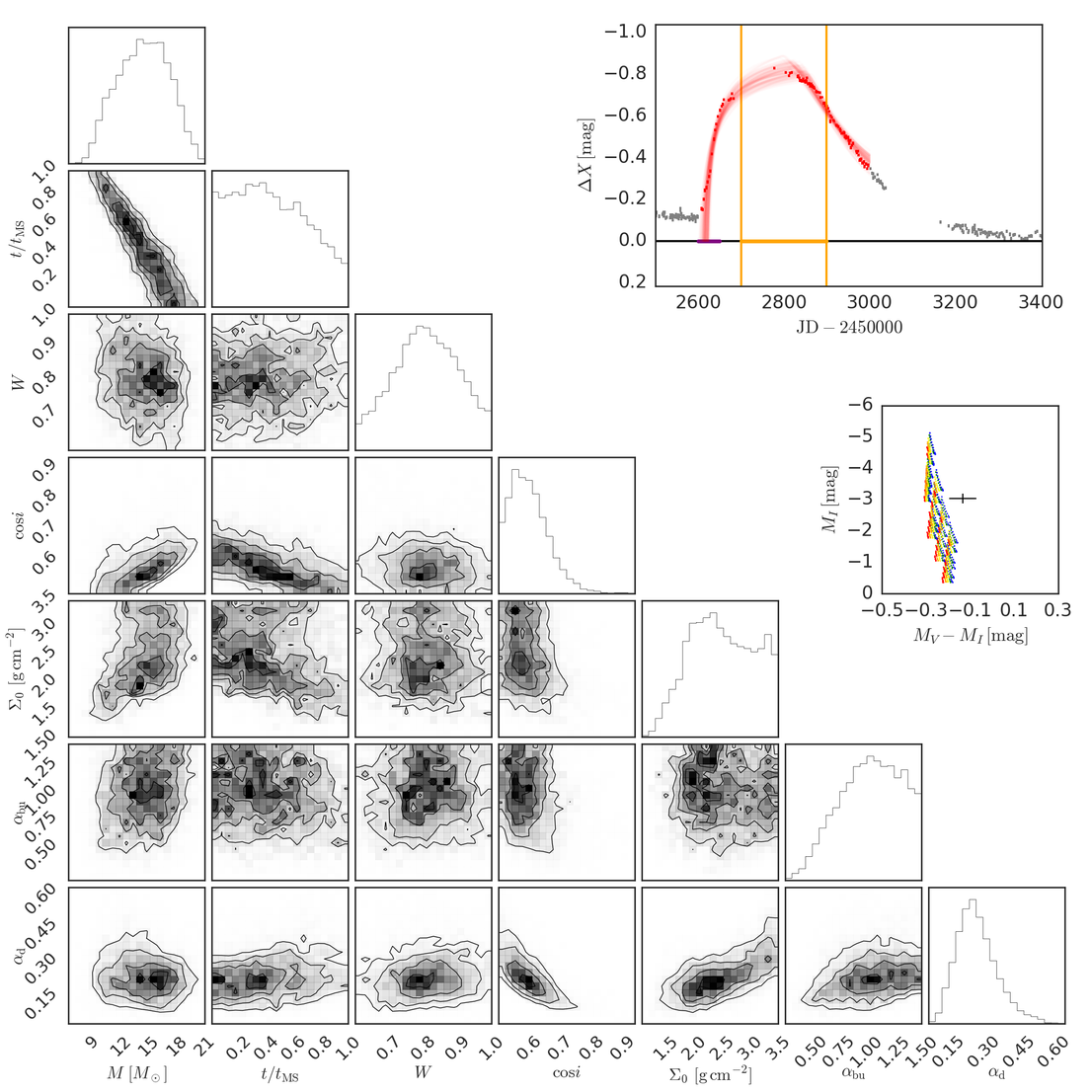}
}
\caption[]
{
Same as Fig.~\ref{example_bb1} for SMC\_SC4 159829 and bump ID 01. 
}
\label{smc_sc4_159829_01}
\end{figure*}
\clearpage

\begin{figure*}
\centering{
\includegraphics[width=1.0\linewidth]{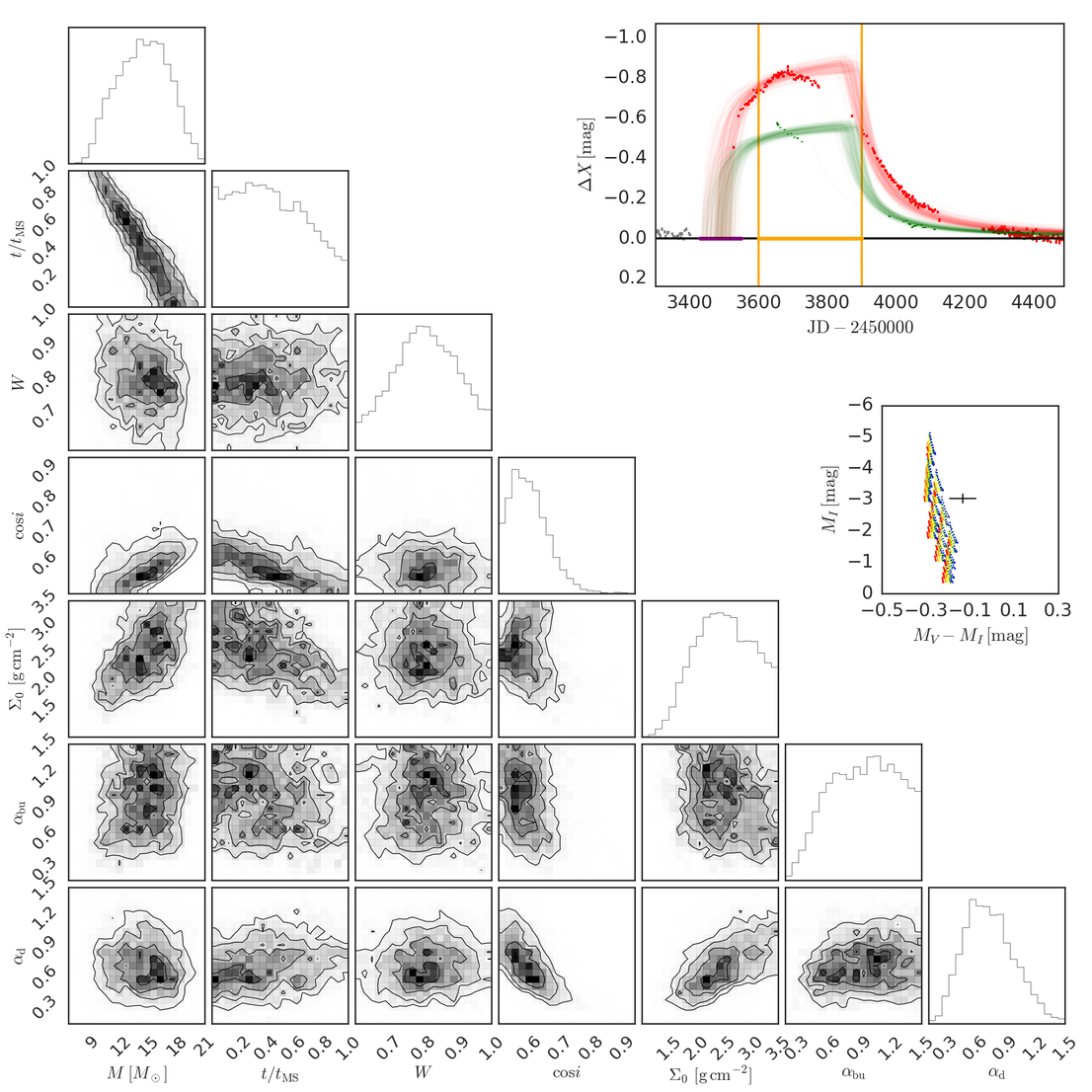}
}
\caption[]
{
Same as Fig.~\ref{example_bb1} for SMC\_SC4 159829 and bump ID 02. 
}
\label{smc_sc4_159829_02}
\end{figure*}
\clearpage

\begin{figure*}
\centering{
\includegraphics[width=1.0\linewidth]{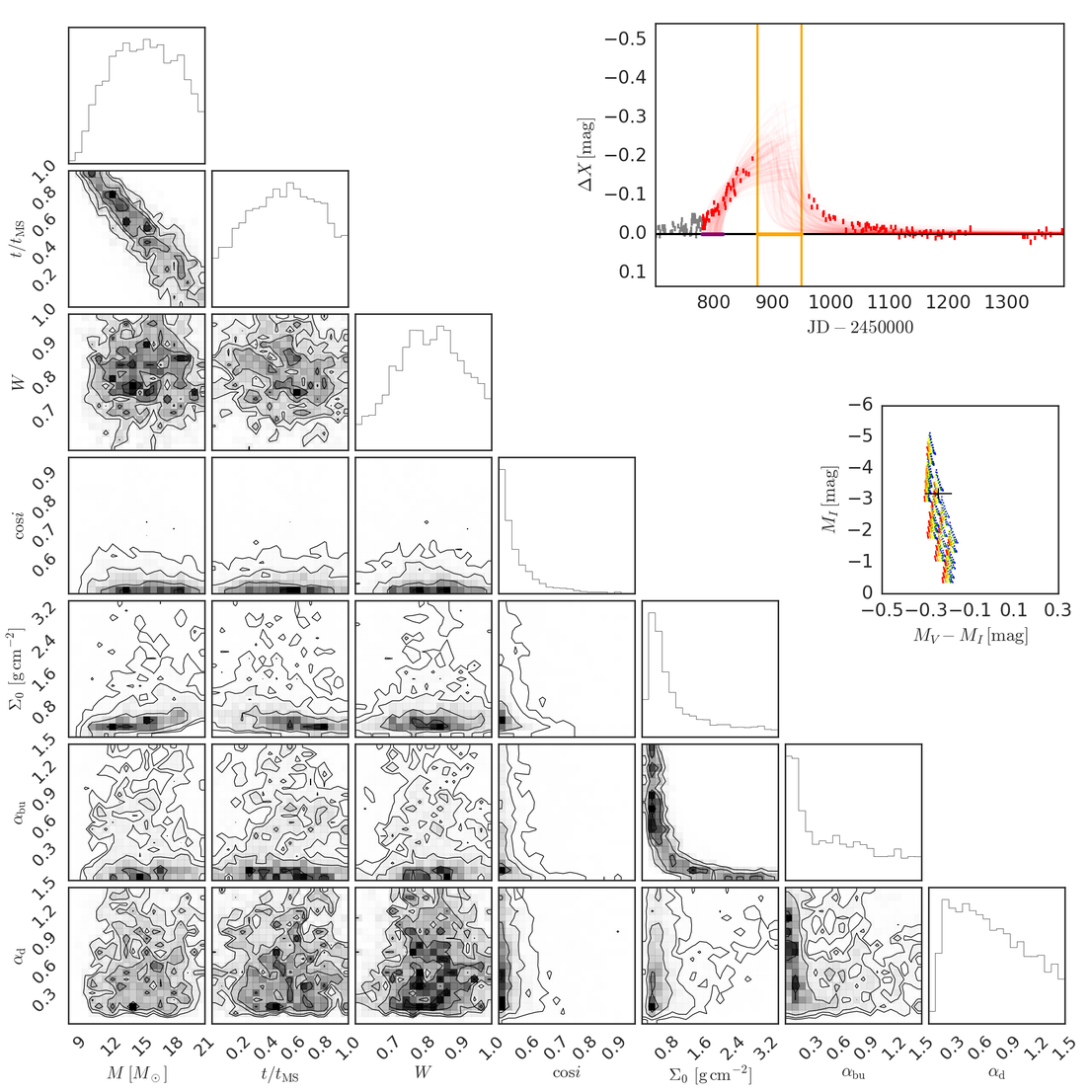}
}
\caption[]
{
Same as Fig.~\ref{example_bb1} for SMC\_SC4 159857 and bump ID 01. 
}
\label{smc_sc4_159857_01}
\end{figure*}
\clearpage

\begin{figure*}
\centering{
\includegraphics[width=1.0\linewidth]{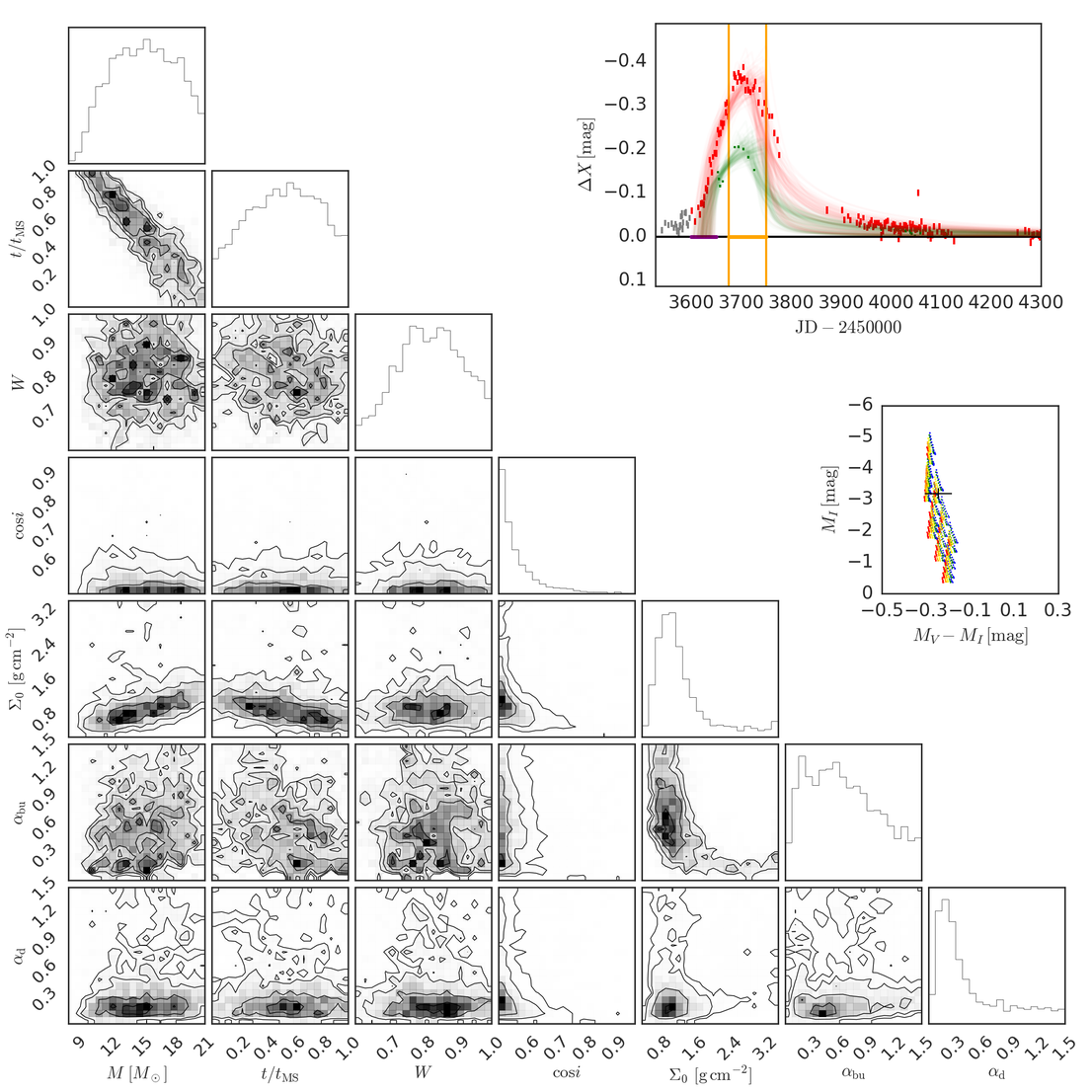}
}
\caption[]
{
Same as Fig.~\ref{example_bb1} for SMC\_SC4 159857 and bump ID 02. 
}
\label{smc_sc4_159857_02}
\end{figure*}
\clearpage

\begin{figure*}
\centering{
\includegraphics[width=1.0\linewidth]{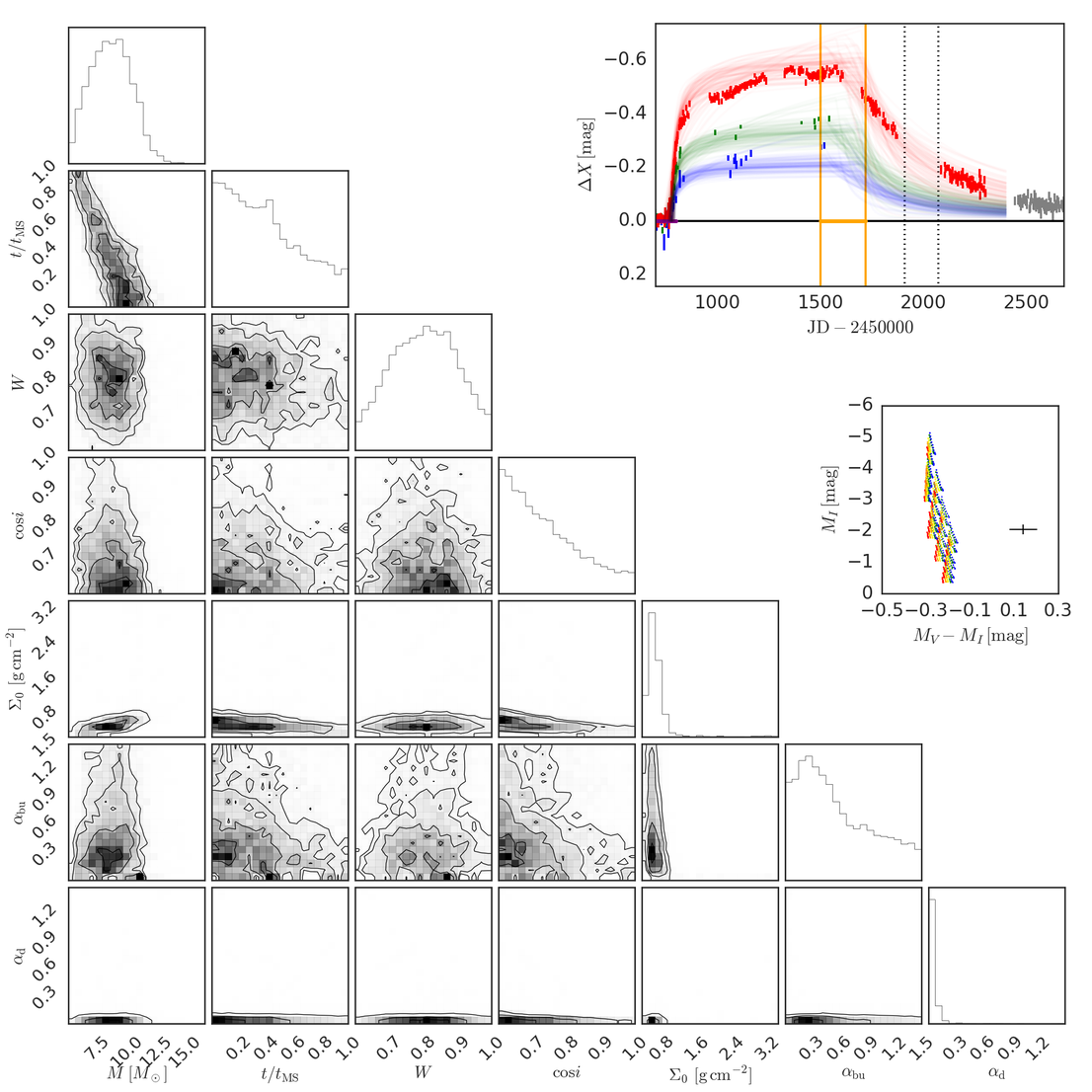}
}
\caption[]
{
Same as Fig.~\ref{example_bb1} for SMC\_SC4 163828 and bump ID 01. 
}
\label{smc_sc4_163828_01}
\end{figure*}
\clearpage

\begin{figure*}
\centering{
\includegraphics[width=1.0\linewidth]{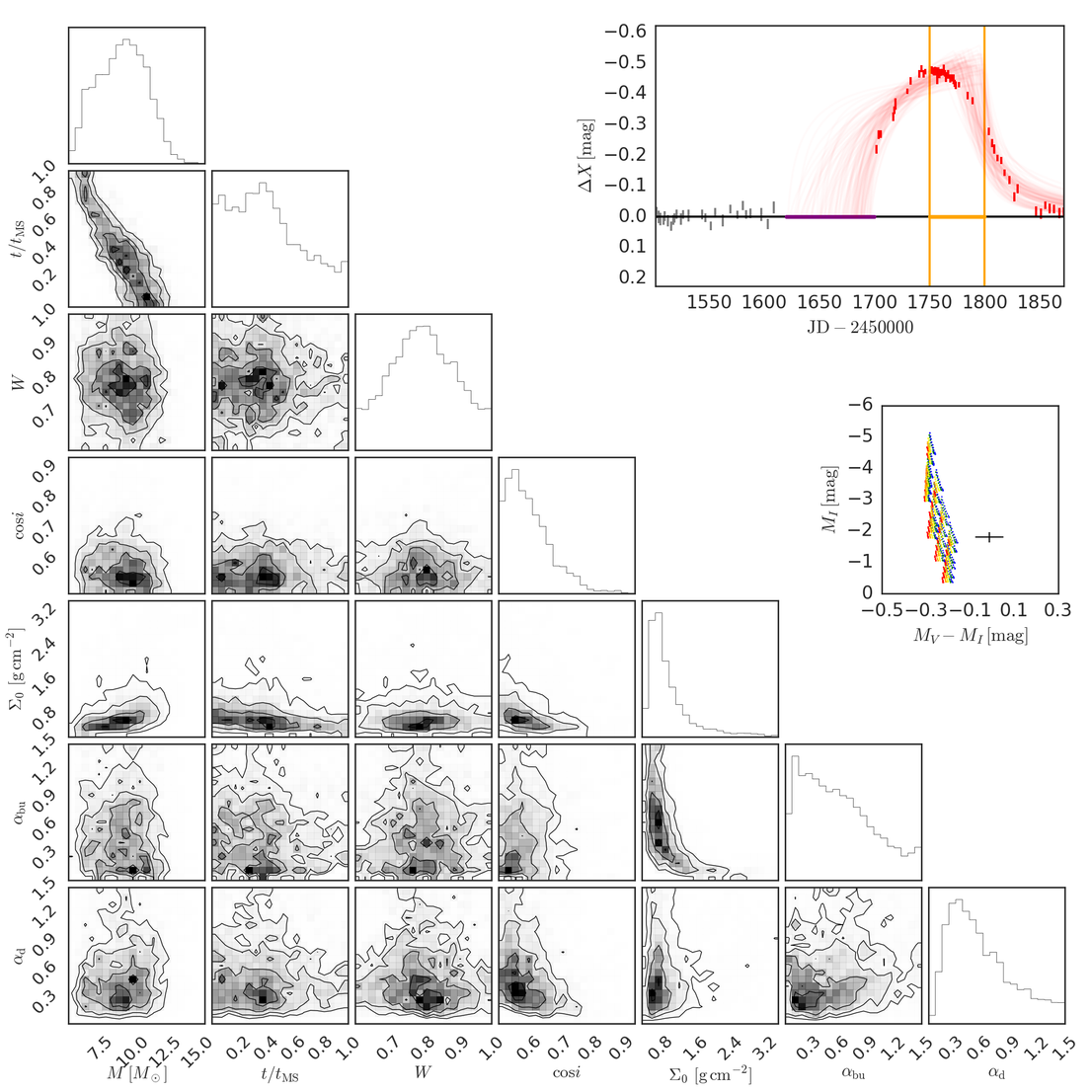}
}
\caption[]
{
Same as Fig.~\ref{example_bb1} for SMC\_SC4 167554 and bump ID 01. 
}
\label{smc_sc4_167554_01}
\end{figure*}
\clearpage

\begin{figure*}
\centering{
\includegraphics[width=1.0\linewidth]{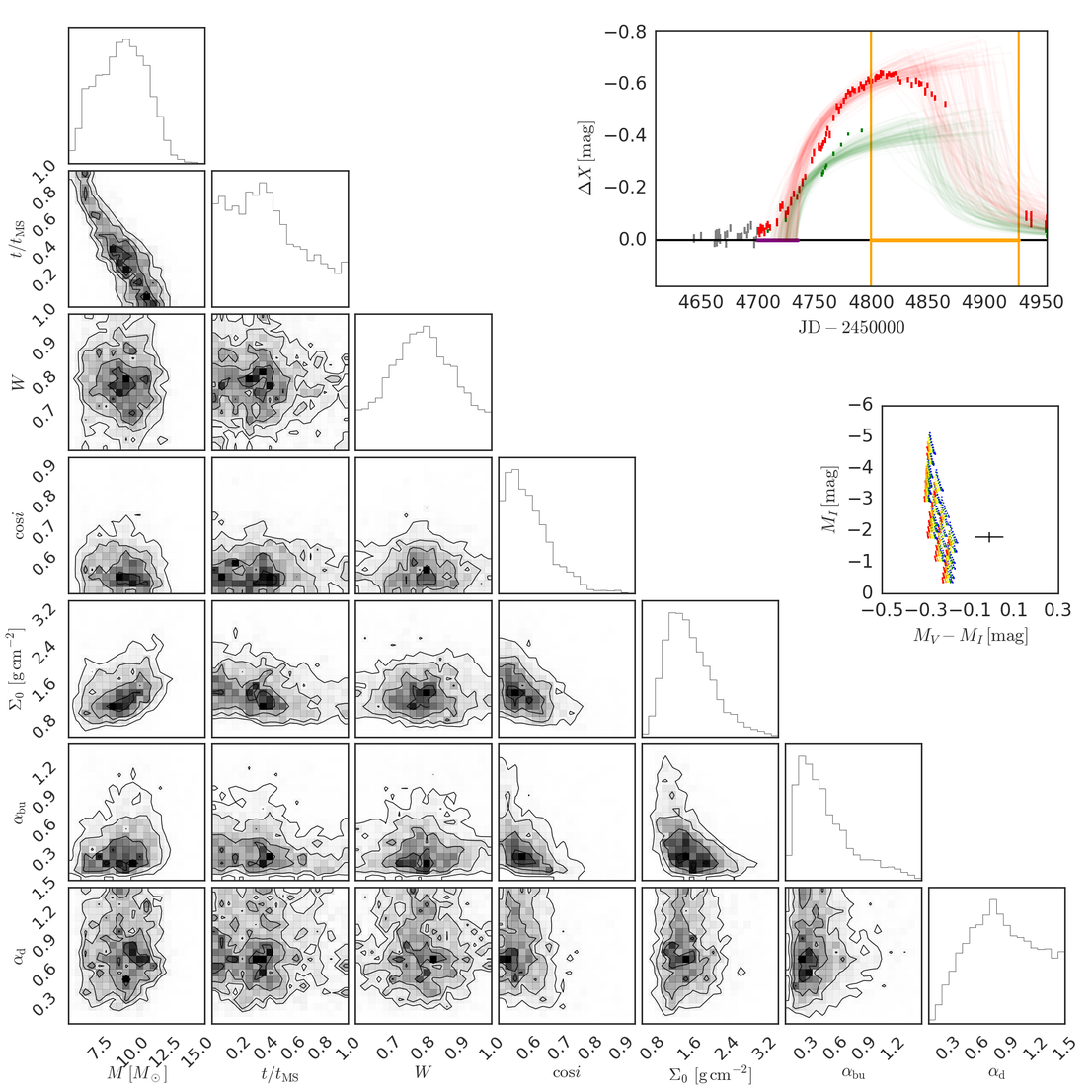}
}
\caption[]
{
Same as Fig.~\ref{example_bb1} for SMC\_SC4 167554 and bump ID 02. 
}
\label{smc_sc4_167554_02}
\end{figure*}
\clearpage

\begin{figure*}
\centering{
\includegraphics[width=1.0\linewidth]{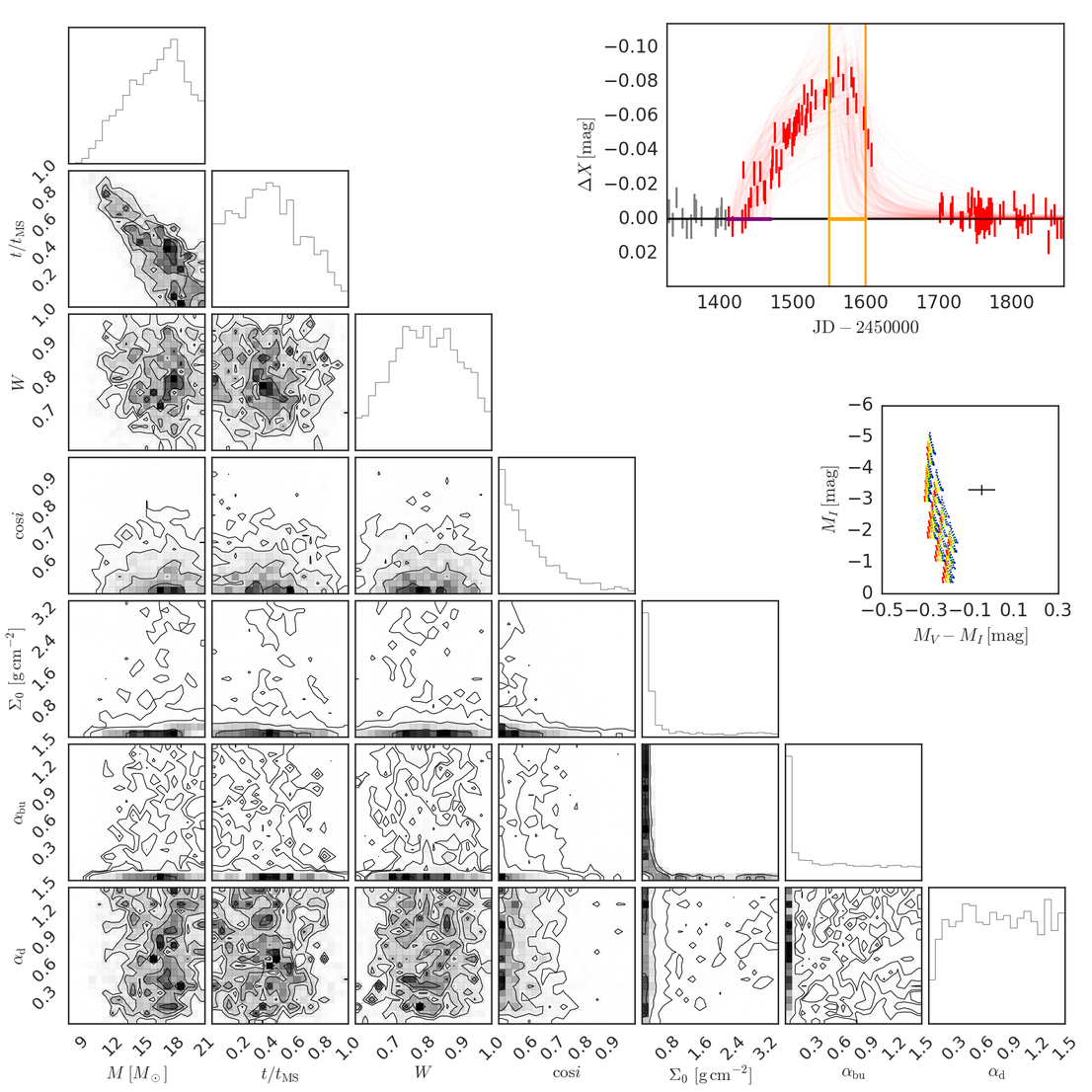}
}
\caption[]
{
Same as Fig.~\ref{example_bb1} for SMC\_SC4 171253 and bump ID 01. 
}
\label{smc_sc4_171253_01}
\end{figure*}
\clearpage

\begin{figure*}
\centering{
\includegraphics[width=1.0\linewidth]{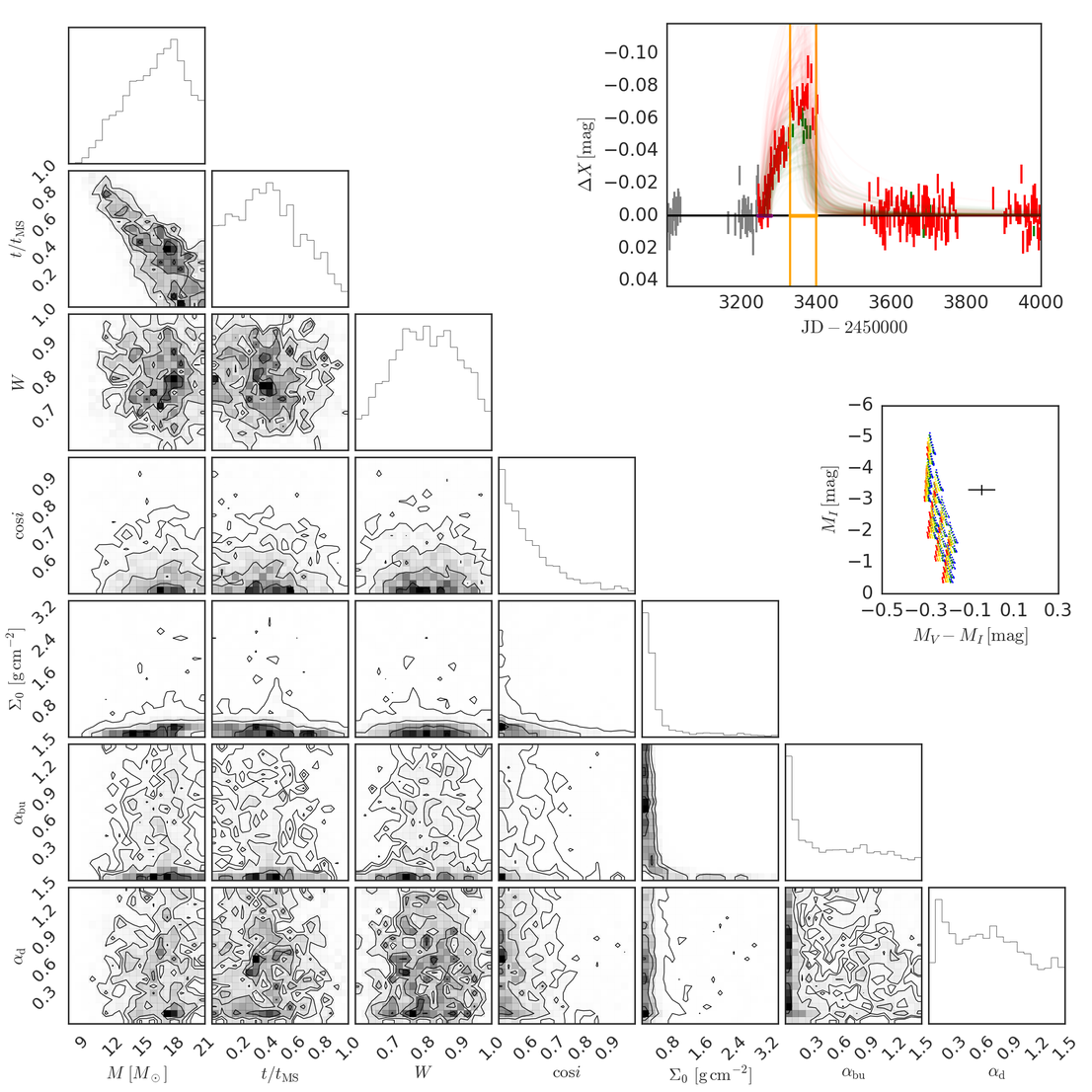}
}
\caption[]
{
Same as Fig.~\ref{example_bb1} for SMC\_SC4 171253 and bump ID 02. 
}
\label{smc_sc4_171253_02}
\end{figure*}
\clearpage

\begin{figure*}
\centering{
\includegraphics[width=1.0\linewidth]{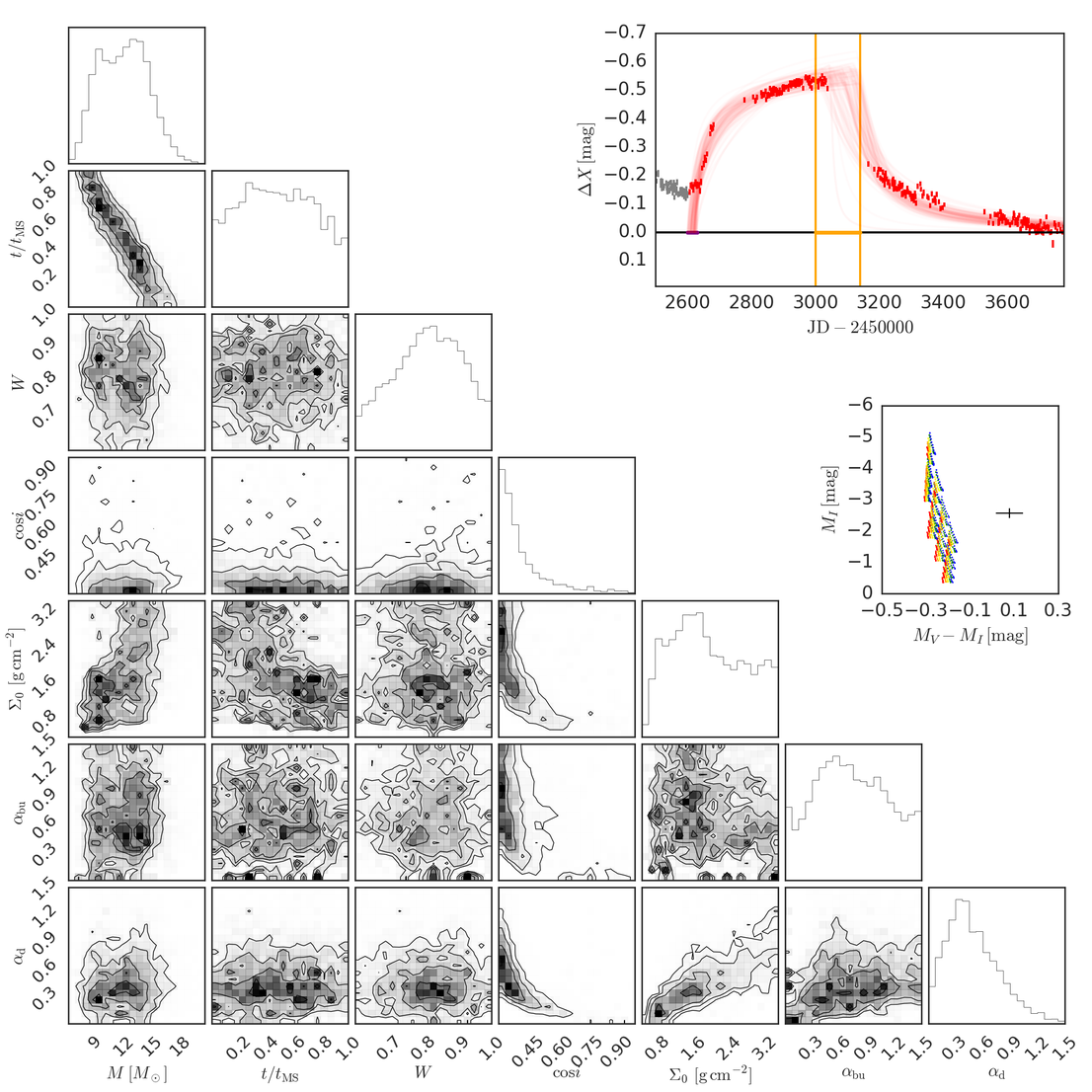}
}
\caption[]
{
Same as Fig.~\ref{example_bb1} for SMC\_SC4 175272 and bump ID 01. 
}
\label{smc_sc4_175272_01}
\end{figure*}
\clearpage

\begin{figure*}
\centering{
\includegraphics[width=1.0\linewidth]{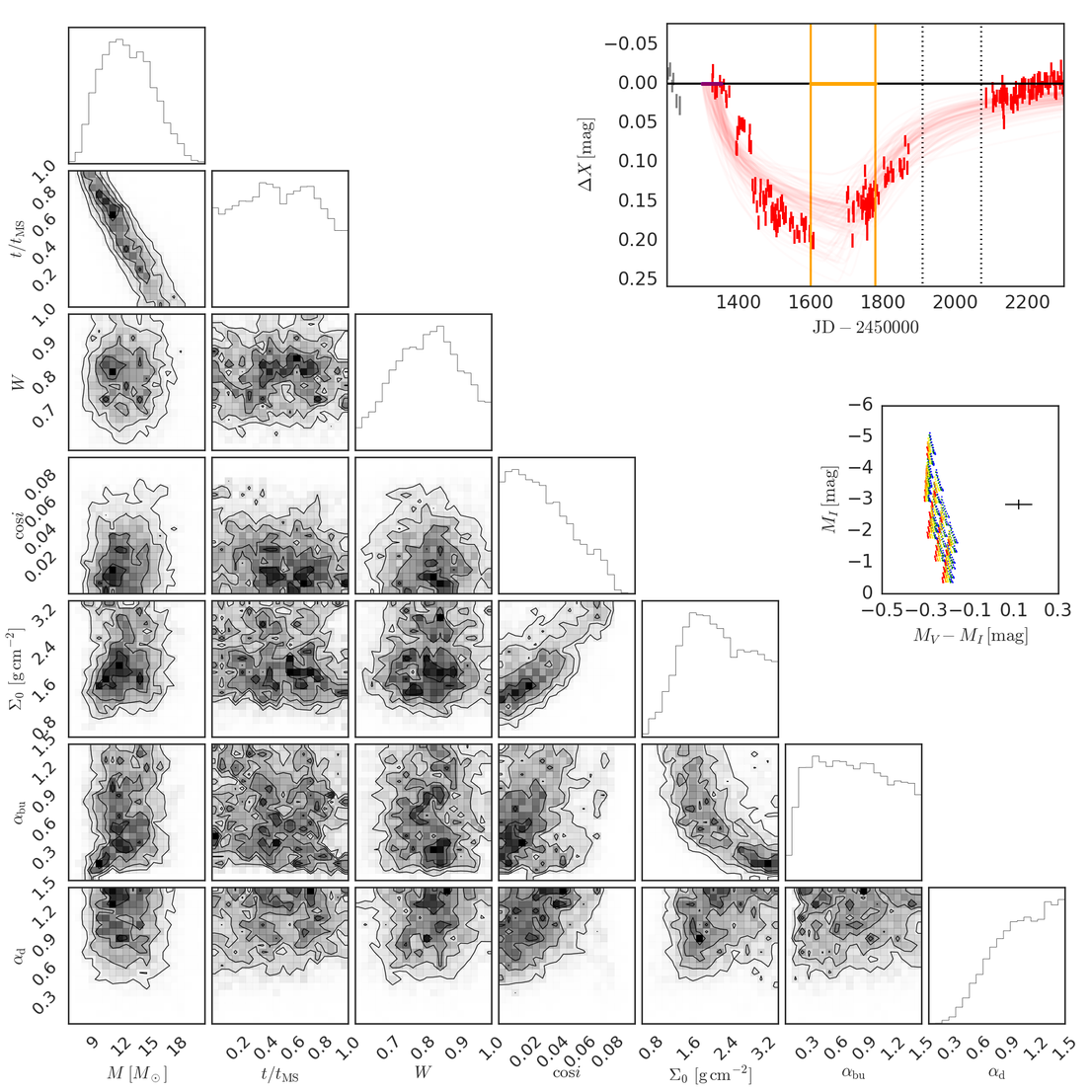}
}
\caption[]
{
Same as Fig.~\ref{example_bb1} for SMC\_SC4 179053 and bump ID 01. 
}
\label{smc_sc4_179053_01}
\end{figure*}
\clearpage

\begin{figure*}
\centering{
\includegraphics[width=1.0\linewidth]{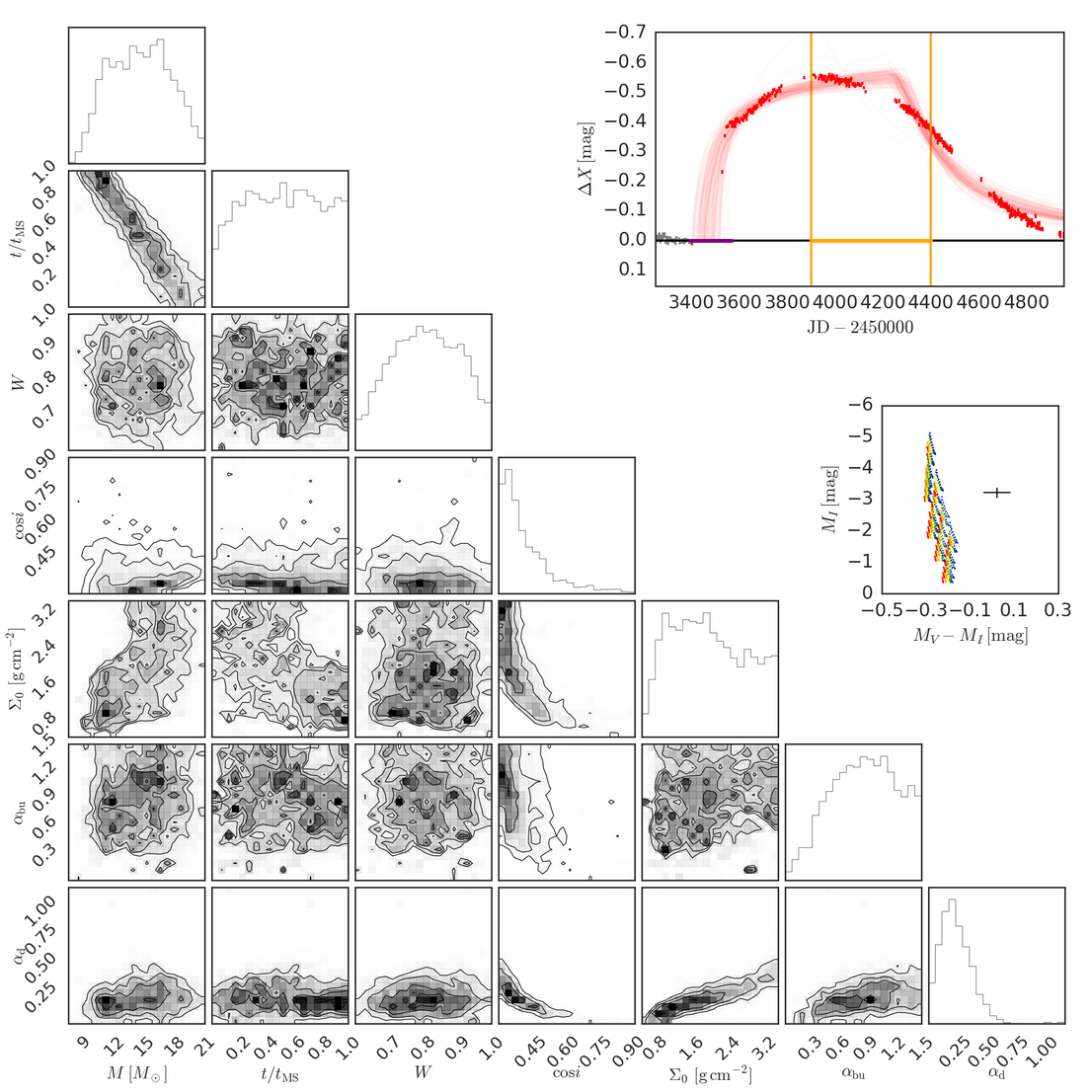}
}
\caption[]
{
Same as Fig.~\ref{example_bb1} for SMC\_SC5 11453 and bump ID 01. 
}
\label{smc_sc5_11453_01}
\end{figure*}
\clearpage

\begin{figure*}
\centering{
\includegraphics[width=1.0\linewidth]{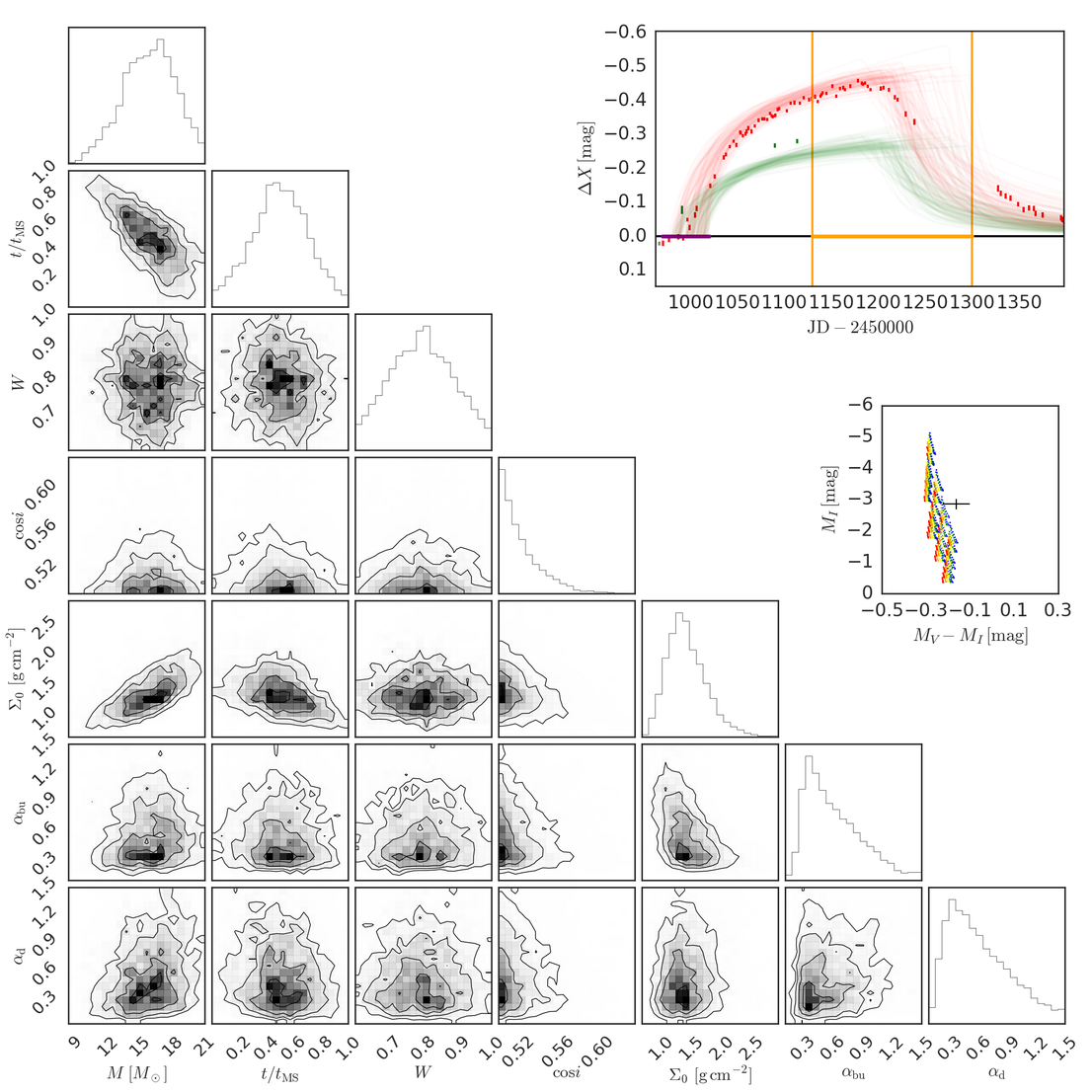}
}
\caption[]
{
Same as Fig.~\ref{example_bb1} for SMC\_SC5 21117 and bump ID 01. 
}
\label{smc_sc5_21117_01}
\end{figure*}
\clearpage

\begin{figure*}
\centering{
\includegraphics[width=1.0\linewidth]{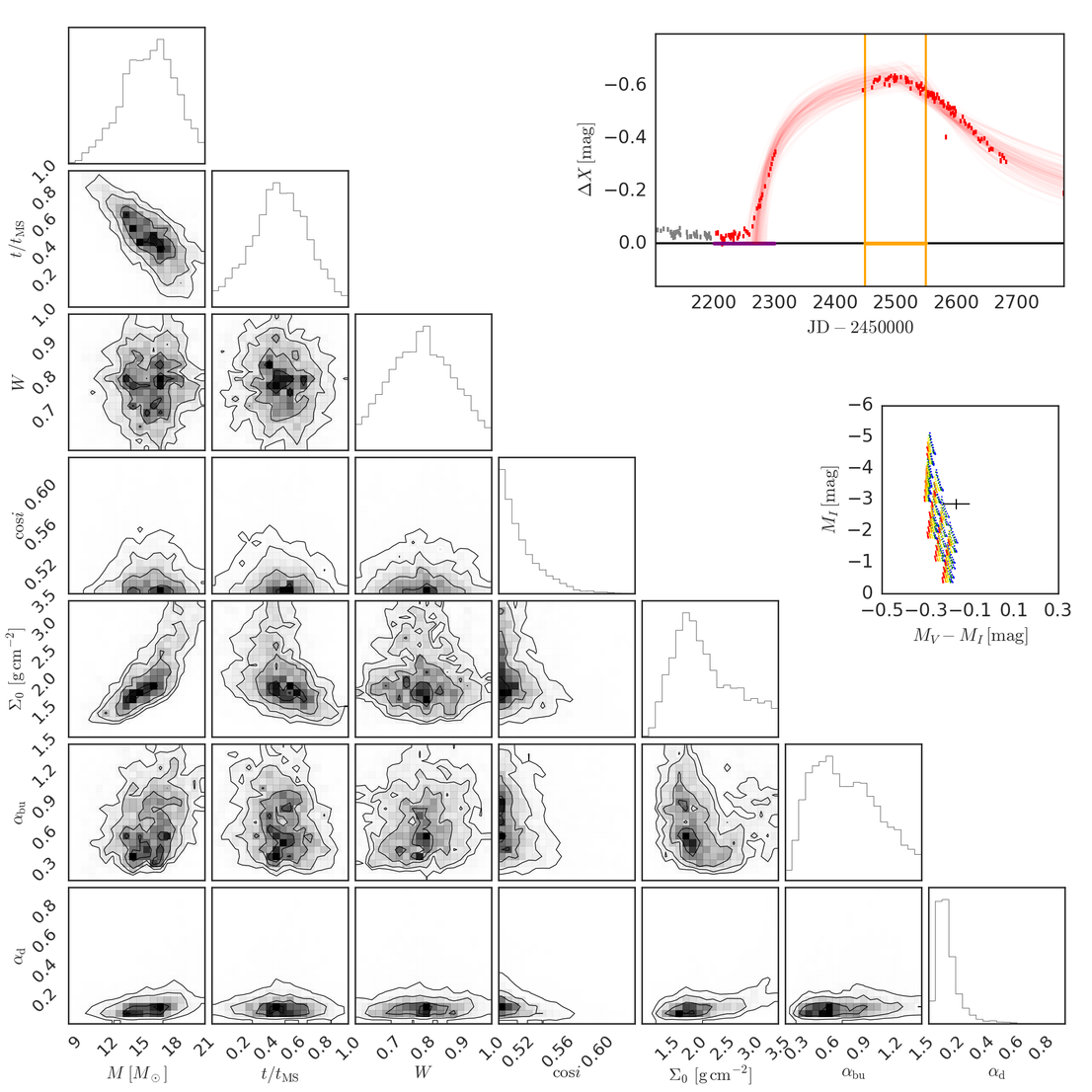}
}
\caption[]
{
Same as Fig.~\ref{example_bb1} for SMC\_SC5 21117 and bump ID 02. 
}
\label{smc_sc5_21117_02}
\end{figure*}
\clearpage

\begin{figure*}
\centering{
\includegraphics[width=1.0\linewidth]{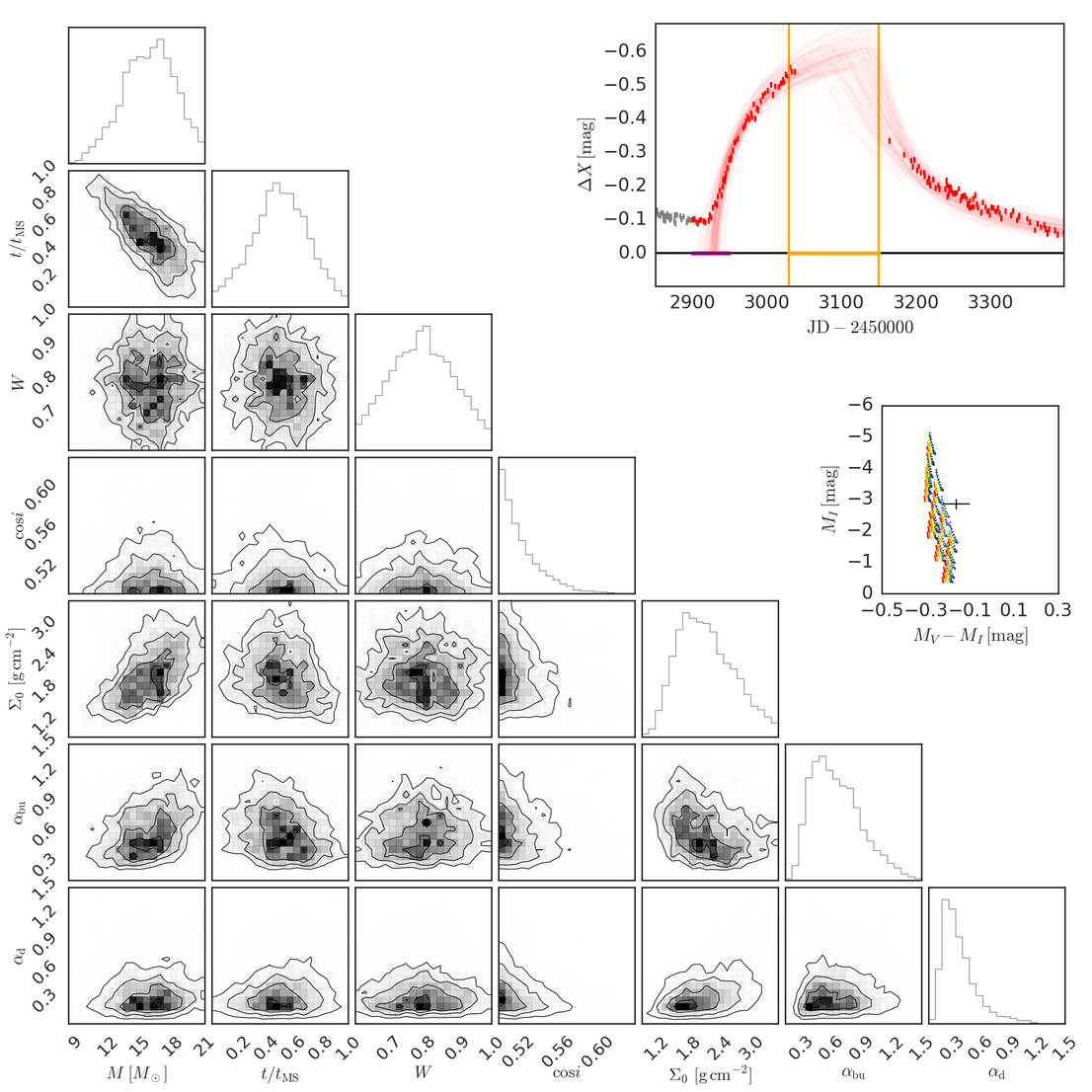}
}
\caption[]
{
Same as Fig.~\ref{example_bb1} for SMC\_SC5 21117 and bump ID 03. 
}
\label{smc_sc5_21117_03}
\end{figure*}
\clearpage

\begin{figure*}
\centering{
\includegraphics[width=1.0\linewidth]{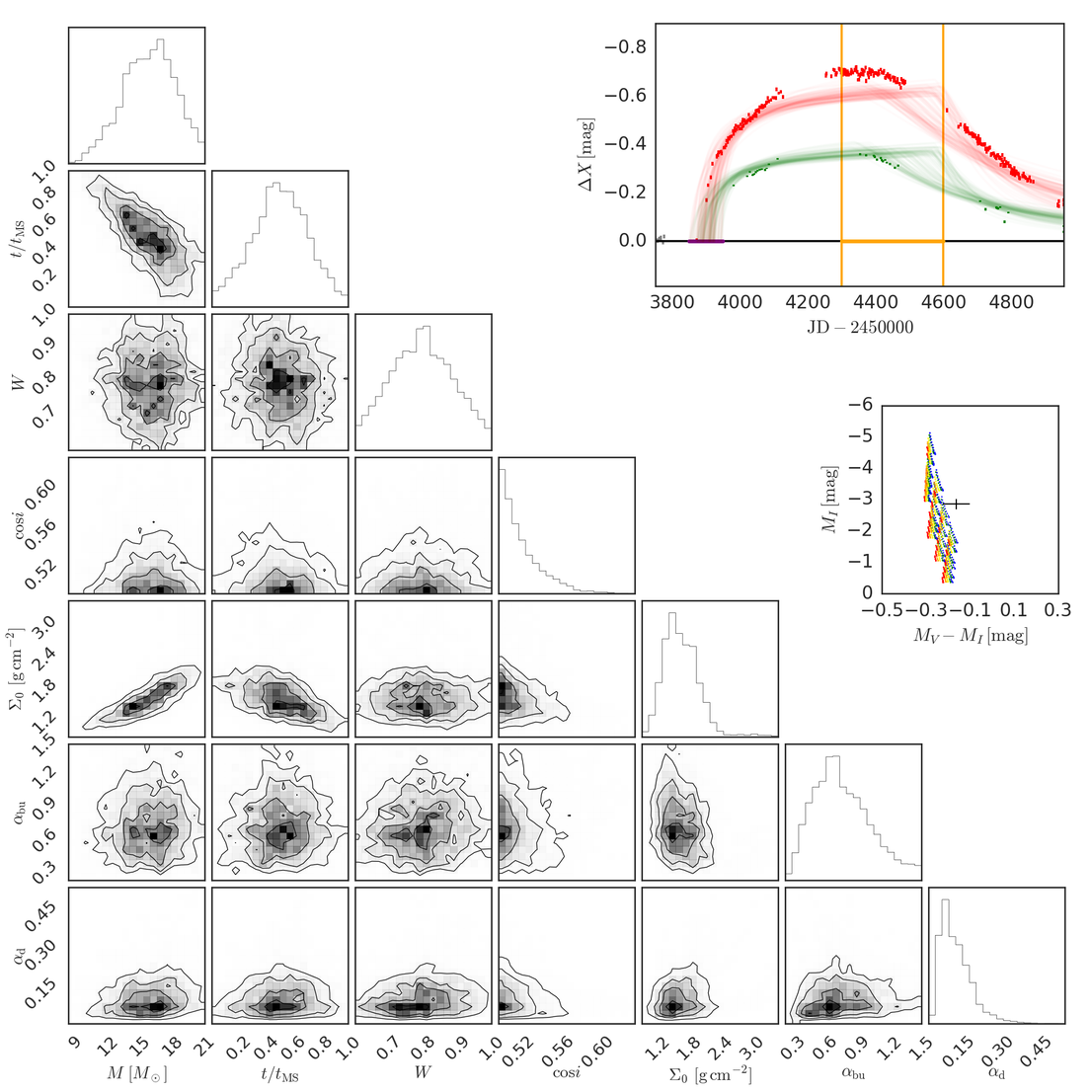}
}
\caption[]
{
Same as Fig.~\ref{example_bb1} for SMC\_SC5 21117 and bump ID 04. 
}
\label{smc_sc5_21117_04}
\end{figure*}
\clearpage

\begin{figure*}
\centering{
\includegraphics[width=1.0\linewidth]{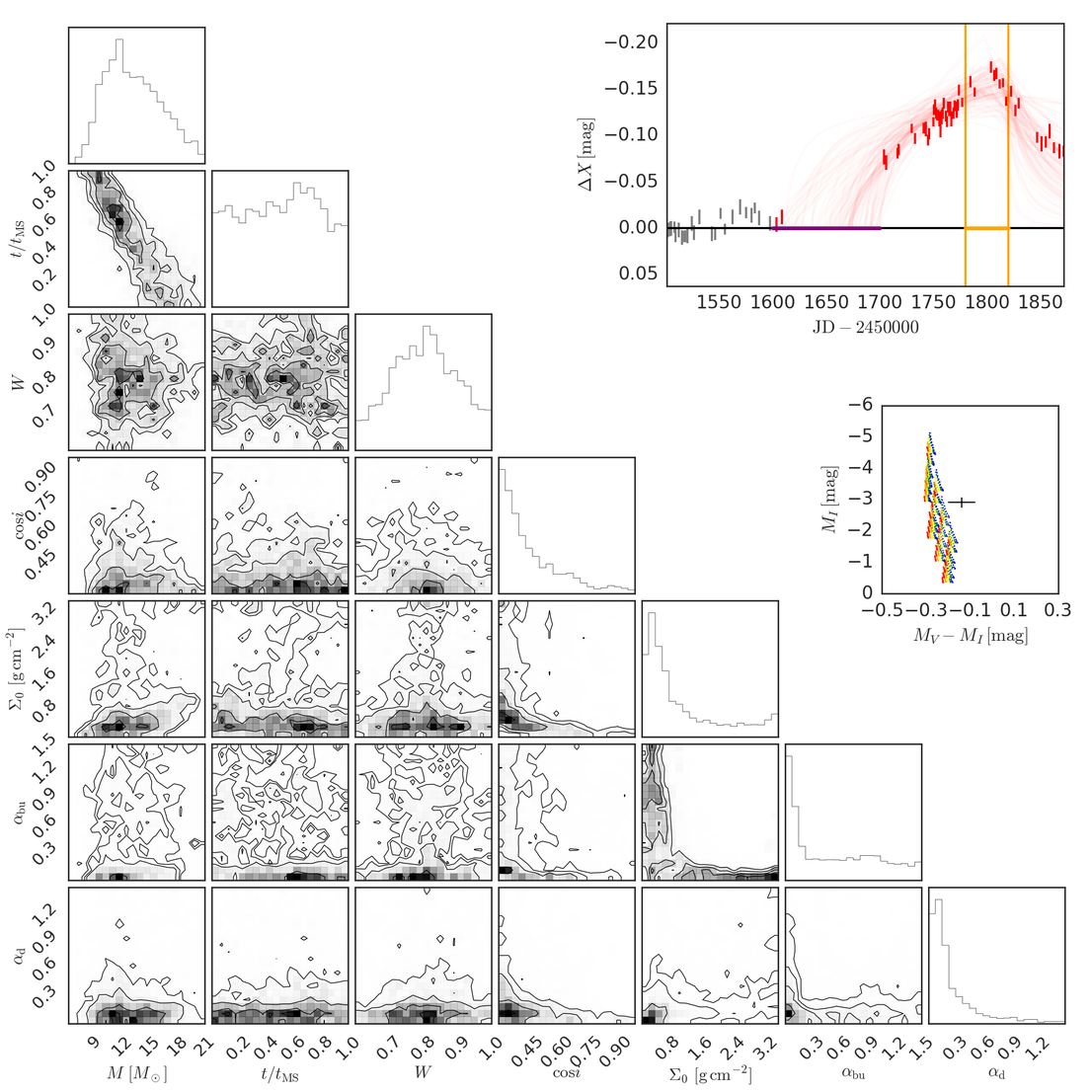}
}
\caption[]
{
Same as Fig.~\ref{example_bb1} for SMC\_SC5 21134 and bump ID 01. 
}
\label{smc_sc5_21134_01}
\end{figure*}
\clearpage

\begin{figure*}
\centering{
\includegraphics[width=1.0\linewidth]{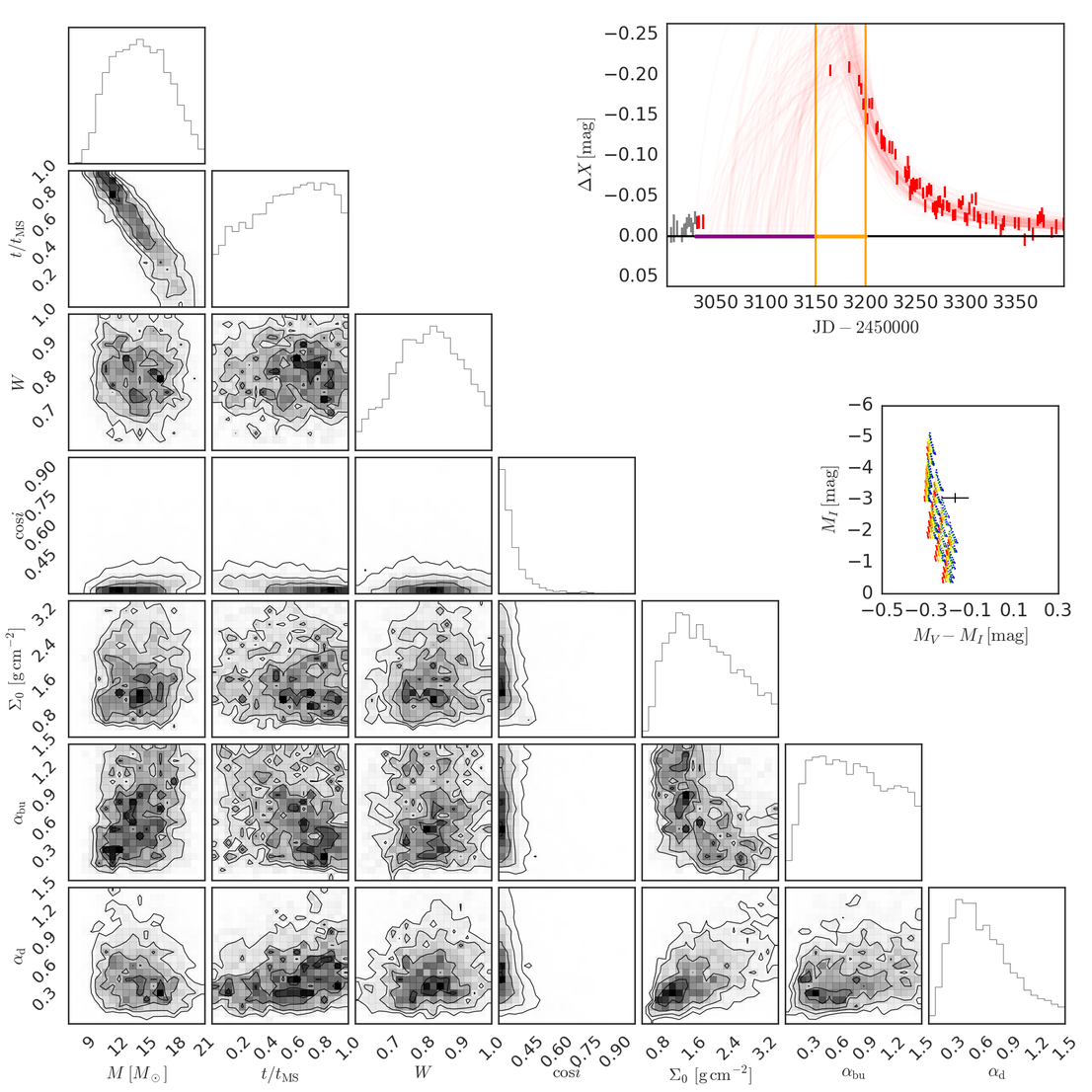}
}
\caption[]
{
Same as Fig.~\ref{example_bb1} for SMC\_SC5 32377 and bump ID 01. 
}
\label{smc_sc5_32377_01}
\end{figure*}
\clearpage

\begin{figure*}
\centering{
\includegraphics[width=1.0\linewidth]{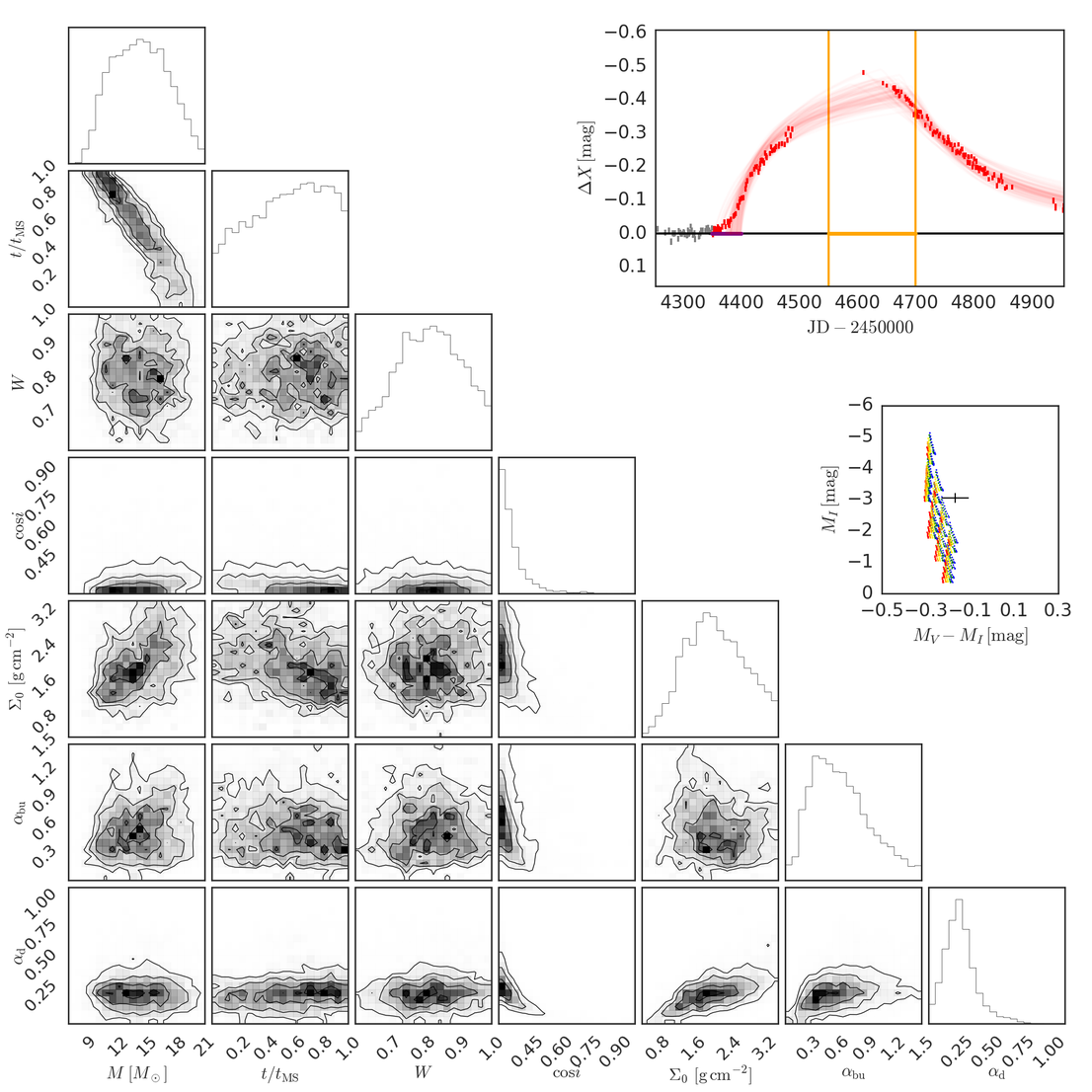}
}
\caption[]
{
Same as Fig.~\ref{example_bb1} for SMC\_SC5 32377 and bump ID 02. 
}
\label{smc_sc5_32377_02}
\end{figure*}
\clearpage

\begin{figure*}
\centering{
\includegraphics[width=1.0\linewidth]{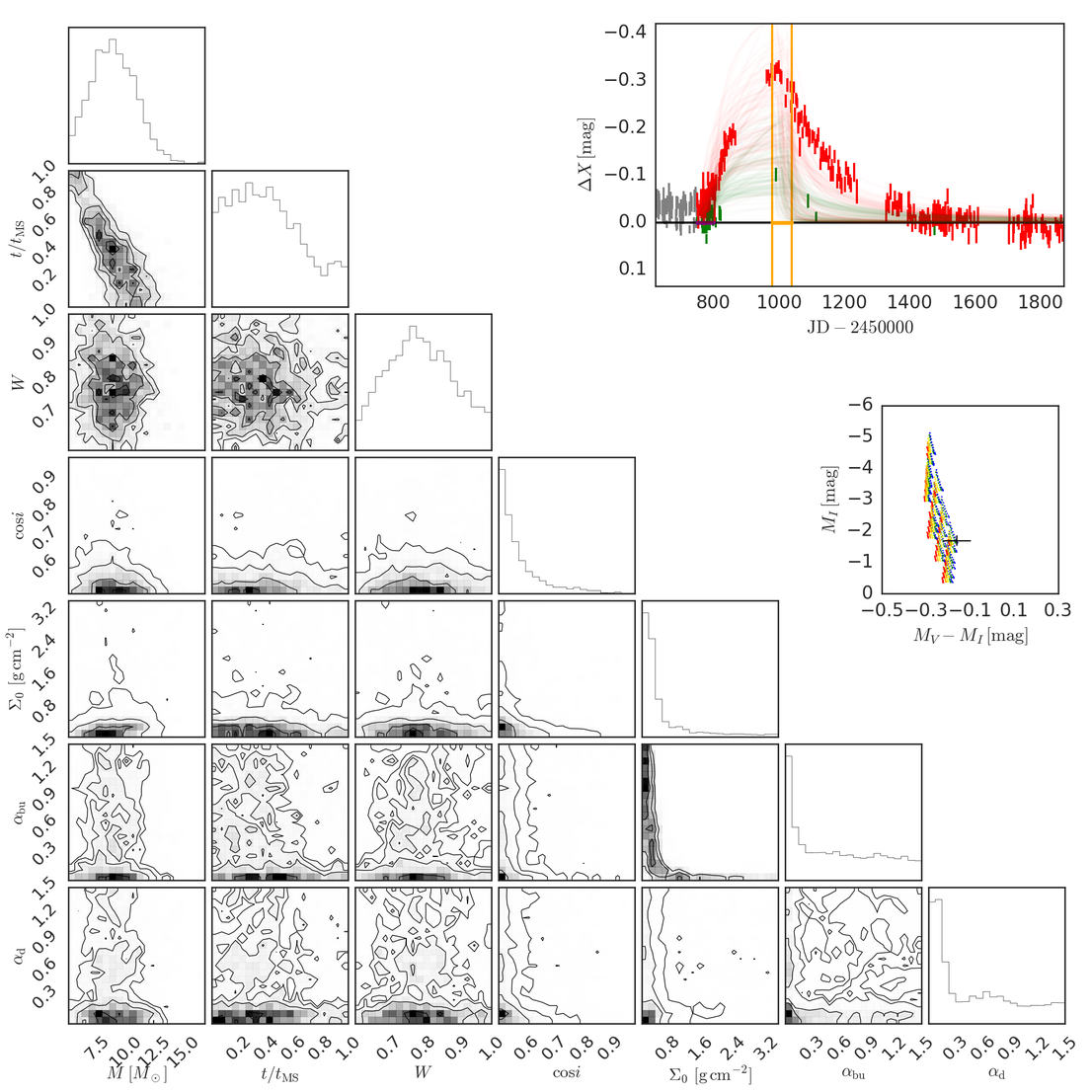}
}
\caption[]
{
Same as Fig.~\ref{example_bb1} for SMC\_SC5 43650 and bump ID 01. 
}
\label{smc_sc5_43650_01}
\end{figure*}
\clearpage

\begin{figure*}
\centering{
\includegraphics[width=1.0\linewidth]{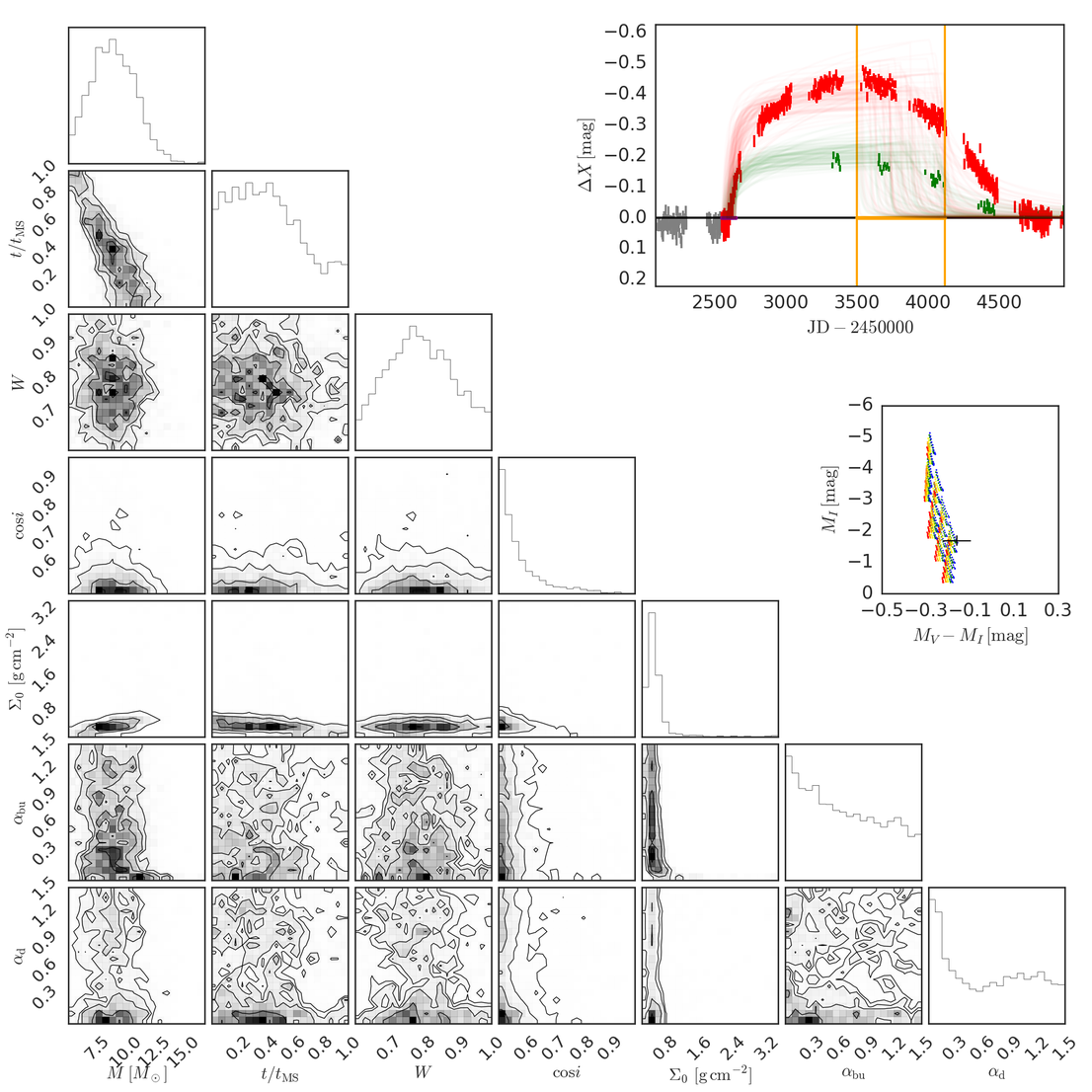}
}
\caption[]
{
Same as Fig.~\ref{example_bb1} for SMC\_SC5 43650 and bump ID 02. 
}
\label{smc_sc5_43650_02}
\end{figure*}
\clearpage

\begin{figure*}
\centering{
\includegraphics[width=1.0\linewidth]{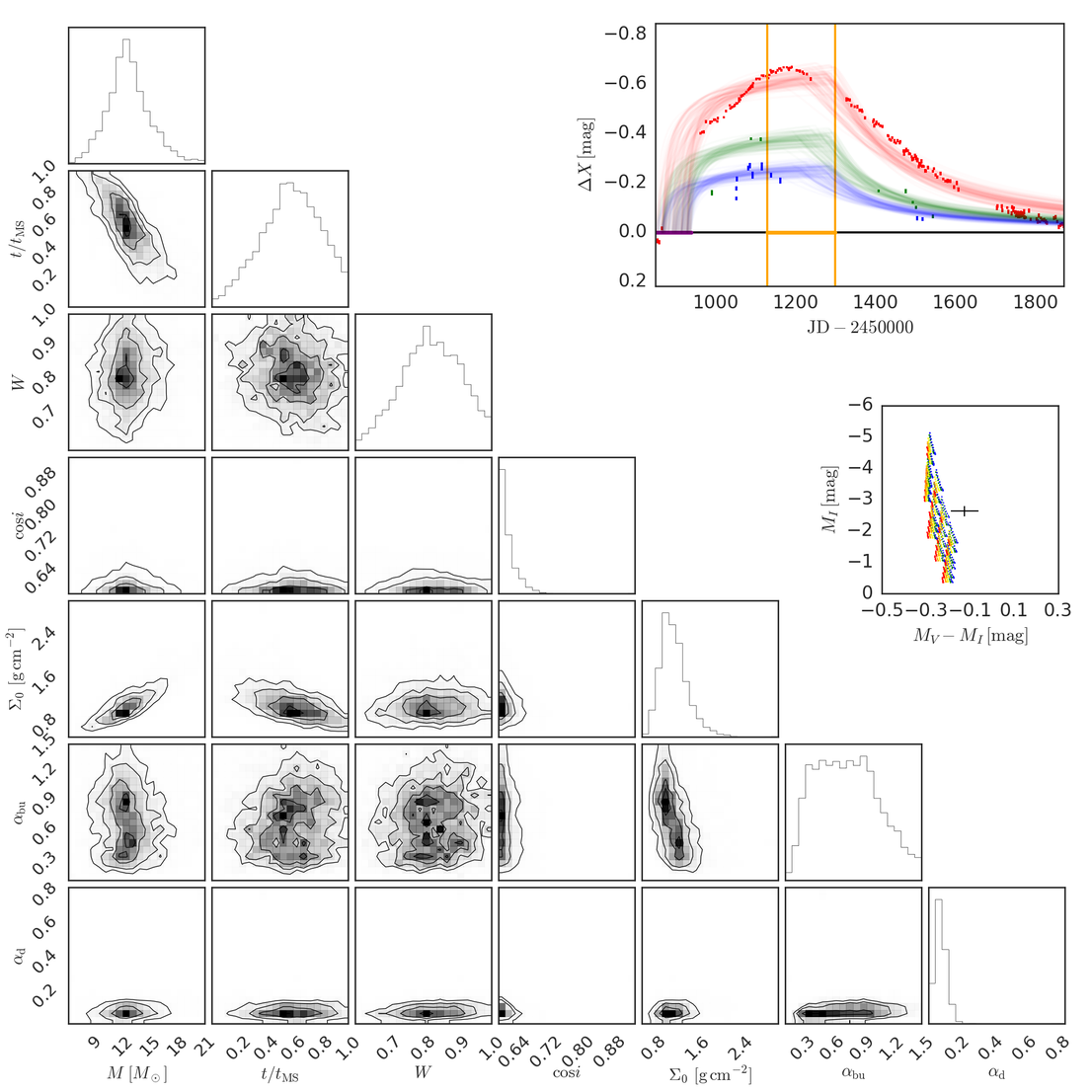}
}
\caption[]
{
Same as Fig.~\ref{example_bb1} for SMC\_SC5 54851 and bump ID 01. 
}
\label{smc_sc5_54851_01}
\end{figure*}
\clearpage

\begin{figure*}
\centering{
\includegraphics[width=1.0\linewidth]{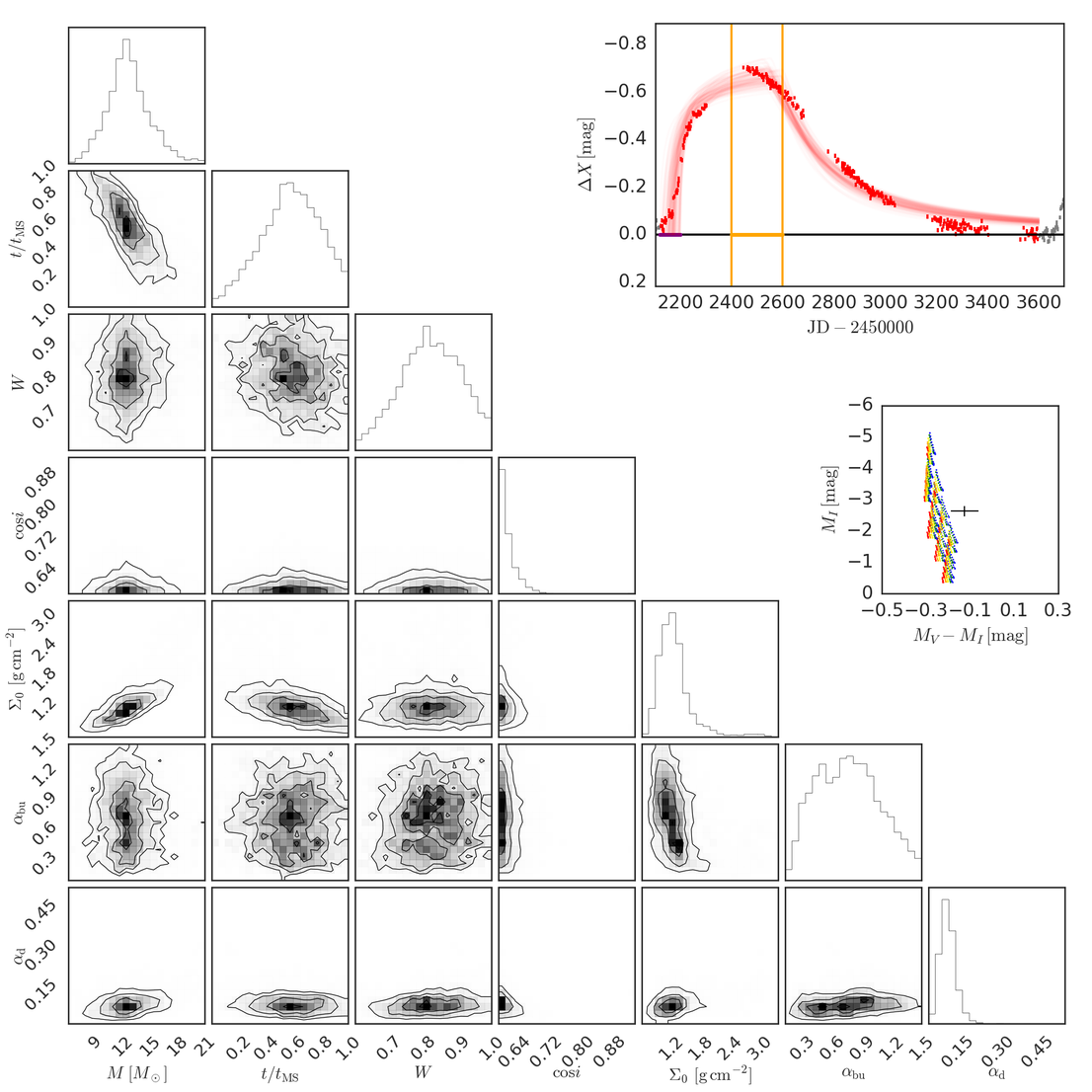}
}
\caption[]
{
Same as Fig.~\ref{example_bb1} for SMC\_SC5 54851 and bump ID 02. 
}
\label{smc_sc5_54851_02}
\end{figure*}
\clearpage

\begin{figure*}
\centering{
\includegraphics[width=1.0\linewidth]{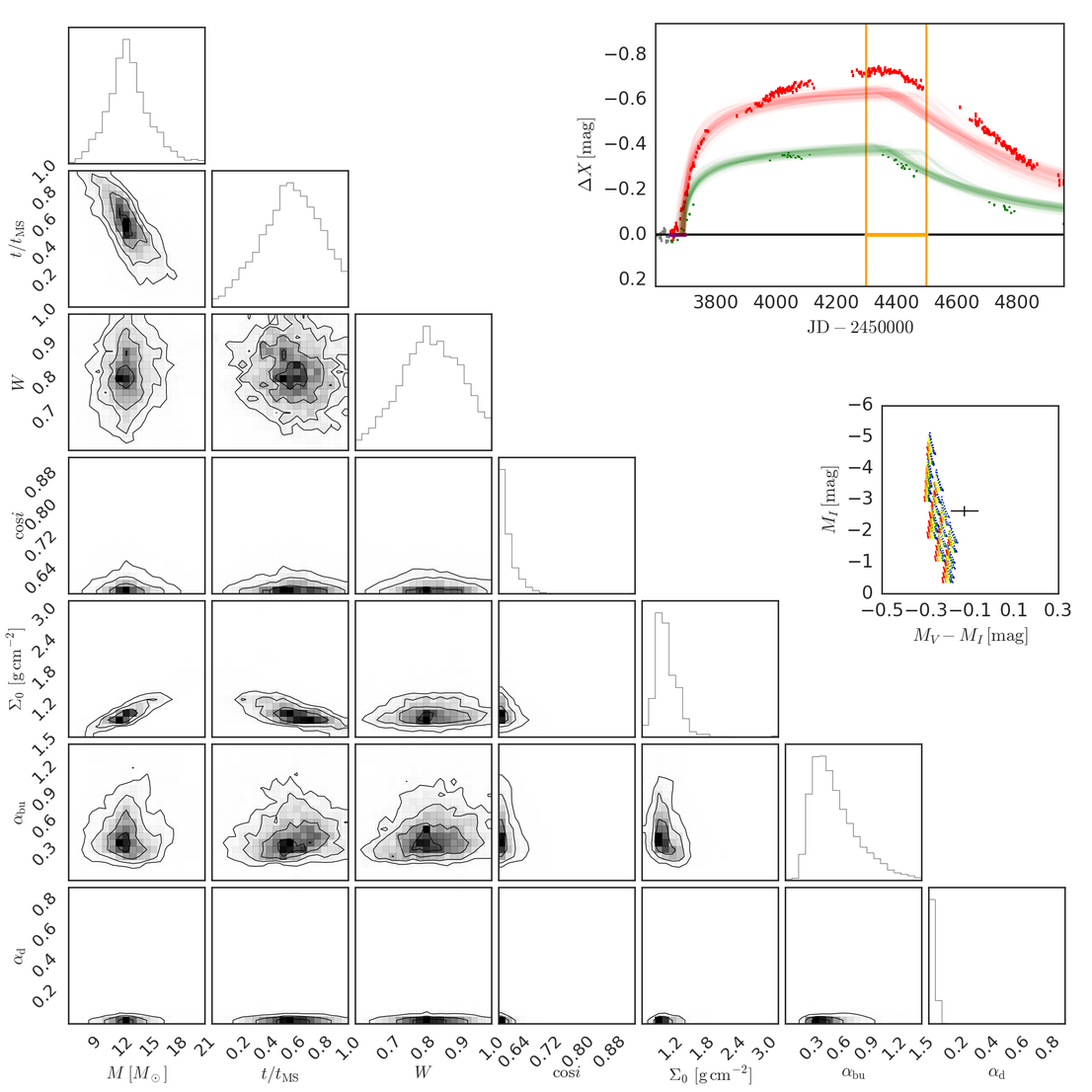}
}
\caption[]
{
Same as Fig.~\ref{example_bb1} for SMC\_SC5 54851 and bump ID 03. 
}
\label{smc_sc5_54851_03}
\end{figure*}
\clearpage

\begin{figure*}
\centering{
\includegraphics[width=1.0\linewidth]{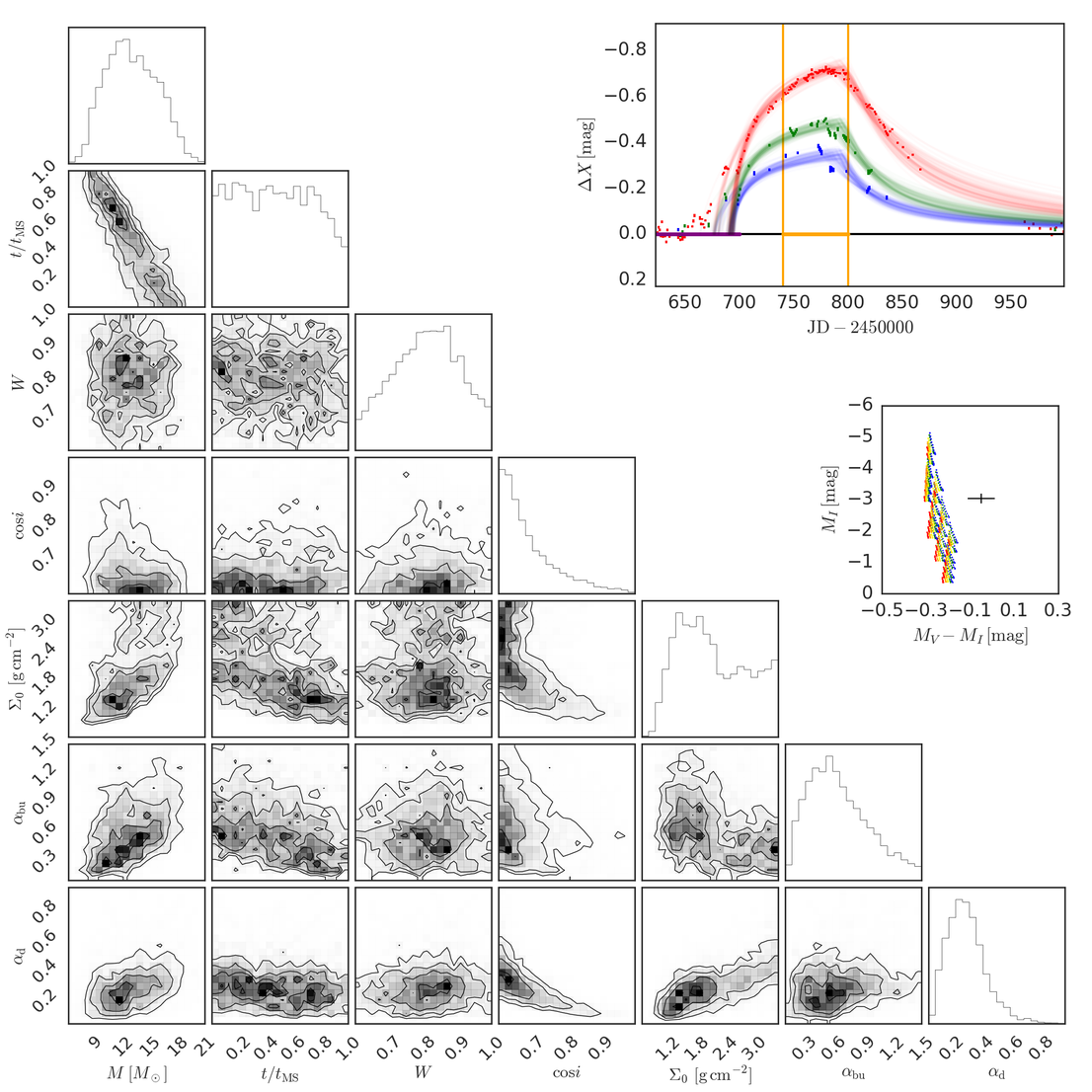}
}
\caption[]
{
Same as Fig.~\ref{example_bb1} for SMC\_SC5 65500 and bump ID 01. 
}
\label{smc_sc5_65500_01}
\end{figure*}
\clearpage

\begin{figure*}
\centering{
\includegraphics[width=1.0\linewidth]{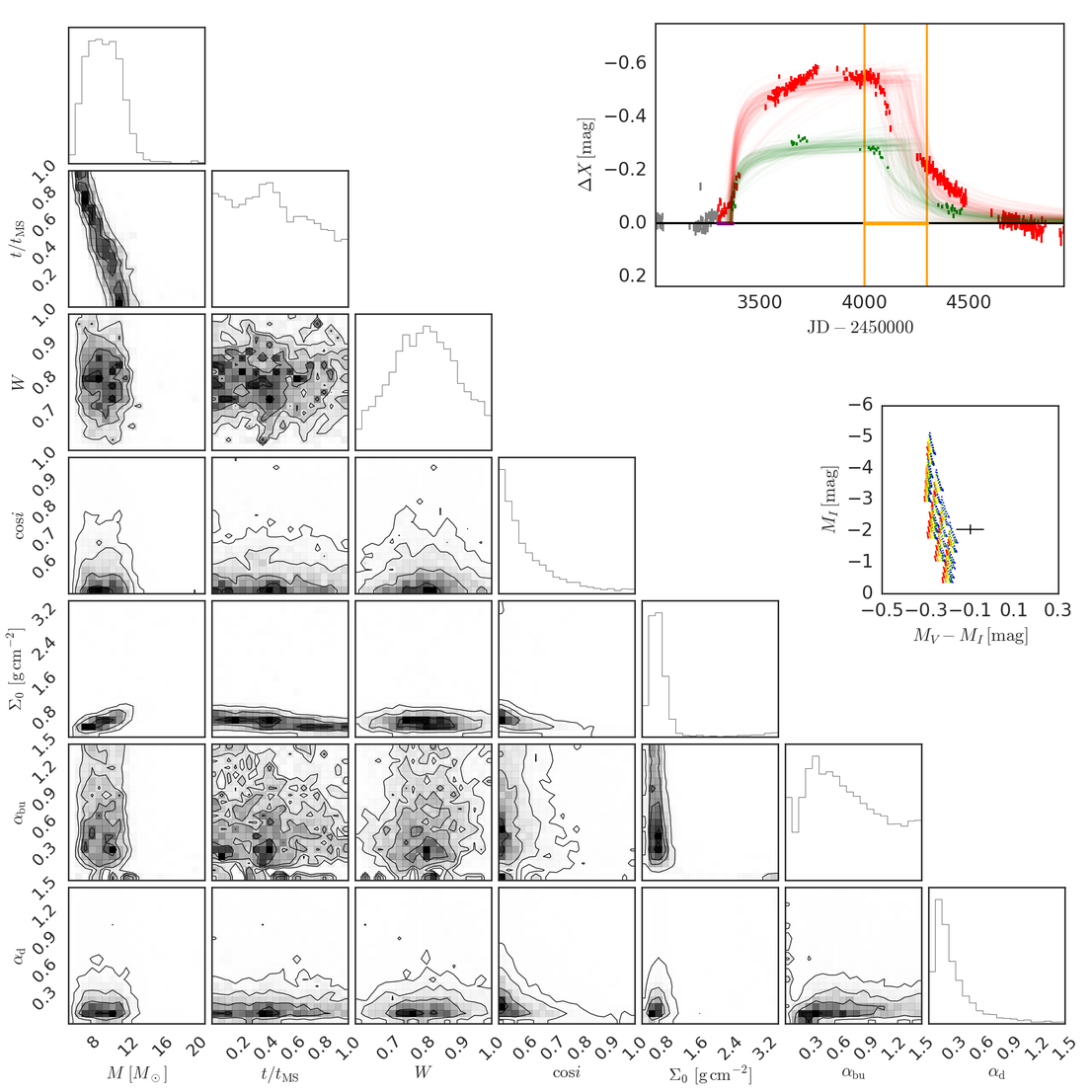}
}
\caption[]
{
Same as Fig.~\ref{example_bb1} for SMC\_SC5 129535 and bump ID 01. 
}
\label{smc_sc5_129535_01}
\end{figure*}
\clearpage

\begin{figure*}
\centering{
\includegraphics[width=1.0\linewidth]{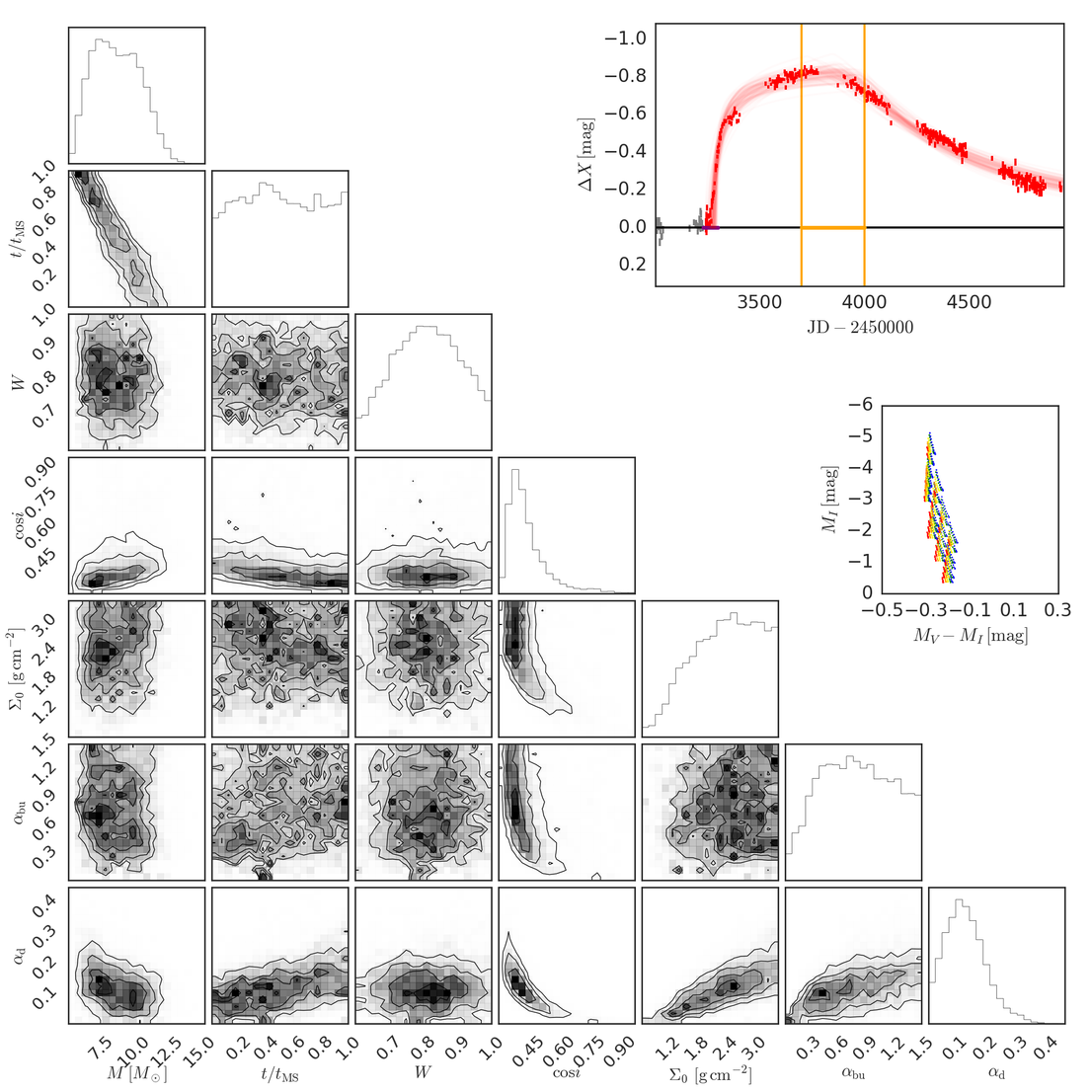}
}
\caption[]
{
Same as Fig.~\ref{example_bb1} for SMC\_SC5 145724 and bump ID 01. 
}
\label{smc_sc5_145724_01}
\end{figure*}
\clearpage

\begin{figure*}
\centering{
\includegraphics[width=1.0\linewidth]{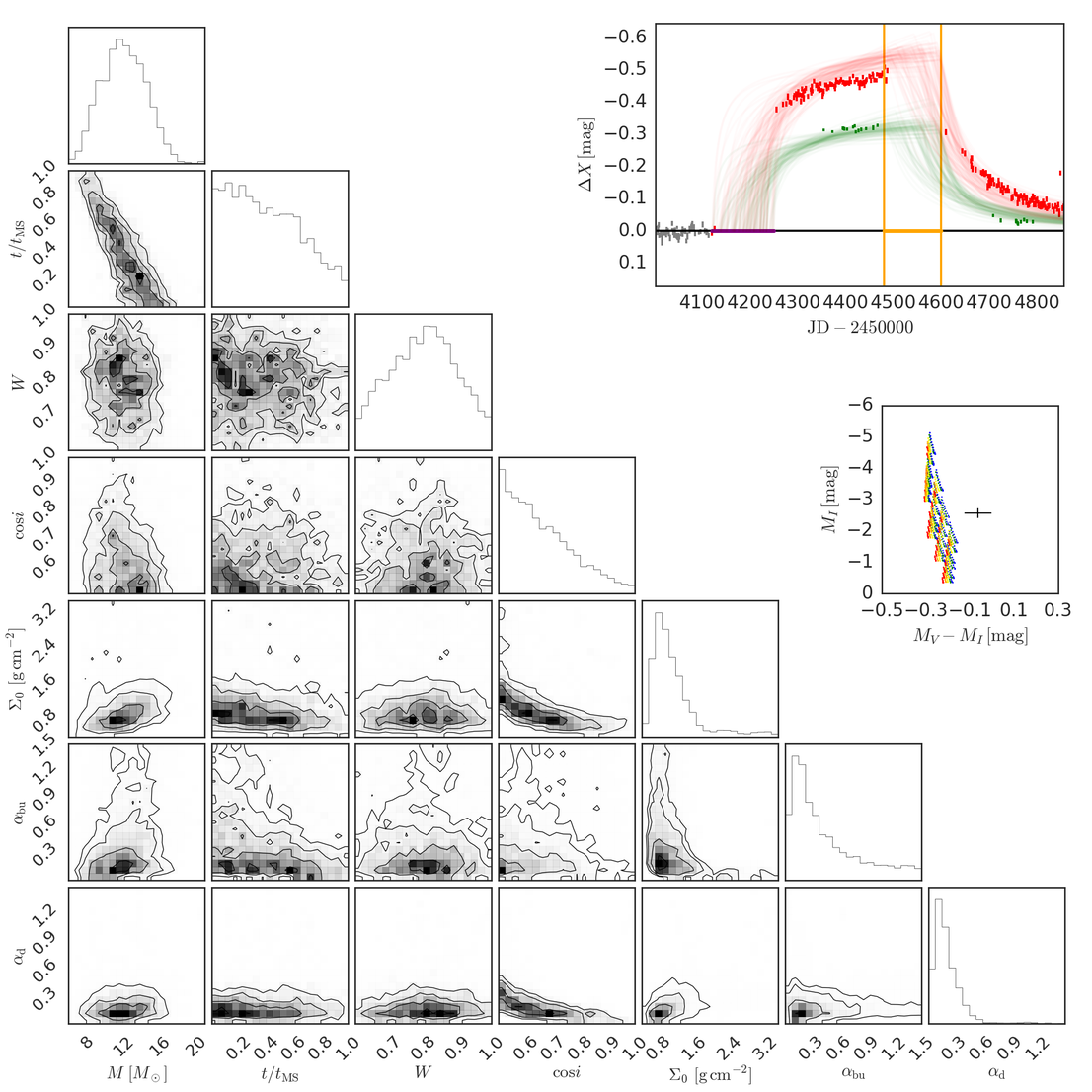}
}
\caption[]
{
Same as Fig.~\ref{example_bb1} for SMC\_SC5 180034 and bump ID 01. 
}
\label{smc_sc5_180034_01}
\end{figure*}
\clearpage

\begin{figure*}
\centering{
\includegraphics[width=1.0\linewidth]{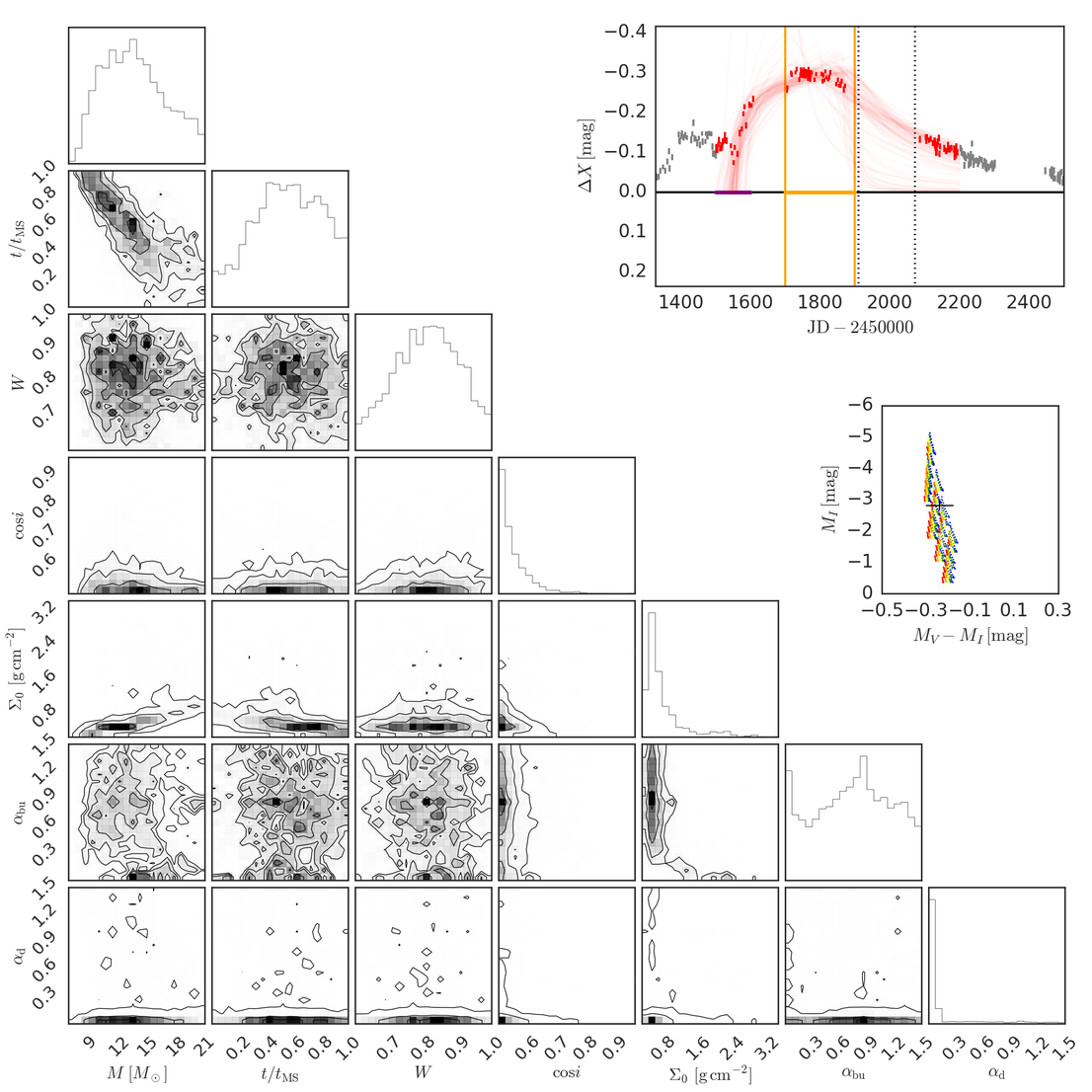}
}
\caption[]
{
Same as Fig.~\ref{example_bb1} for SMC\_SC5 260841 and bump ID 01. 
}
\label{smc_sc5_260841_01}
\end{figure*}
\clearpage

\begin{figure*}
\centering{
\includegraphics[width=1.0\linewidth]{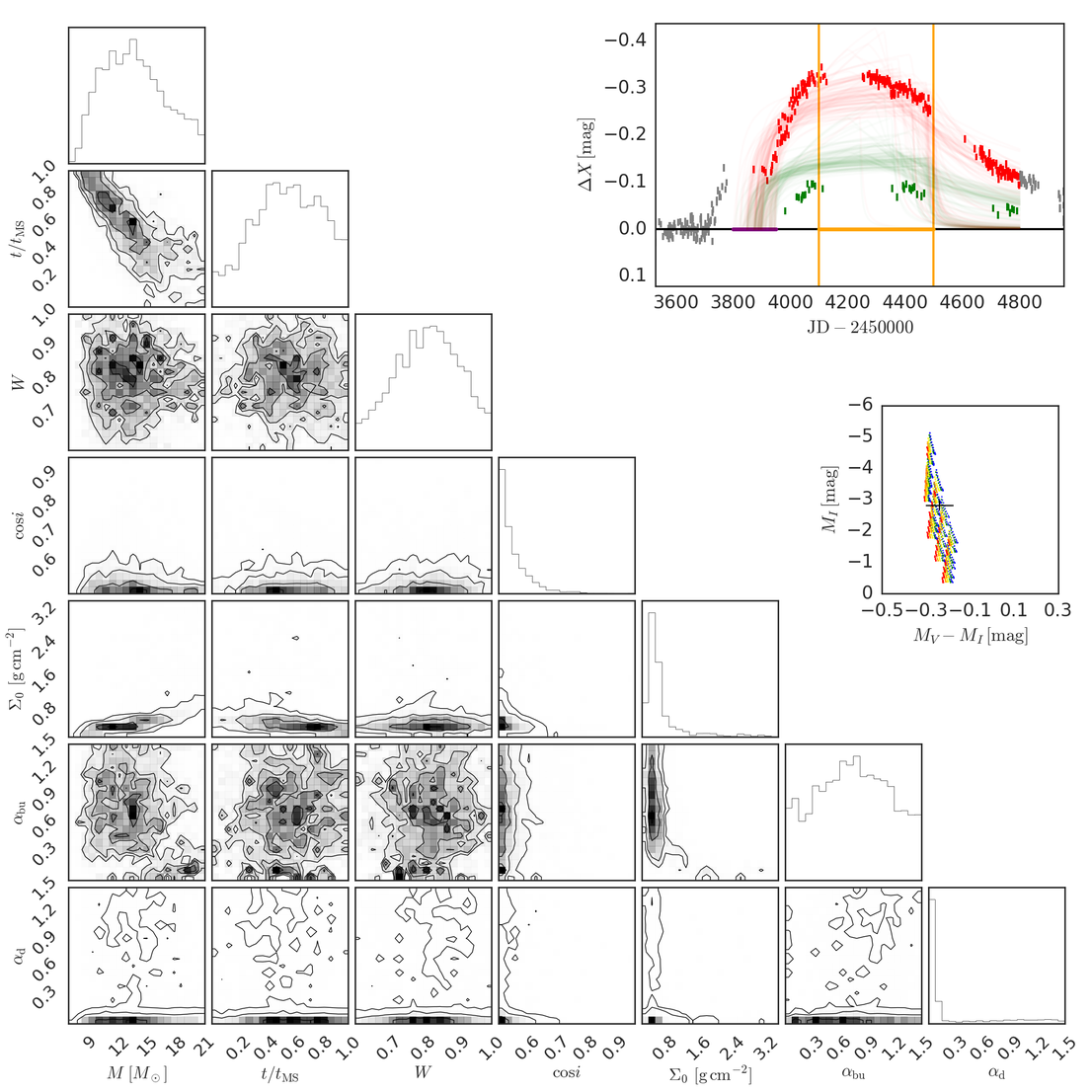}
}
\caption[]
{
Same as Fig.~\ref{example_bb1} for SMC\_SC5 260841 and bump ID 02. 
}
\label{smc_sc5_260841_02}
\end{figure*}
\clearpage

\begin{figure*}
\centering{
\includegraphics[width=1.0\linewidth]{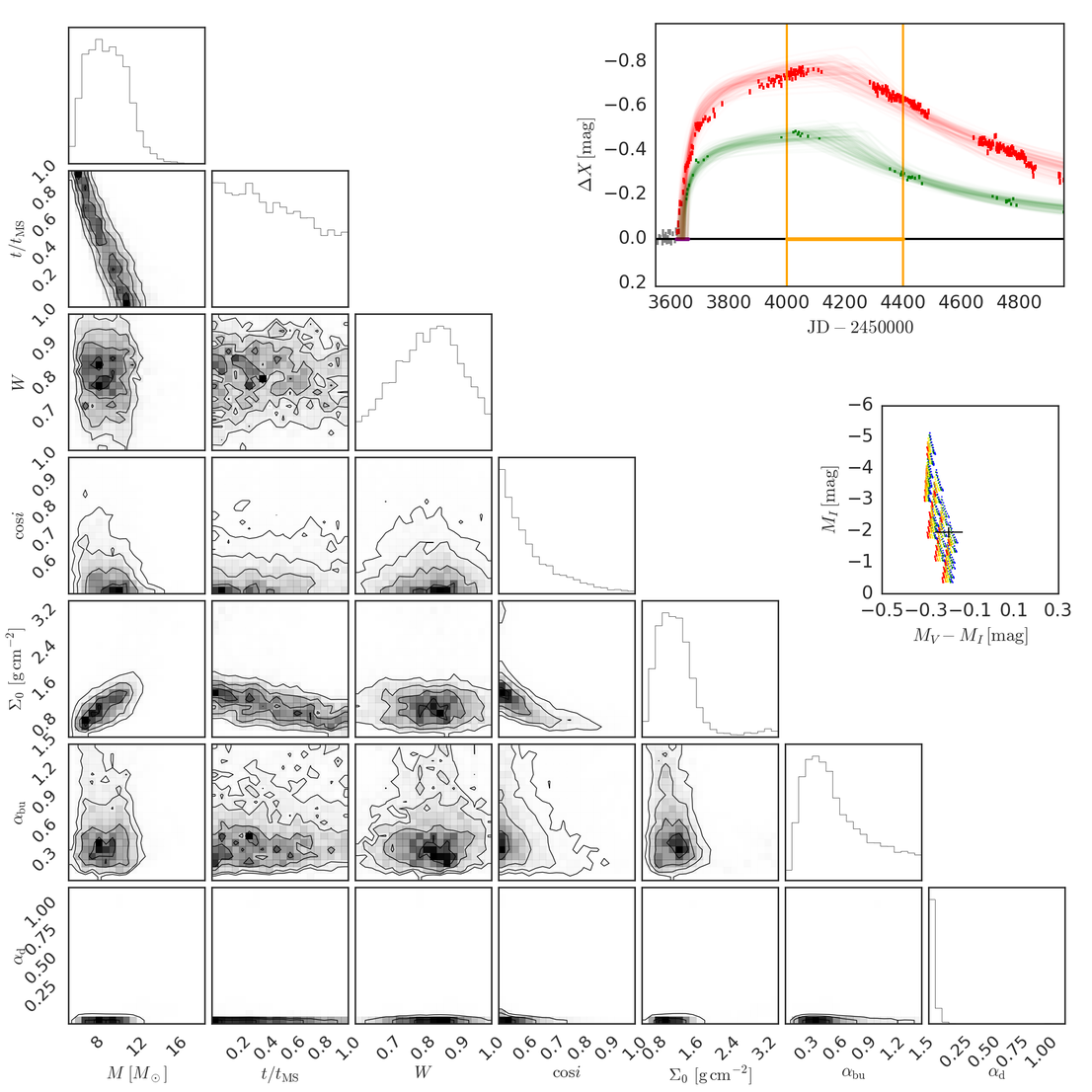}
}
\caption[]
{
Same as Fig.~\ref{example_bb1} for SMC\_SC5 260957 and bump ID 01. 
}
\label{smc_sc5_260957_01}
\end{figure*}
\clearpage

\begin{figure*}
\centering{
\includegraphics[width=1.0\linewidth]{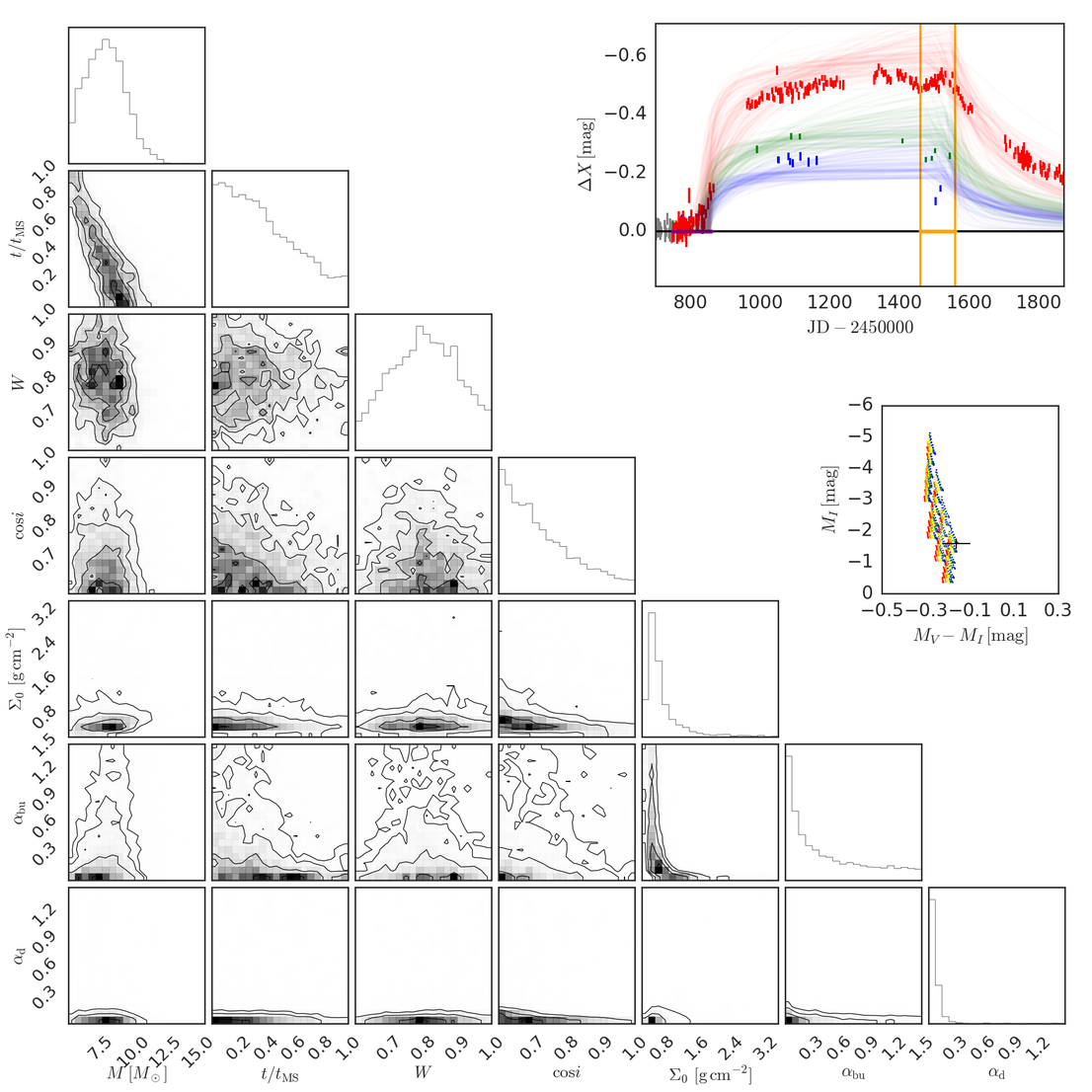}
}
\caption[]
{
Same as Fig.~\ref{example_bb1} for SMC\_SC5 266088 and bump ID 01. 
}
\label{smc_sc5_266088_01}
\end{figure*}
\clearpage

\begin{figure*}
\centering{
\includegraphics[width=1.0\linewidth]{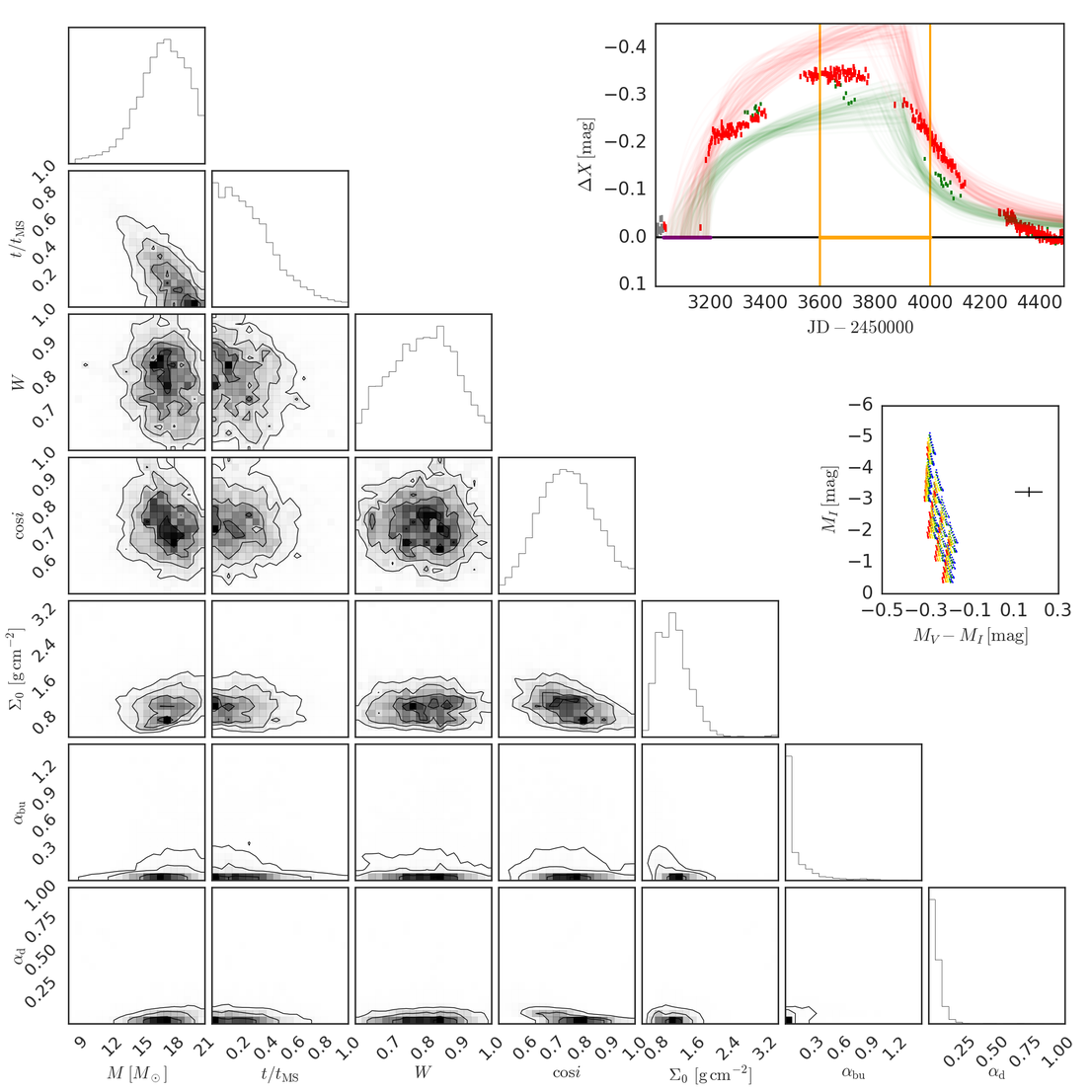}
}
\caption[]
{
Same as Fig.~\ref{example_bb1} for SMC\_SC5 276982 and bump ID 01. 
}
\label{smc_sc5_276982_01}
\end{figure*}
\clearpage

\begin{figure*}
\centering{
\includegraphics[width=1.0\linewidth]{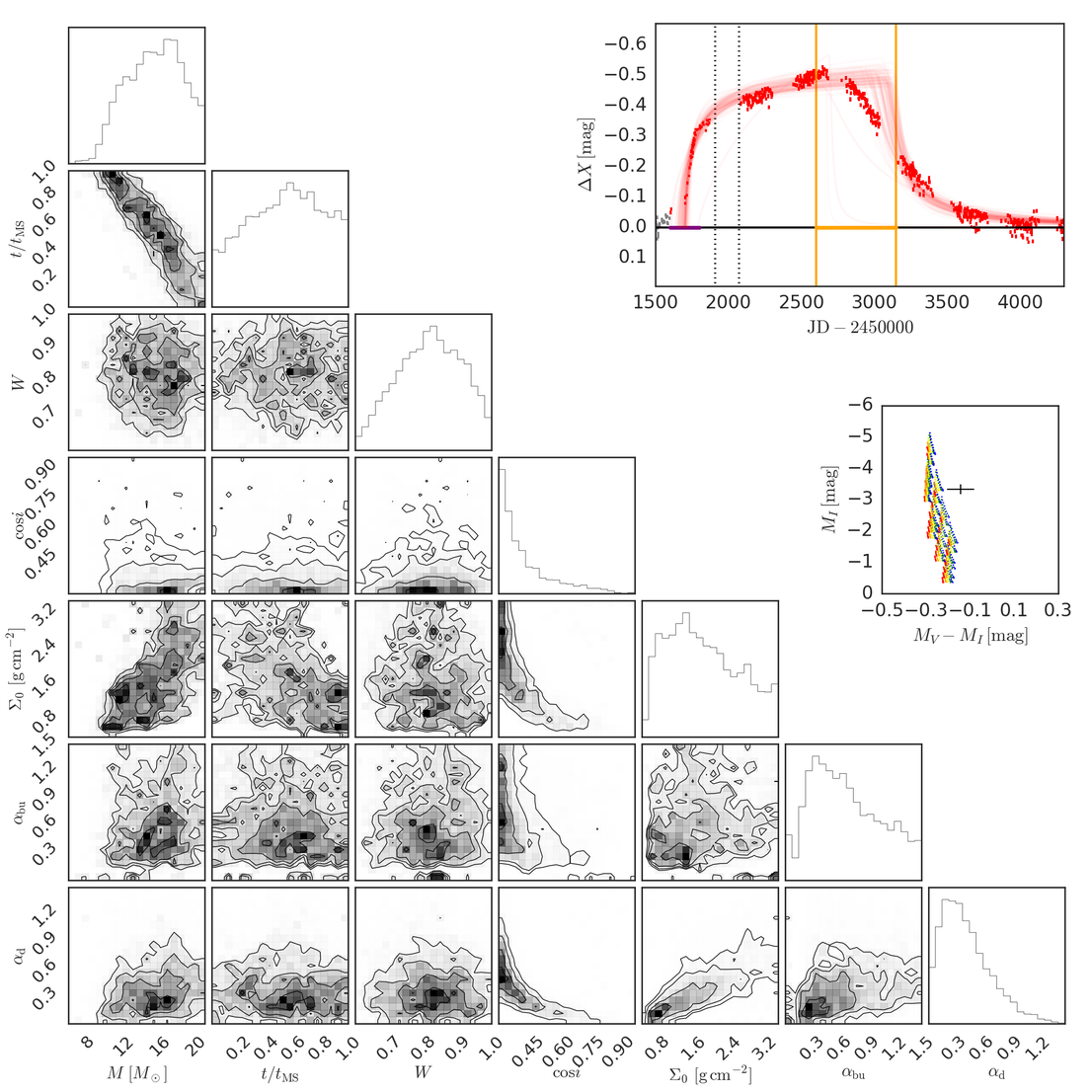}
}
\caption[]
{
Same as Fig.~\ref{example_bb1} for SMC\_SC5 282963 and bump ID 01. 
}
\label{smc_sc5_282963_01}
\end{figure*}
\clearpage

\begin{figure*}
\centering{
\includegraphics[width=1.0\linewidth]{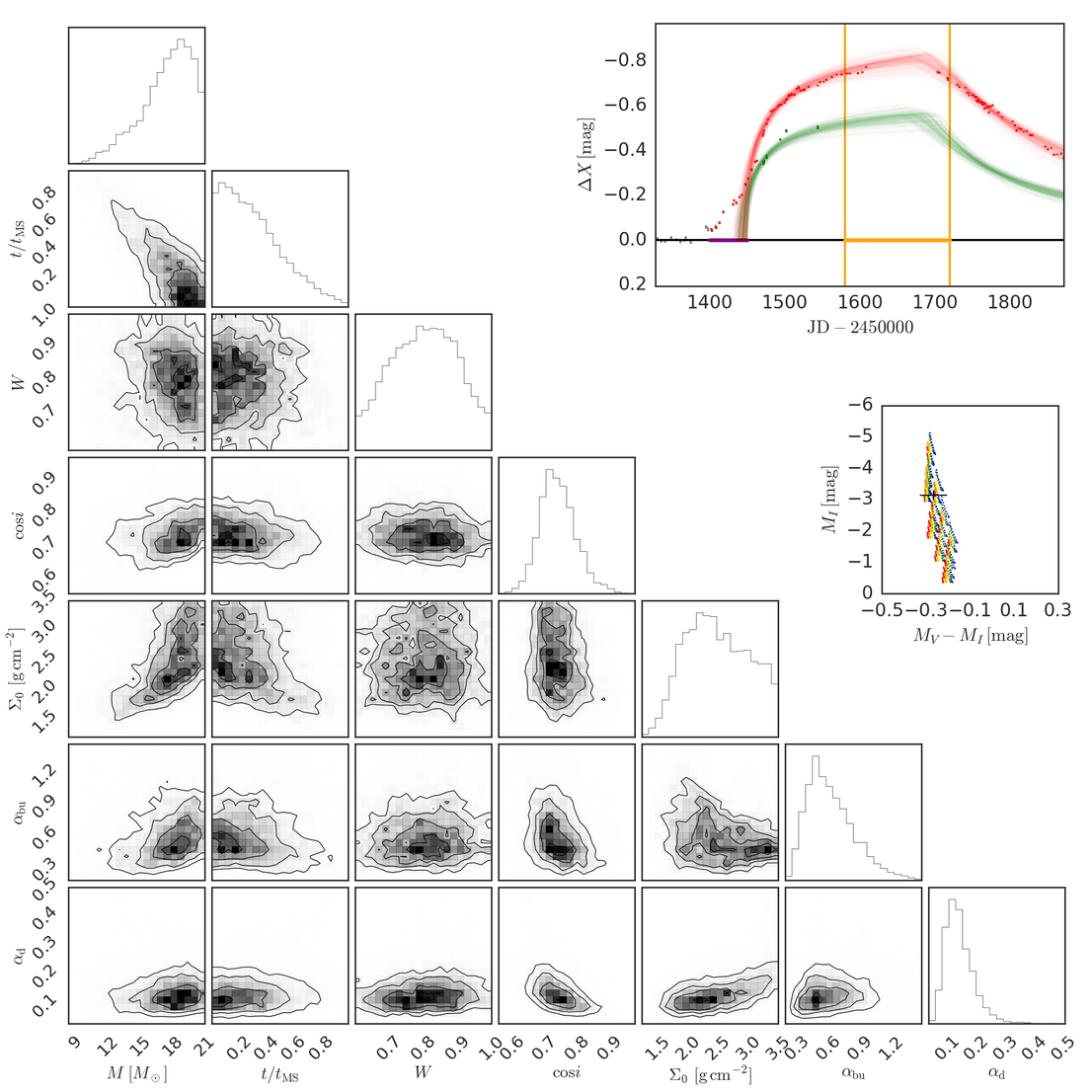}
}
\caption[]
{
Same as Fig.~\ref{example_bb1} for SMC\_SC6 11085 and bump ID 01. 
}
\label{smc_sc6_11085_01}
\end{figure*}
\clearpage

\begin{figure*}
\centering{
\includegraphics[width=1.0\linewidth]{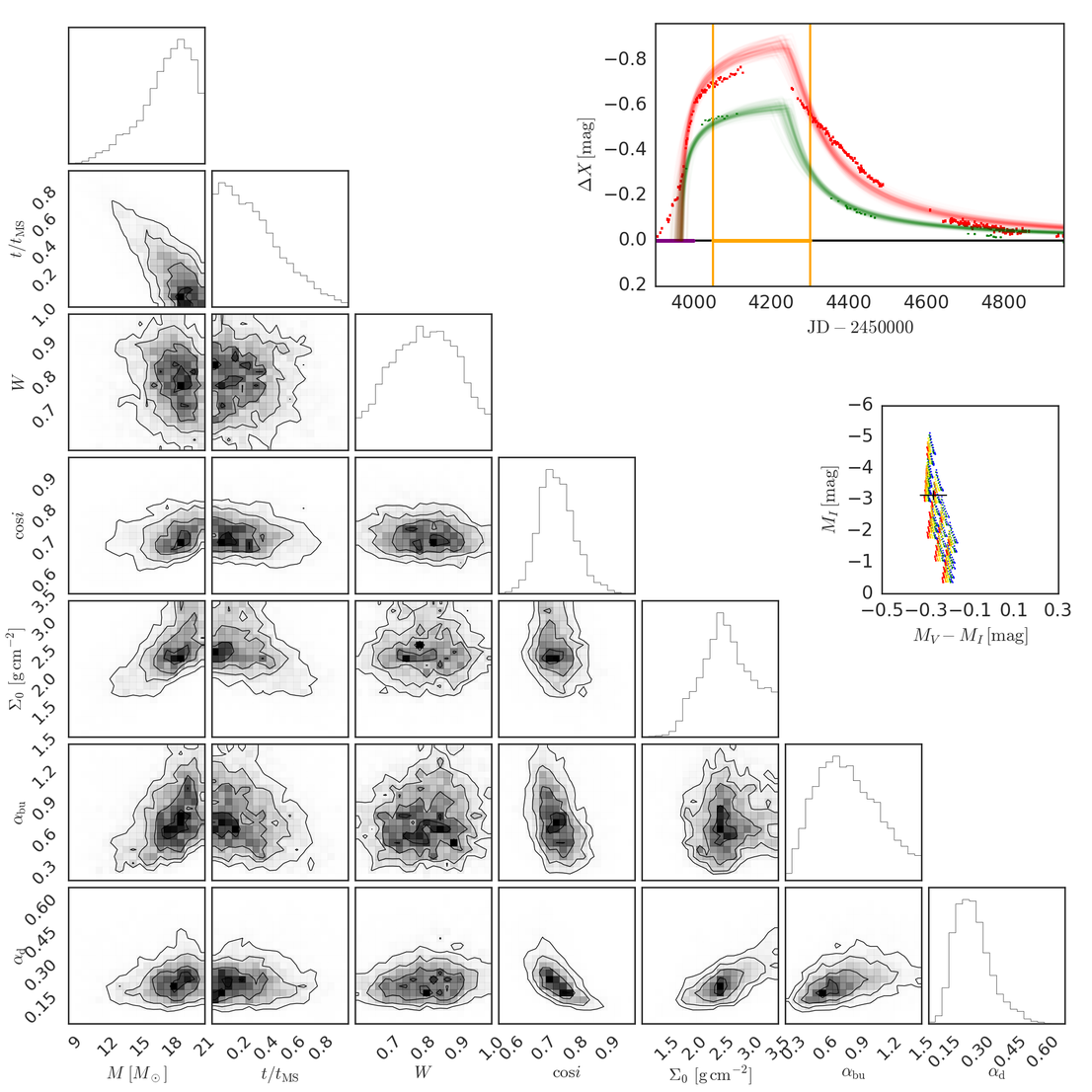}
}
\caption[]
{
Same as Fig.~\ref{example_bb1} for SMC\_SC6 11085 and bump ID 02. 
}
\label{smc_sc6_11085_02}
\end{figure*}
\clearpage

\begin{figure*}
\centering{
\includegraphics[width=1.0\linewidth]{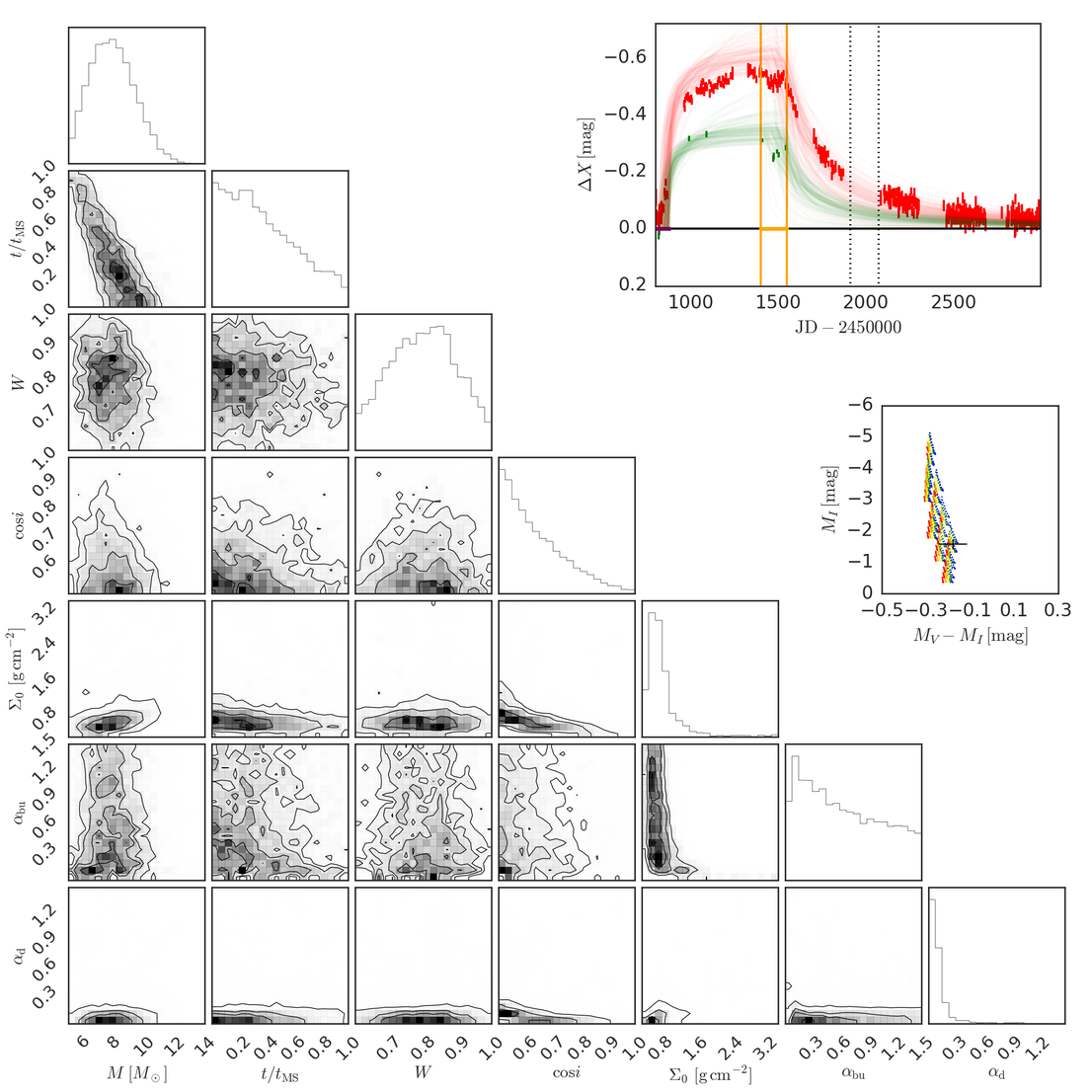}
}
\caption[]
{
Same as Fig.~\ref{example_bb1} for SMC\_SC6 17538 and bump ID 01. 
}
\label{smc_sc6_17538_01}
\end{figure*}
\clearpage

\begin{figure*}
\centering{
\includegraphics[width=1.0\linewidth]{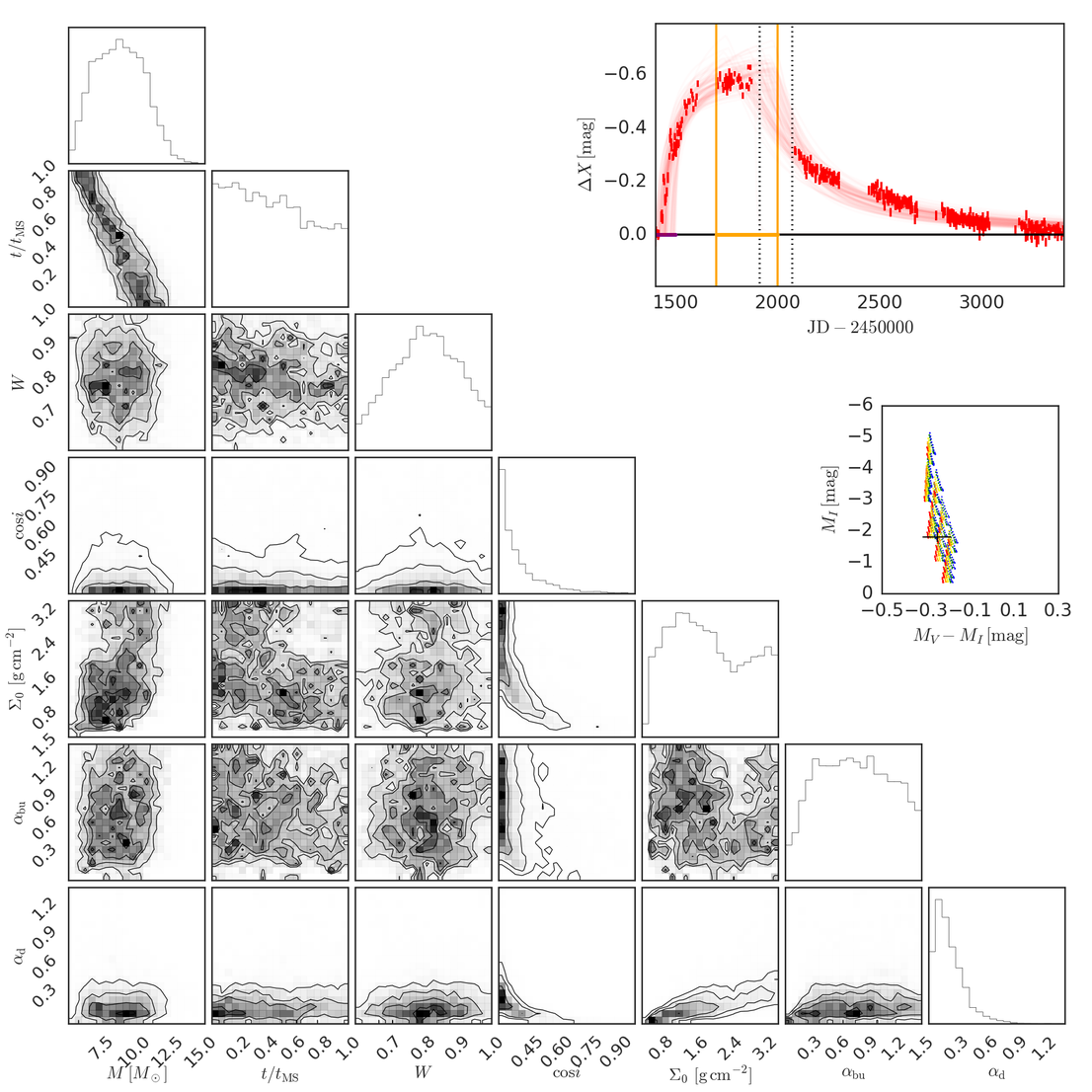}
}
\caption[]
{
Same as Fig.~\ref{example_bb1} for SMC\_SC6 42440 and bump ID 01. 
}
\label{smc_sc6_42440_01}
\end{figure*}
\clearpage

\begin{figure*}
\centering{
\includegraphics[width=1.0\linewidth]{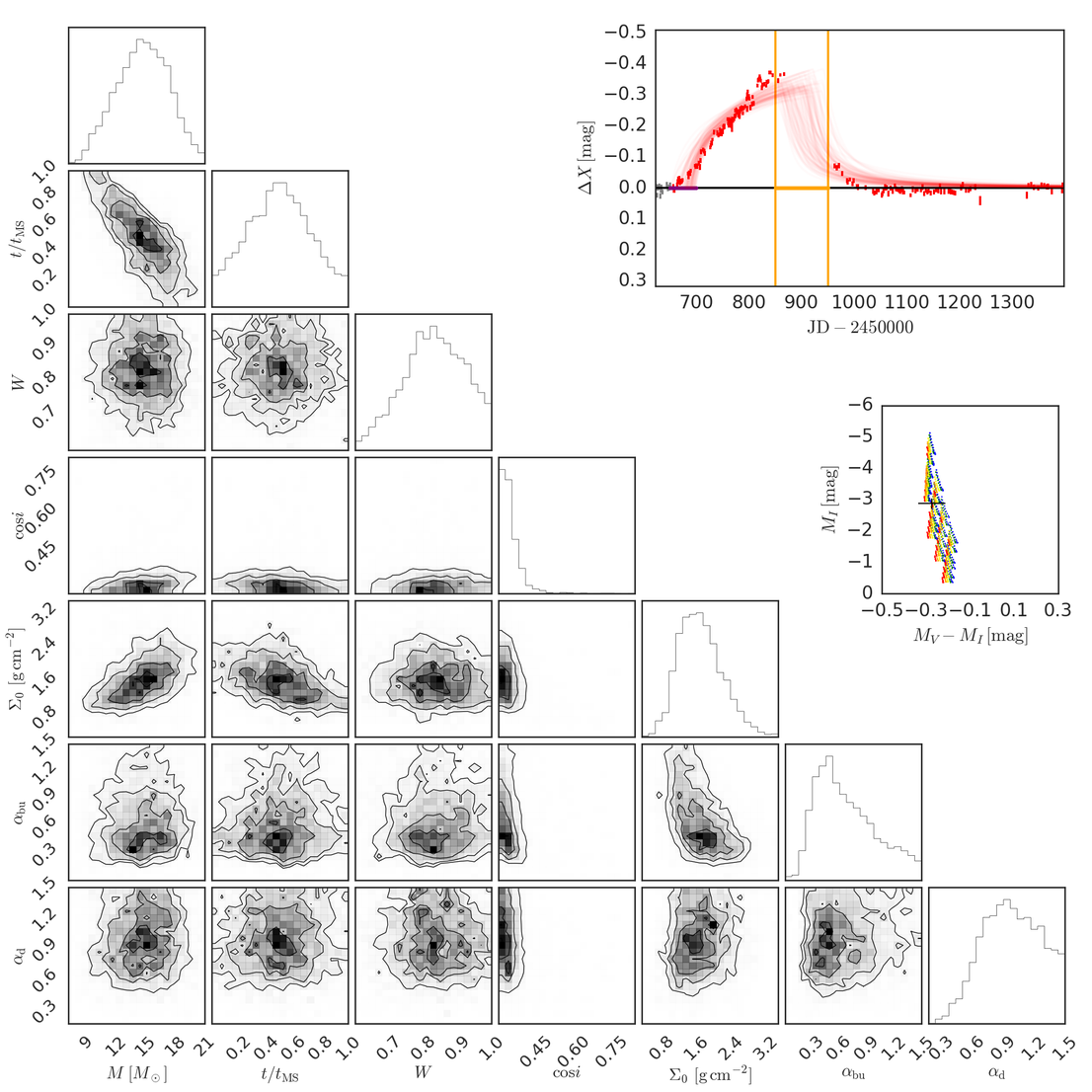}
}
\caption[]
{
Same as Fig.~\ref{example_bb1} for SMC\_SC6 99991 and bump ID 01. 
}
\label{smc_sc6_99991_01}
\end{figure*}
\clearpage

\begin{figure*}
\centering{
\includegraphics[width=1.0\linewidth]{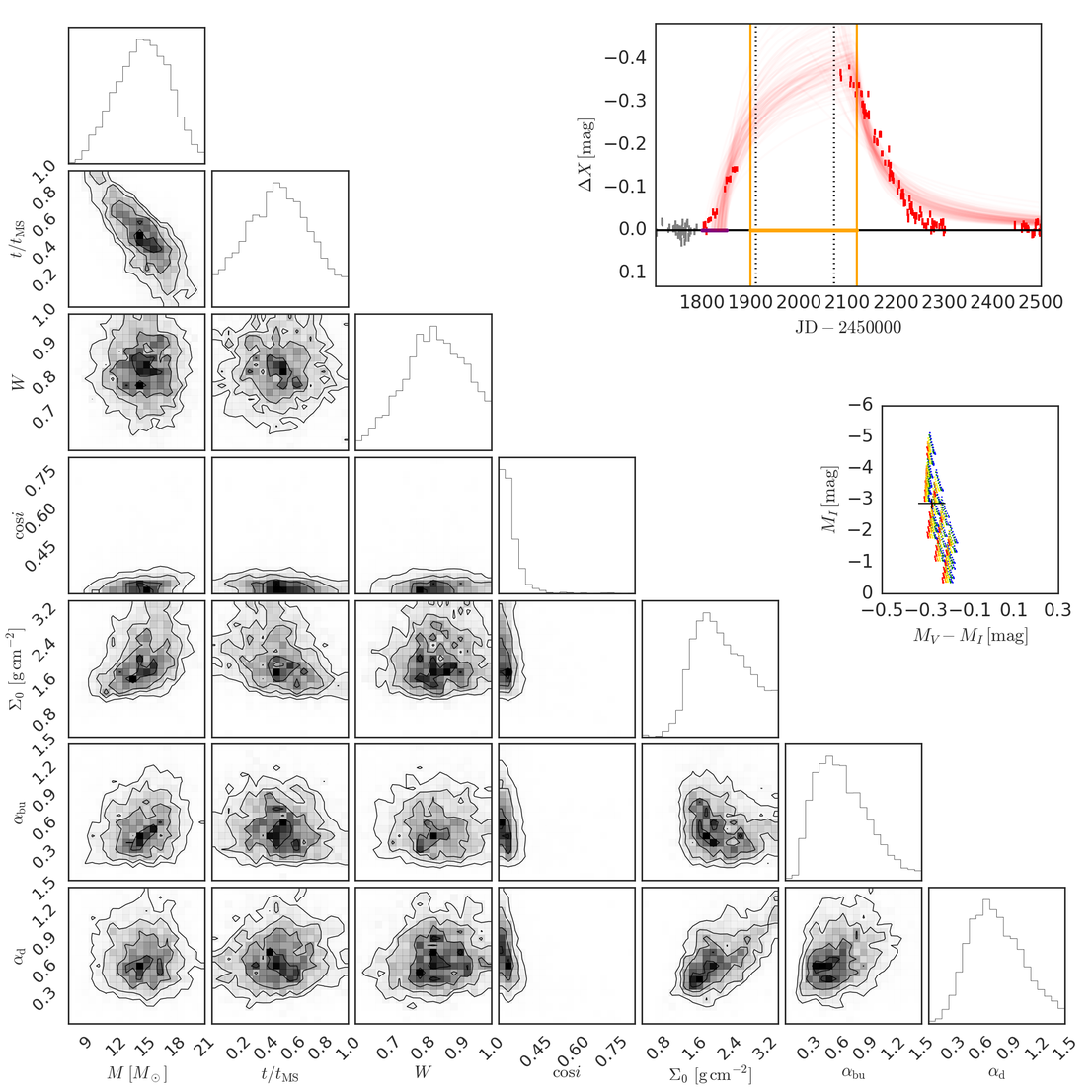}
}
\caption[]
{
Same as Fig.~\ref{example_bb1} for SMC\_SC6 99991 and bump ID 02. 
}
\label{smc_sc6_99991_02}
\end{figure*}
\clearpage

\begin{figure*}
\centering{
\includegraphics[width=1.0\linewidth]{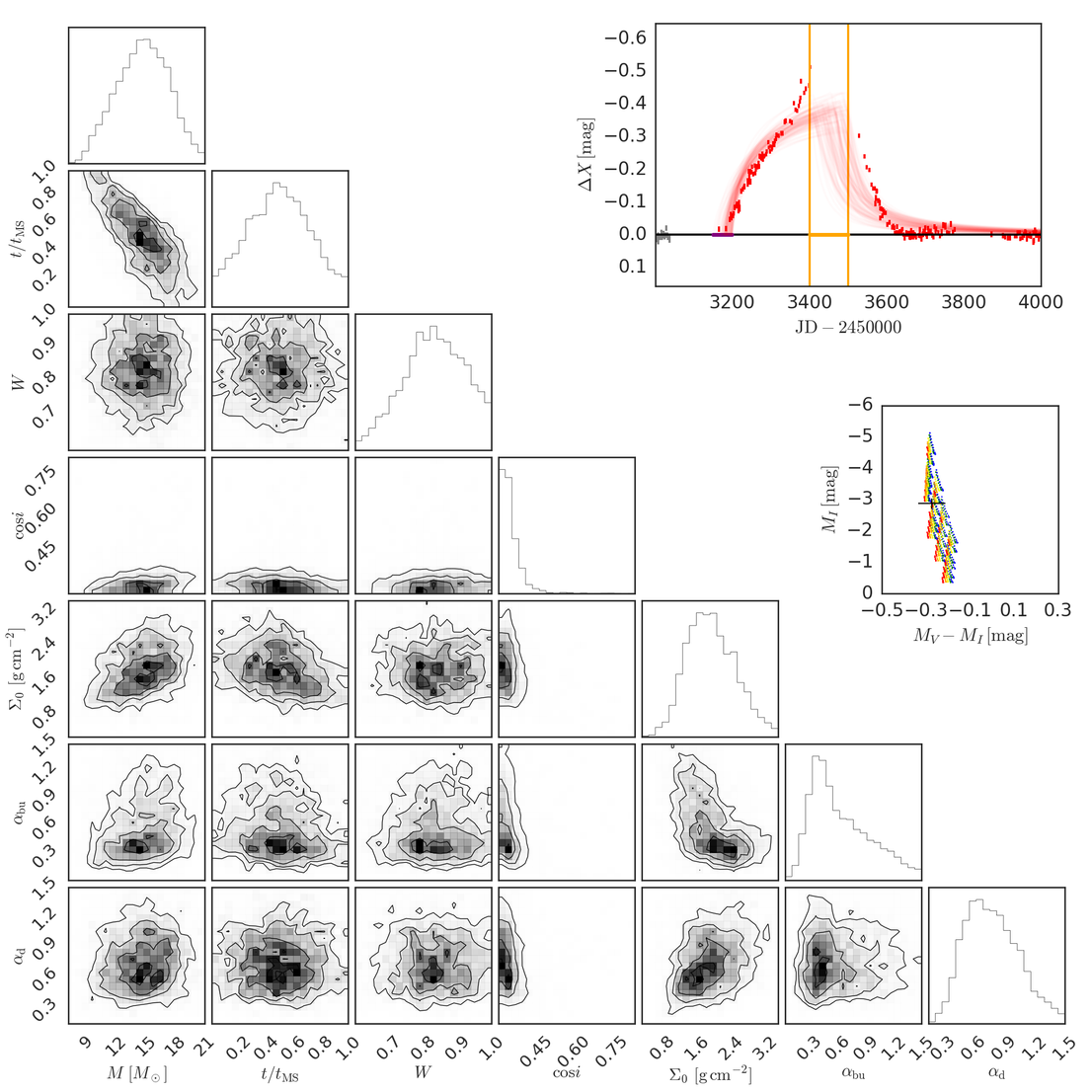}
}
\caption[]
{
Same as Fig.~\ref{example_bb1} for SMC\_SC6 99991 and bump ID 03. 
}
\label{smc_sc6_99991_03}
\end{figure*}
\clearpage

\begin{figure*}
\centering{
\includegraphics[width=1.0\linewidth]{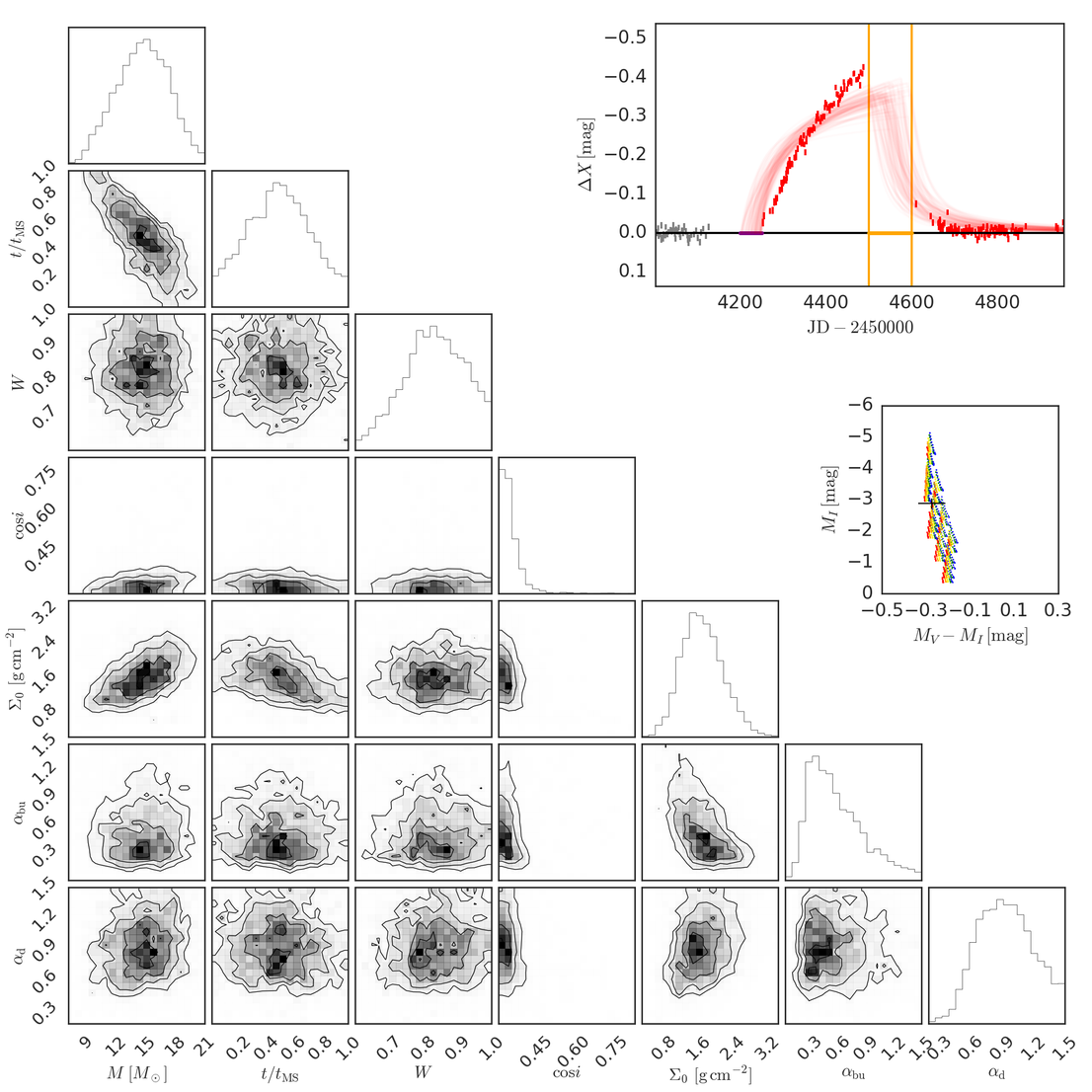}
}
\caption[]
{
Same as Fig.~\ref{example_bb1} for SMC\_SC6 99991 and bump ID 04. 
}
\label{smc_sc6_99991_04}
\end{figure*}
\clearpage

\begin{figure*}
\centering{
\includegraphics[width=1.0\linewidth]{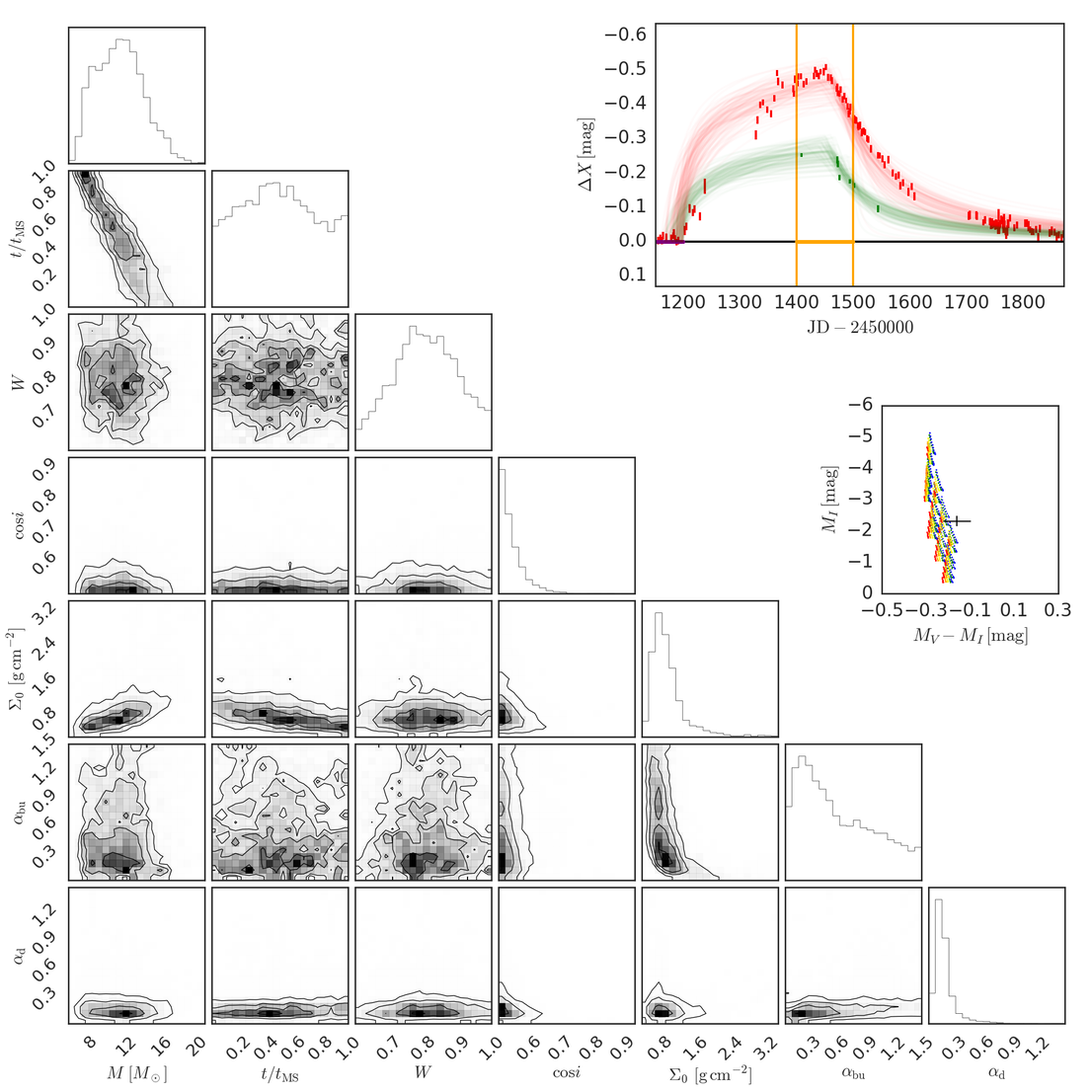}
}
\caption[]
{
Same as Fig.~\ref{example_bb1} for SMC\_SC6 105368 and bump ID 01. 
}
\label{smc_sc6_105368_01}
\end{figure*}
\clearpage

\begin{figure*}
\centering{
\includegraphics[width=1.0\linewidth]{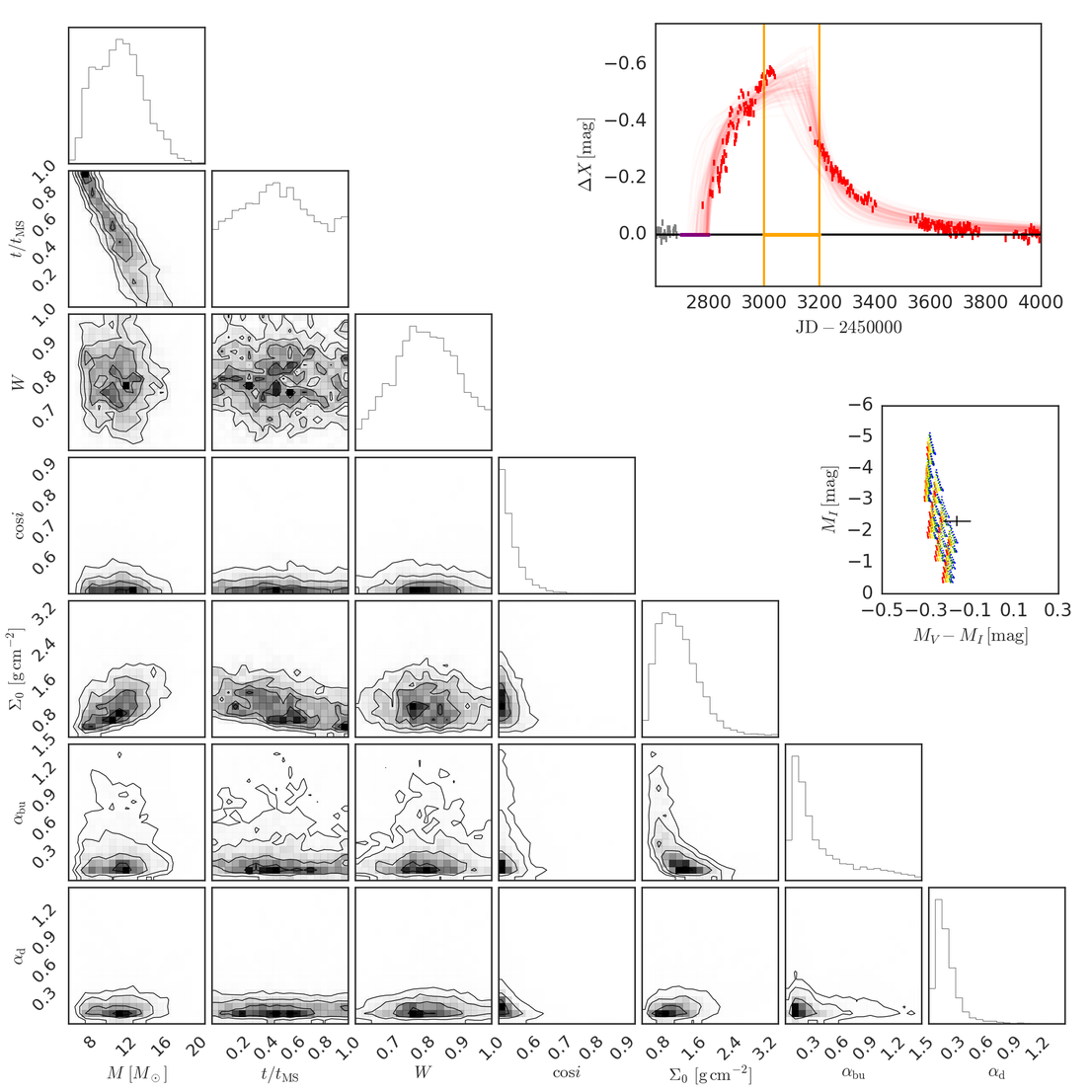}
}
\caption[]
{
Same as Fig.~\ref{example_bb1} for SMC\_SC6 105368 and bump ID 02. 
}
\label{smc_sc6_105368_02}
\end{figure*}
\clearpage

\begin{figure*}
\centering{
\includegraphics[width=1.0\linewidth]{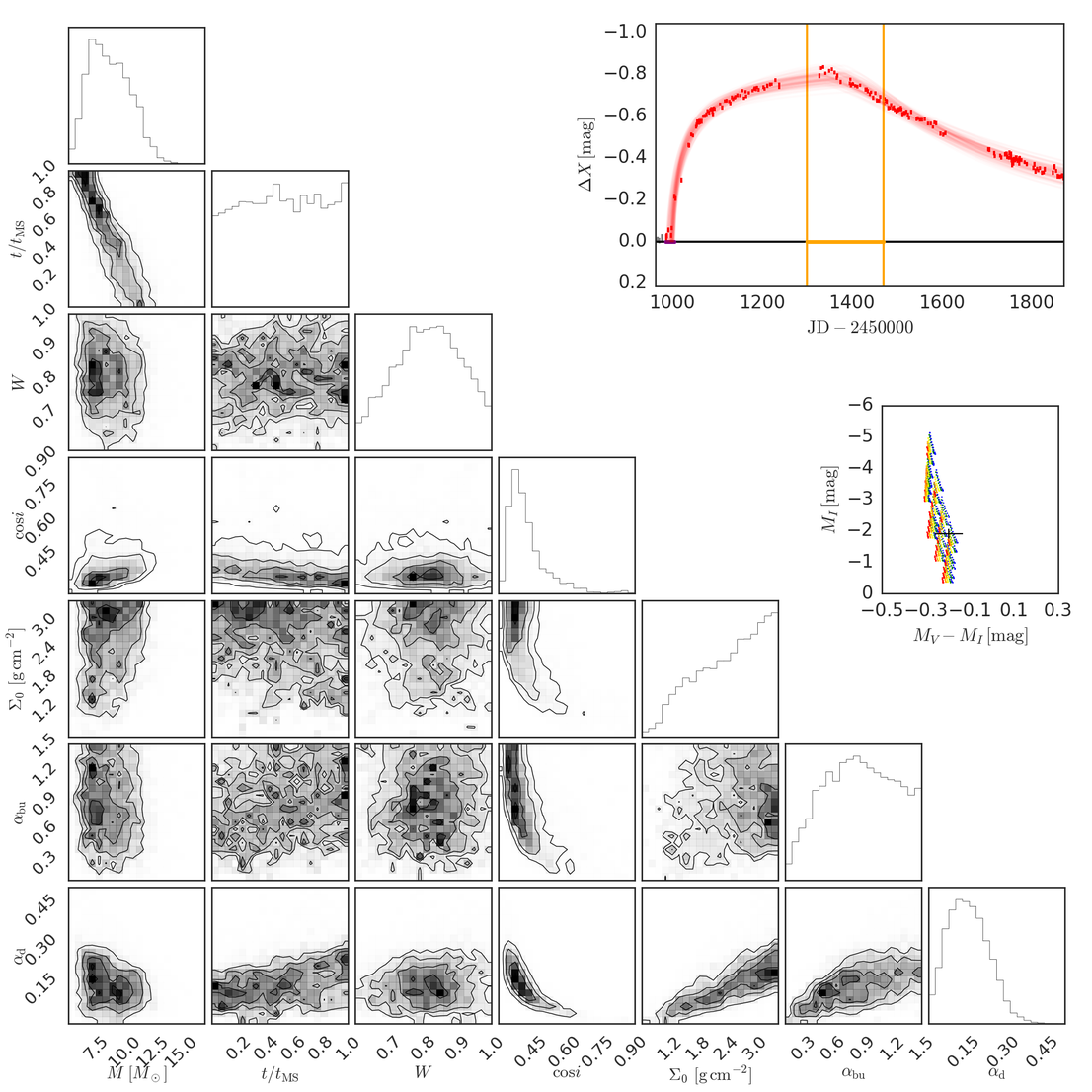}
}
\caption[]
{
Same as Fig.~\ref{example_bb1} for SMC\_SC6 116294 and bump ID 01. 
}
\label{smc_sc6_116294_01}
\end{figure*}
\clearpage


%
\begin{figure*}
\centering{
\includegraphics[width=1.0\linewidth]{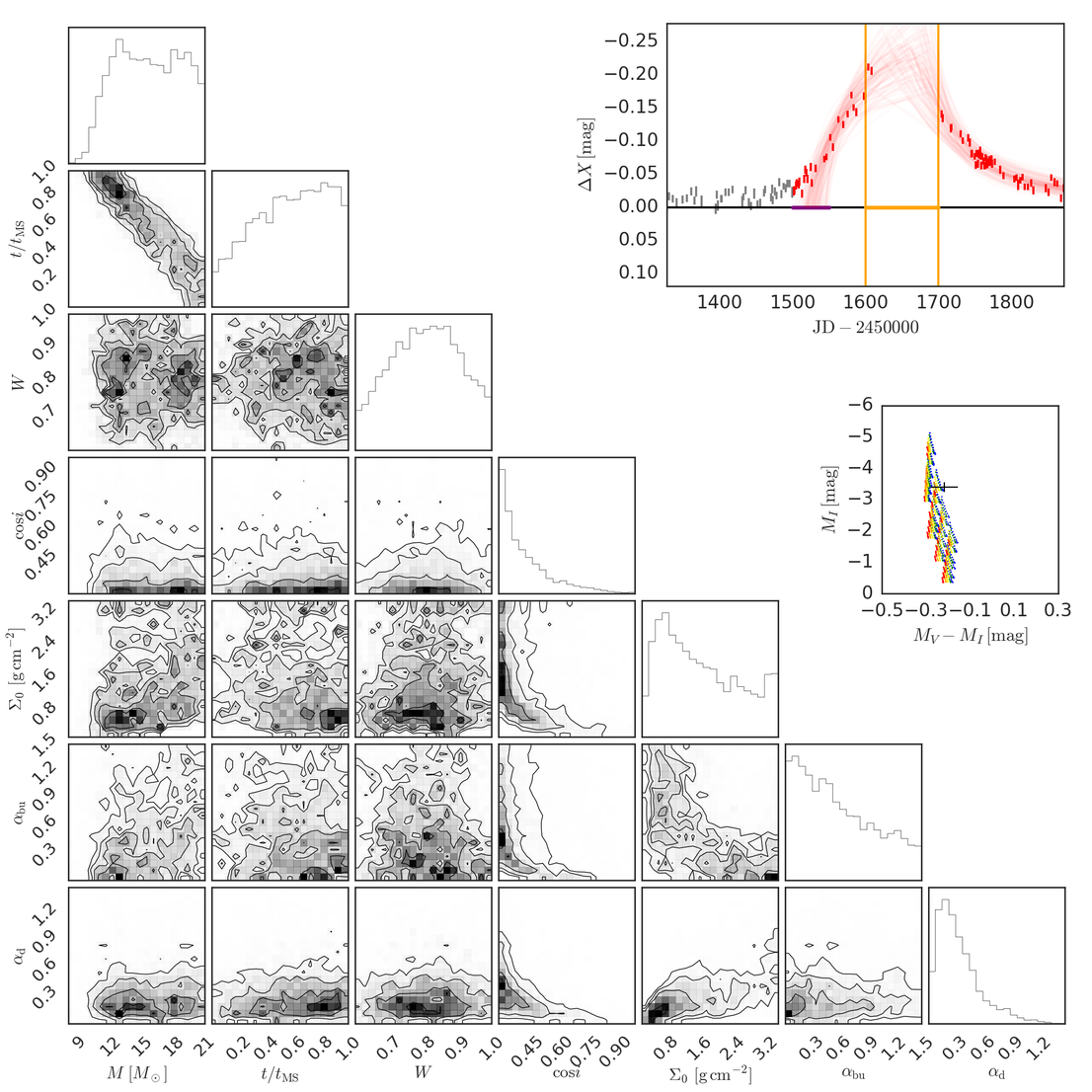}
}
\caption[]
{
Same as Fig.~\ref{example_bb1} for SMC\_SC6 199611 and bump ID 01. 
}
\label{smc_sc6_199611_01}
\end{figure*}
\clearpage

\begin{figure*}
\centering{
\includegraphics[width=1.0\linewidth]{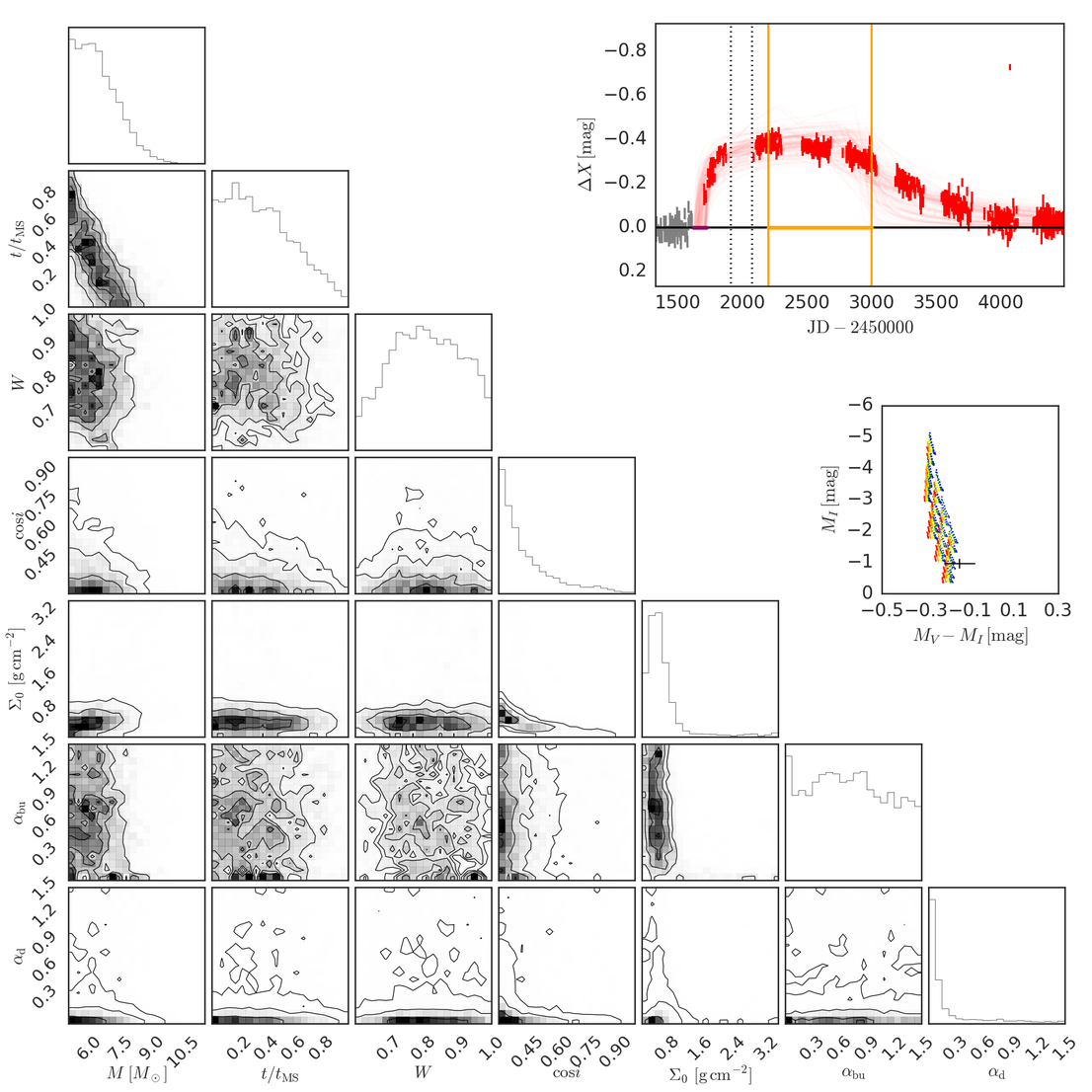}
}
\caption[]
{
Same as Fig.~\ref{example_bb1} for SMC\_SC6 272665 and bump ID 01. 
}
\label{smc_sc6_272665_01}
\end{figure*}
\clearpage

\begin{figure*}
\centering{
\includegraphics[width=1.0\linewidth]{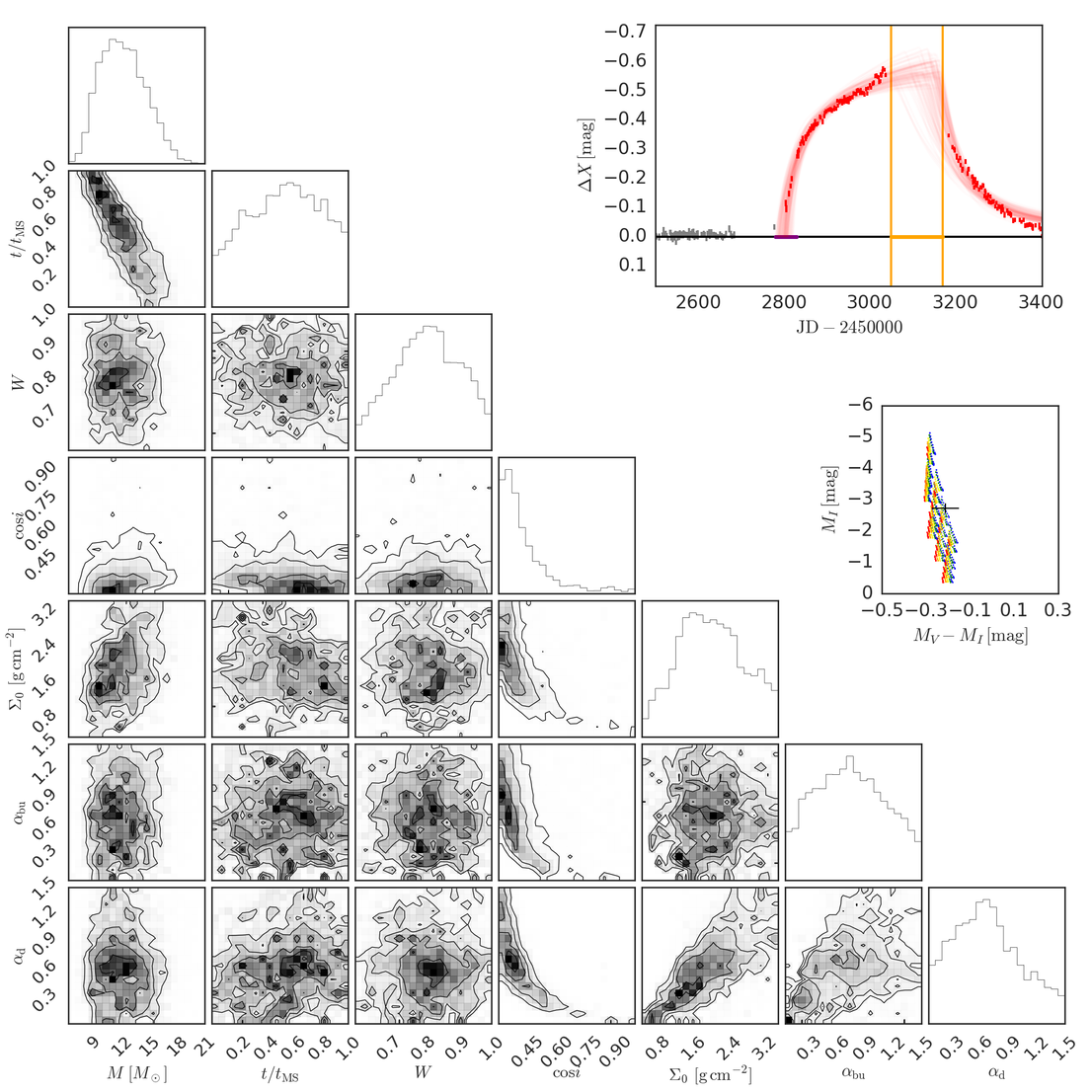}
}
\caption[]
{
Same as Fig.~\ref{example_bb1} for SMC\_SC7 57131 and bump ID 01. 
}
\label{smc_sc7_57131_01}
\end{figure*}
\clearpage

\begin{figure*}
\centering{
\includegraphics[width=1.0\linewidth]{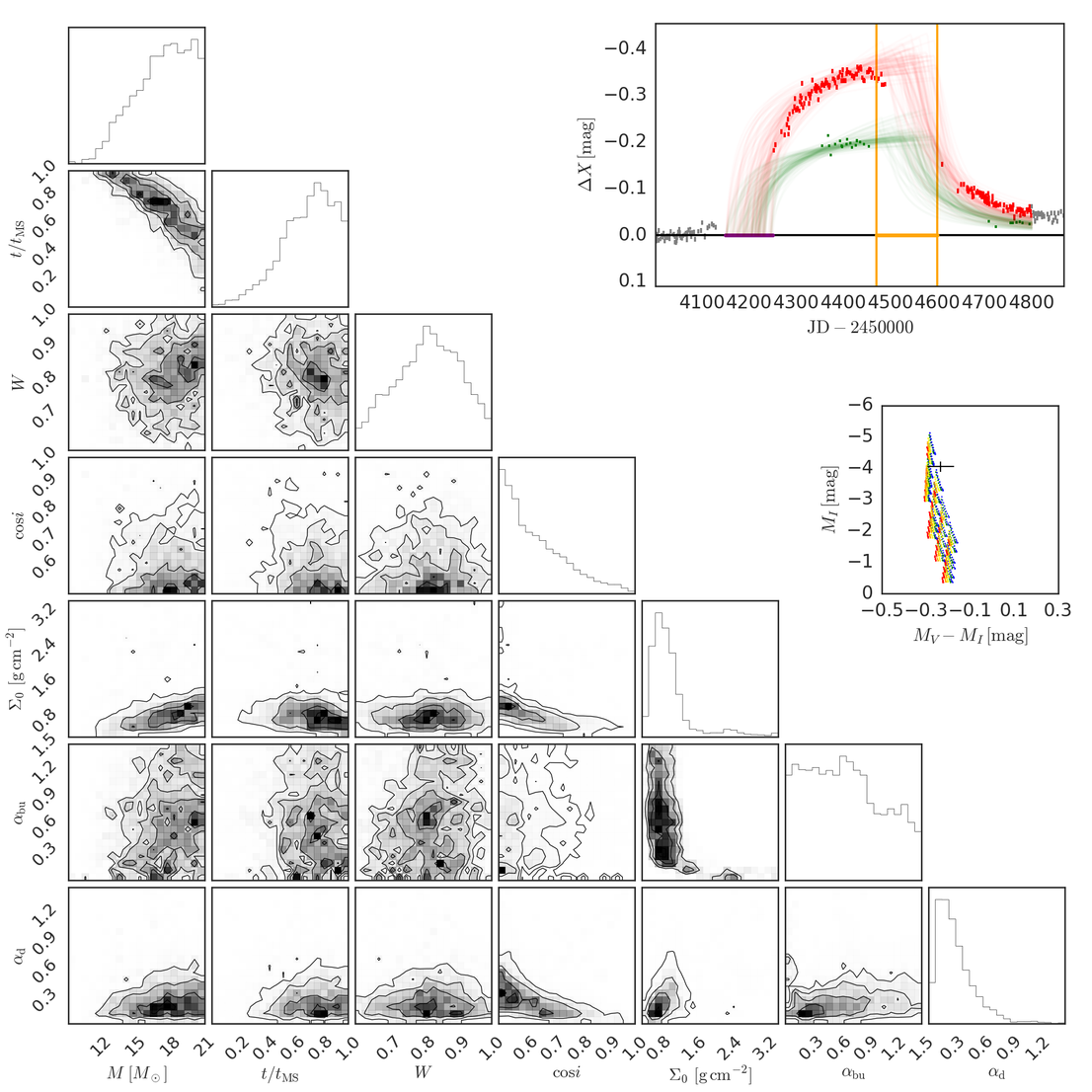}
}
\caption[]
{
Same as Fig.~\ref{example_bb1} for SMC\_SC8 183240 and bump ID 01. 
}
\label{smc_sc8_183240_01}
\end{figure*}
\clearpage

\begin{figure*}
\centering{
\includegraphics[width=1.0\linewidth]{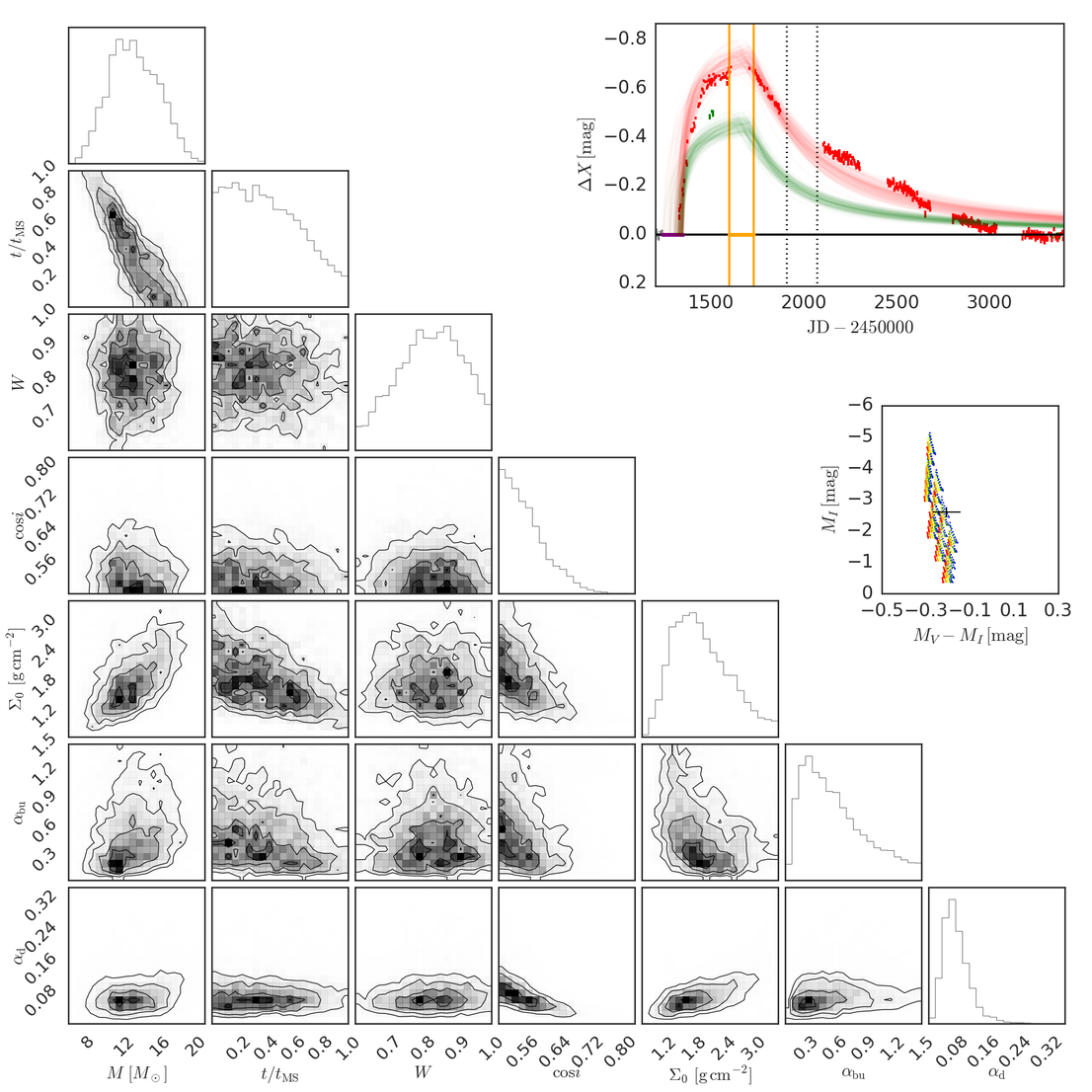}
}
\caption[]
{
Same as Fig.~\ref{example_bb1} for SMC\_SC9 105383 and bump ID 01. 
}
\label{smc_sc9_105383_01}
\end{figure*}
\clearpage

\begin{figure*}
\centering{
\includegraphics[width=1.0\linewidth]{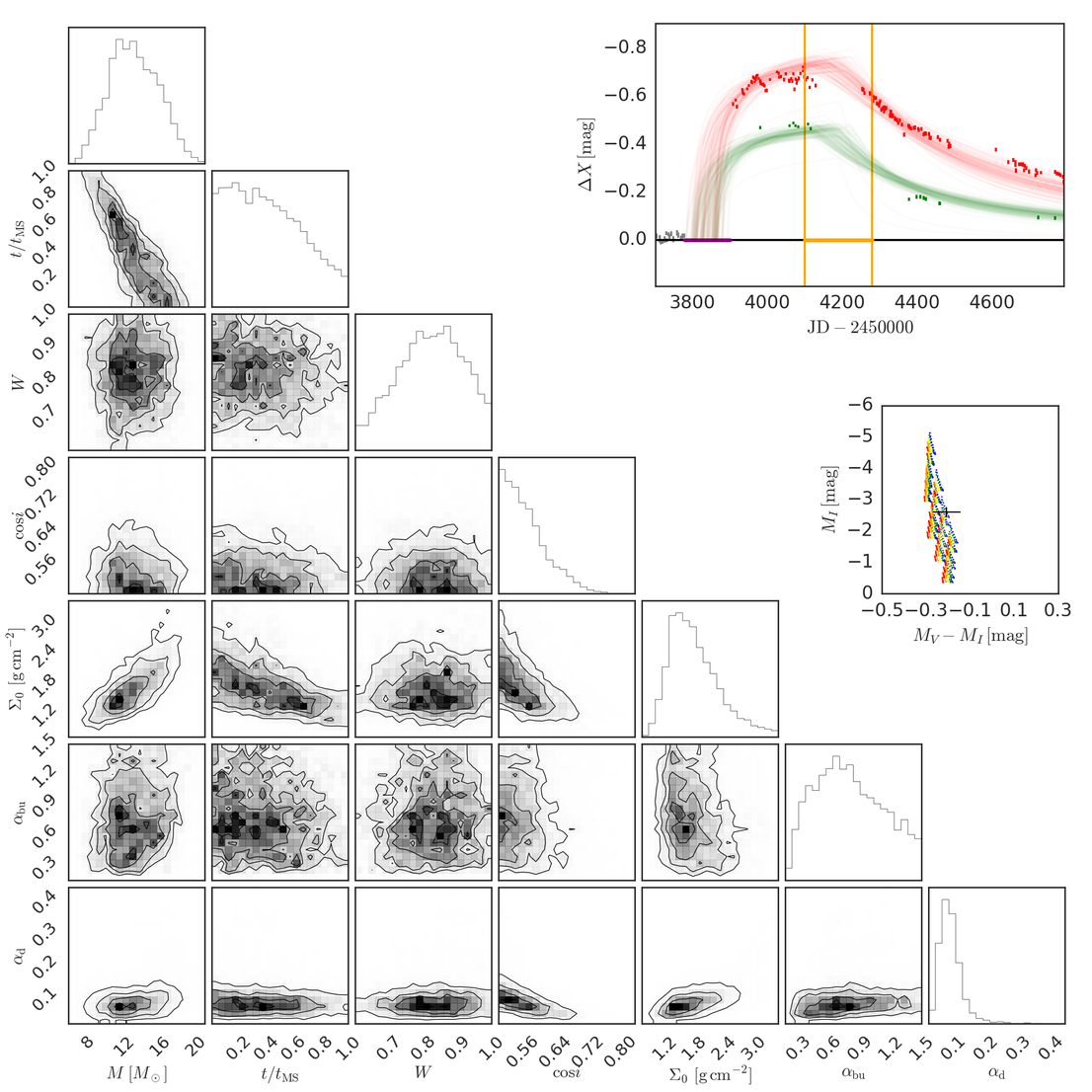}
}
\caption[]
{
Same as Fig.~\ref{example_bb1} for SMC\_SC9 105383 and bump ID 02. 
}
\label{smc_sc9_105383_02}
\end{figure*}
\clearpage

\begin{figure*}
\centering{
\includegraphics[width=1.0\linewidth]{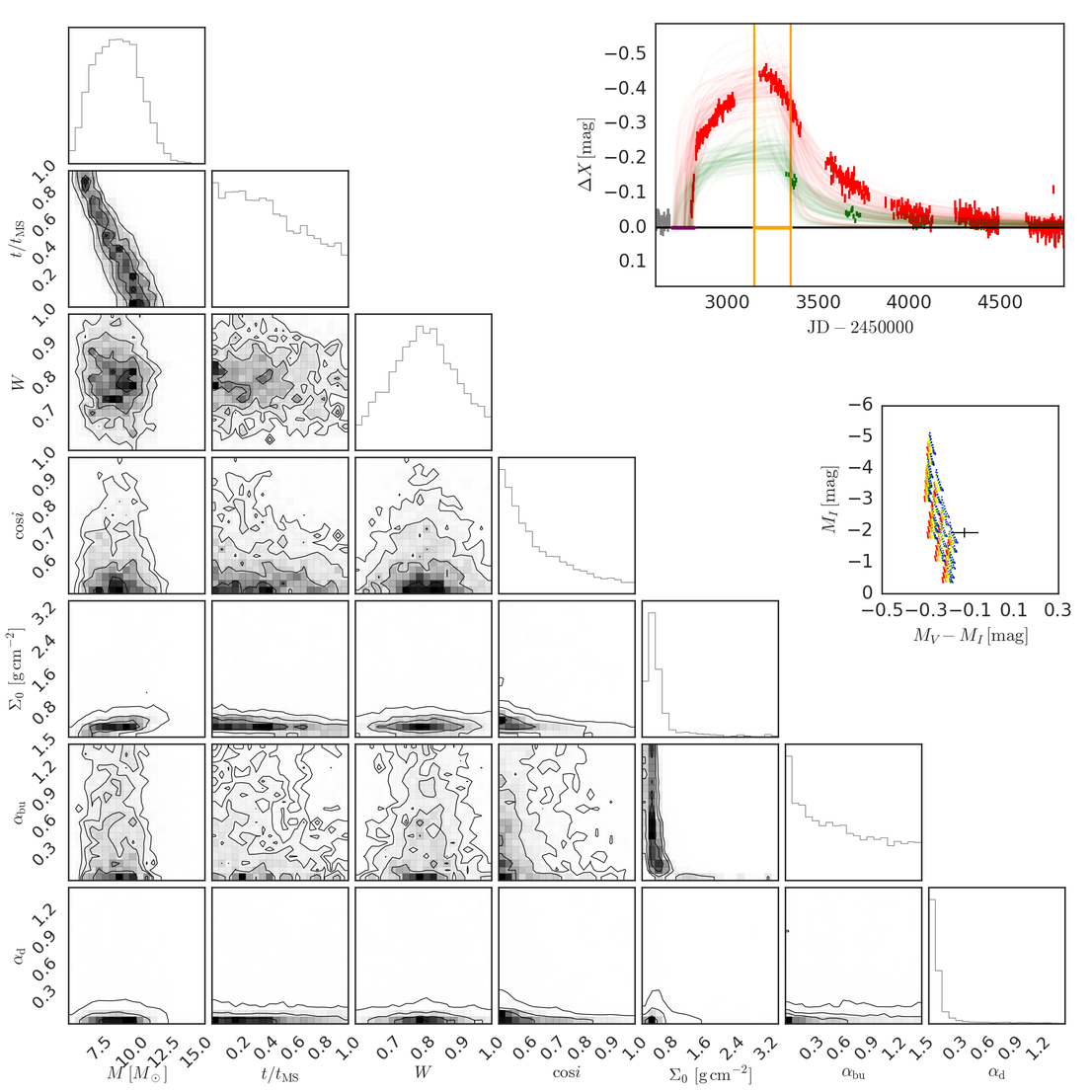}
}
\caption[]
{
Same as Fig.~\ref{example_bb1} for SMC\_SC9 168422 and bump ID 01. 
}
\label{smc_sc9_168422_01}
\end{figure*}
\clearpage

\begin{figure*}
\centering{
\includegraphics[width=1.0\linewidth]{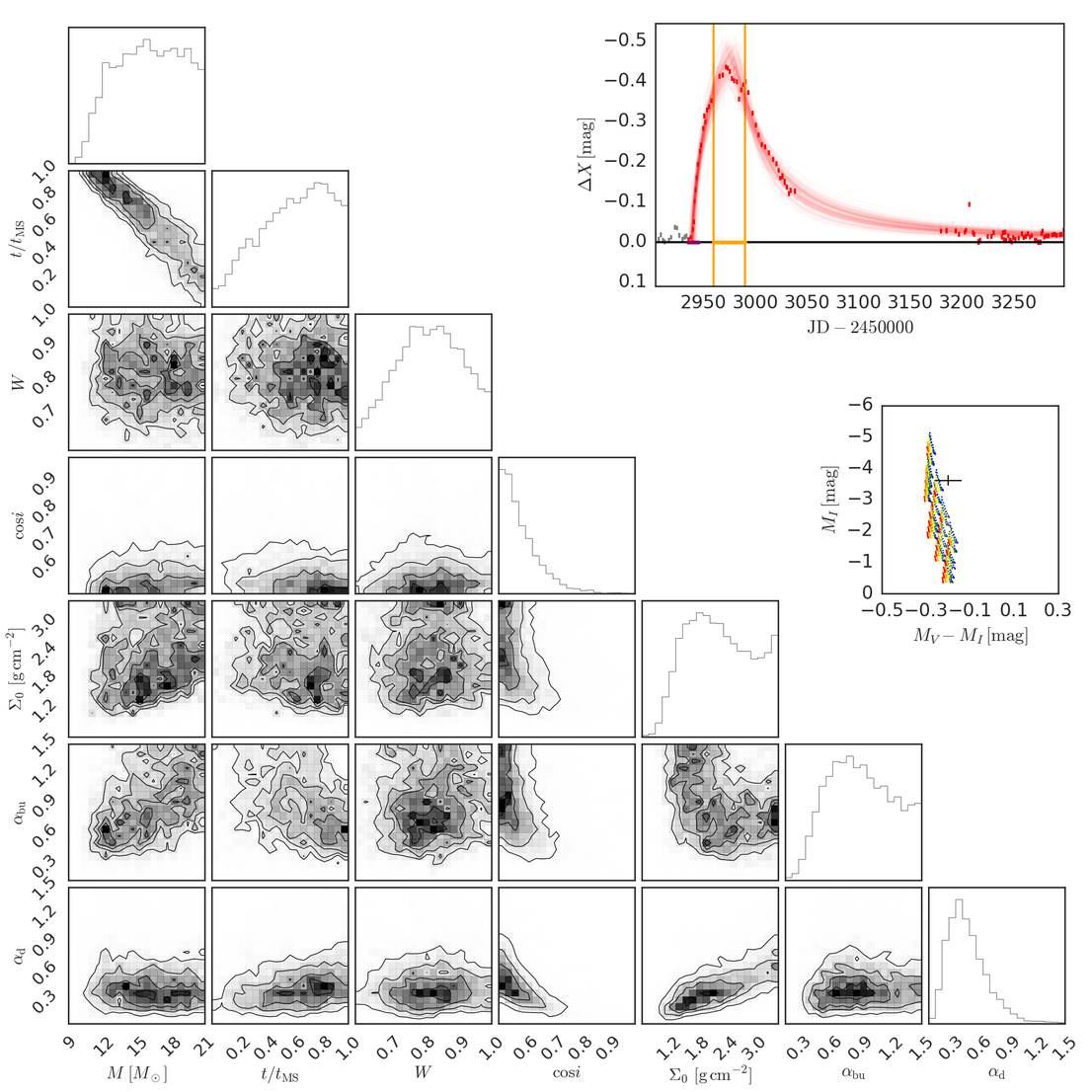}
}
\caption[]
{
Same as Fig.~\ref{example_bb1} for SMC\_SC10 8906 and bump ID 01. 
}
\label{smc_sc10_8906_01}
\end{figure*}
\clearpage

\begin{figure*}
\centering{
\includegraphics[width=1.0\linewidth]{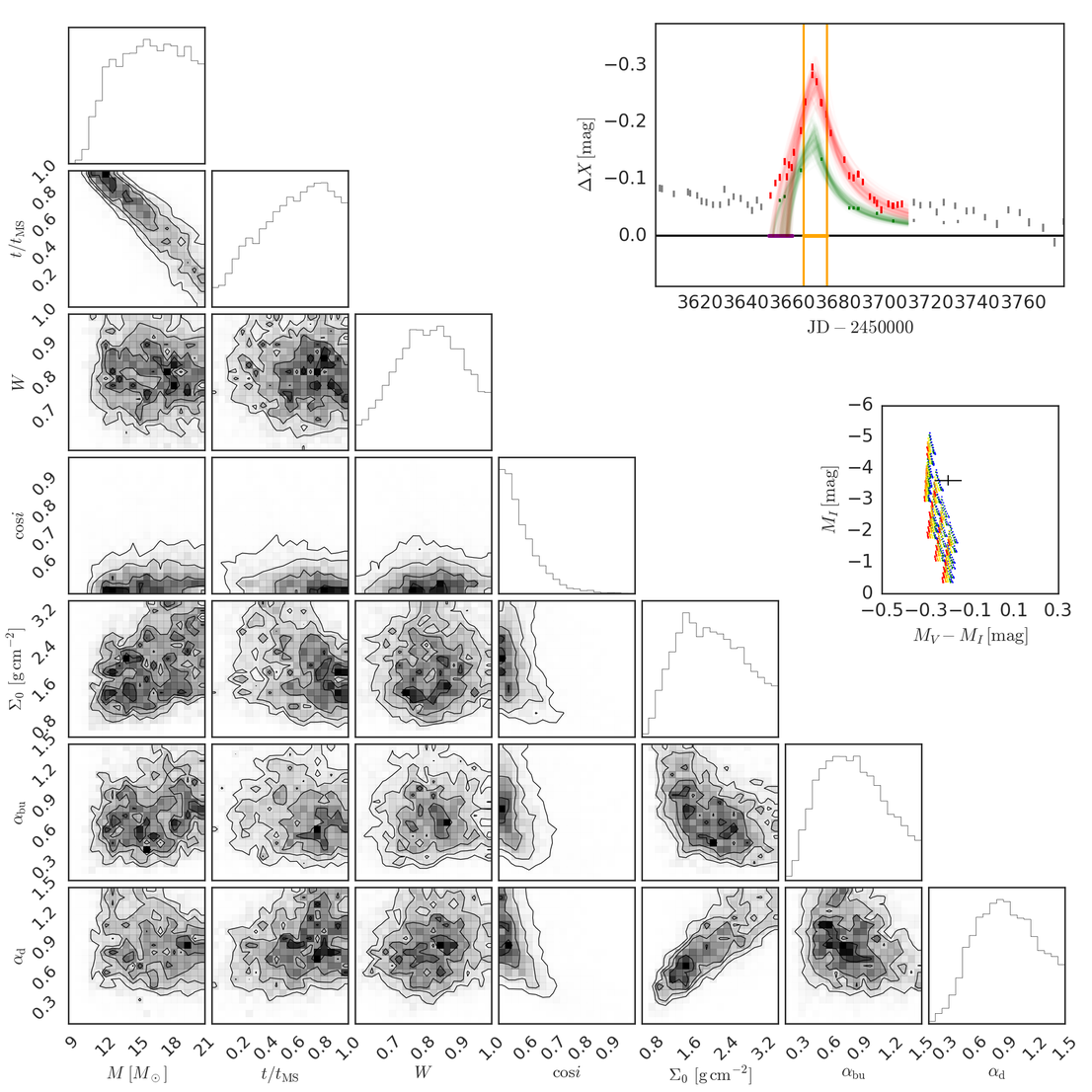}
}
\caption[]
{
Same as Fig.~\ref{example_bb1} for SMC\_SC10 8906 and bump ID 02. 
}
\label{smc_sc10_8906_02}
\end{figure*}
\clearpage

\begin{figure*}
\centering{
\includegraphics[width=1.0\linewidth]{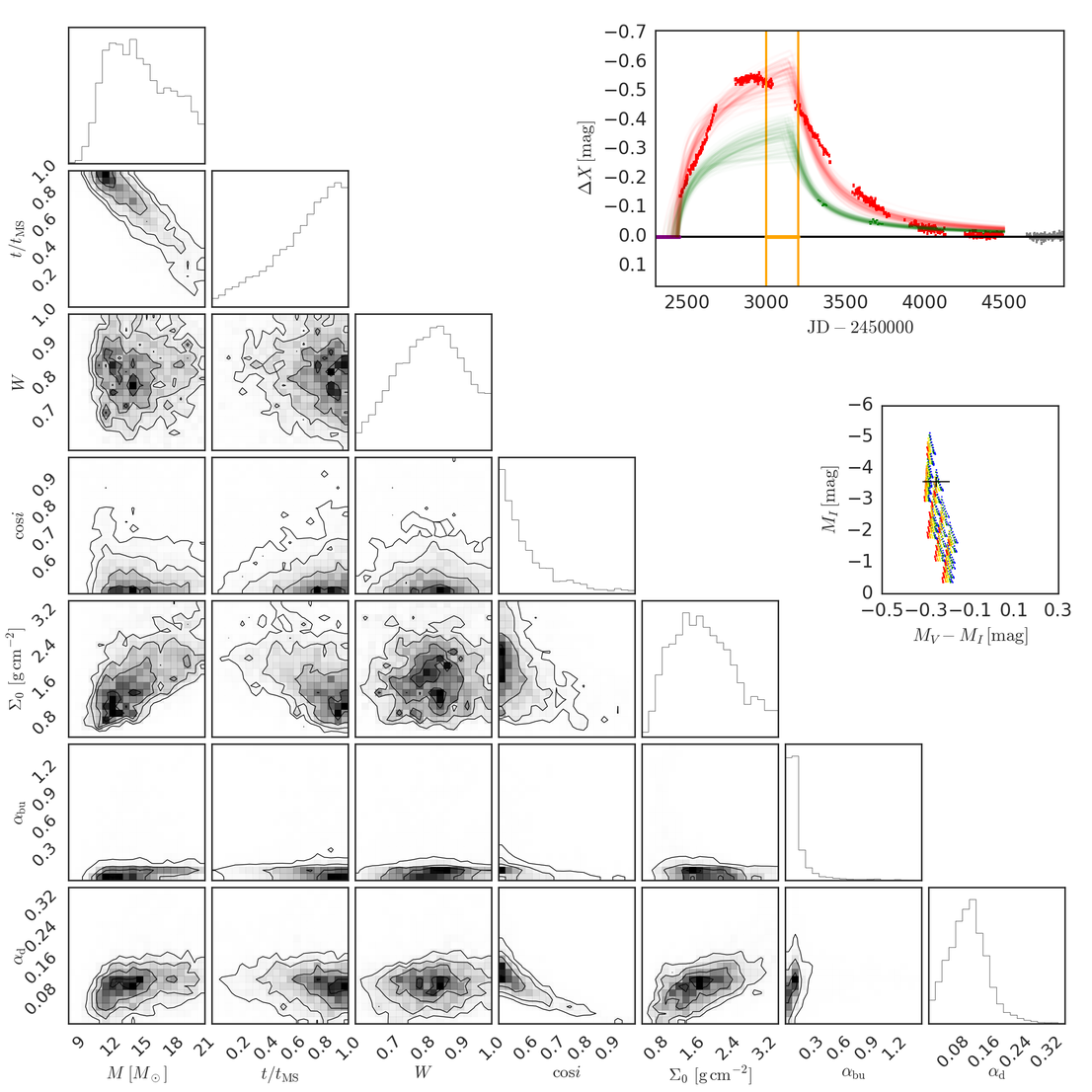}
}
\caption[]
{
Same as Fig.~\ref{example_bb1} for SMC\_SC11 28090 and bump ID 01. 
}
\label{smc_sc11_28090_01}
\end{figure*}
\clearpage

\begin{figure*}
\centering{
\includegraphics[width=1.0\linewidth]{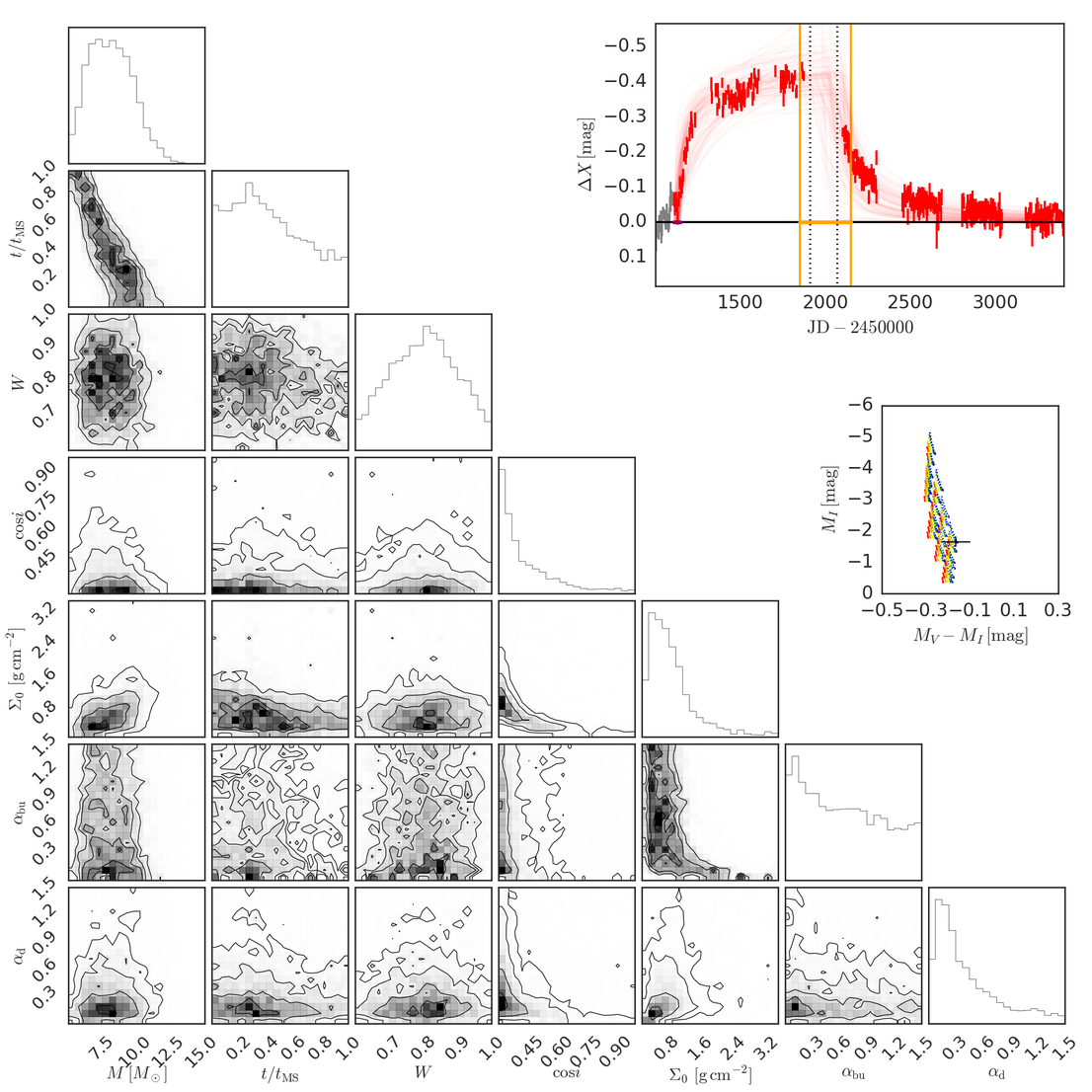}
}
\caption[]
{
Same as Fig.~\ref{example_bb1} for SMC\_SC11 46587 and bump ID 01. 
}
\label{smc_sc11_46587_01}
\end{figure*}
\clearpage


\end{appendix}

\bsp	
\label{lastpage}
\end{document}